\documentclass[bibliography=totoc,twoside,letterpaper,12pt]{report}
\usepackage{bera} 
\usepackage{trajan}
\usepackage[T1]{fontenc}
\usepackage{calligra}
\usepackage{fancyhdr}
\usepackage{float}
\usepackage{graphicx}
\usepackage{fancyhdr}
\usepackage{amsmath}
\usepackage[table]{xcolor}
\usepackage{booktabs}
\usepackage{multirow}
\usepackage{array}
\usepackage{longtable}
\usepackage{adjustbox}
\usepackage{rotfloat}
\usepackage{indentfirst}
\usepackage{comment}

\makeindex
\usepackage{xspace} 
\usepackage{amssymb}
\usepackage{amsmath}

\usepackage[update,prepend]{epstopdf} 
\usepackage{grffile} 
\newcommand{\ChiEFT}{$\chi$EFT\xspace}
\newcommand{\mpi}{\ensuremath{m_\pi}}

\newcommand{\alphae}{\ensuremath{\alpha_{E1}}}
\newcommand{\betam}{\ensuremath{\beta_{M1}}}
\newcommand{\gammaee}{\ensuremath{\gamma_{E1E1}}}
\newcommand{\gammamm}{\ensuremath{\gamma_{M1M1}}}
\newcommand{\gammaem}{\ensuremath{\gamma_{E1M2}}}
\newcommand{\gammame}{\ensuremath{\gamma_{M1E2}}}

\newcommand{\alphaep}{\ensuremath{\alpha_{E1}^{(\mathrm{p})}}}
\newcommand{\betamp}{\ensuremath{\beta_{M1}^{(\mathrm{p})}}}

\newcommand{\alphaen}{\ensuremath{\alpha_{E1}^{(\mathrm{n})}}}
\newcommand{\betamn}{\ensuremath{\beta_{M1}^{(\mathrm{n})}}}
\newcommand{\gammaeen}{\ensuremath{\gamma_{E1E1}^{(\mathrm{n})}}}
\newcommand{\gammammn}{\ensuremath{\gamma_{M1M1}^{(\mathrm{n})}}}

\newcommand{\gammazeron}{\ensuremath{\gamma_{0}^{(\mathrm{n})}}}
\newcommand{\gammapin}{\ensuremath{\gamma_{\pi}^{(\mathrm{n})}}}
\newcommand{\alphaes}{\ensuremath{\alpha_{E1}^{(\mathrm{s})}}}
\newcommand{\betams}{\ensuremath{\beta_{M1}^{(\mathrm{s})}}}

\newcommand{\HIGS}{HI$\gamma$S\xspace}
\newcommand{\threeHe}{${}^3$He\xspace}
\newcommand{\fourHe}{${}^4$He\xspace}

\newcommand{\fm}{\ensuremath{\mathrm{fm}}}

\newcommand{\half}{\frac{1}{2}}

\newcommand{\de}{\partial}

\usepackage{authblk}
\usepackage{atbegshi}
\AtBeginDocument{\AtBeginShipoutNext{\AtBeginShipoutDiscard}}
\begin{document}
\begin{titlepage}
\title{\vspace{15.0cm}International Workshop on Next Generation Gamma-Ray Source}
\author[1,2]{C. R. Howell}
\author[3,2]{M. W. Ahmed}
\author[4]{A. Afanasev}
\author[5]{D. Alesini}
\author[6]{J. R. M. Annand}
\author[7]{A. Aprahamian}
\author[8]{D. L. Balabanski}
\author[9]{S. V. Benson}
\author[10]{A. Bernstein}
\author[11]{C. R. Brune}
\author[12]{J. Byrd}
\author[13]{B. E. Carlsten}
\author[14,2]{A. E. Champagne}
\author[15]{S. Chattopadhyay}
\author[16,2]{D. Davis}
\author[4]{E. J. Downie}
\author[13]{M. J. Durham}
\author[4]{G. Feldman}
\author[1,2]{H. Gao}
\author[17]{C. G. R. Geddes}
\author[4]{H. W. Grie{\ss}hammer}
\author[18]{R. Hajima}
\author[1,2]{H. Hao}
\author[19]{D. Hornidge}
\author[20]{J. Isaak}
\author[14,2]{R. V. F. Janssens}
\author[1,2]{D. P. Kendellen}
\author[21]{M. A. Kovash}
\author[22]{P. P. Martel}
\author[23]{Ulf-G. Mei{\ss}ner}
\author[24]{R. Miskimen}
\author[35]{B. Pasquini}
\author[11]{D. R. Phillips}
\author[20]{N. Pietralla}
\author[25]{D. Savran}
\author[26]{M. R. Schindler}
\author[4,2]{M. H. Sikora}
\author[27]{W. M. Snow}
\author[1]{R. P. Springer}
\author[28]{C. Sun}
\author[29]{C. Tang}
\author[36]{B. Tiburzi}
\author[30]{A. P. Tonchev}
\author[18,3]{W. Tornow}
\author[8]{C. A. Ur}
\author[31]{D. Wang}
\author[1,2]{H. R. Weller}
\author[32]{V. Werner}
\author[1,2]{Y. K. Wu}
\author[1,2]{J. Yan}
\author[37]{Z. Zhao}
\author[33]{A. Zilges}
\author[34]{F. Zomer}
\affil[1]{Department of Physics, Duke University, Durham, NC 27708}

\affil[2]{Triangle Universities Nuclear Laboratory, Durham, NC 27708}

\affil[3]{ Department of Mathematics and Physics, North Carolina Central University, Durham, NC 27707}

\affil[4]{Department of Physics, The George Washington University, Washington, D.C. 20052}

\affil[5]{LNF-INFN, Via E. Fermi 40, 00044 Frascati (Rome), Italy}

\affil[6]{SUPA School of Physics and Astronomy, University of Glasgow, Glasgow, G12 8QQ, UK}

\affil[7]{Department of Physics, University of Notre Dame, Notre Dame, IN 46556, USA}

\affil[8]{Extreme Light Infrastructure Nuclear Physics (ELI-NP), Horia Hulubei National Institute for R\&D in Physics and Nuclear Engineering (IFIN-HH), Magurele, Romania}

\affil[9]{Thomas Jefferson National Accelerator Facility, Newport News, VA 23606}

\affil[10]{Physics Department and Laboratory for Nuclear Science Massachusetts Institute of Technology, Cambridge, MA 02139, USA}

\affil[11]{Department of Physics and Astronomy, Ohio University, Athens, OH 45701}

\affil[12]{Argonne National Laboratory, Argonne, Illinois 60439}

\affil[13]{Los Alamos National Laboratory, Los Alamos, NM 87545}

\affil[14]{Department of Physics and Astronomy, University of North Carolina, Chapel Hill, NC}

\affil[15]{Fermi National Accelerator Laboratory, Batavia, IL 60510}

\affil[16]{Department of Physics, North Carolina State University,
Raleigh, NC 27695}

\affil[17]{Lawrence Berkeley National Laboratory, Berkeley, CA 94720}

\affil[18]{National Institutes for Quantum and Radiological Science and Technology, Tokai, Naka, Ibaraki 3191106 Japan}

\affil[19]{Mount Allison University, Sackville, New Brunswick E4L1E6, Canada}

\affil[20]{Institut f{\"u}r Kernphysik, Technische Universit{\"a}t Darmstadt, 64289 Darmstadt, Germany}

\affil[21]{Department of Physics and Astronomy, University of Kentucky, Lexington, KY 40506}

\affil[22]{Institut f{\"u}r Kernphysik, University of Mainz, D-55099 Mainz, Germany}

\affil[23]{Helmholtz-Institut f{\"u}r Strahlen-und Kernphysik and Bethe Center for Theoretical Physics, Universit{\"a}t Bonn, D-53115 Bonn, Germany}

\affil[24]{University of Massachusetts, Amherst, MA 01003}

\affil[25]{GSI Helmholtzzentrum f{\"u}r Schwerionenforschung GmbH, 64291 Darmstadt, Germany}

\affil[26]{Department of Physics and Astronomy, University of South Carolina, Columbia, SC}

\affil[27]{Department of Physics, Indiana University, Bloomington, IN 47408}

\affil[28]{Lawrence Berkeley National Laboratory, Berkeley, CA, 94720}

\affil[29]{Department of Engineering Physics, Tsinghua University, Beijing 100084, China}

\affil[30]{Nuclear and Chemical Sciences Division, Lawrence Livermore National Laboratory, Livermore, CA 94550}

\affil[31]{Shanghai Institute of Applied Physics, Chinese Academy of Sciences, Shanghai 201800, China}

\affil[32]{Institut f{\"u}r Kernphysik, Technische Universit{\"a}t Darmstadt, 64289 Darmstadt, Germany}

\affil[33]{Institut f{\"u}r Kernphysik, Universit{\"a}t zu K{\"o}ln, Z{\"u}lpicher Stra{\ss}e 77, D-50937 K{\"o}ln, Germany}

\affil[35]{University of Pavia, Pavia, Italy}
\affil[36]{City College of New York, New York, NY}
\affil[37]{University of Virginia, Charlottesville, VA}
\maketitle
\end{titlepage}
{{\bf Disclaimer: } \small \textit {This report was prepared as an account of work sponsored by an agency of the United States
Government. Neither the United States Government nor any agency thereof, nor any of their employees,
makes any warranty, express or implied, or assumes any legal liability or responsibility for the accuracy,
completeness, or usefulness of any information, apparatus, product, or process disclosed, or represents
that its use would not infringe privately owned rights. Reference herein to any specific commercial product,
process, or service by trade name, trademark, manufacturer, or otherwise, does not necessarily constitute
or imply its endorsement, recommendation, or favoring by the United States Government or any agency
thereof. The views and opinions of authors expressed herein do not necessarily state or reflect those of the
United States Government or any agency thereof.
This workshop and the resulting whitepaper are supported by the US. Department of Energy Office of Nuclear Physics under grant number DE-SC0014616 and by the Triangle Universities Nuclear Laboratory (TUNL).}}

\tableofcontents                                                                            

\chapter{Executive Summary}
The photon is a theoretically well-understood probe for investigating the structure of matter over a wide range of distance and energy scales.  The angular-momentum selectivity and unconstrained transfer of isospin in most gamma-ray ($\gamma$-ray) induced reactions enable highly precise strategic investigations of nuclear and nucleon structure and of collective motion responses of the internal degrees of freedom associated with electric charge and current distributions.  In addition, high-energy photons are well suited for non-intrusive material analysis applications in areas such as homeland security, nuclear security, structural integrity assessments, and medical diagnostics.  

Bremsstrahlung radiation produced by an electron beam incident on a high-$Z$ target has been the workhorse photon-beam source used in nuclear physics research and applications for over a century.  The energy spectrum of a bremsstrahlung $\gamma$-ray beam enables measurements of nuclear structure and reaction dynamics over a continuous energy range in a single experiment.  Tagged bremsstrahlung sources provide the capability of associating the energy of the photon inducing a reaction with the photons detected in the final state.  Also, coherent bremsstrahlung beams produced using a crystal target offer researchers a partially polarized $\gamma$-ray beam at energies near the end point of the bremsstrahlung spectrum.  The scientific insight provided by research conducted using these sources is impressive, and facilities with advanced bremsstrahlung photon sources continue to be used in highly productive research programs.   However, the continuous energy nature of bremsstrahlung $\gamma$-ray beams, which facilitates survey measurements over a broad energy range, also has the adverse effect of creating backgrounds that limit the sensitivity of the experiments.  These backgrounds make it difficult to investigate nuclear phenomena with extremely low cross sections or to perform measurements on targets with small sample sizes, as is the case for isotopes with low natural abundances.  

The nearly monoenergetic $\gamma$-ray beams produced by laser Compton scattering offer an alternative to bremsstrahlung beams, providing an enhanced signal-to-background ratio in basic and applied research, and reducing the radiation exposure in material analysis applications. In addition, laser Compton $\gamma$-ray beams can be produced with a beam polarization greater than 95\% for both linear and circular polarization. Linearly polarized laser Compton beams enable unambiguous determinations of the spin and parity of excited nuclear states, and beams with both linear and circular polarization are used in studies of the spin structure of nucleons.

Over the last three decades, laser Compton $\gamma$-ray beam facilities have provided intense polarized and nearly mono-energetic $\gamma$-ray beams for research programs in basic and applied nuclear physics. These facilities include GRAAL at the European Synchrotron Radiation Facility in Grenoble, LEGS at Brookhaven National Laboratory, LEPS at the SPring-8 facility, NewSUBARU at the University of Hyogo, and HI$\gamma$S at the Triangle Universities Nuclear Laboratory. However, HI$\gamma$S and LEPS are the only facilities worldwide that are operated with nuclear physics as the primary research focus. These facilities will soon be joined by two more that are currently under construction, ELI-NP in Romania and the $\gamma$-ray beam line at the Shanghai Synchrotron Radiation Facility (SSRF) in China. These facilities will produce $\gamma$-ray beams with energies below 12 MeV (ELI-NP), and below 20~MeV and above 300~MeV (SSRF). 

Given that the time span between initial planning and the end of construction of an accelerator-driven light source is about a decade, consideration for next-generation laser Compton $\gamma$-ray sources should start now. Technological advances in electron accelerators, lasers, and optics made during the last decade create new options for producing intense polarized $\gamma$-ray beams with narrow energy widths at beam energies from around 1~MeV to several GeV, the energies relevant to nuclear-physics research. 
A workshop on {\em The Next Generation Gamma-Ray Sources}, sponsored by the Office of Nuclear Physics at the Department of Energy, was held November 17--19, 2016 in Bethesda, Maryland.  The goals of the workshop were to identify basic and applied research opportunities at the frontiers of nuclear physics that would be made possible by the beam capabilities of an advanced laser Compton beam facility. To anchor the scientific vision to realistically achievable beam specifications using proven technologies, the workshop brought together experts in the fields of electron accelerators, lasers, and optics to examine the technical options for achieving the beam specifications required by the most compelling parts of the proposed research programs. An international assembly of participants included current and prospective $\gamma$-ray beam users, accelerator and light-source physicists, and federal agency program managers.  Sessions were organized to foster interactions between the beam users and facility developers, allowing for information sharing and mutual feedback between the two groups.  The workshop findings and recommendations are summarized below. \\

\noindent
\underline{\bf Findings of Topical Working Groups:} \\
The topical sessions at the workshop focused on five research areas:  low-energy quantum chromodynamics (QCD), nuclear structure, nuclear astrophysics, fundamental symmetries, and applications.   The advanced accelerator and light source technologies working group recommended laser Compton scattering inside an optical cavity as the primary technology for the next generation $\gamma$-ray sources. The working group concluded that two $\gamma$-ray beam facilities are needed to meet the beam requirements of the research presented in the topical sessions at the workshop (1) a medium-energy source with $\gamma$-ray beams in the range of 25 and 400 MeV, and (2) a low-energy source capable of delivering $\gamma$-ray beams from 1.5 to to about 30 MeV. The topical sessions at the workshop focused on five main subjects:  nucleon structure and low-energy quantum chromodynamics (QCD), nuclear structure, nuclear astrophysics, fundamental symmetries, and applications.  Probing nucleon structure requires medium-energy photons, i.e., $\gamma$-ray beams with energies from about 60 to 350~MeV, while the four other research topics would use $\gamma$-ray beams at energies below about 20~MeV. Descriptions of the research opportunities for all five are in the science sections of this document. Summaries of the findings of the topical working groups are given below.

\noindent {\bf Medium-Energy NGLCGS Facility} \\
Understanding the emergence of hadron structure and the nuclear force in terms of  QCD are key questions at the frontier of nuclear physics. These phenomena are a consequence of quarks and gluons interacting at confinement-scale distances, where color forces are strong. The beams at a medium-energy NGLCGS facility will enable measurements that uniquely probe hadron structure and hadronic interactions in this non-perturbative regime of QCD. The experimental program  will investigate LEQCD phenomena with unprecedented precision in the photon energy range from about 60 MeV to the nucleon-to-Delta(1232) transition. The initial research program at such a facility will two main components: (1) Compton scattering to investigate nucleon structure and (2) photopion production to investigate the QCD origins of isospin symmetry breaking in the strong nuclear force. Both program components will require high-density polarized targets.

\begin{itemize}
	\item{\em Low-Energy QCD and Nucleon Structure}\\
	\noindent
	 The beam capabilities at a medium-energy NGLCGS will enable high-precision measurements of the scalar and spin nucleon polarizabilities by Compton scattering from unpolarized and polarized targets. Such measurements, together with advances in calculations using lattice QCD and QCD-based effective field theories, will explore the QCD origin of nucleon structure associated withe the collective response of nucleons to electromagnetic impulses with unprecedented sensitivity. For example, the experimental programs at the NGLCGS facility are proposed to improve the statistical accuracy of the nucleon spin polarizability measurements by more than a factor of 10 from current values and to impact the dynamical scaler polarizabilities of the neutron and proton with precision from energies below to above the pion production threshold.
	
	\item{\em Low-Energy QCD: Photopion Production}\\
	\noindent
	The high intensity and high energy resolution of beams at a medium-energy NGLCGS facility will enable determination of electromagnetic s-wave and p-wave amplitudes to the $\pi^0$ production with sufficiently high precision to observe charge-symmetry violation. Photoproduction of pions at energies near the reaction threshold provides mechanisms for investigation of the QCD origin of breaking of isospin symmetry in strong nuclear interactions. 
\end{itemize}

\noindent {\bf Low-Energy NGLCGS Facility}
The beams at low-energy NGLCGS facilities will enable high-accuracy measurements in nuclear structure, nuclear astrophysics, fundamental symmetries, and $\gamma$-ray beam applications in nuclear and homeland security. The research opportunities in each area are summarized below:

\begin{itemize}
	\item{\em Nuclear Structure:}\\
	\noindent
	The beams at an advanced $\gamma$-ray source will enable systematic studies of weak collective dipole and quadrupole nuclear excitations with unprecedented precision.  Such studies will provide nuclear structure details and information about the symmetry energy of the nuclear equation of state that are difficult to obtain by other means.  The high $\gamma$-ray beam intensities will enable mapping of states accessed through M1 transitions from the ground state in nuclei with a level of detail and breadth that will contribute to modeling of coherent neutrino-nucleus scattering and to calculating nuclear matrix elements for neutrinoless double-beta decay.   Also, a next-generation $\gamma$-ray beam facility will enable new exclusive measurements of photodisintegration of few-nucleon systems with a precision that provides sensitivity to  three-nucleon interactions.
	
	\item{\em Nuclear Astrophysics:}\\
	\noindent
	New $\gamma$-ray beam capabilities will enable measurements that contribute broadly to open questions in nuclear astrophysics, questions such as big-bang nucleosynthesis, helium burning in massive stars, and synthesis of heavy nuclei.  One of the most important reactions in stellar modeling is $^{12}$C($\alpha$,$\gamma$)$^{16}$O.  The rate of this reaction relative to the carbon forming reaction 3$\alpha \rightarrow ^{12}$C determines the fate of massive stars. The measurement of the rate of the $^{16}$O($\gamma$,$\alpha$)$^{12}$C reaction at energies approaching the temperatures at the core of stars is a grand challenge in nuclear astrophysics.
	
	\item{\em Fundamental Symmetries:}\\
	\noindent
	The beams at an advanced low-energy $\gamma$-ray beam facility will enable measurements of parity violating photodisintegration of few-nucleon systems.  In particular, a measurement of parity violation (PV) in deuteron photodisintegration near threshold is sensitive to a nucleon-nucleon (NN) PV amplitude that is not accessible using other systems.  Such measurements sample the short-range part of the NN interaction, providing unique quantities for comparison with calculations using lattice gauge theory.
	
	\item{\em Applications:}\\
	\noindent
	The intense mono-energetic $\gamma$-ray beams at low-energy NGLCGS facilities will enable photonuclear reaction measurements important for technologies and techniques used in homeland security and nuclear safeguards. Programs to develop field-deployable system will benefit by having a target areas equipped for evaluating concepts for $\gamma$-ray beam interrogation of cargo, nuclear fuel, and special assemblies at these facilities.  The new beam capabilities at advanced $\gamma$-ray sources  will also create opportunities for applications in medicine.
	
\end{itemize}

\noindent
\underline{\bf Findings of Advanced Accelerator and Light-Source Technologies Working Group:} \\
Details of the various technology options considered are presented in the chapter on accelerator concepts for NGLCGS facilities.  A summary of the findings of the  working group is below.

It is unlikely that a single $\gamma$-ray beam source can meet the requirements of both the low-energy (E$_\gamma$ < 20~MeV) and medium-energy (E$_\gamma$ > 60~MeV) parts of the field as described in the working group summaries above and in the science sections of this document. In the options discussed, the $\gamma$ rays are produced by Compton scattering of electrons from photons in an optical cavity that is pumped with an external laser.  Two options for the electron beam accelerators for the low-energy $\gamma$-ray source were considered: a storage ring and an energy-recovery linac with superconducting RF cavities. For the medium-energy $\gamma$-ray source, a storage ring was the primary option.  There is confidence that a high-quality electron beam with low emittance and low energy spread can be maintained in modern storage-ring lattices, thereby enabling production of $\gamma$-ray beams with low energy spread.  The new facility construction cost of the storage-ring option for either a low-energy or medium-energy next-generation Compton $\gamma$-ray source will be about \$150M.  The working group cautions that this estimate is extremely uncertain; it is intended only to set the scale within about a factor of two.  For the low-energy sources, less expensive options, such as upgrades to existing facilities, were also discussed. \\

\noindent
\underline{\bf Recommendations:} \\
The working groups through consensus make the following recommendations.  The order of this listing is not prioritized.  

\begin{itemize}
	
	\item{High intensity $\gamma$-ray beams with circular and linear polarizations will be produced at next-generation $\gamma$-ray sources by Compton scattering of photons from relativistic electrons inside a high finesse optical cavity.  The optical cavity will be pumped by a laser system with high precision control of beam polarization. The electron beam accelerator will use proven technologies, either a storage ring or an energy-recovery linac.  The main technological challenge is the production of reliable optical cavities with the technical specifications required for the next-generation $\gamma$-ray sources. \\

\noindent
{\bf The highest priority R\&D work for the next-generation $\gamma$-ray source should be the development of high finesse optical cavities and the associated laser and optical systems.  It is important for this work to include testing and optimization of the cavity under $\gamma$-ray production conditions.}
}
	
	\item{For photopion production experiments, small angle electron scattering (virtual photon tagging) with intense electron beams was discussed as an alternative to measurements using tagged bremsstrahlung sources or $\gamma$-ray beams produced by Compton scattering.  This alternative technique would be implemented with high-current electron beams, possibly in a storage ring, and with thin targets that allow detection of the low-energy charged particles produced in the reaction. This method promises to be much more effective than conventional photon tagging techniques. \\
	
\noindent	
{\bf R\&D should be supported to develop an alternative to Compton scattering for producing $\gamma$ rays with energies above the pion-production threshold. A system for virtual photon tagging in small-angle electron scattering is a promising candidate for such studies.}
}
	
	\item{Low-energy QCD phenomena will be explored at NGLCGS facilities with unprecedented precision at energies from below the pion-production threshold through the Delta (1232) resonance region. The core experimental programs will involve Compton scattering to study the spin-dependent electromagnetic response of nucleons and near-threshold photopion production.  Both programs require polarized beams and targets.   The measurements enabled by NGLCGS facilities, together with advances in calculations using lattice QCD and QCD-based effective field theories, will explore the QCD origin of nucleon structure and charge-symmetry breaking. \\
	
\noindent	
{\bf Investments in polarized targets are needed to prepare for experiments at the NGLCGS facilities.} \\

\noindent	
{\bf Investments in nuclear theory are needed to support the planning and analysis of low-energy QCD experiments at the NGLCGS facilities.}   

}
\item{To ensure full realization of the scientific potential of a NGLCGS it is crucial that a strong theory effort in this area be maintained as the machine concept is developed and implemented. This will facilitate planning for experiments that optimally realize the scientific goals articulated in this document and continue the strong tradition of synergy between theory and experiment in low- and medium-energy photonuclear physics. Mechanisms that will ensure there is a strong international theory community working on this physics that is fully engaged include workshops with small lead times and durations of up to a month and partial support for postdoctoral researchers and/or graduate students working on theory projects related to the NGLCGS. In addition, computing resources can help address the challenge of solving QCD in this regime and contribute to the 2015 Long-Range Plan’s recommendation of “new investments in computational nuclear theory that exploit US leadership in high-performance computing.}
	\item{The $^{12}$C($\alpha$,$\gamma$)$^{16}$O reaction helps regulate the efficiency of helium burning in massive stars and ultimately determines the mass of the iron core in the incipient supernova. The uncertainty in the measured cross section for this reaction substantially limits our understanding of the late stages of the life of massive stars and the details of the nucleosynthesis under the explosive conditions of supernovae.  The beams at NGLCGS facilities will enable measurements of the $^{16}$O($\gamma$,$\alpha$)$^{12}$C reaction that determine the reaction rate of $\alpha$-particle capture on $^{12}$C at center-of-mass energies lower than have been achieved with other techniques.  Measurements of angular distributions require thin targets to allow detection of the $\alpha$ particles along with charged-particle detectors with wide angle coverage.  Options include time projection chambers with optical and charge readout and silicon strip detectors with thin solid or gas targets. \\

\noindent	
{\bf Investments in active targets, such as time projection chambers, are needed to carry out the highest impact nuclear astrophysics measurements at NGLCGS facilities.}
}

\end{itemize}

\chapter{Introduction} 
\begin{figure}[ht]
	\centering
\includegraphics[width=0.8\linewidth] {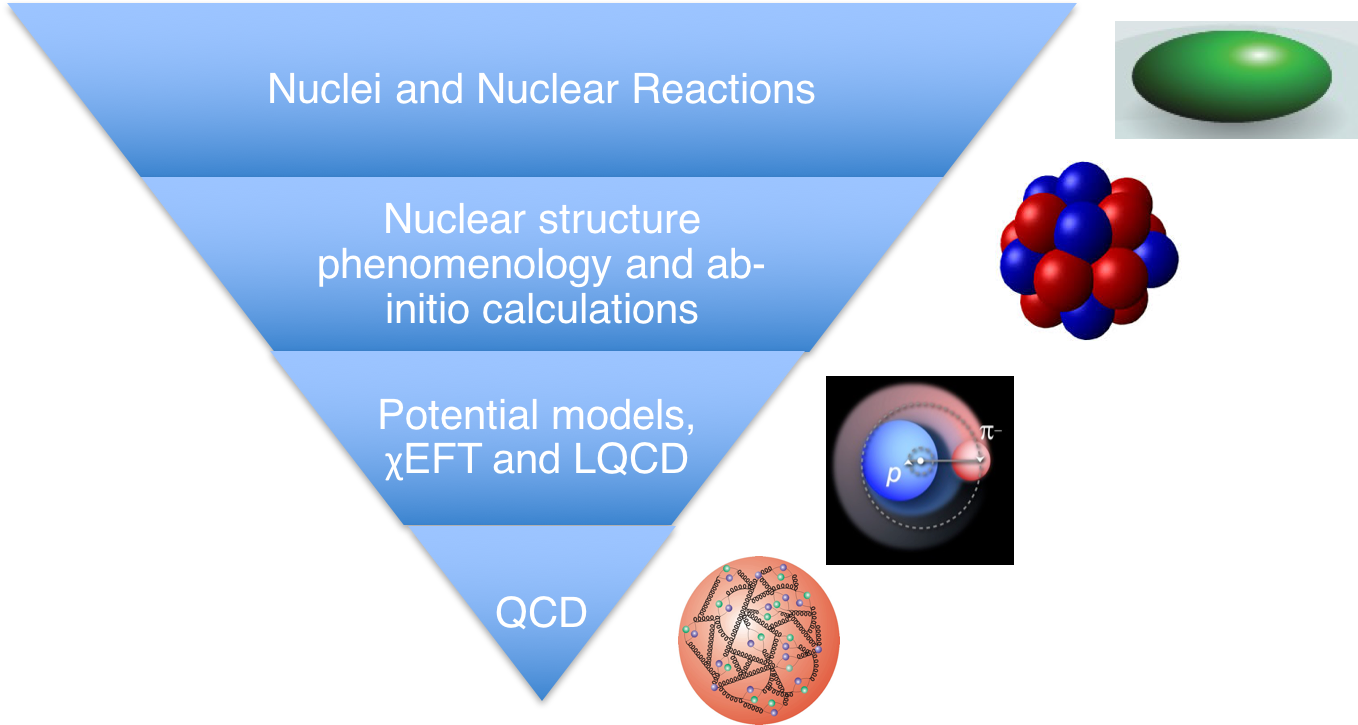}
	\caption{Schematic diagram for a coherent theoretical treatment of nuclear systems starting from high energies (at the bottom), where perturbative QCD can be applied, and ending with low-energy nuclear phenomena, where mean-field potential models are most efficient. This figure is adapted from ref. \cite{How16}. }
	\label{fig:coherent_theory}
\end{figure}

\lhead{}
\chead{}
\rhead{Introduction}
\footnote{References are included at the end of each chapter} 
Over the last decade, substantial progress has been made in developing formalisms and computational methods that contribute to theoretically coherent descriptions of nuclear phenomena with origins in QCD.  An ultimate goal would be to describe nuclear matter, over wide distance and energy scales, with a QCD Lagrangian.  Achieving this aim will likely require the effort of generations of scientists, as is often the case for grand challenges in science and technology.   Figure~\ref{fig:coherent_theory} shows a schematic diagram of a plausible hierarchy for organizing the theoretical treatments of nuclear systems spanning a variety of phenomena.  The scheme starts at high energies, where the most fundamental degrees of freedom are quarks and gluons, and progressively evolves in complexity.  This diagram is intended only to represent gross features that should be included in a coherent picture of strongly interacting matter.  At the top of the diagram, mean-field potentials that describe nuclear structure properties, the collective motion of nuclei, and nuclear reactions should be derived from residual strong interactions between nucleons.  Ab-initio calculations of the structure of light nuclei \cite{Pie01a, Nav07,Nav08,Nog06} and few-nucleon reaction dynamics \cite{Glo96} enable refinement of two-nucleon and multi-nucleon interactions using effective degrees of freedom.  Current theoretical tools for describing the strong nuclear force (two- and three-nucleon interactions) include semi-empirical potential models (see, e.g., Refs.  \cite{Mac01,Wir95,Sto94,Coo75,Pie01b}), effective field theory formulations of two-nucleon (2N) and three-nucleon (3N) interactions \cite{Hammer:2019poc,Epelbaum:2019kcf,Ent03, Epe09}, and lattice-QCD (LQCD) calculations of few-nucleon systems (see, e.g., Refs. \cite{Detmold:2019ghl,Bea15}).  Descriptions of the collective properties of nucleons in terms of effective field theories (see, e.g., Ref.~\cite{Hammer:2019poc,Epelbaum:2019kcf}) and LQCD (see, e.g., Ref.~\cite{Detmold:2019ghl,Bea14}) are steps toward bridging gaps between QCD and theoretical treatments of few-nucleon systems.

Photon beams are a highly-selective probe of the electric-charge and magnetism distributions of nuclei and nucleons.  Measurements of photon scattering and photon-induced reactions provide information on the collective response of the internal degrees of freedom of composite nuclear objects, as depicted in Fig.~\ref{fig:coherent_theory}.  For almost a century, bremsstrahlung $\gamma$-ray beams have been the workhorse for investigating nuclear and nucleon structure and for measuring photon-induced nuclear reactions.  These beams have the advantage of allowing measurements to be performed over a broad energy range in a single experiment, thereby providing information about the energy dependence of phenomena.  However, the continuous energy nature of bremsstrahlung $\gamma$-ray beams has the disadvantage of limiting measurement sensitivity due to backgrounds created by the photons with energies outside the region of interest.  A complementary tool is monoenergetic $\gamma$-ray beams, such as those created by laser Compton photon sources.  The narrow energy bandwidth of laser Compton $\gamma$-ray beams enhances the signal-to-background ratio in comparison to what is achievable using bremsstrahlung beams.  In addition, the high beam polarization available for both linear and circular polarization in the $\gamma$-ray beams produced by laser Compton photon sources enables unambiguous determination of the spin and parity of excited nuclear states and facilitates studies of the spin structure of nucleons.  Over the last several decades, laser Compton $\gamma$-ray beam facilities have provided intense polarized and nearly mono-energetic $\gamma$-ray beams for research programs in basic and applied nuclear physics.  The facilities include GRAAL at the European Synchrotron Radiation Facility in Grenoble, LEGS at Brookhaven National Laboratory, LEPS at the SPring-8 facility, NewSUBARU at the University of Hyogo, and HI$\gamma$S at the Triangle Universities Nuclear Laboratory.  New laser Compton $\gamma$-ray beam facilities under construction include ELI-NP in Romania and the $\gamma$-ray beam line at the Shanghai Synchrotron Radiation Facility (SSRF) in China.   The current generation of laser Compton $\gamma$-ray beam sources are based mostly on either a single-pass light pulse from an external laser scattered off a single-pass electron beam bunch or a single-pass laser pulse scattered off electron beam bunches circulating in a storage ring.  The technique of intracavity Compton scattering of electrons circulating in a storage ring that is employed at HI$\gamma$S is indicative of the technologies to be used in next-generation laser Compton $\gamma$-ray sources.  Combining the high beam current of storage rings with the high photon density inside optical cavities can potentially produce monoenergetic $\gamma$-ray beams with intensities more than three orders of magnitude higher than HI$\gamma$S, which is currently the most intense laser Compton source in the world.

\begin{figure}[ht]
	\centering
\includegraphics[width=0.5\linewidth] {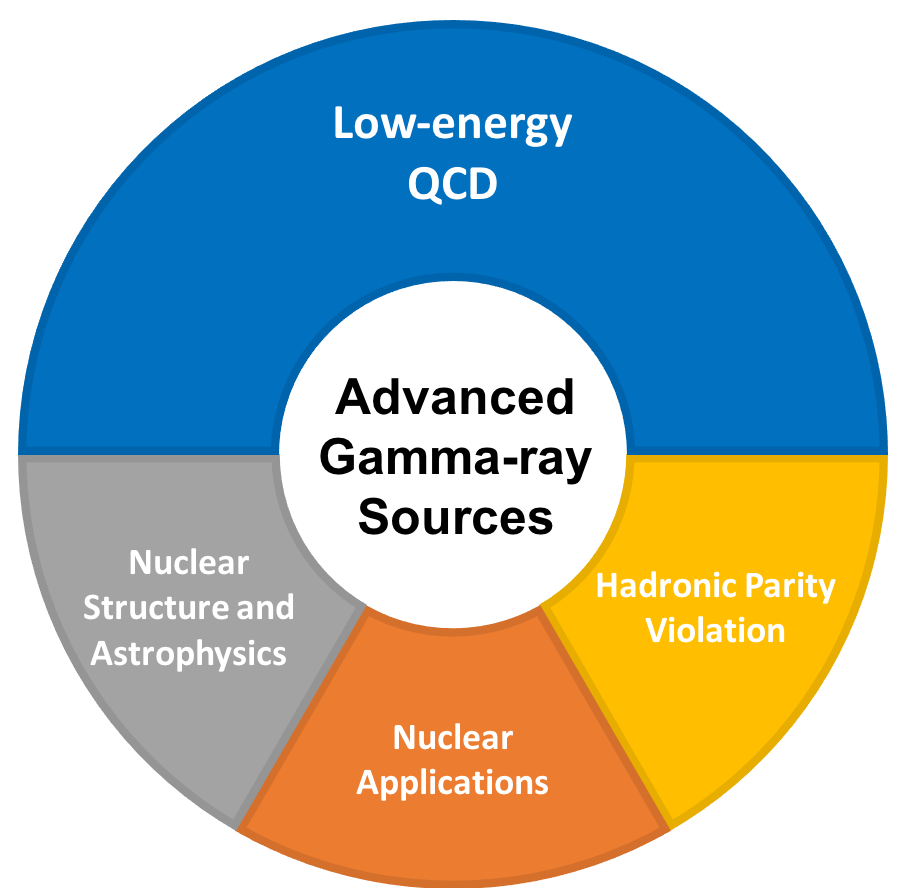}
	\caption{Illustration of the research areas impacted by advanced $\gamma$-ray source technologies.  The upper half of the graphics represent the main research at $\gamma$-ray beam energies from about 60 MeV to 350 MeV.   The lower half represents the research areas covered at energies below 20 MeV. }
	\label{fig:gsources}
\end{figure}

This whitepaper describes research opportunities that would be opened up by the $\gamma$-ray beam capabilities at next-generation laser Compton $\gamma$-ray beam facilities.  Its content is based mainly on discussions and findings of {\it The International Workshop on the Next Generation Laser Compton Gamma-Ray Beam Facility} that was held November 17--19, 2016 in Bethesda, MD.  The workshop brought  together both researchers from the international low-energy and medium-energy nuclear physics communities, along with accelerator physicists and experts in optics.  The accelerator and optics experts anchored this exercise into the constraints of beam performance parameters achievable with technologies that can be implemented in the coming decade.  For both practical reasons and scientific considerations, the upper energy reach of the $\gamma$-ray source was limited to 500 MeV.  Within these boundaries, the broad areas of nuclear physics considered correspond to the upper three panels of Fig.~\ref{fig:coherent_theory}.  It is probably fair to state that most participants began the workshop with the belief that one laser Compton source would be capable of serving both low-energy experiments, with $\gamma$-ray beam energies (E$_{\gamma}$) below about 20 MeV, and medium-energy experiments, with E$_{\gamma} > 60$ MeV.  The questions addressed in low-energy and medium-energy $\gamma$-ray experiments have considerable intellectual overlap and the intense, monoenergertic $\gamma$-ray beams of a NGLCGS would facilitate progress in both regimes. However, the $\gamma$-ray source considerations mean that research at low and medium energies likely must be carried out at two different facilities.  For this reason, the discussion of the research opportunities created by  laser Compton $\gamma$-ray sources is organized by the two energy ranges of the facilities presented at the workshop, E$_{\gamma}$ $\le$ 20 MeV and 60 MeV $\le$ E$_{\gamma}$ $\le$ 350 MeV.  The areas explored in each energy region are illustrated in Fig.~\ref{fig:gsources}.  While low-energy QCD is mentioned explicitly only in the upper half of the circle it is also strongly connected to the proposed research on parity violation in hadronic systems and the studies of few-nucleon systems and light nuclei that compare ab-initio calculations to data. Indeed, since all nuclear dynamics is underpinned by QCD low-energy QCD in fact pervades all the opportunities that would be created by an NGLCGS. These include parity violation in hadronic systems and studies of few-nucleon systems and light nuclei using ab-initio calculations.  Details of the opportunities are presented in the topical chapters that follow.  A synopsis of each topical discussion is given below.

The main opportunities at energies below 20 MeV are presented in chapter 3 and are in the areas of nuclear structure (including photon-induced fission), nuclear astrophysics, and parity violation in few-nucleon systems and light nuclei.  Studies of collective modes of excitation can provide information about short-range correlations between nucleons in nuclei and can reveal features of the nuclear equation of state. The giant dipole resonance (GDR) dominates the nuclear collective response and is well understood. The rotational (rigid rotor) and vibrational (scissor) modes at energies below about 4 MeV have been extensively studied and are described by well-established models. The nuclear excitations on the low-energy tail of the GDR, above about 5 MeV, and below the particle separation energy are not well characterized. Much of the difficulty in understanding the nature of the excitations in this energy region is associated with distinguishing between effects due to the GDR and those due to other mechanisms. The dipole excitation strength in excess of the GDR tail is referred to as the pygmy dipole resonance (PDR). The generally accepted mechanism for the PDR is the vibration of a neutron skin off an isoscalar core. If this picture is correct, a study of its strength as a function of isospin and $A$ should provide information about the density dependence of the symmetry energy term in the nuclear equation of state. The beam intensities available at the next-generation laser Compton $\gamma$-ray sources will enable nuclear resonance florescence (NRF) measurements on nuclei at extremes of isospin where the natural abundance is low and consequently target material is sparse. In addition, the beam at next-generation laser Compton $\gamma$-ray sources will enable thorough measurements of the M1 $\gamma$-ray strength in nuclei.  Special attention will be given to nuclei in the mass vicinity of isotopes used in double-beta decay experiments.  The beams at the next-generation laser Compton $\gamma$-ray source will also enable photon-induced fission measurements at energies near threshold, where the cross section is low, and the study of low-yield fragments.  Such experiments will provide new information about the potential energy surface that governs the evolution of the fission process. Examples of the initial nuclear structure research made possible by the NGLCGS facilities are described in section \ref{Sec:NS}.

The beams at the next-generation laser Compton $\gamma$-ray sources will enable strategic measurements of a variety of NRF and photon-induced nuclear reactions that contribute to modeling $p$-process, $s$-process and $r$-process nucleosynthesis.  However, the highest impact contribution to nuclear astrophysics will be in determining the reaction rate for alpha-particle capture by carbon via the $^{12}$C($\alpha$,$\gamma$)$^{16}$O reaction in massive stars during the alpha-burning stage.  The goal for all measurements of this reaction is to determine the cross section at a center-of-mass energy of 300 keV.  In $\gamma$-ray beam facilities, the time reversed reaction $^{16}$O($\gamma$,$\alpha$)$^{12}$C will be measured, and detailed balance will be applied.  This measurement will be the flagship nuclear astrophysics experiment at advanced $\gamma$-ray sources, and, as such, should be measured using a variety of techniques. Nuclear reaction rate measurements enabled by the NGLCGS facilities that are at key points in stellar nucleosynthesis networks are discussed in section \ref{Sec:NS2}.

In the area of fundamental symmetries, the primary focus will be on studying parity violation in hadronic systems. Such parity violation is a measure of the weak interaction inside systems of strongly interacting particles.  The ultimate goal is to measure parity violation in photodisintegration reactions of few-nucleon systems, such as $^2$H and $^3$He.  Because these measurements must be performed near the reaction threshold energy, and because the parity violating asymmetry is extremely small ($\sim$ 10$^{-7}$), these experiments make the most stringent demands on the source-performance parameters.  That is, a $\gamma$-ray source that delivers beam intensities an order of magnitude below what is required for these experiments will enable all the nuclear structure and nuclear astrophysics discussed in this whitepaper.  Measurements of parity violating photoabsorption asymmetries on parity doublets in light nuclei will be used to assess beam and instrumentation asymmetries in the early stage of developing the capabilities for performing 10$^{-7}$ asymmetry measurements.  For this purpose, nuclei with photoabsorption asymmetries of 10$^{-4}$ to 10$^{-3}$ will be selected.    Hadronic Parity Violatoin studies using photon-induced reactions are described in section \ref{Sec:HPV}.

In addition to the basic science research enabled by the NGLCGS facilities, these $\gamma$-ray sources will provide capabilities that support technology and techniques development work for nuclear security. How such facilities might be used in efforts to advance $\gamma$-ray beam interrogation systems is outlined in Section \ref{Sec:App}.

 Quantum chromodynamics is the fundamental theory of the strong nuclear force. When written in terms of quark and gluon degrees of freedom, it is deceptively simple. Indeed, QCD's asymptotic freedom guarantees that this form of the theory describes strong interactions at sufficiently high energy. However, in that regime, QCD does not bind quarks and gluons into neutrons and protons.  The task of fully and rigorously deriving the presence and properties of neutrons, protons, and nuclei from the Standard Model remains a grand challenge for physics, requiring us to understand the emergence of new degrees of freedom as the theory becomes strongly coupled. Without such an understanding, treatments of the nuclear force based on the fundamental theory will remain elusive, and so too will the ability to model nuclei in a reliable manner that is grounded in QCD.

High-intensity beams of polarized $\gamma$ rays in the energy range of 60--300 MeV provide a unique opportunity to test our understanding of the emergence of neutron and proton structure from the standard model. In this regime, a description of experimentally observed phenomena in terms of neutrons and protons and their low-lying excitations is efficient, thanks to the chiral symmetry of QCD.

The initial research opportunities in nucleon structure and low-energy QCD at NGLCGS facilities are described in Chapter \ref{Ch:QCD}.  One of the main goals at these energies will be to map out the spin-independent and spin-dependent polarizabilities starting at energies below the pion production threshold, extending over the pion production threshold energy, and continuing through the Delta(1232) region.  These measurements will be carried out using Compton scattering from unpolarized and polarized targets.  The other major aim in the medium-energy domain is precision near-threshold pion photoproduction data. Such data would provide the opportunity to extract the s-wave $\pi^0$p scattering length and so gain a new window on the QCD origin of charge-symmetry breaking.. 

In the limit of vanishing up-, down-, and strange-quark masses, the QCD Lagrangian admits a global chiral symmetry:  SU(3)$_L \times$ SU(3)$_R$.  This symmetry is broken spontaneously, which implies the existence of eight pseudoscalar massless Goldstone bosons.  Furthermore, since the quark masses are finite (but small), these Goldstone bosons acquire a small mass and are identified with the pions, kaons, and etas.  The interaction of these Goldstone bosons with themselves or with matter fields such as nucleons is weak.  This allows for a systematic low-energy expansion in terms of small momenta and quark masses: Chiral Perturbation Theory.  Chiral perturbation theory ($\chi$PT) describes meson-meson interactions in this energy domain with high precision. It also describes the interactions of mesons with nucleons at very low energies. Incorporating the $\Delta(1232)$ as an explicit degree of freedom in the theory yields a chiral effective field theory ($\chi$EFT) that describes processes at energies below the chiral symmetry breaking scale ($\Lambda \approx 1$ GeV) via an expansion in a small, dimensionless parameter. This allows reliable quantification of residual theory uncertainties.  Thus $\chi$EFT exploits the chiral symmetry of QCD in order to rigorously connect QCD with the phenomenology of nuclear and particle physics.  Because $\chi$EFT encodes the consequences of the standard model for low-energy processes involving photons, pions, and nucleons, it can be used to test whether observed strong-interaction phenomena are consistent with the standard model.  In particular, $\chi$EFT allows us to elucidate the experimental consequences of the pattern of chiral-symmetry breaking in QCD. 

Accelerator concepts for the NGLCGS are described in Chapter \ref{Ch:Acc}. Two options for the accelerator configuration were discussed for energies below 20 MeV: an energy-recovery linac with superconducting RF cavities or a storage ring.  Only the storage-ring option was presented for energies above 60 MeV.  In all source configurations, the Compton scattering occurs inside an optical cavity that is pumped by an external laser.  The electron source, accelerator structures, and storage-ring lattices that meet the required technical specifications for source performance are robust and can be implemented to order.  The situation is not as well established for the optical cavity.  R\&D is needed to develop an optical cavity design that can meet the power, stability, and reliability requirements of the next-generation laser Compton $\gamma$-ray source.

\bibliographystyle{unsrt}
\renewcommand{\bibname}{References}

\chapter{Nuclear Physics Research and Applications with Gamma-ray Beams Below 20 MeV}
\lhead{}
\chead{}
\rhead{Nuclear Physics Below 20 MeV}
\section{Nuclear Structure} 
\begin{figure}[h!]
	\centering
	\includegraphics[width=0.6\linewidth]
        {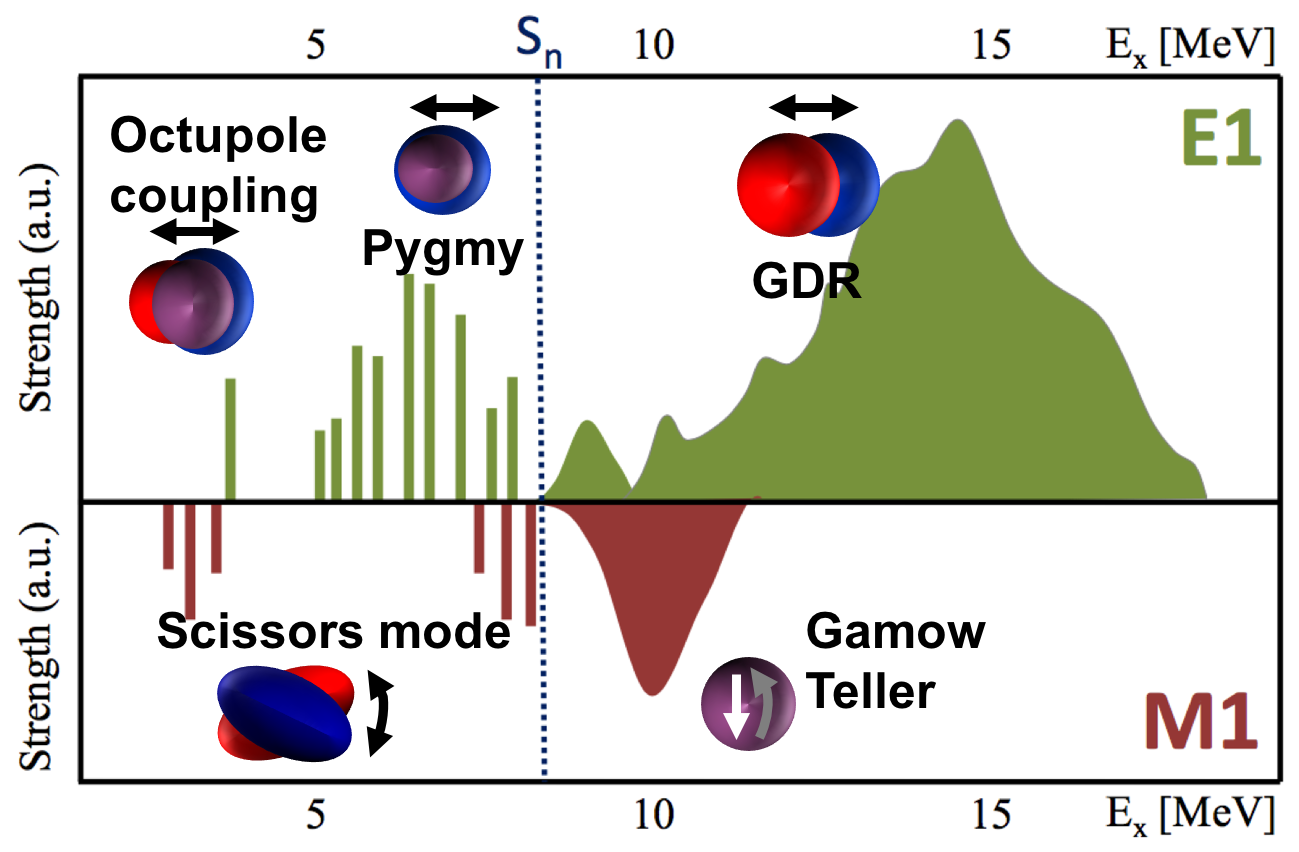}
	\caption{Diagram of the landscape of the response of nuclei to photon absorption.}
	\label{fig:em_landscape}
\end{figure}

\lhead{Nuclear Structure}
\rhead{Nuclear Physics Below 20 MeV}
\label{Sec:NS}
The landscape for the response of nuclei to electromagnetic radiation
as a function of excitation energy is shown schematically in
Fig.~\ref{fig:em_landscape}.  The electric giant dipole resonance
(GDR) is the dominant collective response of nuclei.  This mode can be
modeled as the oscillation of the proton matter distribution against
the neutron distribution.  In the classical interpretation, the width
of the cross-section enhancement associated with the resonance
provides information about the damping forces, i.e., the viscosity of
the nuclear matter.  The nature of the nuclear response at excitation
energies below the GDR around the particle in thresholds is still rather
unclear and is the topic of considerable theoretical and experimental
work \cite{Sav13}. The excitation in this region has been observed to
be largely electric dipole (E1).  This additional structure below the
GDR has been shown to be a common feature in the E1 response of atomic
nuclei, and is generally referred to as the Pygmy Dipole Resonance
(PDR).  The E1 response offers the possibility of studying the
equation of state of neutron matter
\cite{rein10,tami11,piek11,ross13,roca15,bald16}. Also the
$\gamma$-ray absorption cross section in the PDR energy region
influences nuclear reaction rates in stars
\cite{gori04,litv09,daou12,tonc17}. In deformed nuclei, the collective
nuclear dipole response at excitation energies below the PDR region is
mainly due to magnetic dipole (M1) transitions, which can be described
as a scissor motion. The ($\gamma$,$\gamma$') reaction, which is
referred to as nuclear resonance fluorescence (NRF), is the most
effective approach for probing these features of nuclear structure.

Nuclear fission is a complex process in which the collective motion of
nucleons results in a strong change in the shape of the nucleus,
ultimately leading to a breakup into fragments. Photon-induced fission
can provide unique insight into the evolution of the potential energy
surface during the fission process as it progresses from photon
absorption through to the scission point. Measurements of fission
product yields of isotopes with different half lives provide
information on the evolution of the nuclear deformation and breakup of
the nucleus during the fission process.

The high intensity, nearly monoenergetic, and linearly polarized
photon beams at a next-generation laser Compton $\gamma$-ray beam
facility will enable high-precision photonuclear reaction
measurements.  Research opportunities for NRF and photofission
measurements are described in this section.

\subsection{Nuclear Resonance Fluorescence}

The nuclear resonance fluorescence (NRF) reaction is one of the
work horses in the investigation of the dipole response of atomic
nuclei. The method enables the extraction of intrinsic properties of
excited states such as spin, parity and transition widths in a model
independent way. Furthermore, due to the low momentum transfer of
photons, primary dipole transitions are induced, i.e., in even-even
nuclei states with $J=1^{\pm}$ are populated. This makes NRF an
excellent reaction to systematically study the phenomena in the dipole
response of nuclei, such as the PDR and the scissors mode over a large range of
nuclei.

Thus far, mostly the decay branch back to the ground state as been
investigated with the NRF reaction. However, the detailed decay pattern of
excited states is often connected to important details in its nuclear
structure. The decay widths to the individual lower-lying excited
states or the ground state are directly linked to the corresponding
transition matrix elements. Thus, the individual decay channels are
sensitive to different components of the nuclear-state wave
function. For example, in the case of transitions to lower-lying
excited states, the de-excitation takes place via a different
component in the wave function than the excitation from the ground
state. Therefore, the observation of these transitions and the
determination of the branching ratios reveal important experimental
information that tests modern nuclear-structure models.

\begin{figure}[ht]
	\centering
	\includegraphics[width=0.8\linewidth]
        {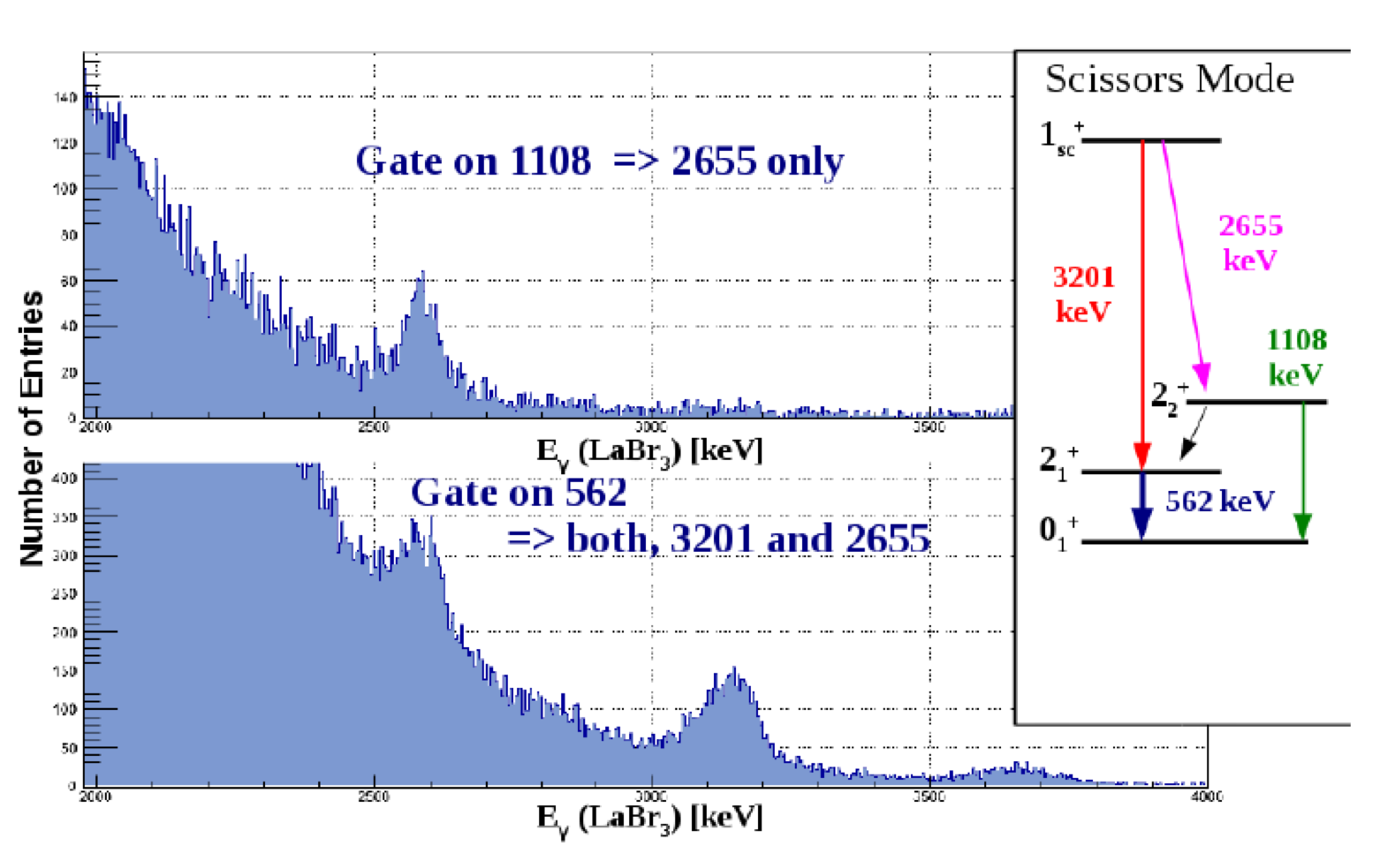}
	\caption{Example of $\gamma\gamma$ coincidence spectroscopy. The spectra show projections of LaBr$_3$ $\times$ LaBr$_3$ matrices after gating on transitions from the first and second 2$^+$ states in $^{76}$Ge. The corresponding level scheme is given on the right-hand side. The scheme is an example for any dipole excited state with arbitrary intermediate levels between the $J = 1$ excited state and the ground state.}
	\label{fig:scissormode}
\end{figure}

The method of $\gamma\gamma$ coincidences in the spectroscopy of the
decay of excited states has been proven to be a powerful tool for
determining even small branching ratios to excited low-lying
states \cite{Loeh13,Loh16}. The principle is illustrated in the left part of
Fig.~\ref{fig:scissormode}. After excitation by photoabsorption, the
high-lying state at excitation energy $E_x$ may de-excite either
directly to the ground state with width $\Gamma_0$ or via an
intermediate state with width $\Gamma_i$. By detecting the two emitted
$\gamma$ rays in coincidence, even small branching ratios
$\Gamma_i$/$\Gamma_0$ can be determined with good precision, since the
background is strongly suppressed. This technique allows the branching
of dipole-excited states to individual lower-lying levels to be
determined, thus greatly aiding in building up complex level and
decay schemes.  An example is shown in Fig.~\ref{fig:scissormode}. The
right-hand side of the figure provide, the level structure for a
measurement on $^{76}$Ge, where the decay of the 1$^+$ scissors mode
has been investigated \cite{Coo15}. The 1$^+$ state at 3.763 MeV is
populated by photo-excitation. Gating on the decays of the 2$^+_1$ and
2$^+_2$ states, as shown on the left side of the figure, allows the
order of the $\gamma$-ray transitions in the decay to be determined.

\begin{figure}[H]
	\centering
	\includegraphics[width=0.6\linewidth]
        {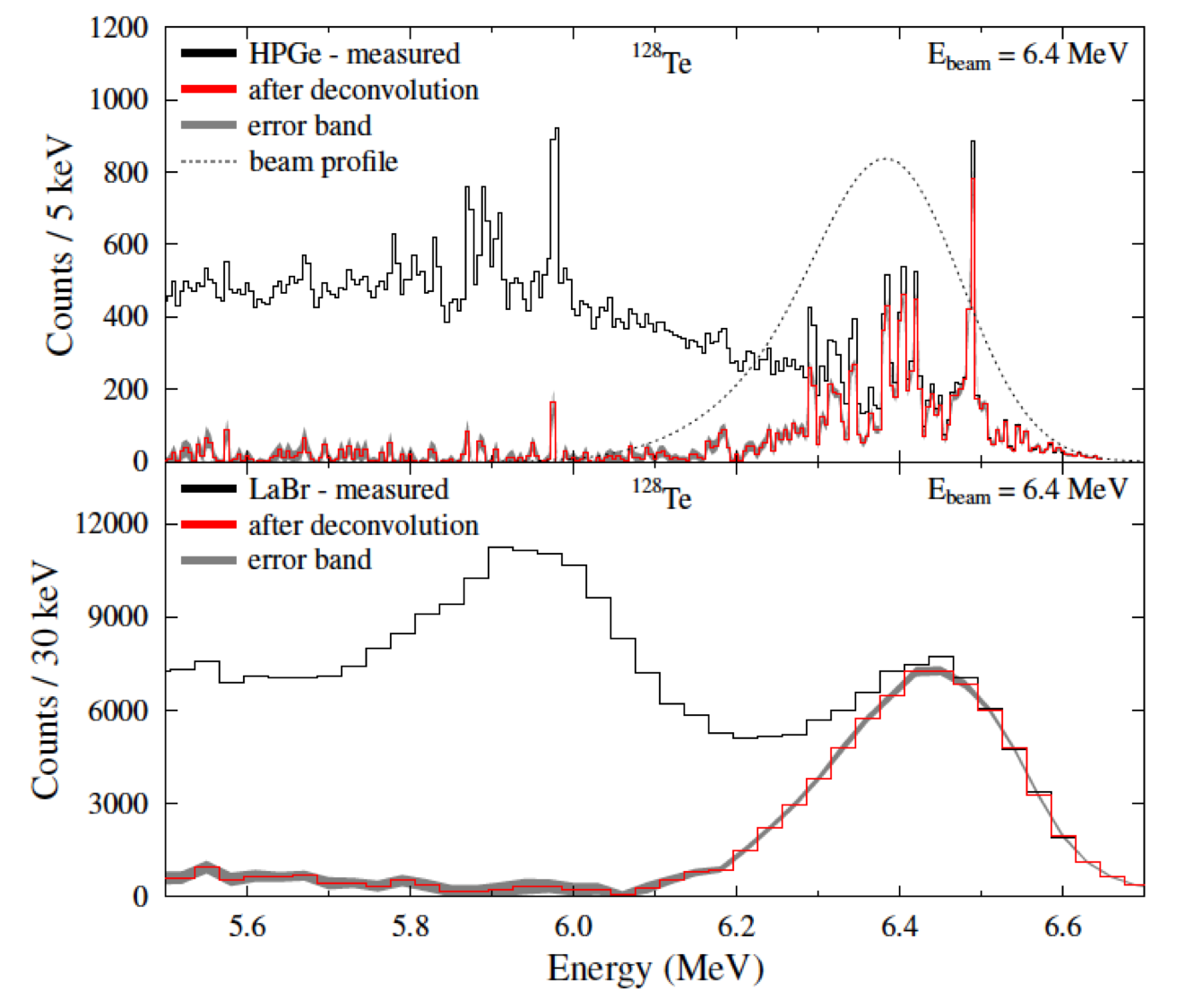}
	\caption{HPGe (top) and LaBr$_3$ (bottom) spectra from singles $\gamma$-ray spectroscopy of $^{128}$Te at a beam energy of 6.4 MeV.  The black histograms show the actual spectra, and the red ones are the result of the deconvolution, which includes correcting for the detector response. The beam profile is indicated by a dashed curve in the upper panel.  This figure is taken from \cite{Isa16}.}
	\label{fig:ggcoincidence}
\end{figure}

The combination of large-volume LaBr$_3$ and high-resolution HPGe
detectors opens the way for $\gamma\gamma$-coincidence measurements
with sufficient statistical accuracy to track $\gamma$-ray decay
cascades. High-resolution HPGe detectors allow for a separation of
individual $\gamma$-ray transitions from dipole excited states in
regions where the level density is not yet too high. Even in regions
with a high level density, individual states can sometimes have
substantial excitation strength and are, thus, separable from the
otherwise rather continuous spectrum. This provides a way to normalize
spectra to complementary data sets obtained using bremsstrahlung
beams. Nevertheless, most dipole-excitation strength in regions of
high level density is so strongly fragmented that individual states
cannot be resolved. In this case, the large-volume LaBr$_3$ detectors
have a significant advantage over HPGe detectors because of their
higher detection efficiency.  This high efficiency allows the
fragmented strength in the beam region to be separated from the
background and enables integrated branching strengths to be
determined. In deformed nuclei, where the first-excited-state energies
are very low, decays to the ground state and the first excited state
are close in energy, posing a challenge to the accelerator team to
deliver beams with high intensity and low energy spread to the target.
Deconvolution methods have recently been developed and applied to aid
in extracting decays out of the detector response 
(Fig.~\ref{fig:ggcoincidence}), see e.g. \cite{Isa16,Isa13}.

With the combination of HPGe and LaBr$_3$ detectors, the high
efficiency of the latter provides sufficient statistics to observe
branching decays from higher-lying states, either in low-resolution
LaBr$_3$ or even in high-resolution HPGe spectroscopy, as demonstrated
in \cite{Loeh13,Loh16}. Especially at low energies, electric quadrupole
excitation probabilities can be sufficiently large for observation,
and can be separated from the dipole strength making use of angular
distribution and correlation measurements using arrays of HPGe and
LaBr$_3$ detectors. In certain cases even the mixing ratio of M1/E2
transitions can be measured.  This recently facilitated a first
measurement of the isovector quadrupole strengths in $^{156}$Gd
\cite{Bec17}, for example.

The capabilities of the next-generation laser Compton $\gamma$-ray
source combined with modern $\gamma$-ray detectors will enable
measurements that address a broad range of physics through 
photo-excitation studies. The topics include multi-phonon structures such as
quadrupole-octupole coupled states and the search for rotational
isovector scissors excitations; the investigation of the pygmy dipole
resonance (PDR) along with its structural evolution and fingerprints
in going from spherical to deformed regions; and detailed
spectroscopic studies of the M1 spin-flip resonance. Therefore, the
physics topics to be addressed embrace shape coexistence, octupole correlations, isovector
excitations, neutron skins, and photon strength functions, often with
strong relevance for the fields of nuclear astrophysics and
weak-interaction physics such as neutrinoless double-beta decay and
neutrino scattering.
 
\subsubsection{Shape Coexistence and Link with Physics with Radioactive beams}
In the context of nuclear structure studies, a direct link between research carried out with $\gamma$-ray beams and with radioactive beams should be briefly discussed. Shell structure is a cornerstone in the description of atomic nuclei as many-body quantum systems. The first data on neutron-rich nuclei far from stability have provided evidence that shell structure evolves with neutron excess. For example, some well-known magic numbers disappear while others appear. Progress in the description of these exotic nuclei is due in no small part to the development of new experimental techniques that have gone hand in hand with the introduction of new theoretical concepts. It has been shown that effective single-particle energies are significantly modified in neutron-rich systems through the action of the monopole component of the proton-neutron tensor force ~\cite{Ots05}. Furthermore, multi-particle multi-hole excitations trigger shell evolution as a function of spin and excitation energy ~\cite{Ots16}.  In this picture, the occupation of specific deformation-driving orbitals leads to changes in nuclear shapes and to the possibility of shape coexistence. The region between Z=20, N=20 $^{40}$Ca and Z=28, N=50 $^{78}$Ni provides a good illustration of the synergy between experiments at the national user facilities and measurements with $\gamma$-ray beams. For example, triple shape coexistence has recently been discovered in neutron-rich $^{66,68,70}$Ni ~\cite{Leo17}. These 0$^+$ states are the result of multi-particle multi-hole excitations. Some of these levels, in particular those associated with prolate deformation, are understood as proton excitations that are predicted to be present in the stable $^{60,62,64}$Ni as well ~\cite{Leo17}, and the evolution of these states' excitation energy with N  is a matter of much theoretical debate. The issue can be addressed with intense $\gamma$-ray beams: in a first phase, all the low-spin excitations can be mapped out. Following this discovery phase, information on the wave function of the states will be obtained from state lifetime measurements, and from the determination of decay branching ratios. Furthermore, excitations built on the excited 0$^+$ levels will be delineated and their properties characterized. Other regions of the nuclear chart lend themselves to similar studies. Specifically, shape changes and shape coexistence phenomena, driven by multi-particle multi-hole excitations are predicted to occur in the Sr-Mo-Zr region as well as in the vicinity of doubly-magic $^{208}$Pb. 

\subsubsection{Multi-Phonon and Rotational Low-Lying States}

\begin{figure}[H]
	\centering
	\includegraphics[width=0.6\linewidth]
        {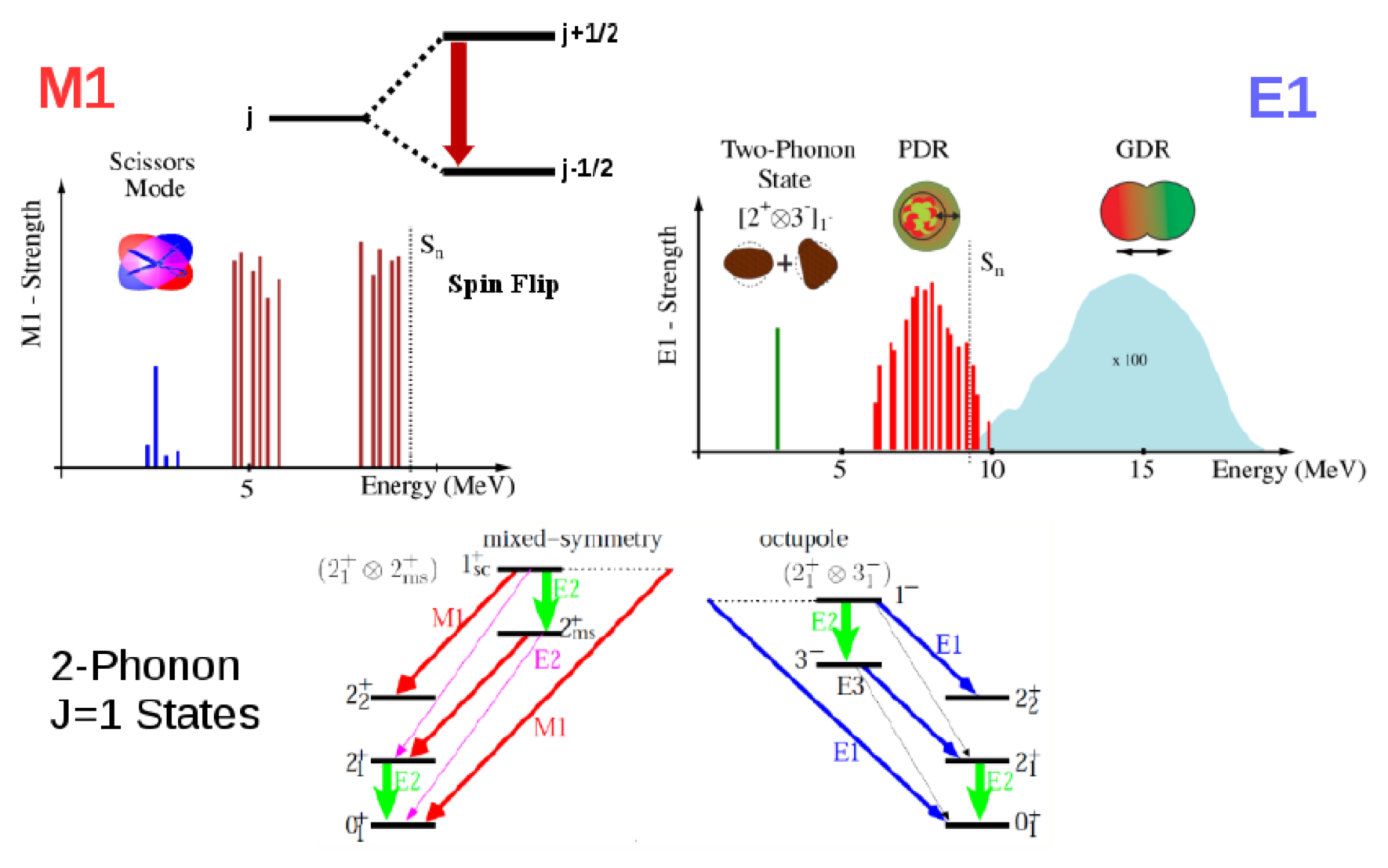}
	\caption{Schematic diagram of nuclear excitations involving multiple phonons and illustrating mixed symmetry and octupole excited states.}
	\label{fig:2phonon}
\end{figure}

At low energies, in spherical or near-spherical nuclei, excited states
are often formed by phonon excitations. For example, the first 2$^+$
state is identified with a quadrupole phonon, and the first 3$^-$
state with an octupole phonon. Such phonons can be coupled, leading to
multiplets of multi-phonon states built from like phonons, or from
mixtures of, e.g., quadrupole and octupole phonons. As such, the
quadrupole-octupole coupled quintuplet of states, 1$^-$ $\cdot \cdot
\cdot$ 5$^-$ contains an E1-excited 1$^-$ state which typically is the
first excited 1$^-$ state, located at roughly the sum energy of its
constituents and carrying most of the E1 strength at low
energies. (See the schematic diagram of $\gamma$-ray transitions
associated with multiple-phonon de-excitation is given in
Fig.~\ref{fig:2phonon}.) In addition to excitations where protons and
neutrons move in phase, there are so-called mixed-symmetry states
where at least one pair of protons/neutrons has a different phase
\cite{Pie08}. This leads to additional states, such as a
quadrupole-excited one-phonon mixed-symmetric 2$^+_{ms}$ state.
Again, the phonon coupling between the latter and the symmetric
2$^+_1$ phonon state leads to a multiplet of mixed-symmetry states,
one of which is an M1-excited 1$^+$ state, which is typically the
strongest M1 excitation at low energies of roughly 3 to 4 MeV. In
rotational nuclei, this state evolves into the well-known scissors
mode, where valence protons and neutrons are counter-oscillating in a
scissors-like fashion.

Different mechanisms forming E1 and M1 excited states at such low
energies can, however, compete with one another. For example,
spin-flip excitations, where a nucleon is moved from an $\ell \pm 1/2$
orbital to its $\ell \mp 1/2$ partner, can generate M1 excitation
strengths similar to those of the scissors mode. In the E1 sector, the
generation of E1 strength in valence spaces containing either no or a
very limited number of opposite-parity orbitals is under much
discussion, and there are other possibilities, such as the formation
of alpha clusters (see below). Therefore, it is not only important to
locate those dipole excited states and to determine their parity and decay
strengths, it is also crucial to measure their decay pattern, which
contains important information on their underlying structure. These
tasks are uniquely facilitated by the capabilities at a
next-generation laser Compton $\gamma$-ray source for measuring decay
branches, parities, and relative excitation strengths. Furthermore, in
rotational nuclei, band structures are expected on $J = 1$ band-heads
such as the scissors-mode state. Indeed, in recent work \cite{Bec17},
a first candidate for the long-sought rotational 2$^+_{sc}$ member of
the scissors-mode band has been discovered.  In addition, measurements
of angular correlations made possible by arrays of $\gamma$-ray
detectors with large angle coverage, high detection efficiency, and
high energy resolution used in combination with linearly polarized
monoenergetic $\gamma$-ray beams will enable measurements of
multipole-mixing ratios. These, in turn, can reveal absolute isovector
E2 strengths. High quality data of this type are crucial in
constraining effective charges in nuclear dynamics calculations.

Another aspect of dipole excited states has recently been found and is
connected to weak-interaction physics. While the strong interaction
dominates in nuclei, nuclei are also laboratories for studying the
weak interaction through the possibility of $\beta$ or $\beta\beta$
decay. Of special interest for many experiments, and for the
characterization of neutrinos in general, is the search for
neutrinoless double-beta (0$\nu\beta\beta$) decay. The existence of
this decay mode would identify the neutrino as a Majorana particle,
one which is its own anti-particle.  It would provide information
about the absolute mass of the neutrino involved in the decay through
the ratio of the measured decay rate and the calculated nuclear matrix
element for the decay.  Since many candidate isotopes (mother and/or
daughter) for 0$\nu\beta\beta$ decay occur in mass regions near
transitions from spherical to deformed shapes, their model description
is challenging. Furthermore, isovector parameters are seldom known. NRF
measurements can be used to extract information on phenomena such as
shape coexistence and mixed-symmetry states from the observation of
the scissors mode and its decays, especially its decays to the
potentially mixed-configuration 0$^+_{1,2}$ states (see
Ref. \cite{Bel13}).

\subsubsection{The E1 Pygmy Dipole Resonance}

The strongest and best-known E1 mode in atomic nuclei is the giant
dipole resonance (GDR). It is a broad resonance structure peaking well
above the particle-separation threshold energy. Therefore, in NRF
experiments, only the low-energy tail of the GDR below the particle
threshold energy is directly observable. Often, this tail is
parameterized by a Lorentzian, eventually with modifications such as
temperature dependence, e.g., parameterizations that directly use the
so-called photon strength functions (PSFs). PSFs have a direct
relation to nuclear astrophysics, since they give a handle on the
balance between particle-capture and photodisintegration processes in
stellar nuclear synthesis. Over the last decade, observations have
given rise to additional strength, on top of the
low-energy tail of the GDR.  This added strength is the so-called
pygmy dipole resonance (PDR) \cite{Sav13}. This strength, typically a
few percent of the GDR strength, appears around the neutron-separation
threshold. It would directly influence PSFs and would, therefore, impact
nucleosynthesis, which ultimately shapes the abundance patterns of
elements \cite{Gor98}.

Various studies of the PDR have involved magic or near-magic spherical
nuclei.  Usually, the origin of the PDR is thought to be an
oscillation of a neutron skin against the proton-neutron saturated
core. This picture is supported by several microscopic calculations,
showing an excess of neutron transition densities near the nuclear
surface. However, other possibilities exist. For example, the
occurrence of alpha clusters on the surface, which would yield
enhanced E1 strength, has been suggested \cite{Iac85}. Additional
problems lie in the differentiation of PDR strength from the
low-energy tail of the GDR.  Alpha scattering seems to be a possible
method for accessing this problem \cite{Sav06,End10,Sav18}, because,
compared to NRF data, low-energy E1 strength appears to split into
isoscalar and isovector parts. 

Other challenges lie in the determination of the total strength. For
example, proton-scattering data partly point to significantly more
observed E1 strength than photon-scattering data \cite{krum15}. This
may, at least in part, be due to branching transitions from
dipole-excited states and many weak unresolved transitions, which
would have been missed in earlier NRF experiments using bremsstrahlung
beams. At the next-generation $\gamma$-ray source, the focus will be
on identifying the strength distribution (state-by-state or average),
measuring the parities as well as branching behavior of the excited
states. From these data, more stringent conclusions can be drawn on
the nature of the PDR, its overall strength, and therefore also on the
determination of PSFs. In addition, the Brink hypothesis (the
excitations that are built on the ground state should also be built on
any excited state), which is part of the basis of the statistical
model, can be tested in a model independent way using
$\gamma$-$\gamma$ decay spectroscopy, as has been recently demonstrated
\cite{Isa18}.

A further challenge lies in the deformation degree of freedom. In the
end, most nuclei on the chart of the nuclides are deformed, but our
information on the evolution of the PDR toward deformed nuclei is
sparse at best. Therefore, the experimental program should extend to
address this problem. Indications of a potential splitting of the PDR
into two parts with different $K$ quantum number have recently been
observed in proton-scattering on $^{154}$Sm \cite{Kru14}.  This would
be similar to the splitting of the GDR with respect to the two (or
even three) axes of the nuclear body. However, the high degree of
fragmentation in deformed nuclei renders a state-by-state analysis
meaningless. Therefore, measurements that track the cascade of
$\gamma$-ray decays are most useful. In addition, the very low energy
of the first excited state in deformed nuclei requires a reduced
energy spread of the photon beam compared to existing facilities in
order to enable a clean separation of the decay intensity to the first
excited state and the ground state.  Such measurements over sizable
isotope chains will be made possible by the beam capabilities at the
next-generation laser Compton $\gamma$-ray beam facilities.

\subsubsection{Alpha-Cluster Excitations}

Below the particle-separation threshold, two underlying structures of
E1 excitations have been intensively studied in the last decade, the
octupole modes \cite{But96,Kne06} and the PDR \cite{Sav13}. For both
modes, a non-uniform distribution of protons and neutron generates E1
transitions at lower energies. As pointed out in earlier sections,
these two modes could have substantial impact on nuclear structure
models and nuclear synthesis calculations.  Another mode giving rise
to enhanced E1 strength at lower energies is the oscillation of an
alpha cluster relative to the remaining bulk nuclear matter
\cite{Iac85}. For light nuclei, alpha clustering is well-established
\cite{Oer06}, and its implications for the E1 strength have been
discussed. Also, in heavy isotopes such as $^{212}$Po, strong
indications of a $^{208}$Pb + $\alpha$ system have been observed
\cite{Ast10}. Recently, the existence of alpha clusters in heavy
nuclei was supported by adding four-particle correlations to
shell-model calculations \cite{Roe14}.  Because the formation of alpha
clusters provides interesting insights into the formation of bosonic
clusters in strongly coupled fermionic systems, identifying new
signatures of alpha clustering in heavier nuclei is of general
scientific interest.

Several enhanced E1 transitions were observed in the neodymium chain
by ($\gamma$,$\gamma$') experiments up to 4~MeV
\cite{Pit90,Fri92,Eck97}. To shed light on the origin of the low-lying
$J = 1^-$ states, {\it spdf} interacting-boson-model (IBM)
calculations have been performed \cite{Spi15}. It has been proposed
that the $p$-boson is related to $\alpha$-cluster configurations. By
identifying the basis states by means of their boson contents
$\mid$[n$_s$], [n$_p$], [n$_d$], [n$_f$]$\rangle$, the ratio
n$_p$/n$_f$ permits an assignment of quadrupole-octupole or $p$-boson
character to be associated with the excited states. The results of
these calculations suggest the presence of alpha-cluster modes
occurring at the surface of nuclei with atomic masses just above magic
numbers. For example, in IBM calculations for $^{144}$Nd, a remarkable
increase of $p$-boson contributions has been observed for energies up
to the neutron-separation threshold.

\subsubsection{The M1 Spin-Flip Resonance}

Although, overall, much less M1 strength is expected in the region
where the PDR occurs, there should be significant strength due to the
spin-flip resonance. The exact energy of this resonance depends on the
underlying shell structure, but typically it coincides with the onset
of the major E1 strength distribution. Knowledge about this M1
resonance is sparse, and present parameterizations are based on little
available data. The challenges at hand are the identification of M1
strength within the sea of E1 excited states. In cases of low
fragmentation, this can be done with a state-by-state analysis of
excited-state parities, but at high fragmentation, one has to choose an
average approach. The best way to filter out M1 strength in those
energy regions is the use of photon-scattering with a linearly
polarized beam, such as those available at laser Compton $\gamma$-ray
sources.

The impact of such higher-lying M1-excited states has different
facets. On the one hand, they will provide constraints to microscopic
calculations, since the occurrence of spin-flip transitions involving
particles moving between $\ell \pm 1/2$ partner orbitals, usually
across a major nuclear shell. In addition, states excited by a
spin-flip transition are those which are populated through
Gamov-Teller transitions in $\beta$ decay.  On the other hand, such
1$^+$ states will directly influence PSFs in even-even nuclei, since
decays between 1$^-$ (GDR/PDR) states and the spin-flip states are
allowed, changing the overall shape of the PSFs. More reliable data on
the M1 response can, therefore, serve as a test of parity
asymmetry. (Usually the same number of positive and negative parity
states is assumed.)

Again, M1 strength, this time at energies above the scissors mode, can
prove important for neutrino physics. Specifically, there are
detectors, such as the molybdenum-based MOON detector \cite{Eji08},
which are in the planning and construction phases and are expected to
serve, not only for the detection of potential 0$\nu\beta\beta$ decay
events, but also for the direct detection of neutrinos through inverse
``$\beta$ decay'' by the capture of neutrinos. Another facet of
neutrino detection lies in the excitation of 1$^+$ states by neutrinos
\cite{Ydr12}. Hence, the M1 response of relevant nuclei plays an
important role in the characterization of such detectors and also in
rate estimates using nuclear models such as the quasiparticle
random-phase approximation.

The beam requirements for carrying out the nuclear structure research
using NRF described in this section are given in
Table~\ref{tab:nsbeam}.

\begin{table}[h]
	\caption{Suggested beam parameters for NRF experiments.}
	\centering
\begin{tabular}{c|c}
	\hline  Parameter&Value  \\ 
	\hline  Energy& 2-20 MeV \\ 
	\hline  Flux ($\gamma$/s) & 10$^9$ at 1\% FWHM  \\ 
	\hline  Polarization& Linear  \\ 
	\hline  Diameter& 10 mm on target \\ 
	\hline  Beam Repetition Rate & few MHz  \\
	\hline  Beam Pulse Width & $<$ 1.0 ns  \\ 
	\hline 
\end{tabular} 
	\label{tab:nsbeam}
\end{table} 

\subsection{Photon-Induced Nuclear Fission}

Nuclear fission is a highly exothermic and strongly collective nuclear
process in which most of the energy is released through the kinetic
energy of the ejected fission fragments.  The evolution of a
fissioning system proceeds from the initial impact of the incident
particle through the intermediate saddle point(s), then through
scission, and finally to the configuration of separated fission
fragments. This evolution is governed by a multi-dimensional
potential-energy surface (PES) and by the shell structure of the
fragments.  Development of reliable theoretical models of nuclear
fission is important for basic research and for the development of
nuclear energy and nuclear security technologies.  For these purposes,
data for different types of observables are needed, including fission
product yields over wide ranges of masses and half lives, as well as the
kinetic energy and angular distributions of the emitted fragments and
neutrons.

To first approximation, the fission barrier of a heavy nucleus can be
studied by measuring the fission probability as a function of the
excitation energy. By comparing the experimentally determined fission
probabilities to the results obtained with the WKB approximation, the
shape of the fission barrier (its height and curvature parameter) can
be determined.

\begin{figure}[H]
	\centering
	\includegraphics[width=0.4\linewidth]
        {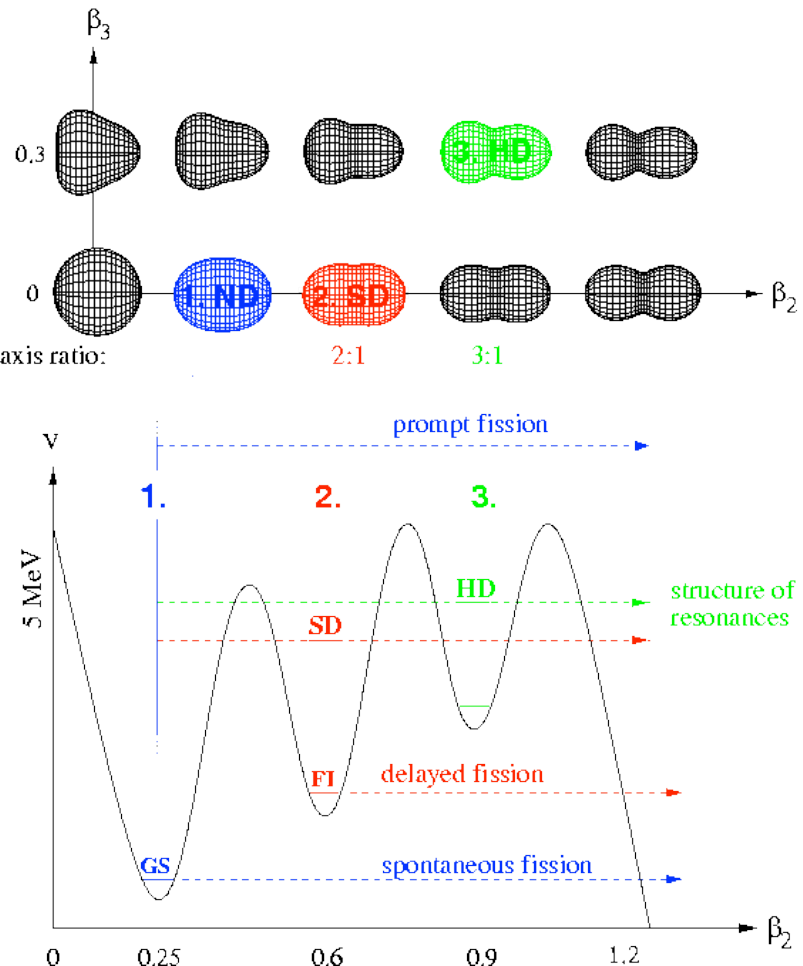}
	\caption{Schematic overview of the multiple-humped fission barrier in actinide isotopes, along with the corresponding nuclear shapes. The lower part shows a cut through the potential energy surface along the fission path, revealing an SD second minimum at an axis ratio of 2:1 and an HD third minimum at an axis ratio of 3:1. The energies and locations of the saddle points and minima were taken from Ref. \cite{Thi02}. In the upper part, the corresponding nuclear shapes are displayed as a function of the quadrupole and octupole degrees of freedom.}
	\label{fig:fissionbarrier}
\end{figure}

Significant progress has been made in studying the multiple-humped fission
barrier landscape of actinides, where a strongly deformed deep third
minimum in the potential landscape was established \cite{Thi02,
Kra11}. This is illustrated in Fig.~\ref{fig:fissionbarrier}, where
horizontal dashed lines show transmission resonances in the
superdeformed (SD) second minimum and in the hyperdeformed (HD)
third minimum. One expects that the HD minimum in a
cluster description consists of a rather spherical $^{132}$Sn-like
component with magic neutron and proton numbers $N = 82$ and $Z =
50$. Recent theoretical results \cite{Mcd13,Jac13} highlight the role
of shell corrections in the prominence of the third minima for thorium
and light uranium isotopes. There is a clear trend suggesting that
lower $N$ values produce larger neutron shell corrections and, thus,
more prominent third minima. This is in stark disagreement with
experimental results. Using the $^{231}$Pa($^3$He,$d$f) reaction,
Csige {\it et al.} \cite{Csi09} studied sub-barrier fission in
$^{232}$U and inferred a fission barrier with a well-formed third
minimum that does not agree with the theoretical
predictions. Pronounced third minima have also been inferred to exist
through observation of fine structure in the cross section of
transfer-reaction-induced fission or through sub-barrier photofission
cross-section measurements. There is particular interest in
identifying resonances in the third minimum, where very large
deformations cause the GDR to split into two components with a
low-lying oscillation along the long symmetry axis with a typical
excitation energy of four to five MeV. It is expected that these
resonances may have a significantly enhanced $\gamma$-ray width
$\Gamma_\gamma$ of about 100 eV in the population of the third
minimum, while, for the second minimum, one expects a $\gamma$-ray width
of only about 1 eV.  E1 resonances excited from the ground state will
have about the same absolute excitation strength as resonances in the
third minimum, but resonances in the third minimum will be much
broader, with a total width of about 1 keV.

Photofission measurements enable the selective investigation of
extremely deformed nuclear states in the light actinides and can be
utilized to better understand the landscape of the multiple-humped PES
of these nuclei. The selectivity of these measurements stems from the
low and reasonably well-defined amount of angular momentum transferred
during the photoabsorption process. High-resolution studies can be
performed on the mass, atomic number, and kinetic energy distributions
of the fission fragments following the decay of well-defined initial
states in the first, second and third minima of the PES in the region
of the light actinides.  The beams available at the next-generation
laser Compton source facilities will enable high fidelity studies of
the PES using transmission resonance spectroscopy, as well as studies
of heavy clusterization in the actinides, and of rare fission
processes such as ternary fission. Each of these opportunities is
discussed in this section.

\subsubsection{Transmission Resonance Spectroscopy in Photofission}

The approach to investigating extremely deformed collective and single-particle nuclear states of the light actinides is based on the
observation of transmission resonances in the prompt fission cross
section. Observing transmission resonances as a function of the
excitation energy allows the identification of the excitation energies
of the SD and HD states. Moreover, the observed states
can be ordered into rotational bands, with moments of inertia proving
that the underlying nuclear shape of these states is indeed an SD or
an HD configuration. For the identification of the rotational bands,
the spin information can be obtained by measuring the angular
distributions of the fission fragments. Furthermore, the PES of the
actinides can be parameterized very precisely by analyzing the overall
structure of the fission cross section, and by fitting it with the results of nuclear
reaction code calculations.


So far, transmission resonances in the light actinides have been
studied primarily in light-particle-induced nuclear reactions. These
studies do not benefit from the same selectivity found in photonuclear
excitation and, consequently, they are complicated by a statistical
population of the states in the second (and third) minimum. These
measurements have also suffered from a dominant prompt-fission
background. By contrast, the next-generation laser Compton
$\gamma$-ray sources will enable the identification of sub-barrier
transmission resonances with very low integrated cross sections down
to $\Gamma \sigma \approx 0.1$ eVb and in so far unexamined
nuclei. The narrow energy bandwidth of the next generation
$\gamma$-ray sources will enable a significant reduction of the
presently dominant background from non-resonant processes.

Besides exploring the level structure in the second and third minima
of the fission barrier of the light actinides, the harmonicity of the
potential barrier can also be examined, and the parameters of the
fission barrier can be extracted. Such fission barrier parameters are
crucial inputs for cross-section calculations in the thorium-uranium
fuel cycle of fourth-generation nuclear power plants. The selectivity
of the photofission measurements allows high-resolution investigation
of fission resonances in photofission in the second and third minima
of the fission barrier. Detailed studies of SD and HD states via
transmission resonance spectroscopy is relevant also for achieving
much cleaner energy production by an efficient transmutation of the
long-lived, and most hazardous radioactive components of nuclear waste,
and by controlling the fission process through using entrance channels
via HD states.

\subsubsection{Photofission as a Probe of Heavy Clusterization in the Actinides}

Theoretical considerations suggest that in a cluster description, the
HD configuration of a light actinide consists of a
spherical $^{132}$Sn-like component with magic neutron and proton
numbers $N=82$ and $Z=50$, respectively, complemented by an attached,
elongated second cluster of nucleons. Since the fission-product mass
distribution is distinctly determined by the configuration at the
scission point, and the third minimum is very close to the scission
configuration, it is expected that the mass distribution, following the
decay of an HD nucleus, will exhibit a pronounced asymmetric
structure. However, such a dramatic effect of the shell structure has
not been observed so far due to the fact that available studies used
particle-induced fission.

Brilliant quasi-monoenergetic $\gamma$-ray beams will enable
high-resolution investigations of the mass, atomic number, and kinetic
energy distributions of the fission fragments following the decay of
states in the first, second and third minima of the PES in the region
of the light actinides. In these measurements, the heavy
clusterization and the predicted cold valleys of the fission potential
can be studied for the first time. Data on the heavy cluster formation
will provide valuable information on the fission dynamics.

\subsubsection{Investigation of Rare Fission Modes: Ternary Photofission}

So far, information on ternary and more exotic fission modes has been
deduced from neutron-induced and spontaneous fission
experiments. Ternary particles are released very close to the scission
point, thus providing valuable information on both the scission-point
configuration of the fissioning nucleus and the dynamics of the
fission process. However, ternary photofission has never been studied
due to the very low cross section of the reaction channel.

In such studies, the geometry can be fixed by using polarized
$\gamma$-ray beams, which is a clear advantage over neutron-induced or
spontaneous fission experiments. Moreover, by employing
quasi-monoenergetic $\gamma$-ray beams, excitation-energy correlations
of the ternary fission process can be explored with good
resolution. These experiments investigate open problems such as the
mechanism of ternary particle emission, the role of the deformation
energy and of the spectroscopic factor, and the possible formation of
heavier clusters. It will be  interesting to measure 
light-particle decay in photofission and to search for the predicted
enhanced $\alpha$ decay of SD and HD states of the
light actinides.

The availability of brilliant quasi-monoenergetic $\gamma$-ray beams
will make it possible to study the angular distribution of the ternary
particles. These, in turn, will provide important spectroscopic
information on the fissioning system.  The beam requirements to carry
out the photofission research program described in this section are
given in Table~\ref{tab:nfbeam}.

\begin{table}[h]
	\caption{Suggested beam parameters for the photofission research program.}
	\centering
\begin{tabular}{c|c}
	\hline  Parameter&Value  \\ 
	\hline  Energy& 5--20 MeV \\ 
	\hline  Flux ($\gamma$/s) & 10$^9$ at 3\% FWHM  \\ 
	\hline  Polarization& Linear and Circular \\ 
	\hline  Diameter& 10 mm on target \\ 
	\hline  Beam Repetition Rate & few MHz  \\
	\hline  Beam Pulse Width & $<$ 1.0 ns  \\
	\hline  Macropulse Width & 1 $\mu$s to 10 s   \\ 
	\hline  Macropulse Repetition Rate & 0.1 Hz to 500 kHz   \\ 
	\hline 
\end{tabular} 
	\label{tab:nfbeam}
\end{table}

\section{Nuclear Astrophysics} 
\lhead{Nuclear Astrophysics}
\rhead{Nuclear Physics Below 20 MeV}
\label{Sec:NS2}
Many questions in astrophysics require a detailed understanding of stars and stellar properties, thus challenging stellar models to become more sophisticated, quantitative, and realistic in their predictive power. This in turn requires more detailed input, such as thermonuclear reaction rates and opacities, and a concerted effort to validate models through systematic observations. Consequently, the study of nuclear reactions in the universe remains at the forefront of nuclear physics and astrophysics research.

Nuclear reactions generate the energy in stars and are responsible for the synthesis of the elements. When stars eject part of their matter through various means, they enrich the interstellar medium with their nuclear ashes and thereby provide the building blocks for the birth of new stars, of planets, and of life itself.  Element synthesis and nuclear energy generation in stars are the two primary research topics in nuclear astrophysics. Both require accurate knowledge of charged-particle- and neutron-induced nuclear reactions that take place in the hot stellar plasma. It is remarkable how the quantum mechanical nature of atomic nuclei influence the macroscopic properties of stars.

The DOE report {\it The 2015 Long Range Plan for Nuclear Science: Reaching for the Horizon} organizes nuclear astrophysics into five broad topical areas: (1) the origin of the elements, (2) the life of stars, (3) the death of stars, (4) the matter of neutron stars, and (5) connections: dark matter, QCD phase diagram, weak interactions and neutrinos.  The beams available at the next-generation laser Compton $\gamma$-ray source facilities will enable measurements that contribute to the first four topical areas.  The main opportunities are for cross-section measurements of ($\gamma$,$\gamma$') nuclear resonance florescence (NRF) processes and ($\gamma$,particle) reactions.   The NRF measurements provide important information for determining photon strength functions (PSFs), $\gamma$-ray transition probabilities, and nuclear structure spectroscopic information, all of which are inputs to nuclear astrophysics reaction-network calculations. The ($\gamma$,particle) reaction measurements provide data that are important input for $\gamma$-ray-induced reactions on stable nuclei in stars and also for the time reverse of particle capture on unstable nuclei.  These measurements are particularly relevant for $p$-process, $s$-process, and $r$-process nucleosynthesis.  The most important contribution to nuclear astrophysics that will come from opportunities created by the beam capabilities of a next-generation laser Compton $\gamma$-ray source is the measurement of the $^{16}$O($\gamma$,$\alpha$)$^{12}$C reaction cross section as a means for determining the cross section for the ${12}$C($\alpha$,$\gamma$)$^{16}$O reaction at center-of-mass energies important for carbon burning in massive stars using time reversal invariance. Some examples of research opportunities at the next-generation laser Compton $\gamma$-ray source are outlined below.  Table \ref{tab:napbeam} gives the suggested beam parameters for the nuclear astrophysics program.

\begin{table}[ht]
	\caption{Suggested beam parameters for the nuclear astrophysics program.  The beam parameters required to carryout the $^{16}$O($\gamma$, $\alpha$)$^{12}$C measurements are the most demanding in this research area.  In addition to the listed beam attributes, another important requirement is that the beam on target have a very small bremsstrahlung component, in order to minimize backgrounds caused by $\gamma$-ray-induced particle production.}
	\centering
\begin{tabular}{c|c}
	\hline  Parameter&Value  \\ 
	\hline  Energy& 2--20 MeV \\ 
	\hline  Flux ($\gamma$/s) & 10$^{11}$ at 1\% FWHM  \\ 
	\hline  Polarization& Linear and circular  \\ 
	\hline  Diameter& 10 mm on target \\ 
	\hline  Beam Repetition Rate & few MHz  \\
	\hline  Beam Pulse Width & $< 1.0$ ns  \\ 
	\hline 
\end{tabular} 
	\label{tab:napbeam}
\end{table} 

\subsection{The Origin of the Elements: $p$-Process and $s$-Process Nucleosynthesis}  
Models of $p$-process and $s$-process nucleosynthesis require reliable ($\gamma$,$n$), ($\gamma$,$p$) and ($\gamma$,$\alpha$) reaction cross sections on hundreds of stable and unstable nuclei. The proton-rich nuclei cannot be produced by neutron capture reactions. Complete network calculations on $p$-process nucleosynthesis include several hundred isotopes and the corresponding reaction rates. Theoretical predictions of the rates, normally in the framework of the Hauser-Feshbach theory, are necessary for modeling the nucleosynthesis reaction network. The reliability of these calculations should be tested experimentally, especially at rate-limiting paths in the network.  Different approaches are available and necessary to improve the experimental data base for the $p$-process. The ($\gamma$,$n$) cross sections in the energy regime of the GDR have already been measured extensively (see, e.g., \cite{Ber75}). More recently, substantial effort has be devoted to using beams with a continuous bremsstrahlung spectrum to determine the reaction rates without any assumptions on the shape of the cross section's energy dependence in the astrophysically relevant energy region, close to the reaction threshold \cite{Vog01,Son04,Erh06}.  A determination of the reaction rates by an absolute cross section measurement is also possible using monoenergetic photon beams produced by a laser Compton $\gamma$-ray source \cite{Uts03}.  The beam capabilities at next-generation laser Compton $\gamma$-ray sources will open the possibility of higher accuracy ($\gamma$,$n$) cross-section measurements and measurements on nuclei with low natural abundances, for which the amount of target material will be small.

In contrast, the experimental knowledge about the ($\gamma$,$p$) and ($\gamma$,$\alpha$) reactions in the corresponding Gamow window is smaller. In fact, the experimental data is based on the observation of the time reversal ($p$,$\gamma$) and ($\alpha$,$\gamma$) cross sections, respectively \cite{Sau97,Bor98,Ozk02,Rap02,Gyu05} for the proton-rich nuclei with mass numbers around 100. Due to the difficulties concerning the experimental accessibility of the ($\gamma$,$\alpha$) reaction rates, a method using elastic $\alpha$ scattering has been established \cite{Ful01,Gal05}. It would be a tremendous improvement in the quality of the database to measure these rates directly using photon beams. Having a significant impact on this database will require a program of systematic measurements on a broad range of nuclei.  This type of program would be made possible by a next-generation laser Compton $\gamma$-ray source.

\begin{figure}[H]
	\centering
\includegraphics[width=0.6\linewidth] {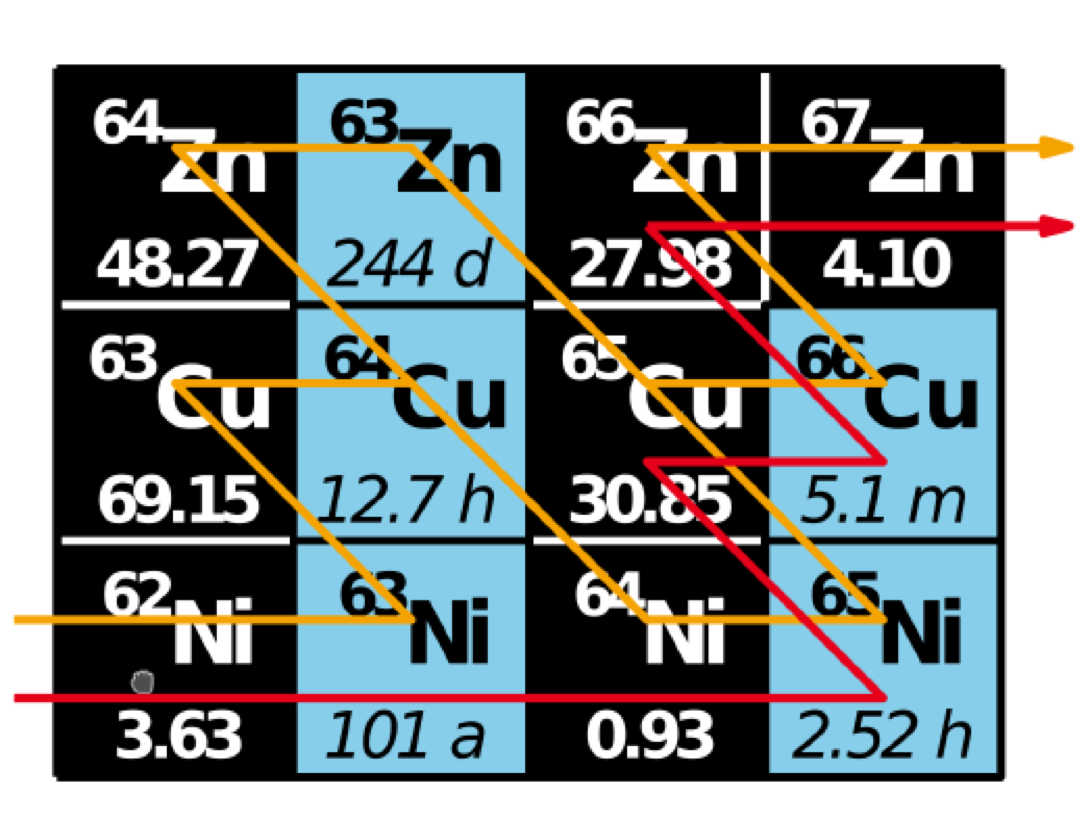}
	\caption{The $s$-process reaction path in the Ni-Cu-Zn region. During He-core burning (yellow), the temperature and neutron density are low compared to those during C-shell burning (red). Therefore, the unstable nucleus $^{63}$Ni either experiences decay or radiative neutron capture and acts as a branch-point nucleus. The isotopic abundance pattern of copper and zinc depends strongly on the stellar neutron-capture rate of $^{63}$Ni.  This figure is adapted from Ref.~\cite{Led13}. }
	\label{fig:64ni}
\end{figure}

The heavy elements above the so-called iron peak are mainly produced in neutron capture processes: the $r$ process (r: rapid neutron capture) deals with high neutron densities, well above 10$^{20}$ cm$^{-3}$, and temperatures of around 2 to 3 GK. It is thought to occur in explosive scenarios such as supernovae \cite{Kra93,Wall97} and was recently verified in neutron star merger via gravity wave observations \cite{Abb17}. In contrast, the average neutron densities during $s$-process nucleosynthesis (s: slow neutron capture) are rather small (around 10$^8$ cm$^{-3}$), so that the neutron capture rate $\lambda_n$ is normally well below the $\beta$-decay rate $\lambda_\beta$, and the reaction path is therefore close to the valley of $\beta$ stability \cite{Kap99,Kap06, Dill08}. However, during the peak neutron densities, branching occurs at unstable isotopes with half-lives as low as several days. The half-lives of these branch points are normally known with high accuracy, at least under laboratory conditions, and rely on theory only for the extrapolation to stellar temperatures \cite{Tak83}. However, their neutron capture cross sections are accessible to direct experiments only in special cases. For this reason, nuclear astrophysics reaction-network simulations currently rely heavily on Hauser-Feshbach (HF) model calculations of the critical neutron capture cross sections.  The difficulty is that there are many examples where the results of HF model calculations differ substantially from measurements (see, e.g., Ref. \cite{Bao00}).   Thus, experimental constraints on the theoretical predictions of these branch points are needed.  Laser Compton $\gamma$-ray sources enable measurements of the inverse ($\gamma$,$n$) reaction, which can also provide information about the stellar enhancement factors (SEFs).  The SEF accounts for the difference in the neutron-capture cross section measured in the laboratory and the effective cross section in the stellar environment.  Because nuclei in stars spend much of the time in excited states, capture reactions mostly occur on excited states, not the ground state.  In addition to determining the cross sections and the SEFs directly from ($\gamma$,$n$) cross-section measurements, the SEFs can be calculated using photon strength functions determined from NRF measurements.  The importance of each nucleus and reaction in the $s$-process network must be analyzed on its own merits. An example of the type of information that can be obtained using linearly polarized mono-energetic photon beams is illustrated in the proposed NRF measurement on $^{64}$Ni at HI$\gamma$S.

The $^{64}$Ni nucleus is the product of neutron capture on $^{63}$Ni, which is a branch-point nucleus. Figure~\ref{fig:64ni} illustrates the situation during He-core burning (yellow) and C-shell burning (red) of a massive star, where the weak component of the $s$-process originates \cite{Kap11}. As the branching is very sensitive to the $^{63}$Ni($n$,$\gamma$) cross section, a measurement was performed at CERN using a radioactive $^{63}$Ni target \cite{Led13,Led14}. However, the reaction rate in the hot environment of the stellar plasma differs significantly from the measured value because excited states are populated \cite{Rau12}. In the case of $^{63gs}$Ni($n$,$\gamma$), the stellar rate is still around 90\% at He-core burning temperatures, but it drops to around 40\% at the higher temperature in the C-shell burning phase \cite{Led13}. Hauser-Feshbach model calculations are required to account for the stellar enhancement \cite{Hau52}. These calculations rely on the photon strength function in $^{64}$Ni, which can be deduced from photoabsorption cross sections and the decay properties of low-spin states.  The linearly polarized monoenergetic beams available at laser Compton $\gamma$-ray sources enable model-independent determination of photoabsorption cross sections as a function of the excitation energy, and consequently, a determination of the photon strength functions (see, e.g., \cite{Ton10,Loh13,Rom13,Isa13,Isa16}).
\subsection{The Life of Stars: The $^{12}$C($\alpha$,$\gamma$) Reaction Cross Section}
\begin{figure}[H]
	\centering
\includegraphics[width=0.4\linewidth] {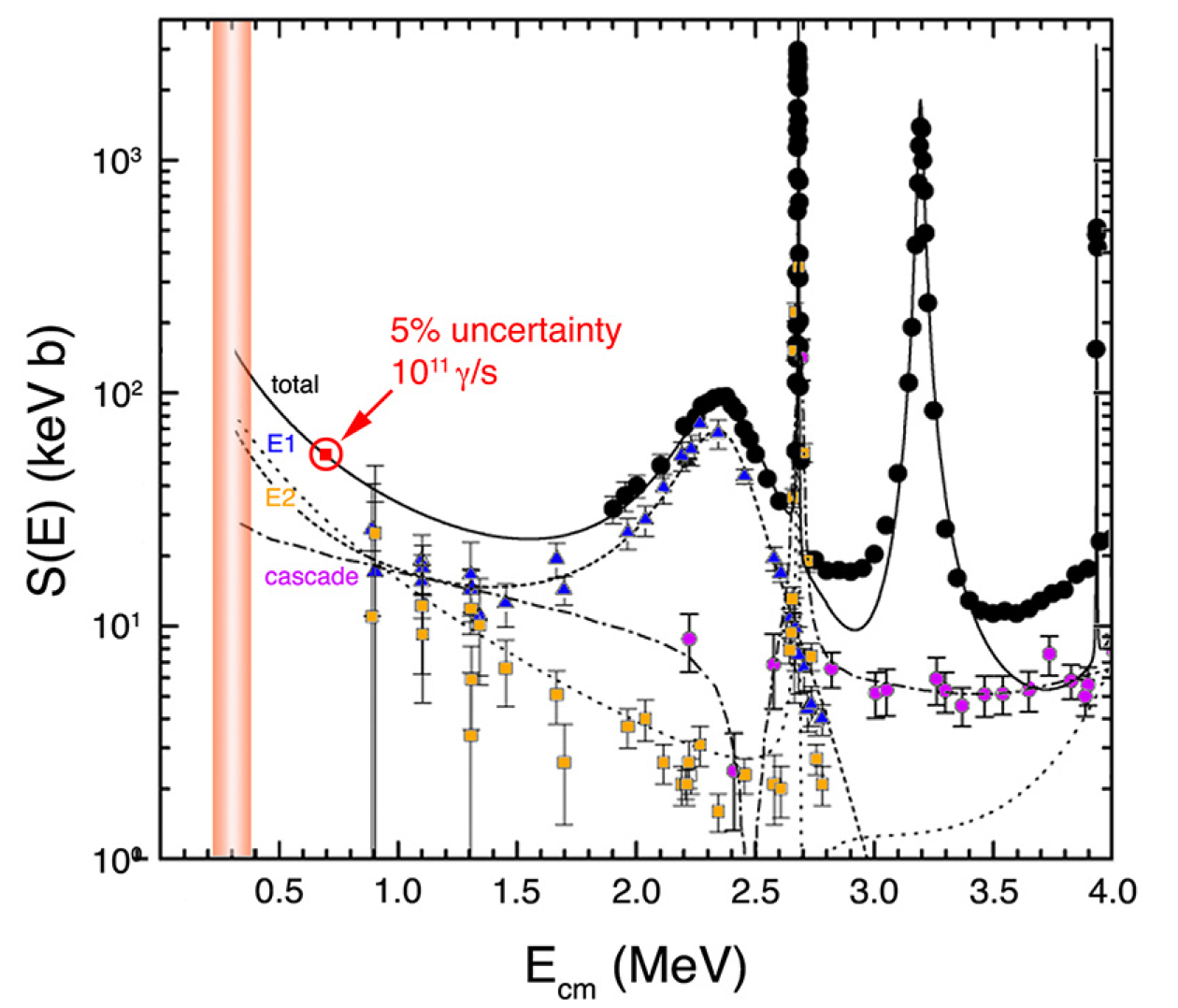}
	\caption{Plot of total S-factor data (filled-in circles) \cite{Sch05} for $^{12}$C($\alpha$,$\gamma$)$^{16}$O compared with E1 (open triangles) and E2 (open squares) $\gamma$-ray measurements \cite{Ass06} and the $E_x = 6.05$ MeV cascade data (open circles) \cite{Mat06}.  The solid line represents the sum of the single amplitudes of an R-matrix fit \cite{Kun02}, while the dotted and dashed lines are the E1 and E2 amplitudes, respectively. In addition, the R-matrix fit of \cite{Mat06} to their cascade data is shown as the dot-dashed line. The latter component is not included in the sum and might explain the high yield in the S-factor data between the resonances. This figure is adapted from Ref.~\cite{Str08}.}
	\label{fig:12cag}
\end{figure}

Late-stage red giant stars produce energy in their interiors via helium burning. In first generation stars, helium burning proceeds through the 3$\alpha$ process and then $^{12}$C($\alpha$,$\gamma$)$^{16}$O, while in later generation stars, $\alpha$ captures also involve the various CNO seed nuclei.  In both cases, the $^{12}$C($\alpha$,$\gamma$)$^{16}$O reaction helps to regulate the efficiency of helium burning in massive stars (those with masses greater than the suns). It also determines core mass, temperature and density during the latter stages of stellar evolution, and ultimately the mass of the iron core in the incipient supernova. In addition, the carbon-to-oxygen ratio (C/O) influences the abundances of elements produced in the ensuing explosion.  After several decades of effort, the uncertainty in the rate of the $\alpha$-particle capture reaction on $^{12}$C is still large enough to substantially limit our understanding of the latter stages of stellar evolution. To resolve this problem, the astrophysical S factor for the $^{12}$C($\alpha$,$\gamma$)$^{16}$O reaction must be determined to an accuracy better than about 10\% in the Gamow window ($E_{cm} = 300$ keV), which is indicated by the vertical band in Fig.~\ref{fig:12cag} \cite{Fow84}.

\begin{figure}[H]
	\centering
\includegraphics[width=0.4\linewidth] {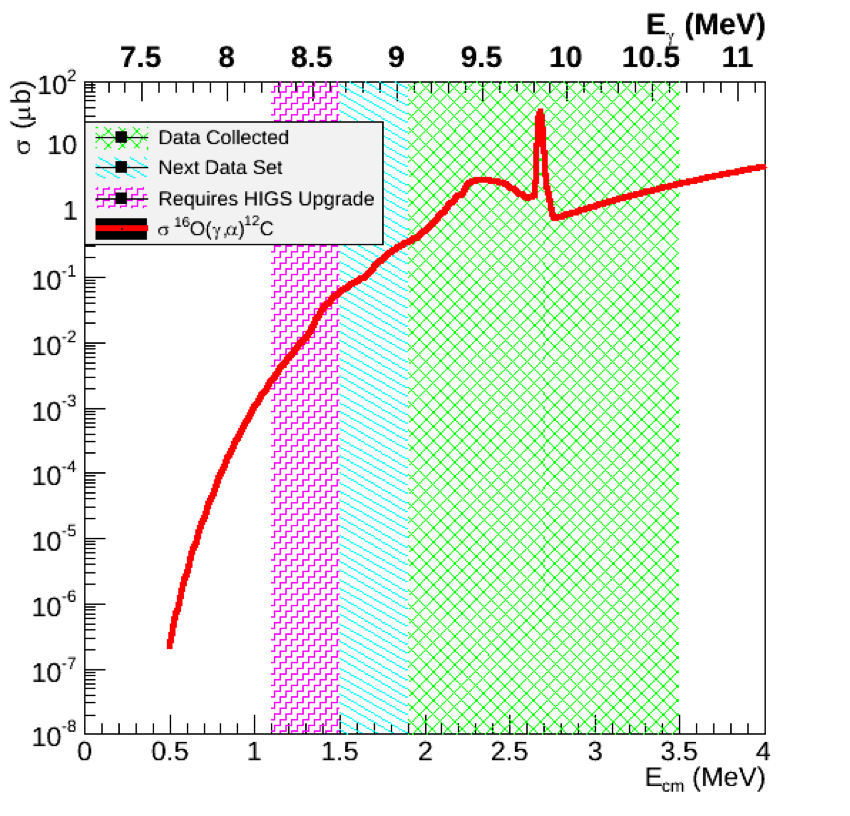}
	\caption{Plot of the cross section of the $^{16}$O($\gamma$,$\alpha$)$^{12}$C reaction as a function the system center-of-mass energy (bottom axis) and the incident $\gamma$-ray beam energy (top axis). The green hash shows the region where data have been collected at HI$\gamma$S using the optical time-projection chamber (TPC). The blue filled bar is the lowest energy region experimentally accessible at HI$\gamma$S using current target and detector technologies. The energy region indicated by the red brick filled area is what would be possible with an intensity upgrade of HI$\gamma$S.
 }
	\label{fig:16oga}
\end{figure}

As shown in Fig.~\ref{fig:12cag}, the cross section has been measured at various levels of precision down to a center-of-mass (cm) energyof 1.2 MeV; then it must be extrapolated down to 300 keV, the energy needed for stellar reaction-rate calculations.  One of the major uncertainties in performing the extrapolation arises from the presence of several resonances which contribute to the cross section at  energies around $E_{cm} = 0.75$ MeV.  Above $E_{cm} = 1$ MeV, the elastic scattering and capture reactions are dominated by a broad 1$^-$ resonance at an excitation energy in $^{16}$O of $E_x = 9.59$ MeV ($E_{cm} = 2.43$ MeV) and a narrow 2$^+$ state at $E_x = 9.85$ MeV ($E_{cm} = 2.70$ MeV). However, a 1$^-$ state at $E_x = 7.12$ MeV, just 42 keV below threshold, determines the capture cross section in the astrophysically relevant energy region, including its interference with the higher lying 1$^-$ and 2$^+$ states.  In addition, broad high-lying states and direct processes produce a coherent background that affects the energy dependence of the cross section and thereby the extrapolation.

Direct measurements of the $^{12}$C($\alpha$,$\gamma$)$^{16}$O cross section at energies below $E_{cm} = 2$ MeV have been attempted for over 30 years. The major difficulty encountered in these experiments is the intense neutron background arising from the $^{13}$C($\alpha$,$n$) reaction. This background tends to swamp the $\gamma$-ray detector. The $\gamma$-ray beams produced by laser Compton sources have a narrow energy spread and thus offer an alternative technique for measuring the cross section for $\alpha$ capture on $^{12}$C at energies important to astrophysics as discussed above. The principle of detailed balance allows the determination of the ($\alpha$,$\gamma$) cross section from the measurement of the cross section for the time reversed ($\gamma$,$\alpha$) reaction.  An added advantage of using the ($\gamma$,$\alpha$) reaction over the direct reaction is that the cross section in the Gamma window is enhanced by about a factor of 50 due to detailed balance.  This experimental concept has been demonstrated at HI$\gamma$S using a CO$_2$ gas-filled optical time projection chamber (TPC) \cite{Gai10,Zim13}. The calculated cross section for the $^{16}$O($\gamma$,$\alpha$)$^{12}$ reaction is shown in Fig.~\ref{fig:16oga}.  An upgrade of HI$\gamma$S to increase the $\gamma$-ray beam flux on target by a factor of about fifty in the energy region important for this reaction,  would enable meaningful measurements in the red shaded energy range ($\langle E_{cm}\rangle = 1.3$ MeV) in the figure.  The next-generation laser Compton sources could potentially deliver more than an additional factor of 500 in beam flux on target relative to HI$\gamma$S, thereby enabling measurements down to about $\langle$E$_{cm}\rangle = 700$ keV. This measurement should be performed at next-generation laser Compton $\gamma$-ray source facilities using a variety of experimental techniques, such as gas TPCs (both optical and charge readout), silicon strip detectors with thin targets, and total cross-section measurements using super-heated high-purity water detectors (e.g., as in Ref.~\cite{DiG15}).

\subsection{The Death of Stars: $r$-Process Nucleosynthesis}

More than half of the heavy nuclei with $A > 120$ are produced by $r$-process nucleosynthesis.  The $r$-process involves nuclear reactions driven by rapid neutron capture, where the neutron capture rate is faster than the competing beta decay.  The likely environments for the $r$-process are type-II supernovae and merging neutron stars.  Simulations of $r$-process nucleosynthesis require reliable ($n$,$\gamma$) reaction cross sections on hundreds of stable and unstable nuclei.  Theoretical predictions of the rates, normally in the framework of Hauser-Feshbach theory, are necessary for modeling neutron capture on unstable nuclei. The reliability of these calculations should be tested experimentally, especially at rate-limiting paths in the network. The ($\gamma$,$n$) time-reversed photodisintegration reaction offers a mechanism for determining the neutron capture cross section and level density of radioactive nuclei.  Next-generation laser Compton $\gamma$-ray sources will enable measurements on isotopes with low natural abundances and thus very limited sample sizes.
\subsection{Neutron Stars: the EOS of Neutron-Rich Matter}  
Obtaining information about the detailed structure of the crust of a neutron star is an important open challenge in astrophysics.  The crust is composed of non-uniform neutron-rich solid matter that is about 1 km thick and located above a liquid core \cite{Hae07,Lat00}. The inner crust comprises the region from the density at which neutrons drip from nuclei to the inner edge separating the solid crust from the homogeneous liquid core. While the density at which neutrons drip from nuclei is well determined, the transition density at the inner edge is much less certain because of insufficient knowledge of the equation of state (EOS) of neutron-rich nuclear matter.  Measurements of collective responses of neutron-rich nuclei provide constraints on the nuclear EOS.  The monoenergetic and linearly polarized beams at laser Compton $\gamma$-ray sources enable unique measurements of dipole and quadrupole excitations. Those most relevant to exploring the EOS are studies of the pygmy dipole resonance (PDR) on nuclei at the neutron rich end of an isotope chain using nuclear resonance florescence (NRF) and determination of the centroid energy and width of the isovector giant quadrupole resonance (IVGQR) via Compton scattering \cite{Hen11}. The underlying structure of the PDR is often interpreted as an oscillation of less bound valence neutrons forming a neutron skin \cite{Paa07,Lit08}. However, the true nature of the PDR is a matter of ongoing discussion. Furthermore the PDR and the E1 strength in general have a direct connection to the neutron skin of nuclei and the symmetry energy of nuclear matter \cite{Kli07,Pie06}.
\begin{figure}[H]
	\centering
\includegraphics[width=0.35\linewidth] {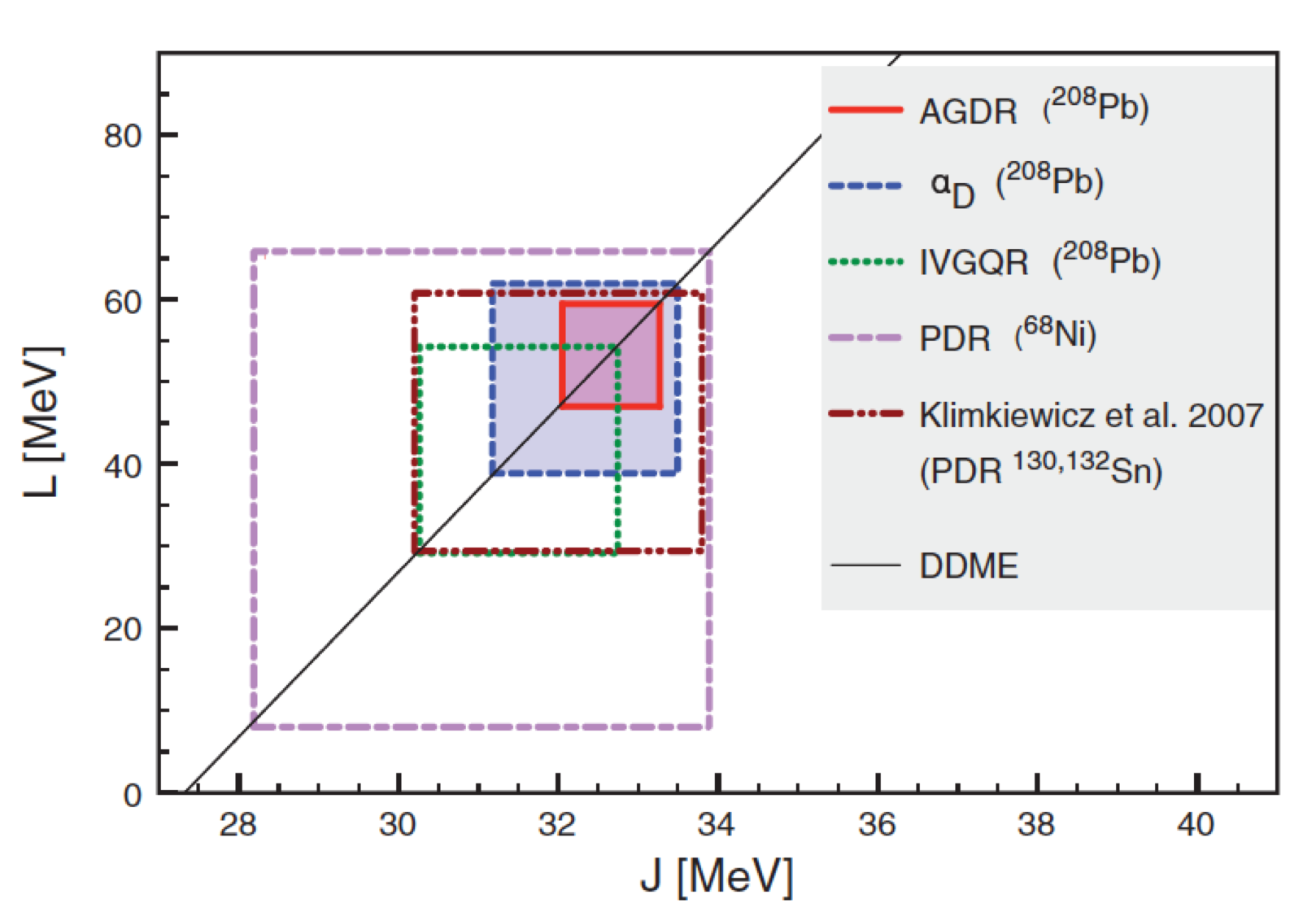}
	\caption{Constraints of the symmetry energy at saturation $J$ and slope parameter $L$, obtained from a comparison of relativistic nuclear energy density functional results and data on anti-analog giant dipole resonance (AGDR) \cite{Kra01} and IVGQR \cite{Hen11} excitation energies in $^{208}$Pb; the dipole polarizability $\alpha$D of $^{208}$Pb \cite{Tam11}; and the PDR energy-weighted strength in $^{68}$Ni \cite{Wie09} and $^{130,132}$Sn \cite{Kli07}. This figure is taken from Ref.~\cite{Paa14}.}
	\label{fig:eos}
\end{figure}

\section{Hadronic Parity Violation} 
\lhead{Hadronic Parity Violation}
\rhead{Nuclear Physics Below 20 MeV}
\label{Sec:HPV}
The capabilities of a next generation laser Compton $\gamma$-ray source, combined with current and expected advances in theory and lattice calculations, provide the opportunity to make significant progress in our understanding of hadronic parity violation (PV). Parity violation in nuclear processes can probe the standard model (SM) and point towards possible SM extensions.  In the SM sector, hadronic PV provides an important probe of two phenomena that are not well understood: neutral-current nonleptonic weak interactions and nonperturbative strong dynamics. While PV  is well understood in quark-quark weak interactions, it is ultimately the interplay of different forces at different length scales in the SM that is responsible for hadronic PV phenomena. Neutral-current interactions are suppressed in flavor-changing hadronic decays, making hadronic PV  (HPV) between nucleons the best place to study neutral-current effects. Because parity violating nucleon-nucleon (NN) interactions are the manifestation of the interplay of nonperturbative strong effects and the short-range weak interactions between quarks, they are sensitive to short-distance quark-quark correlations inside the nucleon. Quark-quark and NN weak interactions also induce parity-odd effects in electron scattering~\cite{Bec01, Beck01, Beck03, Beise05}, nuclear decays~\cite{Ade85}, compound nuclear resonances~\cite{Bow1, Tom}, and atomic structure, where they are the microscopic source for nuclear anapole moments~\cite{Zel57, Fla80, Woo97, Tsi09}. The comparison between NN weak amplitudes in few-nucleon systems and heavy nuclei 
can also test the statistical theory of parity violation in compound nuclei, which will become useful in the future for the interpretation of time reversal violation experiments in the transmission of polarized neutrons through polarized nuclear targets on p-wave compound nuclear resonances~\cite{Bowman2014}. 

The NN weak interactions are therefore worth understanding both for their own sake and for other areas of nuclear, particle, and atomic physics. Agreement between theory and experiment in this area pushes the ``complexity'' frontier of the SM, as it requires a quantitative understanding of strong, electromagnetic, and weak interactions in the strongly-interacting regime of QCD all acting in the same system. This research program will also stretch the technology of lattice gauge theory calculations~\cite{Wasem12} to its computational limits.  

This section explores possibilities for measuring hadronic PV in few nucleon-systems and in the mixing of parity doublet states in light nuclei using circularly polarized high-intensity photon beams.  

\subsection{Hadronic Parity Violation in Few-Nucleon Systems}

Weak NN amplitudes at low energy are suppressed by six to seven orders of magnitude compared to strong NN amplitudes and are therefore difficult to observe. While these effects can be amplified by several orders of magnitude by nuclear dynamics in complex systems, the unambiguous extraction of the weak interactions among nucleons requires parity violating measurements in very light nuclei, where theory is under good control. 
At the moment this includes the deuteron, the triton, and $^{3}$He, with the four-nucleon system in the pipeline.
These very light systems can be calculated using two- and three-nucleon interactions, which in turn are understood in terms of effective field theories (EFTs) that systematically incorporate the symmetries of QCD. The EFT approach can consistently be applied to strong and weak nucleon interactions as well as to external currents. There is also growing optimism that parity-odd effects in $A > 4$ nuclei can be calculated. We are at a juncture in the development of theory where PV may soon be understood in terms of QCD dynamics even in many body nuclei.  Ongoing programs~\cite{KITP} are pursuing the matching of lattice QCD to EFTs, which are used as input into many-body calculations of heavier nuclei. Both SM and beyond-the-SM physics can be treated in this way.

At leading order in the EFT power counting, and at very low photon energies (below 10 MeV), there are five parity violating low-energy constants (LECs) that parameterize the short-distance physics \cite{Dan,Zhu05,Girlanda:2008ts}.\footnote{At higher energies, pions have to be treated as active degrees of freedom and this increases the number of LECs. Parity violation  in chiral EFT has been considered recently in Refs.~\cite{deV13,Viv14,deV14,deV15}}.
The values of these LECs cannot be determined within the EFT framework but have to be extracted from experiment or ultimately from calculations involving nonperturbative QCD. Recent work shows that upon combining the EFT approach with the large-$N_c$ expansion of QCD, the number of independent parity-violating LECs reduces to two at leading order~\cite{Phillips:2014kna}. One of these is the $\Delta I=2$ isotensor coupling, which is poorly constrained from existing measurements. Combining the existing data for $\vec{p}+p$ scattering (which constrains the remaining leading-order isoscalar piece) with the results of a measurement of the $\Delta I=2$ contribution would therefore constrain all terms at leading order in this combined EFT and large-$N_c$ expansion~\cite{Schindler2016}.

An important experimental constraint on the isovector partity violating interactions comes from  the parity violating asymmetry from polarized neutron capture on the proton, $\vec{n}+p \to D+ \gamma$, recently measured by the NPDGamma collaboration at the Spallation Neutron Source (SPS) at ORNL~\cite{Bly18}. In the next few years we expect to possess two additional pieces of experimental information on the NN weak interaction in few-nucleon systems. 
The analysis of parity violation in $\vec{n}+{}^3\text{He} \to {}^3\text{H}+p$ completed recently at the Spallation Neutron Source at ORNL will be published, and another experiment on parity violation in $\vec{n}+{}^4\text{He}$ spin rotation~\cite{Snow2015} will be performed at the new NG-C neutron beam line at the NIST Center for Neutron Research (NCNR). 
However, these will yield scant experimental information on the elusive $\Delta I=2$ component of the NN weak interaction. Fortunately the $\Delta I=2$ operator is the easiest target for a quantitative lattice calculation from the SM because it does not receive contributions from disconnected diagrams.  The isotensor operator is scheduled to be calculated on the lattice in the next few years~\cite{INCITE, Kurth:2015cvl}.  Lattice constraints on other NN weak amplitudes are many years out, at least at physical quark masses.

The cleanest experimental channel for extracting the $\Delta I=2$ component in few-body PV is P-odd deuteron photodisintegration near threshold using the helicity-dependent photodisintegration cross section for circularly polarized photons.  
Limits from previous attempts exist from Chalk River \cite{Earle88} and from the reverse reaction~\cite{Knyaz'kov:1984zz}, but did not resolve a nonzero value. Experiments to measure this observable have been proposed in the past at JLab \cite{Sinclair00,Woj00}, SPRING-8, and the Shanghai Synchrotron, but they have not yet been realized.  
It was therefore listed in the last NSAC Long-Range plan as a priority NN weak interaction measurement for the future. In addition, it is possible to contemplate other parity violation experiments involving helicity-dependent reactions with circularly polarized photons in few-nucleon systems, such as the breakup of three-nucleon systems. 
We emphasize the importance of a $\vec \gamma d \rightarrow p n$ parity-violation measurement because of the unique information it will provide about the weak interaction in hadronic systems and the state of EFT and lattice calculations.

\begin{figure}[h!]
	\centering
\includegraphics[width=0.6\linewidth] {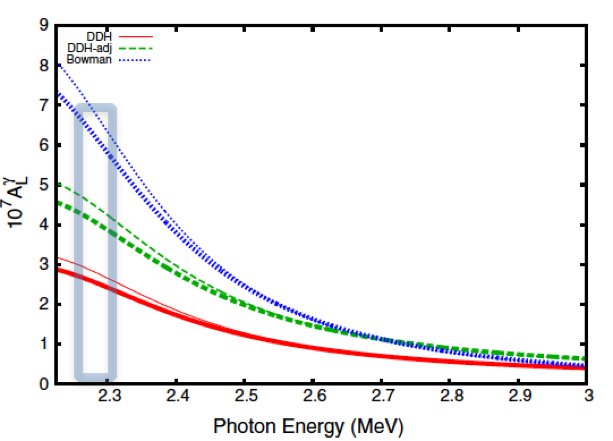}
	\caption{Plot of the calculated $A_L^\gamma$ for parity violating photodisintegration of the deuteron as a function of photon energy. The lowest photon energy on the plot is the breakup threshold value corresponding to the deuteron binding energy. This figure is adapted from Ref.~\cite{Vanasse2014}.  The curves represent next-to-leading-order EFT calculations using three different sets of hadronic weak coupling constants, which are labeled DDH, DDH-adj, and Bowman.  DDH represents the DDH "best values" \cite{Des80}.  DDH-adj refers to a set in which two combinations of $\rho$ and $\omega$ couplings are fit to data on the $\vec{p}p$ longitudinal asymmetry, while the remaining couplings take the DDH "best values" \cite{Sch04}. The set labeled Bowman is obtained by fitting the parity violating couplings to a variety of available data \cite{Bow07}.  The optimum energy window for the measurement is indicated by the rectangle.
 }
	\label{fig:hpvay}
\end{figure}

\begin{figure}[h]
	\centering
\includegraphics[width=0.6\linewidth] {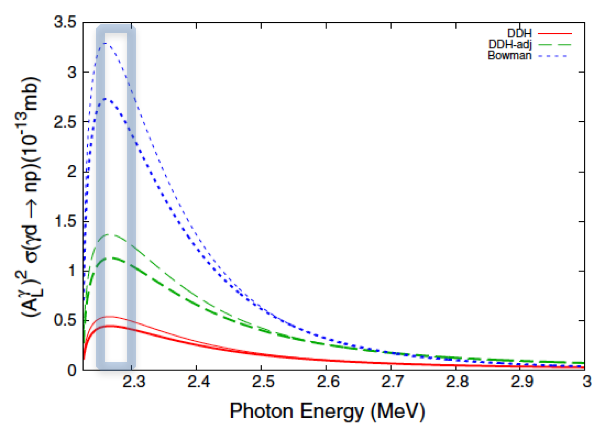}
	\caption{Plot of the calculated figure of merit for PV photodisintegration of the deuteron as a function of photon energy. The lowest photon energy on the plot is the breakup threshold value corresponding to the deuteron binding energy. This figure is adapted from Ref. \cite{Vanasse2014}.  The curves follow the legend given in Fig.~\ref{fig:hpvay}.  The optimum energy window for the measurement is indicated by the rectangle.
 }
	\label{fig:hpvfom}
\end{figure}

The parity-violating asymmetry in deuteron photodisintegration, including its energy dependence, has been considered in pionless EFT in Refs.~\cite{Schindler2010,Vanasse2014}. The next-to-leading order results of Ref.~\cite{Vanasse2014} were combined with different model estimates for the LECs to determine the expected size of the parity-violating asymmetry $A_L^\gamma$, to estimate a figure of merit (FOM), and to extract the best energy regime for performing the experiment. 
$A_L^\gamma$ is computed as the ratio of the difference in the photodisintegration cross section measured with positive ($h = +1$) and negative ($h = -1$) beam helicity divided by the unpolarized cross section, which is the sum of the positive and negative helicity cross sections.  The beam helicity is $h = +1$ when the polarization vector of the circularly polarized beam is in the direction of the beam momentum vector.  
$A_L^\gamma$ for the photodisintegration of the deuteron computed using various parameterizations of the hadronic weak coupling constants is plotted as a function of photon energy in Fig.~\ref{fig:hpvay}.  The FOM quantity that is optimized for obtaining the best statistical accuracy per hour of beam time is FOM = ($A_L^\gamma$)$^2\sigma$, where $\sigma$ is the unpolarized photodisintegration cross section.  The FOM computed for various parameterizations of the hadronic weak coupling constants are plotted as a function of photon energy in Fig.~\ref{fig:hpvfom}.  This analysis indicates that the experiment should ideally be performed in the photon energy region between 2.26 and 2.30 MeV. In this energy window, the average cross section and $A_L^\gamma$ are 600 $\mu$b and 4 $\times$ 10$^{-7}$, respectively.  

\begin{figure}[h]
	\centering
\includegraphics[width=0.6\linewidth] {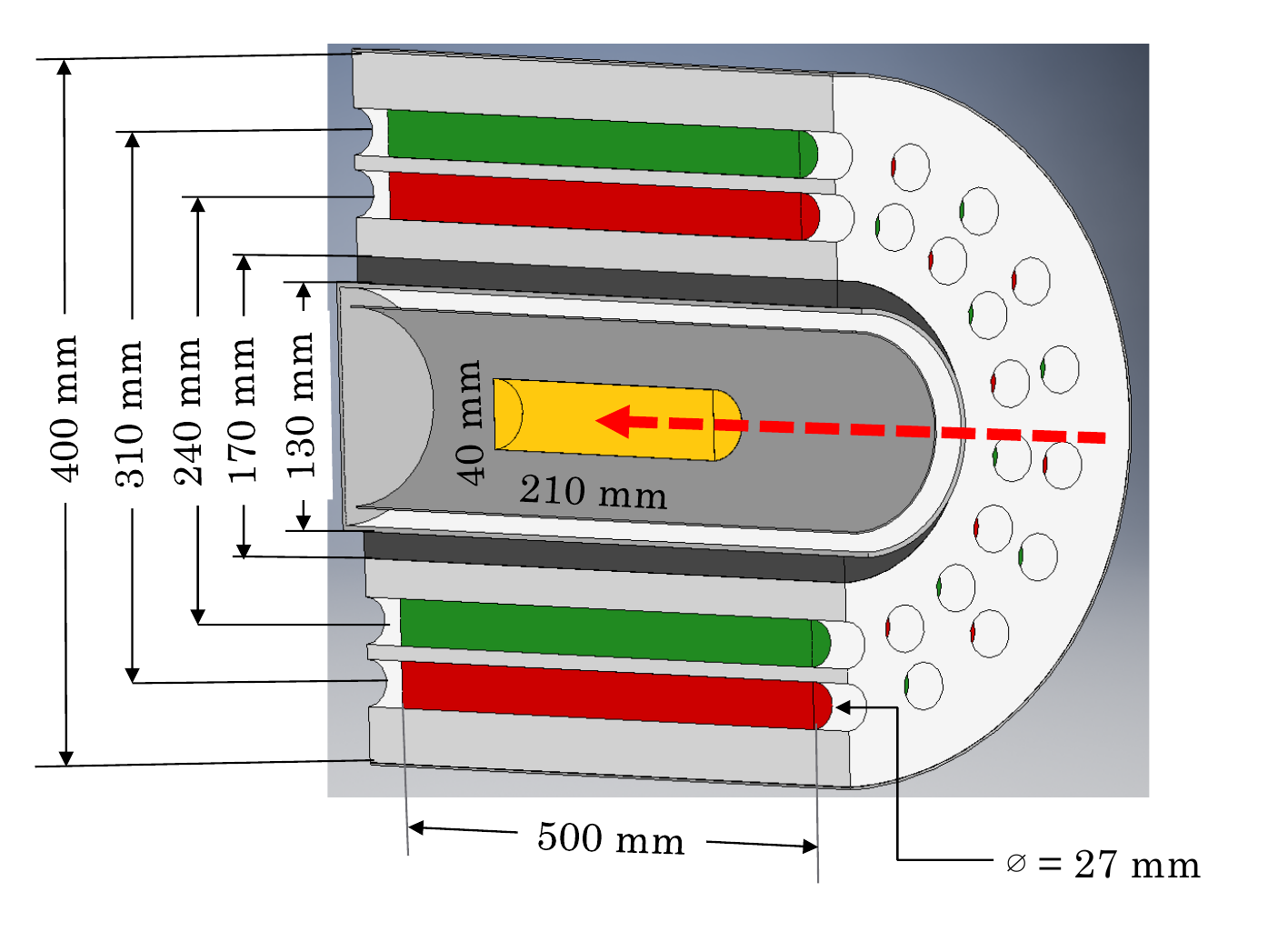}
	\caption{Schematic diagram of the experimental concept for the $A_L^\gamma$ measurement for PV photodisintegration of the deuteron.  
 }
	\label{fig:hpvsetup}
\end{figure}

 A schematic diagram of the concept for the experiment setup is shown in Fig.~\ref{fig:hpvsetup}.  The neutrons from the photodisintegration reaction in the liquid deuterium (LD) target are detected in an array of $^3$He ionization tubes embedded in a cylindrical polyethylene moderator that surrounds the target.  The background due to interactions of scattered $\gamma$ rays in the $^3$He ionization tubes is determined using an array of $^4$He ionization tubes located outside the inner layers of $^3$He ionization tubes.  About 75\% of the $\gamma$ rays incident on the 21-cm long LD target are transmitted through the target.  The $\gamma$-ray beam is collimated to a diameter of 10 mm, and the inner diameter of the $^3$He ionization tube array is 6 cm.  The ionization tubes are 50 cm long and centered about the LD target. With this target thickness, detector geometry, and a $\gamma$-ray beam intensity of 10$^{10}$ $\gamma$/s on target in the above energy window, a statistical accuracy of $\pm$10$^{-7}$ can be obtained with about 9000 hours of beam time.

 The technical demands to minimize $\gamma$-ray helicity-dependent changes in the phase space of the $\gamma$-ray beam are comparable to those encountered at the polarized electron injector section at JLab in the parity violating  electron scattering. Developing the capability for delivering polarized $\gamma$-ray beams to targets with the stability and precision required to perform 10$^{-7}$ asymmetry measurements will demand focused and dedicated resources for R\&D work and implementation of the supporting systems. Suggested beam parameters for studying PV in deuteron photodisintegration are shown in Table \ref{tab:beamd}.

\begin{table}[h]
	\caption{Suggested beam parameters for  the $A_L^\gamma$ measurement in PV deuteron photodisintegration.}
	\centering
	\begin{tabular}{c|c}
		\hline  Parameter&Value  \\ 
		\hline  Energy& 2.25-2.30 MeV \\ 
		\hline  Flux & $10^{10}$  \\ 
		\hline  $\Delta$E/E & 0.001 to 0.01 FWHM  \\
		\hline  Polarization& Circular  \\ 
		\hline  Diameter& 10 mm on target \\ 
		\hline  Time Structure& 10 Hz polarization flip \\ 
		\hline 
	\end{tabular} 
	\label{tab:beamd}
\end{table}

\subsection{Hadronic Parity Violation in Light Nuclei}

The size of the expected parity violating asymmetries from reactions in few-nucleon systems induced with photons at MeV energies are generally around several times 10$^{-7}$. For such experiments, achieving the required systematic accuracy is a formidable challenge, especially at the initial ramp-up stage of a NGLCGS facility. This motivates us to look for parity violating asymmetries that are larger, and therefore easier to measure, even if the interpretation is not as theoretically clean as in few-nucleon systems.
Certain light nuclei possess doublets with narrow energy spacings in the 10 to 100 keV range between levels of opposite parity. Parity-odd asymmetries can be amplified by several orders of magnitude by the interference of such parity doublets states involved in photon-induced reactions. There are a half dozen such nuclei (some of which were investigated using charged-particle beams a few decades ago)~\cite{Ade85} whose parity-odd asymmetries could be measured using the ultra-high intensity mono-energetic polarized photon beam at the NGLCGS. To develop the technical infrastructure and scintific expertise among the staff, a parity violation program at such facilities should start with an asymmetry measurement on parity doublets in nuclei, where the parity violating assymetry is several orders of magnitude larger than that in few-nucleon systems. The experiment could measure either the helicity dependence in the total cross section using transmission or the asymmetry of the fluorescence from the doublet using a HPGe detector array. Examples of candidate nuclei and estimated amplification factors for parity violating asymmetries in nuclear resonance fluorescence are described by Titov, Fuji, Waral, and Kawase \cite{Tit06}.

Either parity-odd asymmetry can be calculated in a simple way in terms of the energies and widths of the states and the parity-odd matrix elements of interest. Parity violation in states which are particle unstable will be difficult to handle theoretically. However, as the theoretical tools to calculate parity violation in these nuclei improve, we can identify one or more candidate nuclei for a focused effort involving both theory and experiment.

\begin{table}[h]
	\caption{Suggested beam parameters for parity violiaying asymmetry  measurements in parity doublets.}
	\centering
	\begin{tabular}{c|c}
		\hline  Parameter&Value  \\ 
		\hline  Energy& 1-20 MeV \\ 
		\hline  Flux & 10$^9$ at 0.1-1\% FWHM  \\ 
		\hline  Polarization& Circular  \\ 
		\hline  Diameter& 10 mm on target \\ 
		\hline  Time Structure& 10 Hz polarization flip \\ 
		\hline 
	\end{tabular} 
	\label{tab:beam}
\end{table} 

Although the parity violating asymmetry is enhanced in the parity doublet systems, the photonuclear cross section is small compared to Compton and pair production cross sections in the energy range of interest below the giant dipole resonance. The beam requirements are therefore still challenging, as shown in Table~\ref{tab:beam}.  

\section{Nuclear and Homeland Security} 
\lhead{Applications}
\rhead{Nuclear Physics Below 20 MeV}
\label{Sec:App}
A next-generation Compton $\gamma$-ray beam facility can contribute to the R\&D of techniques and technologies for applications in homeland security, nuclear safeguards, and medicine.  The facility should have the infrastructure that supports measurements that contribute to the nuclear databases important in supporting the development of technologies and techniques in the above areas.  Equally important, the facility should have a target area that is equipped for evaluating concepts for $\gamma$-ray beam interrogation of cargo, nuclear fuel, and special assemblies.

Safeguarding the nation against evolving threats that involve the use of special nuclear materials (SNM) requires continued innovations in the procedures and technologies used to inspect cargo at ports of entry into the United States. Research frontier areas include the development of systems for $\gamma$-ray and neutron-beam interrogation of cargo for shielded SNM, new cost-effective materials for $\gamma$-ray and neutron detection, and improved techniques and supporting technologies for applications of nuclear forensics in the analysis of interdicted materials. The beam capabilities and technical infrastructure of next generation $\gamma$-ray beam facilities should support the following research: (1) evaluation of $\gamma$-ray beam interrogation techniques and technology concepts, (2) filling gaps in the photonuclear reaction database needed in methods for identification of SNM, including the development and maintenance of online databases, and (3) testing, characterizing, and calibrating $\gamma$-ray and neutron detectors.  The infrastructure for photonuclear data measurements should be optimized for nuclear resonance fluorescence (NRF), photo-neutron reactions, and photofission.

\begin{figure}[h]
	\centering
\includegraphics[width=0.4\linewidth] {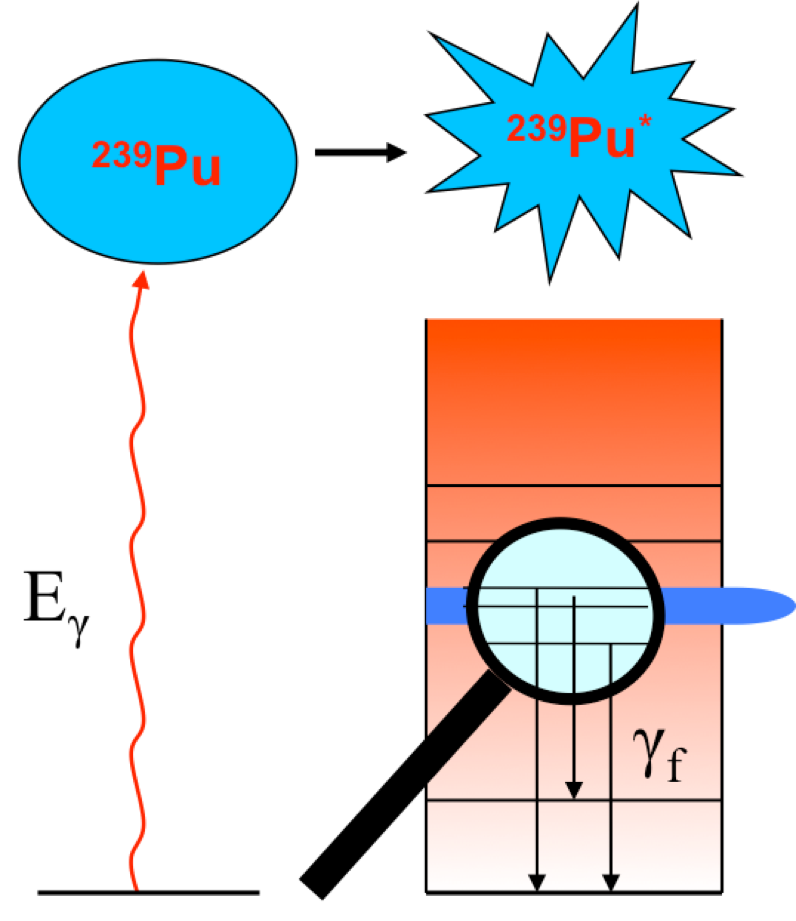}
	\caption{Schematic diagram of the NRF technique. First the incident photon is absorbed by a nucleus at certain resonance energies characterized by the parameters $E_i$, J$_\pi$, $\Gamma_i$. Then the excited state quickly re-emits $\gamma$-rays at the same or a different energy in a decay pattern that is unique for each type of nucleus. The MeV energy scale is high enough to penetrate heavily shielded containers, thus permitting non-destructive, unambiguous isotope identification and quantitative mass determination. }
	\label{fig:nrf}
\end{figure}

Each nucleus with $Z > 2$ is characterized by unique excited states. These states have very narrow widths (rarely more than a fraction of an eV) and can be populated by absorption of photons of the appropriate energy. When such an excited state decays, characteristic $\gamma$-rays are emitted with unique and well-defined energies. The NRF method of isotope identification, shown in Fig.~\ref{fig:nrf}, is based on these observations. The main advantage of this method over other beam-based techniques is that both the excitation and the de-excitation processes proceed via the electromagnetic interaction, which is well understood.  Because of the low angular momentum transferred in photonuclear reactions, photon-induced nuclear excitations are mostly electric dipole (E1), magnetic dipole (M1), and to a lesser extent electric quadruple (E2) transitions from the ground state. If the excited state is unbound to emission of strongly interacting particles (e.g., neutrons, protons, or alpha particles), the photon decay is not strong enough to be observed in competition with particle decay channels. Hence, most of the useful states are positioned below the particle separation energies of around 6 to 8 MeV, and they generally have ground state radiative widths ($\Gamma_\gamma$) of 10 to 100 meV. 

There are two main approaches to identifying nuclei using NRF: (1) detection of scattered characteristic $\gamma$-rays, and (2) resonance absorption or the observation of flux removal from the $\gamma$-ray beam by resonance nuclear scattering.   In the latter method, resonance absorption creates notches in the energy spectrum of the transmitted beam at energies characteristic of the nuclei encountered by the beam.  Bertozzi and Ledoux proposed this method for applications in scanning seagoing cargo containers \cite{Ber05}.  Detection of these notches provides the signature for identifying the isotopes through which the beam has passed.  The application of the notch technique with nearly mono-energetic $\gamma$-ray beams has been studied recently \cite{Pru06}, and a recent measurement performed using the $\gamma$-ray beam at HI$\gamma$S demonstrates that for $\gamma$-ray beams with a small energy spread, such as the 5\% FWHM at HI$\gamma$S, the probability of notch refilling due to small-angle scattering is likely to be negligible for most cargo inspection scenarios \cite{Hag09}.  Beams at a next-generation laser Compton $\gamma$-ray source will enable expansion of the experimental studies of the notch technique reported by Hagmann et al.\cite{Hag09}.  The goals of such work would include: (1) comparison of signal-to-background ratio of the notch technique to that for detection of scattered $\gamma$-rays for high $Z$ nuclei and a variety of cargo configurations; (2) measurements of effective NRF cross sections for selected isotopes; (3) assessments of performance quantities for the notch technique for simple cargo arrangements, in order to benchmark computer models of the setup; and (4) evaluations of the performance of the notch technique implemented with low-resolution high-efficiency $\gamma$-ray detectors, such as LaBr$_3$ scintillators.  The performance quantities include the signal-to-background ratio, the number of signal counts per 10$^9$ incident $\gamma$-rays, and the radiation levels in the proximity of the cargo.

The driving principle of the notch technique is the analysis of the energy distribution of the beam transmitted through the sample material.   The resonance absorption of the $\gamma$-rays by the nuclei in a sample material creates notches in the transmitted beam at the resonance energies.  Because the width of the notch (or absorption peak) in the beam-energy profile is typically less than 1 eV, direct detection of the notch is not possible with current $\gamma$-ray-detector technology, but the natural line width of $\gamma$-ray transitions in nuclei is used to overcome this technical challenge \cite{Ber05}.  The comparison of the beam-flux normalized NRF spectrum on a target measured using the beam transmitted through the cargo sample with that measured with the sample removed gives a measure of the resonance absorption of the cargo material for the nuclei in the downstream target.  The $\gamma$-ray beam flux transmitted through the sample at the energy $E_r$ of a resonance in a nucleus in the sample is given by 
 
 \begin{equation}
\phi = \phi_0\,exp(-\sigma_{eff}\,\rho x).  \\
\label{eqn:notch}
\end{equation} 

In this equation $\phi_0$ is the $\gamma$-ray beam flux incident on the sample at energy $E_r$, $\phi$ is the transmitted beam flux at energy $E_r$, $\sigma_{eff}$ is the effective nuclear resonance absorption cross section, $\rho$ is the density of the nuclei associated with the resonance in the sample, and $x$ is the thickness of the sample.  The quantity $\sigma_{eff}$ includes effects from Doppler broadening caused by the thermal motion of the nuclei in the sample.

\begin{figure}[h]
	\centering
\includegraphics[width=0.6\linewidth] {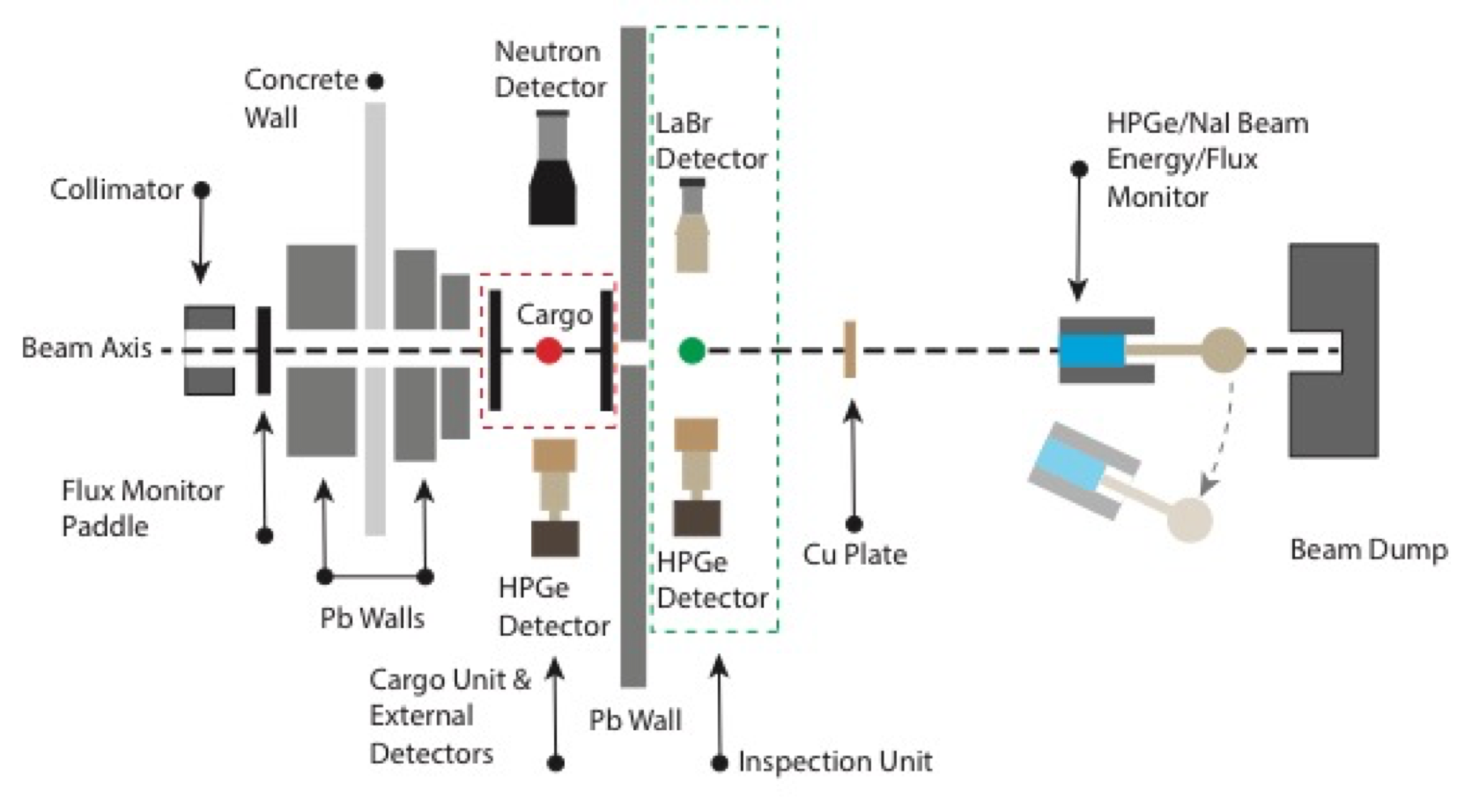}
	\caption{Schematic diagram of the experimental setup for evaluating techniques for remote identification of nuclei using $\gamma$-ray beams.}
	\label{fig:notchsetup}
\end{figure}

A schematic diagram of the experimental setup for the notch-technique evaluation measurements is shown in Fig.~\ref{fig:notchsetup}. 
The time structure of the beam enables the backgrounds uncorrelated with the beam to be measured while collecting beam-induced reaction data and determining the velocity of neutrons emitted from the target via time-of-flight measurements.  
The beam-shaping collimator shields the experimental setup from the radiation produced by the beam using lead and concrete walls indicated in the drawing.  The standard collimator diameter in these measurements will be around 25 mm (1 inch).   These studies will be performed using circularly polarized $\gamma$-ray beams.   
At $\gamma$-ray beam energies below 7 MeV, the collimator material should be lead, and at higher energies the collimator should be made of aluminium to reduce neutron production via the ($\gamma$,$n$) reaction on the collimator.  
The sample materials being studied are placed between the steel plates in the area labeled cargo in the diagram. 
The energy spectrum of the beam transmitted through the sample 
should be analyzed using the NRF measurement setup in the area labeled inspection unit.
 The cargo arrangement will nominally contain targets that are one to several cm thick.  The analysis foils will normally be a few mm thick.  The collimated beam flux before the cargo will be monitored e.g., with a thin plastic scintillator paddle.  
The absolute beam flux can be measured by Compton scattering from a copper plate into an HPGe detector.   The $\gamma$-rays emitted by the cargo sample are detected by HPGe detectors positioned at scattering angles of at least 90 degrees.  At $\gamma$-ray beam energies above the neutron separation energy, the fast neutrons emitted from the sample will be detected.  The energies of the neutrons are determined by time-of-flight analysis.  Both HPGe and LaBr$_3$ detectors are used to detect the $\gamma$-rays emitted by the witness target in the inspection unit.  The LaBr$_3$ detectors offer a possible alternative to HPGe detectors in field application.

Two distinct types of measurements can be performed using this system: (1) determination of $\sigma_{eff}$ for several states in key nuclei such as $^{238}$U, $^{239}$Pu, $^{235}$U, $^{208}$Pb, $^{12}$C, $^{14}$N, $^{16}$O, and $^{24}$Mg; and (2) evaluation of techniques for remote isotope analysis of cargo.  The same nuclei can be studied in both measurements. In addition to determining signal-to-background ratios for the notch measurements (inspection station) and detection of the scattered $\gamma$-rays (cargo setup), the radiation levels inside and just outside the target area should be recorded during measurements, especially when evaluating beam-based interrogation concepts.  In addition,  the facility should have the capability for measuring the energy spectra of the fast neutrons and $\gamma$-rays  at several locations in the target area during data collection under various conditions. 

\begin{figure}[H]
	\centering
\includegraphics[width=0.3\linewidth] {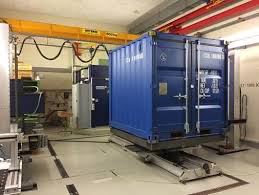}
	\caption{Movable support structure for concept evaluation.  Its movement degrees of freedom allow 3D $\gamma$-ray beam scanning for ``real-world'' scenarios. This figure is adapted from a paper by kolokytha et al. \protect{\cite{kol17Calvin}} of the CEA Saclay in France. }
	\label{fig:support}
\end{figure}

The target area for nuclear and homeland security applications should be equipped with a  support system for imaging and scanning cargo-size containers, e.g., as the system shown in Fig.~\ref{fig:support}, which is used in the European project for developing a system for Automated Comparison of X-ray Images for Cargo Scanning (ACXIS) \cite{kol17Calvin} .  This system should  allow ``real-world'' setups to be moved along multiple axes and rotated to allow beam scans.  The target room should be equipped with detectors and instrumentation with the capabilities for large-area measurements of \begin{itemize}
\item NRF
\item Photofission: prompt and delayed neutrons and $\gamma$-rays
\item Imaging: transmission, backscattering, and NRF with linearly polarized $\gamma$-ray beams  
\end{itemize} 

\renewcommand{\bibname}{References}

\chapter{Nucleon Structure and Low-Energy QCD}
\lhead{}
\chead{}
\rhead{Nuclear Structure and Low-Energy QCD}
%
\label{Ch:QCD}
The nucleon sits roughly halfway on the ladder of distance scales of relevance
for contemporary nuclear physics. Going up the ladder, {\it ab initio}
calculations that use
protons and neutrons and the forces between them as  building blocks predict
the properties of nuclei as large as $^{100}$Sn. Going down the ladder,
lattice simulations that employ the fundamental QCD degrees of freedom---quarks and gluons---are now performed at the physical quark mass, reproduce a variety of hadron properties, and provide an increasingly detailed picture of its internal structure. 

However, much of that internal structure is not resolved inside a typical nucleus under terrestrial conditions. The calculations of nuclei and nuclear matter described in the first paragraph use emergent QCD degrees of freedom: protons, neutrons, pions, and the Delta resonance.
The interactions of these particles are not always well constrained by, e.g., nucleon-nucleon scattering data. Understanding the properties, structure, and interplay of these low-energy QCD degrees of freedom is  crucial to continued progress in our field.

The $\gamma$-ray beam of a next-generation laser Compton source is
ideal for the examination of nucleon structure in this ``low-energy QCD"
domain. Its electromagnetic field couples to the distribution of charge and
magnetization inside the nucleon. At the lower end of the NGLCGS energy range,
$\omega \lesssim 50$ MeV, the photons will not resolve the nucleon's  internal
structure. But as $\omega$ is increased beyond that, different degrees of
freedom become important in determining the photon response of neutrons and
protons. There are two regions of particular note: 1) around $\omega \approx 150$~MeV, the pion-production channel opens; and 2) once $\omega \approx 300$ MeV the photon field is resonant with the lowest excited state of the nucleon, the Delta(1232) resonance.

In this chapter, we discuss two different NGLCGS
experimental programs that probe this physics. In Compton scattering, the
incoming photons of the $\gamma$-ray beam induce oscillations in the nucleon's
internal charge and magnetization distributions. These, in turn, cause the
re-emission of radiation, with a strength that is proportional to the
frequency-dependent coupling between the photon and the internal
oscillation. Compton scattering is thus unique amongst probes of nucleon
structure: it can activate pions and other effective degrees of freedom inside
the nucleon even at energies where these cannot be liberated as physical
particles in the final state.

Indeed, nucleon Compton scattering involves a fascinating interplay of
different QCD mechanisms. The pions' special status as (pseudo-)Goldstone
bosons of QCD means that the nucleon polarizabilities which parameterize the
low-frequency nucleon response to the photon should bear the imprint of this
Goldstone-boson physics. Those pion effects are complemented by those of the
Delta excitation---especially for magnetic properties. Again, since photons
activate virtual excitations, the Delta affects the Compton response even at
energies well below the resonance. 
At higher energies, nucleon Compton scattering is 
sensitive to the still-mysterious scalar sector of QCD and 
to higher-energy nucleon resonances. 

\begin{figure}[H]
	\centering
	\includegraphics[width=0.45\linewidth]{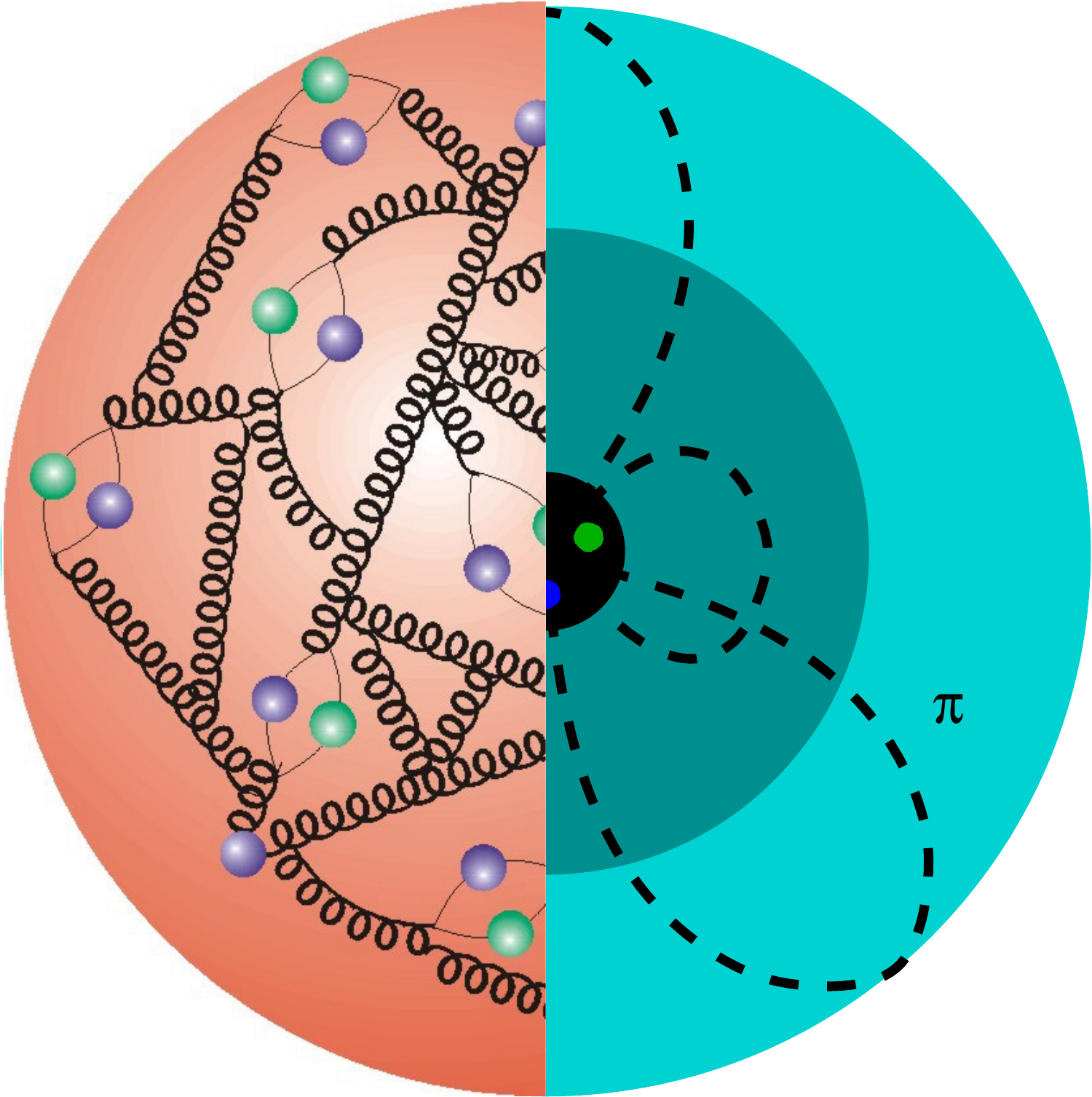}
	\caption{The appropriate degrees of freedom with which to describe
          nucleon structure depend on the distance scale at which it is
          probed: quarks and gluons at high energies (left); or nucleons,
          their pion clouds and nucleonic excitations at low energies
          (right).}
	\label{fig:nucleonuncloaked}
\end{figure}

The impact of virtual degrees of freedom on Compton scattering
 makes it an excellent place to learn how different
hadronic degrees of freedom become active as the photon beam energy is
increased. NGLCGS will be uniquely positioned to be a scale-scanning facility,
examining the entire photon-energy range of relevance for low-energy
QCD. Beyond these energies, where the wavelength decreases further, there is a
transition to the physics examined by high-energy facilities, including a
future EIC, where quarks and gluons become a more efficient way to think about electromagnetic response. But the fundamental QCD degrees of freedom are not efficient to model nuclei, whose physics is dominated by hadron dynamics at distance scales of 1-2 fm, see Fig.~\ref{fig:nucleonuncloaked}. 

Once $\omega \gtrsim 150$ MeV, pions can be produced as physical particles in the final state. We therefore plan that Compton-scattering studies at NGLCGS be complemented by an aggressive program to measure this reaction. While much data in this regime already exists from LEGS, Mainz, and other facilities, important questions remain. In particular, the flux and monochromaticity of the NGLCGS beam will make possible pion-production experiments that can test the limits of isospin symmetry in pion-nucleon and nucleon-nucleon interactions. A precise understanding of isospin violation is becoming a key input for models of nuclei, as facilities like FRIB and RIKEN push ever further into the neutron-rich frontier.

\section{Compton Scattering} 
\lhead{Compton Scattering}
\rhead{Nuclear Structure and Low-Energy QCD}


Compton scattering from hadrons provides unique access to the symmetries and
dynamics of the charges and currents constituting the low-energy degrees of
freedom inside the nucleon~\cite{comptonreview, Hagelstein:2015egb,
  Wissmann:2004, Schumacher:2005an, Report,Pasquini:2018wbl}. This information can be encoded in
the proportionality factors between the incident field and the induced
multipole: the dynamical, energy-dependent polarizabilities of the
nucleon. Intuitively speaking, they measure the response of the low-energy
degrees of freedom in the nucleon in transitions $Xl\to Yl^\prime$ of definite
multipolarity for a real photon with non-zero frequency $\omega$, and
$l^\prime=l\pm\{0;1\}$. After subtracting the effects of a point-like nucleon,
one expands the amplitude as ($X,Y=E,M$; $T_{ij}=\half (\de_iT_j + \de_jT_i)$;
and $T=E,B$):
\begin{equation}
\begin{array}{ll}
\label{polsfromints}
2\pi\!\!\!\!\!&
\big[{\alphae(\omega)}\,\vec{E}^2+
{\betam(\omega)}\,\vec{B}^2 
+{\gammaee(\omega)}\,
\vec{\sigma}\cdot(\vec{E}\times\dot{\vec{E}})
+{\gammamm(\omega)}\,
\vec{\sigma}\cdot(\vec{B}\times\dot{\vec{B}})\\[1ex]&
-2{\gammame(\omega)}\,\sigma^iB^jE_{ij}+
2{\gammaem(\omega)}\,\sigma^iE^jB_{ij} 
+\dots(\mbox{higher multipoles}) \big]
\end{array}
\end{equation} 
For $\omega$ up to about 300 MeV, 
just six dynamical polarizabilities characterize the nucleon response: two
scalar polarizabilities, $\alphae(\omega)$ and $\betam(\omega)$, for electric
and magnetic dipole transitions, and four spin polarizabilities
$\gamma_i(\omega)$ which are addressed in more detail below. The zero
frequency values, $\alphae(\omega=0)$ etc., are often called the
``polarizabilities".

As explained above, Compton scattering has a special role as a scale-scanning
probe of the nucleon's internal degrees of freedom. In the NGLCGS energy
range, the six dynamical polarizability functions above serve as benchmarks of
our understanding of the way in which short-distance QCD dynamics and the
consequences of chiral symmetry breaking---especially as encoded in pionic
excitations---combine to produce nucleon structure. Subtle alterations in this
interplay then produce differences between proton and neutron values, and
precise polarizability numbers (including nucleon differences) are crucial
ingredients in attempts to understand the neutron-proton mass difference and the
extraction of nuclear radii from the Lamb shift in muonic atoms that produced
the ``proton-radius puzzle"~\cite{Gasser:2015dwa, WalkerLoud:2012bg,
  Pohl:2013yb, Birse:2012eb,Pasquini:2018wbl}.

\begin{figure}[!htb]
	\centering
\includegraphics[width=0.55\linewidth]
{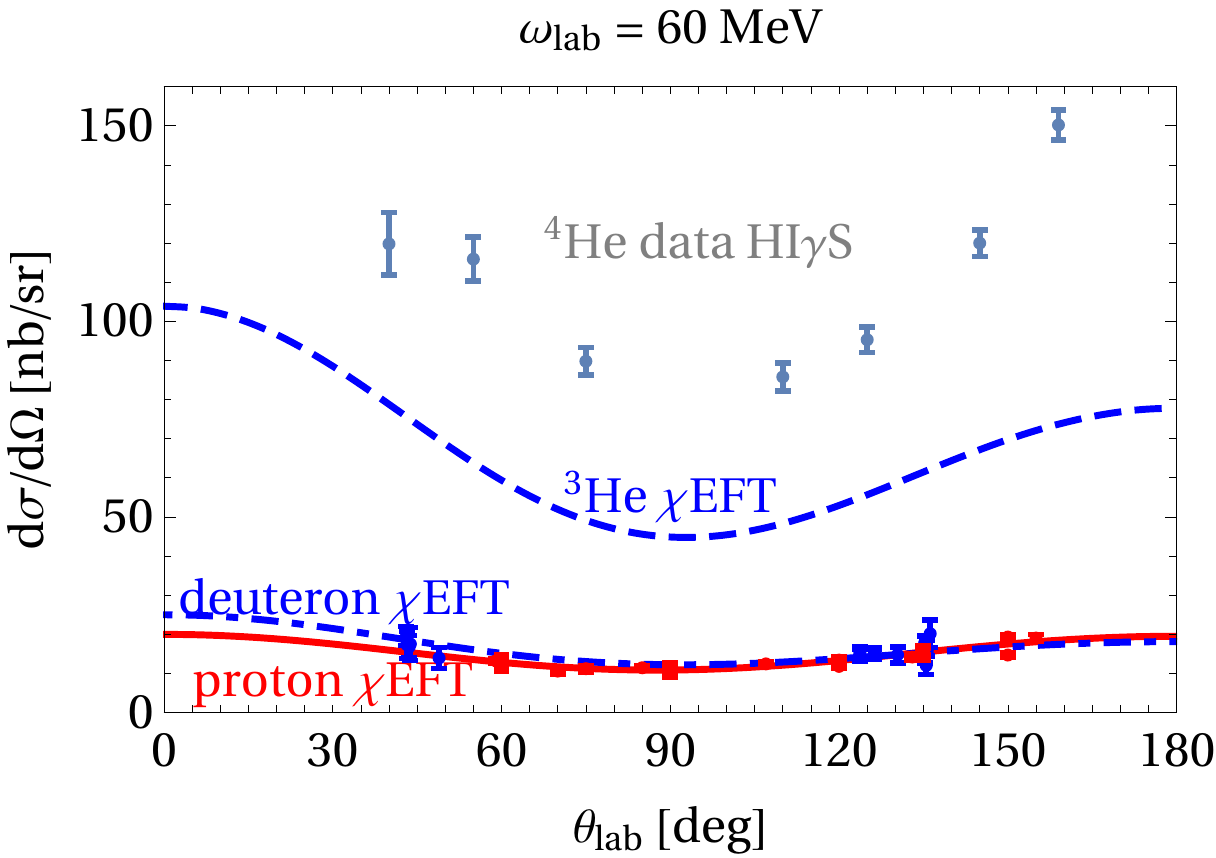}
	\caption{Predicted (curves) and measured (data points) elastic Compton cross sections at $\omega_{\rm lab}=60$ MeV for the different targets discussed in this chapter.}
	\label{fig:crosssectionplot}
\end{figure}

Our goal is to map out the interplay of low-energy QCD mechanisms that is encoded in the dynamical polarizabilities of the proton and neutron.
To this end, we advocate a series of NGLCGS Compton-scattering experiments with
both polarized and unpolarized beams and targets. The targets are the proton,
deuteron, \threeHe and \fourHe, see fig.~\ref{fig:crosssectionplot}. This
means NGLCGS will do challenging experiments on targets where nuclear
effects are absent or clearly under control, while also examining nuclei with
larger Compton cross sections and relying on theory for a reliable extraction
of nucleon-level information. This interplay of experimental signal strength
and calculational complexity across different targets will test the unified picture
of Compton scattering from light nuclei that has been built up over the last twenty years.

As discussed in the 2015 NSAC/DOE and 2017 NuPECC Long-Range
Plans~\cite{LRP15, NAS,nupecc2017}, there has been significant progress in
our understanding of nucleon polarizabilities over the last decade; see also
Sec.~\ref{sec:progress} below. This has occurred because of a synergistic
community effort that employed diverse nuclear-physics tools so that Compton
data on protons and light nuclei now confronts the emerging lattice QCD
computations described in Sec.~\ref{sec:lattice}, with chiral effective-field
theory (\ChiEFT) and dispersion-relation calculations acting as a bridge
between the lattice and the laboratory (see Sec.~\ref{sec:theorybackground}).

\begin{figure}[!htbp]
	\centering
	\includegraphics[width=0.45\linewidth]{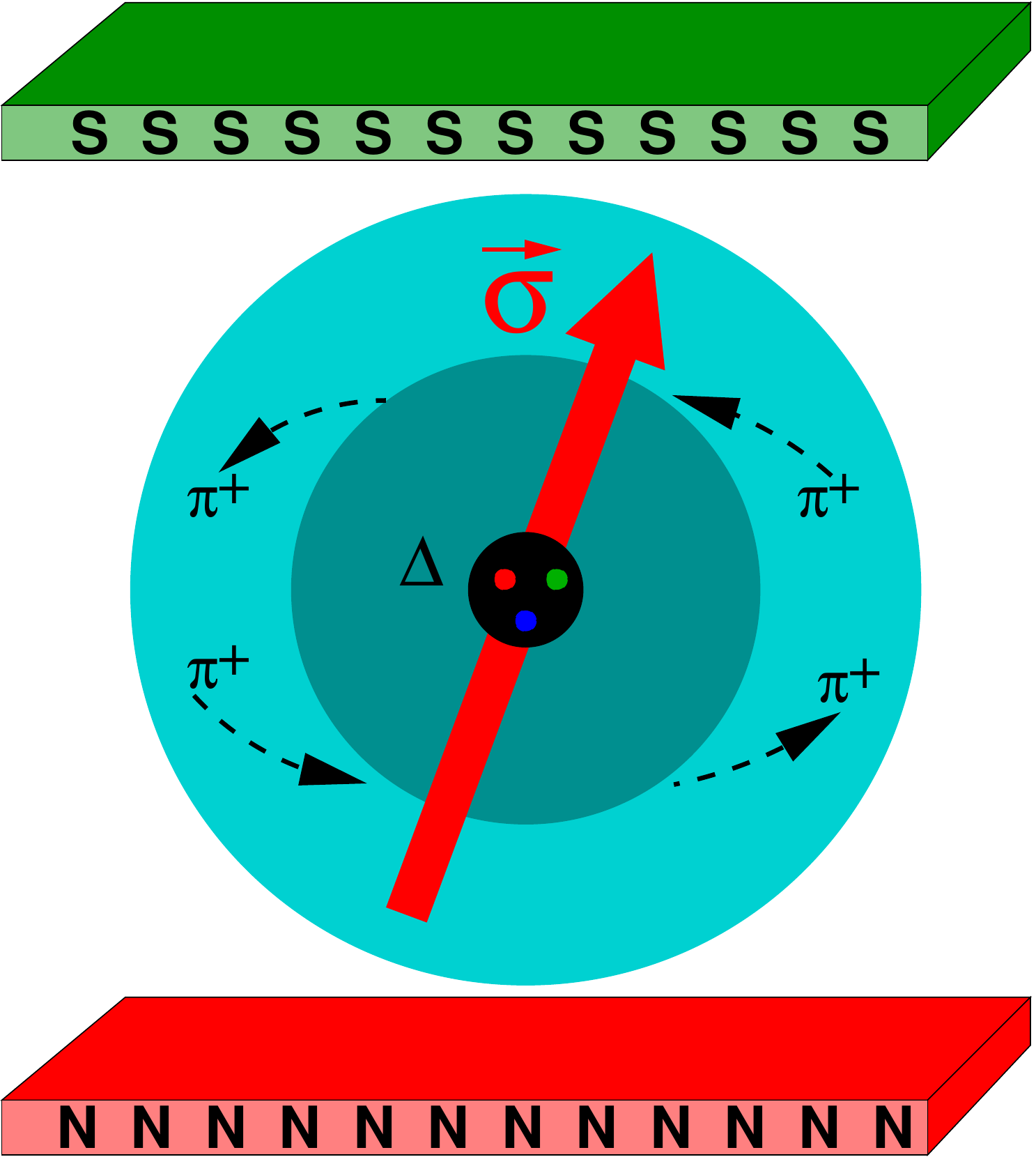}
	\caption{A cartoon that illustrates how the response of the nucleon, including its spin, to an applied
	magnetic field provides information on the nucleon's composition in terms of low-energy degrees of freedom.}
	\label{fig:spincartoon}
\end{figure}

In the next decade, attention will turn to the spin polarizabilities, which so
far are largely unexplored. These parameterize the optical activity of the
nucleon---the stiffness of the nucleon's spin degrees of freedom
(fig.~\ref{fig:spincartoon}).  They therefore complement experiments at
higher-energy facilities that illuminate how the nucleon spin is built up from
quark and gluon spin and orbital angular momentum.  In an intuitive picture,
$\gammaee(\omega)$ and $\gammamm(\omega)$ encode how the electromagnetic field
produced by the nucleon spin leads to bi-refringence, as in the classical
Faraday effect. Until the recent pioneering experiment of
Refs.~\cite{Martel15, Paudyal:2019mee}, only the linear combinations $\gamma_0$ and $\gamma_\pi$
of spin polarizabilities that enter scattering at $0^\circ$ and $180^\circ$
had been determined~\cite{Report,Schumacher:2005an}. For the proton,
even these had substantial uncertainties, with conflicting results from MAMI
and LEGS. For the neutron, even less is known. We will discuss the current
status of
all scalar and spin polarizabilities in Sec.~\ref{sec:progress}.

\subsection{Insights from Lattice QCD}

\label{sec:lattice}

First though, we address how such measurements are related to QCD.  Polarizabilities can be computed directly on the lattice. But since they are second-order in electromagnetic fields and furthermore diverge near the chiral limit, they provide a considerable challenge, and thereby a stringent test of lattice methods. Similar techniques can
then be trusted for QCD input to double-beta decay, dark-matter detection, and
beyond-the-standard-model currents in nucleons and nuclei.

\begin{figure}[!htbp]
	\centering
	\includegraphics[width=0.47\linewidth]{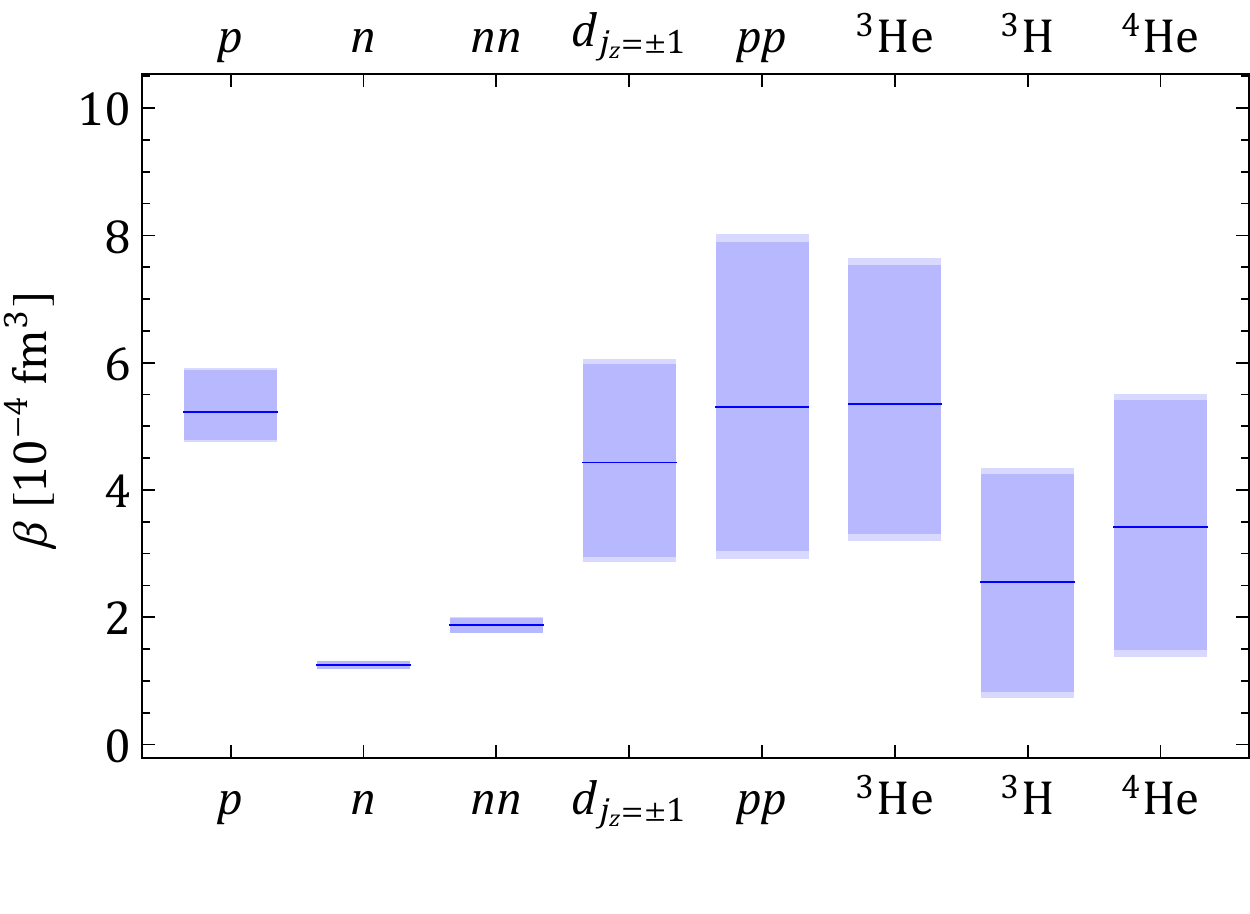}\hspace*{2ex}
	\includegraphics[width=0.47\linewidth]{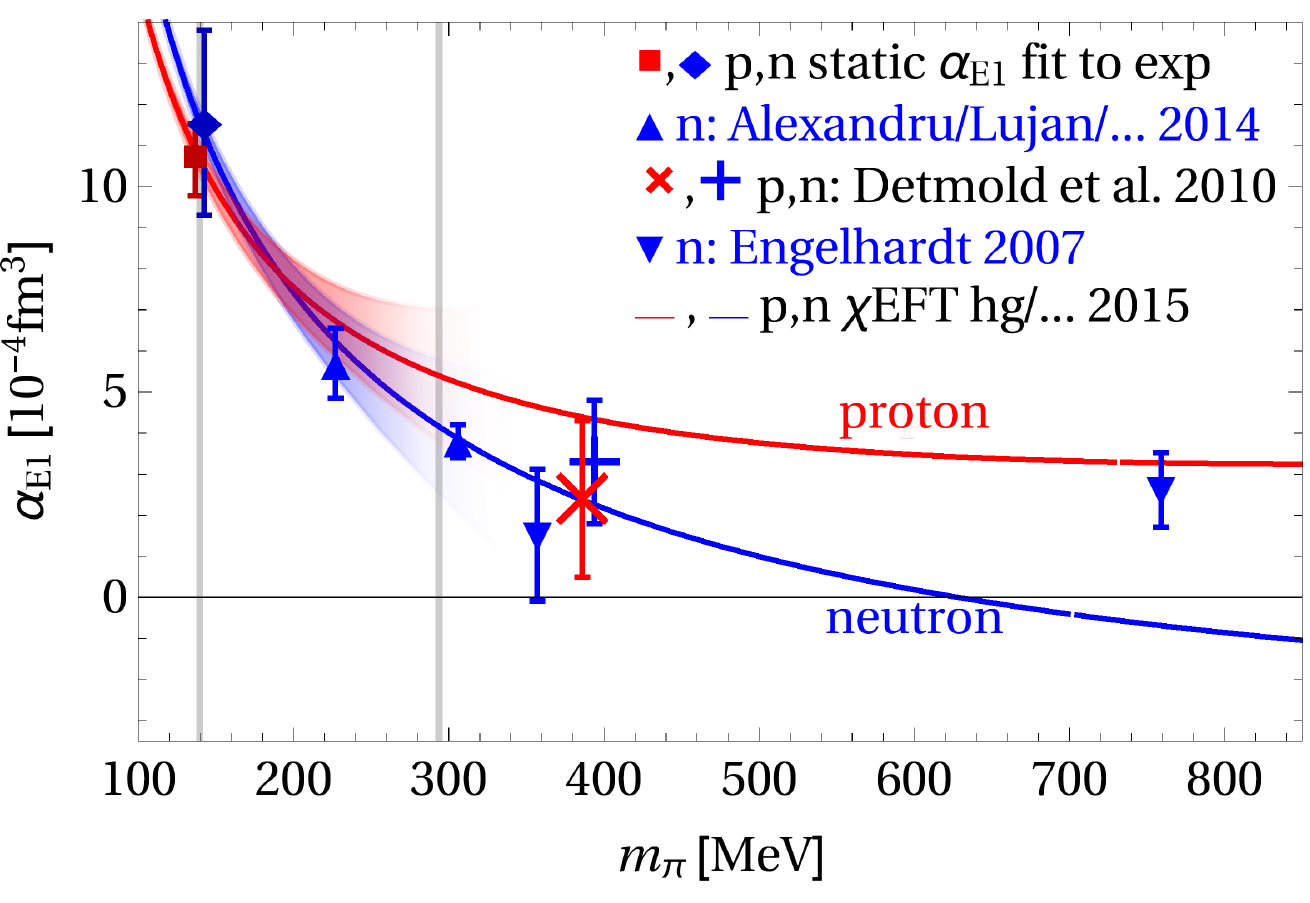}
	\caption{Left: Magnetic polarizabilities of nucleons and light
		nuclei extracted from lattice QCD computations performed at $m_\pi =
		800$ MeV%
		~\cite{Chang:2015qxa}.  Right: Pion-mass dependence of the
		electric polarizabilities in \ChiEFT, with experimental extractions
		at the physical point and available lattice results~\cite{Griesshammer:2015ahu}.}
	\label{fig:beta800}
\end{figure}

The recent success of the background-field technique used for these calculations has led to a strong
effort to refine computations of nucleons and nuclei in uniform
electromagnetic fields~\cite{Lujan:2014kia, Freeman:2014kka, Lujan:2016ffj,
  Primer:2013pva, Beane:2014ora, Beane:2015yha, Chang:2015qxa,
  Detmold:2015daa, Detmold:2019ghl}. Results from the NPLQCD collaboration shown in
fig.~\ref{fig:beta800} even include the first computations of magnetic
properties of light nuclei. There is also a pioneering computation of
$\gammaee$ by Engelhardt~\cite{Engelhardt:2011qq}. Both show that lattice and
experimental uncertainties on these quantities are commensurate, and have the
opportunity to be reduced in parallel.

Currently, the most significant systematic error is that computations are performed at unphysically large pion masses, altering the dynamical response to applied electromagnetic fields. However, recent cross-fertilization between lattice QCD and phenomenology furthers the goal of exposing the role of chiral symmetry in low-energy QCD~\cite{Griesshammer:2015ahu}.
For magnetic polarizabilities, the pion mass provides a dial that alters the
relative weight of diamagnetic and paramagnetic contributions, leading to a
deeper understanding of the underlying effective degrees of freedom. Analysis
of lattice QCD results with \ChiEFT, moreover, provides a consistency check
that is independent of using Compton data. In turn, high-accuracy experiments
in single and few-nucleon systems provide crucial constraints on lattice QCD
computations. 


\subsection{
	Extracting Polarizabilities from Data}

\label{sec:theorybackground}

Equation~\eqref{polsfromints} shows that, near $\omega=0$, the effects of the scalar polarizabilities in cross sections are $\propto \omega^2$, and  $\propto \omega^3$ for spin polarizabilities. 
With increasing energy, polarizability contributions become more prominent. Resonances and particle-production thresholds lead to further enhancements. When inelastic channels are open, the dynamical polarizabilities become complex, and their imaginary parts are directly related to pion photoproduction. That allows, for example, an indirect exploration of some currently ill-determined $n\pi$ photoproduction multipoles~\cite{bernstein-idea}.

Since data are taken at (ideally several) nonzero energies, the ``(static)
polarizabilities'' $\alpha_{E1}(\omega=0)$ etc.\ follow from an extrapolation
of results at nonzero energies and compress the richness of such information
into just a few numbers. A sophisticated and reliable understanding of the
energy dependence of these functions is required, so that the static values
can be extracted from data taken at energies from 70 to 250 MeV. A good
understanding of the energy dependence induced by pion-production and the
$\Delta(1232)$ resonance is therefore mandatory.


Recently, an open letter of theorists with backgrounds in several variants of
dispersion relations and effective field theories summarized the theory status
as follows~\cite{theoryletter}:
\begin{enumerate}
\item Static scalar polarizabilities can be extracted from future data well
  below the pion-production threshold with high theoretical accuracy. 
\item Data around and above the pion production threshold show increased
  sensitivity to the spin polarizabilities and will help to understand and
  resolve some discrepancies between different approaches.
\item All theoretical approaches resort to well-motivated but not fully controlled approximations around and above the $\Delta(1232)$ resonance. Concurrently, sensitivity to the static polarizabilities decreases substantially. Instead, one studies details of the $\Delta(1232)$ resonance, as well as 
the degrees of freedom exchanged between photons and the nucleon in the
t-channel. 
\end{enumerate}
The transition from one regime to another is, of course, gradual rather than
sudden.

Neutron polarizabilities are less well determined than proton ones, see
eq.~\eqref{eq:polsresults}. Since both are equal
at leading order in \ChiEFT, their precise measurement could reveal subtle
differences in the pion dynamics around the neutron and proton.  

Such effects have fundamental implications: Ref.~\cite{Gasser:2015dwa} pointed
out that the subtraction function in their dispersive treatment of the
neutron-proton mass difference can be integrated to yield
$\alphaep - \alphaen$. Accurate measurements of the differences of the
electric polarizabilities will therefore illuminate the interplay of
quark-mass and electromagnetic contributions to $M_\mathrm{n} - M_\mathrm{p}$.


The neutron dynamical polarizabilities can be determined from Compton
scattering on light nuclei.  For photon energies below 100 MeV, elastic
scattering is favored. While the Compton response of the neutron itself is
quite small there, it interferes constructively with large pieces of the
nuclear amplitude, including the proton Thomson term and the sizable
pion-exchange currents of bound systems. The importance of the latter for
Compton scattering in light nuclei means that measurements of the elastic
reaction also provide indirect, non-trivial benchmarks
for the theoretical description of mesonic contributions to nuclear
binding. The different systematics in theory calculations of elastic and
quasi-free Compton scattering on light nuclei make experiments in neutron
quasi-free kinematics an attractive complement to measuring the coherent process.

In each case, \ChiEFT provides the necessary, model-independent
parameterizations of Compton scattering on the various targets described here,
including consistency between single-nucleon and pion-exchange
currents~\cite{comptonreview}. And crucially, it does so in a framework where
the residual theoretical uncertainties can be reliably estimated. A \ChiEFT
that is well-matched to the experiments outlined below exists for the proton
up to about 250 MeV; for the deuteron, up to 120 MeV; and for \threeHe, at 50
to 120 MeV.  Accuracies are 2\% or better up to $\omega\sim\mpi$, and 20\% or
better around the $\Delta$ resonance. For each of these nuclei, comprehensive
studies of the sensitivities of Compton observables with polarised beams
and/or targets, and even for some recoil observables, are available in these
regions~\cite{Babusci:1998ww,Griesshammer:2017txw, Pasquini:2007hf,
  Griesshammer:2013vga, Margaryan:2018opu}.  With present efforts and new
experiments on the horizon, a consistent \ChiEFT description up to 250 MeV for
all light nuclei, including \fourHe, is being pursued vigorously. Results can
be expected within the next few years.



\subsection{Review of Recent Accomplishments}

\label{sec:progress}

\begin{figure}[!htbp]
	\centering
		 \includegraphics[width=0.55\linewidth]
		{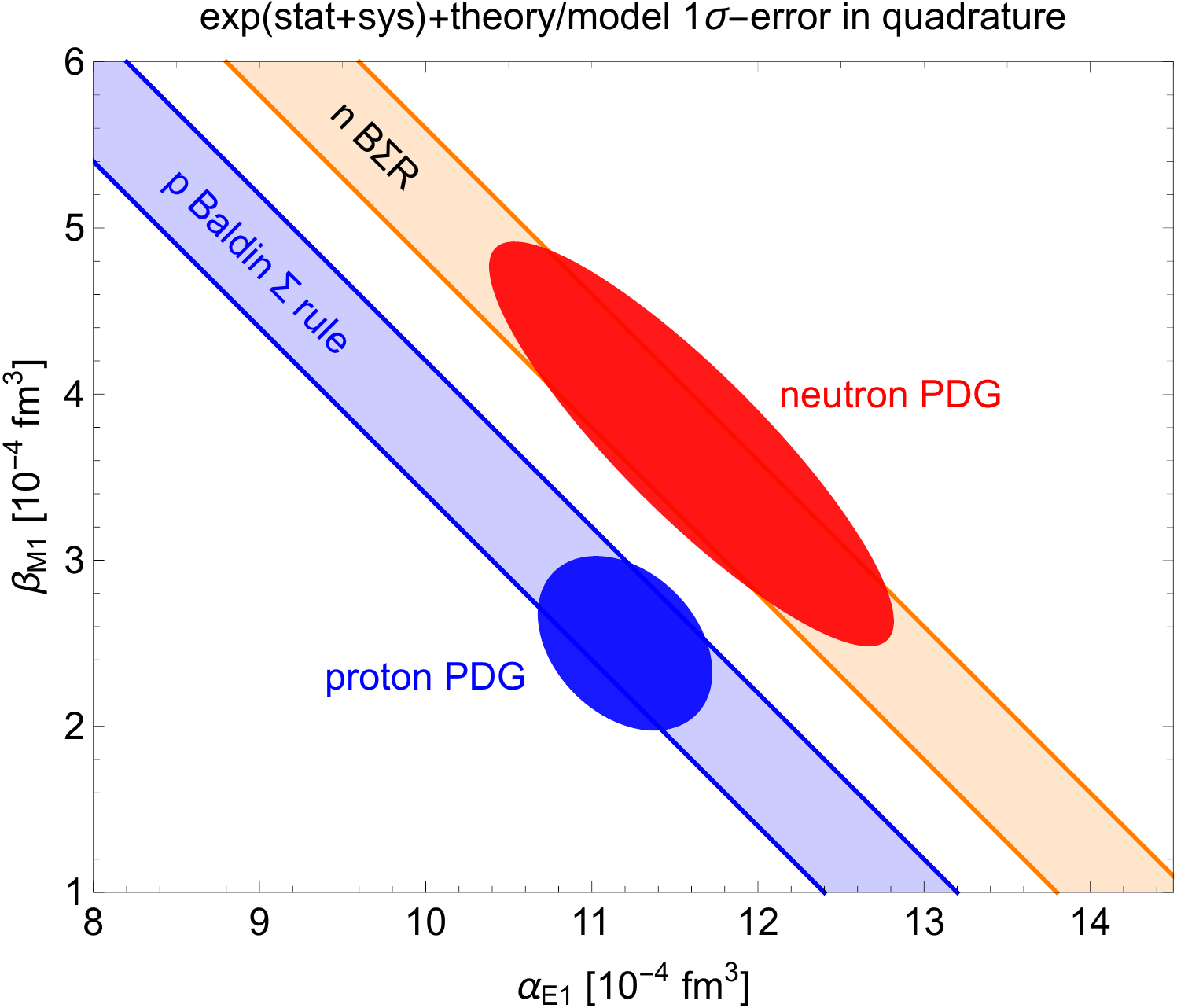}%
		\caption{\label{fig:comptonnucleonPDG} 
		Static scalar polarizabilities
		(blue: proton; red: neutron); most recent PDG listings. 
                Ellipses represent $1\sigma$ errors, with statistical, systematic and theoretical errors
		added in quadrature.} %
\end{figure}

The 2015 U.S.~long range plan states: ``Great progress has been made in
determining the electric and magnetic polarizabilities.'' 
The 2018 Particle Data Group numbers for the proton and neutron scalar polarizabilities (in the canonical units of $10^{-4}$~fm$^3$) are:
\begin{equation}
\begin{split}
\alphaep=11.2 \pm 0.4
 &\;,\; 
\betamp = 2.5 \pm 0.4;
\\[0.5ex]
\alphaen =11.8\pm 1.1&\;,\;
\betamn =3.7 \pm 1.2;
\end{split}
\label{eq:polsresults}
\end{equation}
see fig.~\ref{fig:comptonnucleonPDG}.  The sums, $\alphaep+\betamp=13.8\pm0.4$
and $\alphaen+\betamn=15.2\pm0.4$ are quite accurately known from the Baldin
sum rule and therefore used as constraints in scalar-polarizability
determinations~\cite{Olmos:2001,Levchuk:1999zy, Gryniuk:2015eza}. This implies
that a portion of the error bars are anticorrelated.
Particularly notable in the proton case is that efforts by several theory groups, in different frameworks, have led to compatible results, 
in each of which theory and experimental uncertainties are of similar size~\cite{higherorderpols,Pasquini:2017ehj,Krupina:2017pgr}. For example, the extraction of Ref.~\cite{higherorderpols} quotes a theory error of $\pm 0.3$ while the uncertainty stemming from the $\gamma$-proton data itself is $\pm 0.35$. In spite of this positive development, there is a disconcerting residual dependence of extractions on the  $\gamma$-proton data used~\cite{comptonreview,higherorderpols,Pasquini:2017ehj,Krupina:2017pgr, Pasquini:2019nnx}. This has led to an ongoing discussion about the appropriate criteria for a statistically consistent database. Ultimately, this matter can only be definitively decided by additional high-quality $\gamma$-proton data.
If NGLCGS were to start running tomorrow, this would be a
pressing problem for it to address. But approved experiments at \HIGS and MAMI will
greatly expand the proton Compton database within the next few years;
we expect they will resolve it.

As for the neutron results, a U.S.-led collaboration at MAX-IV added 22 points in 2015,
effectively doubling the deuteron's world dataset and reducing the statistical
error for the neutron values by 30\%~\cite{Myers:2014ace, Myers:2015aba}.
Since \HIGS is now adding precision data, we anticipate that proton and
neutron errors will soon be small enough to enable quantitative exploration of
the extent to which the proton and neutron deformations in electromagnetic
fields differ.

\begin{figure}[!htbp]
	\centering
\vspace*{-.5cm}
  \includegraphics[width=0.85\linewidth]{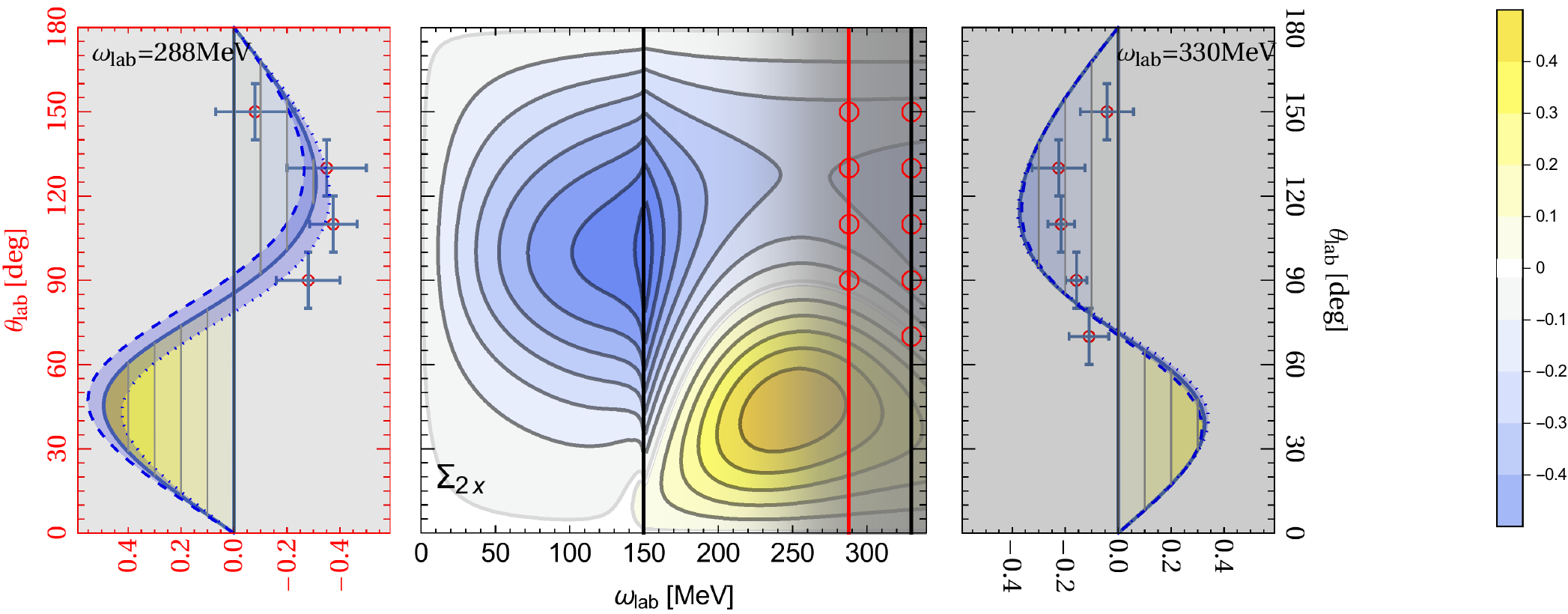}
  \caption{\label{fig:clover} Prediction of the proton double-polarization
    observable $\Sigma_{2x}$ with circularly polarized beam and transverse
    polarized target in \ChiEFT~\cite{Griesshammer:2017txw} compared to the
    data from MAMI~\cite{Martel15, Paudyal:2019mee,Sokhoyan:2016yrc,Martel:2017pln,
        Collicott:2015}.}
\end{figure}

Another U.S.-led collaboration conducted the pioneering measurements of doubly
polarized Compton-scattering observables at MAMI and provided the first
extraction of proton spin polarizabilities~\cite{Martel15, Paudyal:2019mee,
  Sokhoyan:2016yrc,Martel:2017pln, Collicott:2015}. They compare
favorably with predictions from two different implementations of $\chi$EFT
and from DR; see  fig.~\ref{fig:clover}. In units of $10^{-4}\;\fm^4$:
\begin{equation}
\label{eq:spinpols}
\hspace*{-1ex}
\begin{tabular}{lllll}
& {$\gammaee$} &{$\gammamm$}
&{$\gammaem$}&{$\gammame$}\\[0.5ex]
MAMI~\cite{Martel15, Paudyal:2019mee,Sokhoyan:2016yrc,Martel:2017pln, Collicott:2015}&$-3.5\pm1.2$&$3.2\pm0.9$&$-0.7\pm1.2$&
$2.0\pm0.3$\\[0.5ex]
\ChiEFT variant 1~\cite{Griesshammer:2015ahu} 
&$-1.1\pm1.9_\mathrm{th}$&
${2.2\pm0.5_\mathrm{stat}\pm0.6_\mathrm{th}}$ &
$-0.4\pm0.6_\mathrm{th}$& $1.9\pm0.5_\mathrm{th}$\\[0.5ex]
\ChiEFT variant 2~\cite{Lensky:2015awa} 
&$-3.3\pm0.8_\mathrm{th}$&
${2.9\pm1.5_\mathrm{th}}$ &
$+0.2\pm0.2_\mathrm{th}$& $1.1\pm0.3_\mathrm{th}$\\[0.5ex]
DR~\cite{Babusci:1998ww,Drechsel:1999rf,Pasquini:2018wbl} &
 $-[3.4\dots4.5]$&
$[2.7\dots3.0]$& $[-0.1\dots0.3]$& $[1.9\dots2.3]$
\end{tabular}
\end{equation}
But more work is clearly needed to reduce the experimental uncertainties of
$20\%$ to $170\%$ and obtain more accurate information on the low-energy
response of the nucleon's spin degrees of freedom to electromagnetic fields.





\subsection{Overall Goals of a Compton Program}


\label{sec:goals}

The existing unpolarized proton database, while large in size, is noisy, and
the data between 190 MeV and 250 MeV are not of particularly
high quality~\cite{comptonreview}. The deuteron data were markedly improved by
Myers et al.~\cite{Myers:2014ace,Myers:2015aba}, but the uncertainties are not on a par with
the proton ones. Spin polarizabilities are best extracted from data with both
polarized beams and polarized targets. For the proton, pioneering
double-polarization data are now available~\cite{Martel15, Paudyal:2019mee,
  Martel:2017pln,Collicott:2015, Sokhoyan:2016yrc}; but there is no
corresponding data set for a light-nuclear target from which neutron values
can be inferred. Indeed, no data, polarized or unpolarized, have been
published for \threeHe, which seems to be a promising target in this regard.

Further progress in our understanding of the nucleon's electromagnetic
response will thus come on three fronts: 
\begin{itemize}
\item very accurate extractions of neutron scalar polarisabilities that
  will illuminate differences in neutron and proton internal structure;
\item measurements of the largely unexplored neutron spin
  polarizabilities;
\item marked improvement of the proton spin-polarizability values, enabling issues requiring
  precision knowledge of low-energy spin structure to be addressed.
\end{itemize}
We therefore propose a suite of high-accuracy experiments on protons and light
nuclei that, together, will determine the static polarizabilities
and map out the dynamical polarizabilities of the neutron and proton over
a broad energy range. The result will be an understanding of the role that
different QCD effects play in nucleon structure. The experiments are described
in Sec.~\ref{sec:experiments}, where each has its own subsection, with
individual specifications of beam parameters. We close with sections on the
target (Sec.~\ref{sec:targets}) and beam (Sec.~\ref{sec:detectors}) technology
needed to make these experiments a reality.

Several interrelated experiments are necessary
since it is unlikely that a single experiment will determine a particular
static polarizability with the requisite precision. 
The program is designed to provide a coherent data set which will dramatically improve
extractions and enable unprecedented checks of both experimental systematics
and theoretical bias.  For all targets, high-quality data with carefully
formulated correlated and point-to-point systematic errors are needed.
 
The scalar polarizabilities are best extracted from both unpolarized and polarized experiments where the effect of spin polarizabilities is small, namely at photon energies below 100 MeV. 
Proton targets provide a direct avenue for proton polarizabilities. The
deuteron and \fourHe, as isoscalar targets, are sensitive to the average of
proton and neutron polarizabilities, $\alphaep+\alphaen$ etc. In \threeHe, the
sensitivity is to $2\alphaep+\alphaen$ etc. Experiments on different light
nuclei provide important checks of theory systematics, and in particular of
the unified \ChiEFT description of meson-exchange currents in these
nuclei. These binding effects must be subtracted in a model-independent theory
approach.


The spin polarizabilities can be extracted at energies around or above 120 MeV
by combining 
precise data and theory with a carefully selected
set of double-polarized observables whose sensitivities are dominated by only
one or two proton and neutron spin polarizabilities measured over a range of
energies; see e.g.~Refs.~\cite{Pasquini:2007hf,Griesshammer:2017txw}.  Again, proton
polarizabilities are directly accessible with proton targets, but the
isoscalar values from deuteron targets provide valuable cross checks. Since
polarized \threeHe is an effective polarized-neutron target, neutron values
can also be determined without knowing their proton
counterparts~\cite{Margaryan:2018opu}.


Inelastic Compton scattering on the quasi-elastic ridge is dominated by the impulse approximation. Measurements of the neutron-photon interaction in quasi-free kinematics therefore provide another avenue to extract neutron polarizabilities. High-quality theory is needed to control the corrections to the quasi-free approximation (for first steps see ref.~\cite{DemissieThesis}).

Lastly, we point out that a series of measurements over a wide kinematic range
is not only necessary for high-accuracy determinations of scalar and spin
polarizabilities. It also allows reconstruction of the functions
$\alphae(\omega)$, $\betam(\omega)$ and $\gamma_i(\omega)$, yielding the
energy dependence of the dynamical polarizabilities, see recent progress in
Refs.~\cite{Pasquini:2017ehj,Krupina:2017pgr}. This provides
stringent tests of our understanding of different low-energy QCD mechanisms
and of the way they are manifest in Compton scattering.


\subsection{Experiments}

\label{sec:experiments}

\subsubsection{Proton Spin Polarizabilities} 
\begin{figure}[!htbp]
	\centering
		\includegraphics[width=0.49\linewidth]{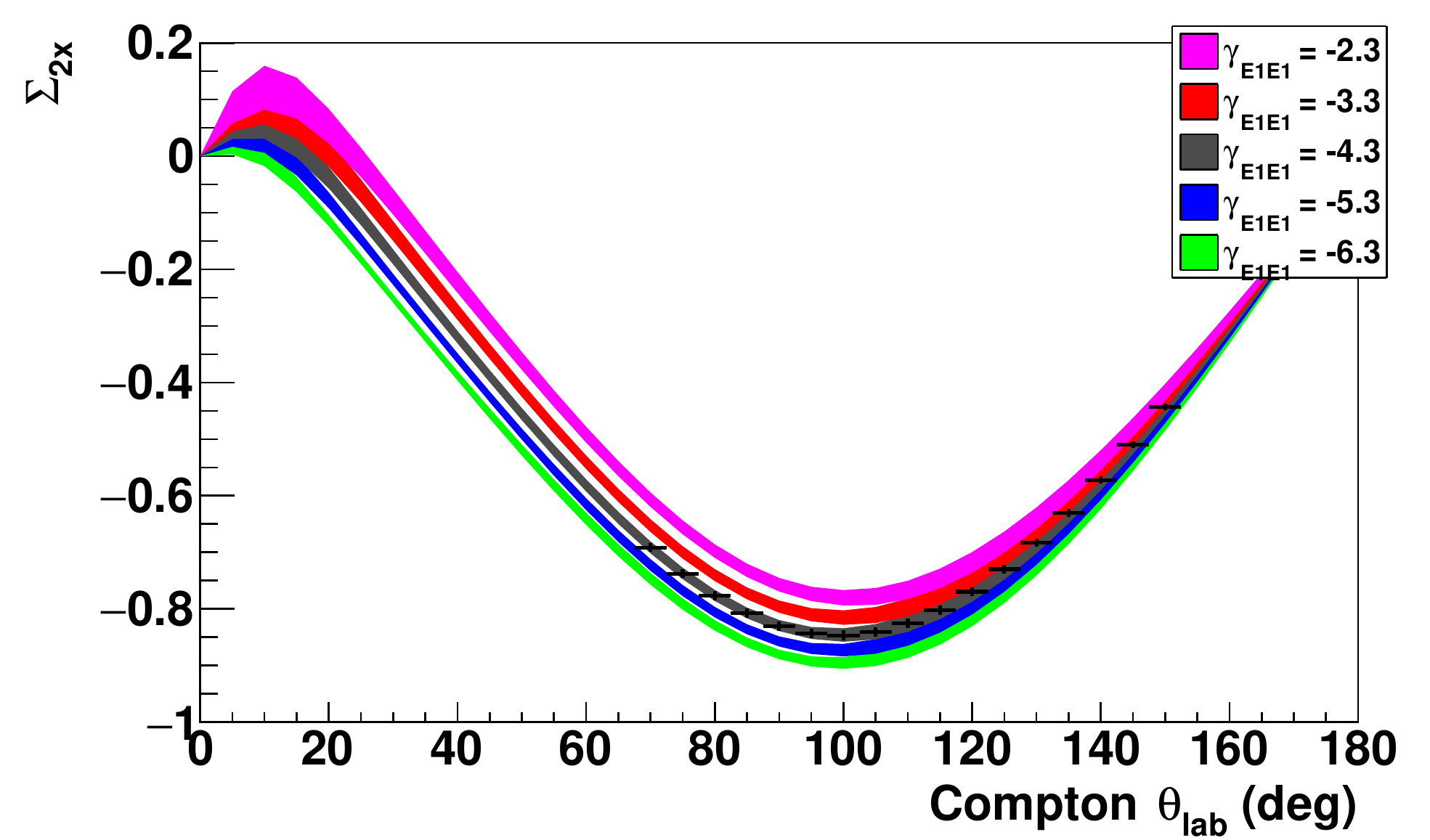}
		\includegraphics[width=0.49\linewidth]{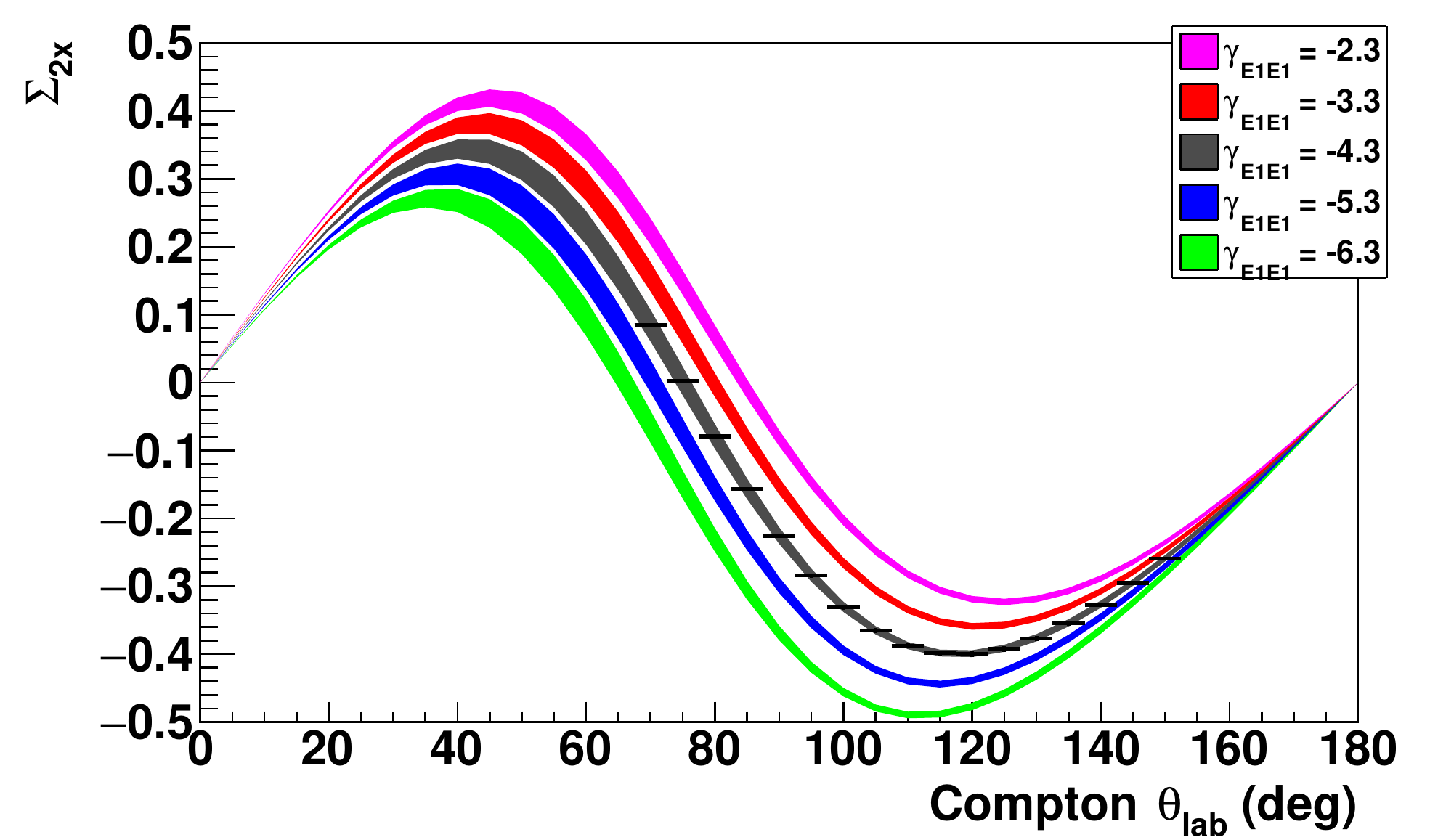}
	\caption{Projected error bars for Compton scattering asymmetries with
		a transversely polarized proton target at 150 MeV (left) and 290 MeV
		(right). The colored bands show the
		predicted asymmetries~\cite{Pasquini:2007hf} with $\gammamm, \gammaem$, and
		$\gammame$ held fixed, and $\gammaee$ varied as shown in
		the figure legend.}
	\label{fig:sigma2x}
\end{figure}

\begin{figure}[!htbp]
	\centering
		\includegraphics[width=0.49\linewidth]{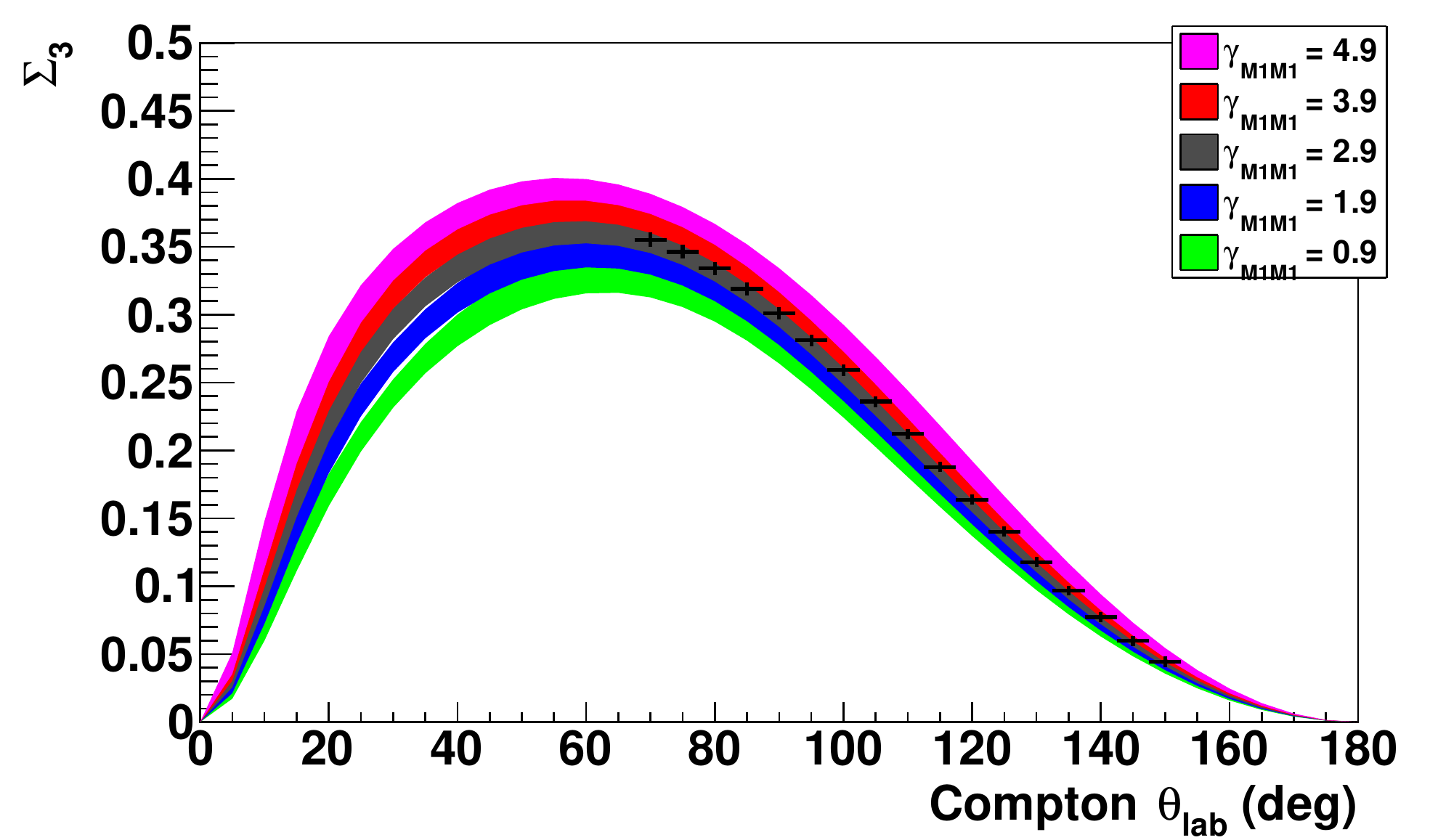}
		\includegraphics[width=0.49\linewidth]{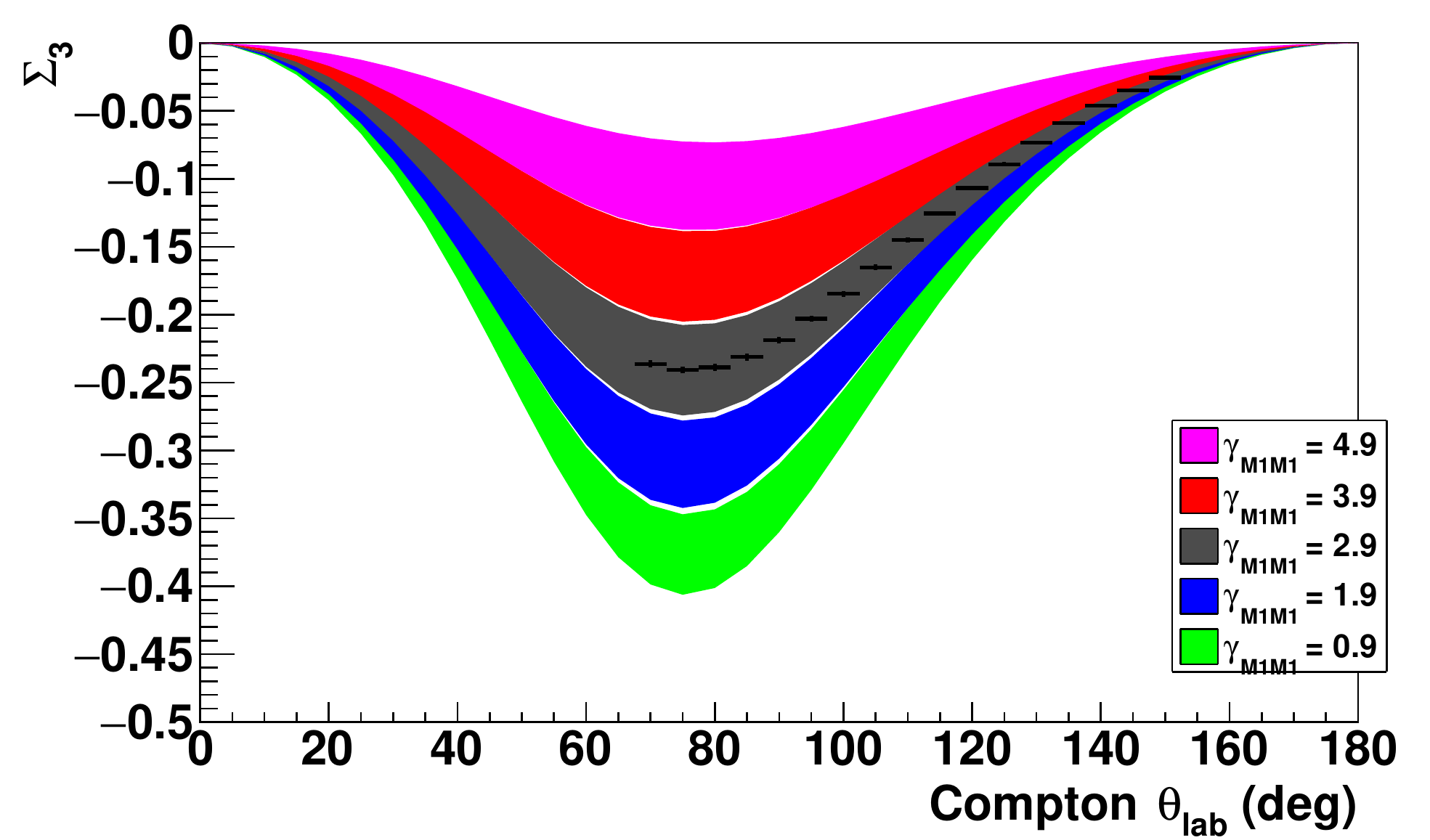}
	\caption{Projected error bars for Compton scattering asymmetries with
		an unpolarized proton target and linearly polarized photons at 150
		MeV (left) and 290 MeV (right). The colored bands show the
		predicted asymmetries~\cite{Pasquini:2007hf} with $\gammaee, \gammaem$, and
		$\gammame$ held fixed, and $\gammamm$ varied as shown in
		the figure legend.}
	\label{fig:sigma3}
\end{figure}

A monochromatic, high flux photon source with $10^9$ $\gamma$/s, with linear
and circular polarizations of approximately 100\%, can provide the definitive
measurement of the proton spin polarizabilities.  Figure \ref{fig:sigma2x}
shows representative asymmetry data points for the $\Sigma_{2x}$ asymmetry
using a transversely polarized target and circularly polarized photons with
200 hours of running at both 150 MeV and 290 MeV. This assumes a frozen-spin
target similar to the Mainz design and a $4\pi$ $\gamma$-ray detector similar
to the Mainz Crystal Ball. The beam conditions are shown in Table
\ref{tab:proton_spin_pol_beam}.  Figure \ref{fig:sigma3} shows representative
asymmetry data points for the $\Sigma_3$ asymmetry assuming an unpolarized
target and linearly polarized photons.  A conservative estimate is that
statistical uncertainties in the polarizabilities can be reduced by a factor
of ten or more relative to those in the Mainz results. 

Comprehensive discussions of theory predictions for the dependence of all
double-polarized observables on the spin polarizabilities are provided in
Refs.~\cite{Babusci:1998ww,Griesshammer:2017txw, Pasquini:2007hf}. To test
the theory dependence of the analysis
 it will also be important to take data at lower energies, around
150 MeV.  At incident energies below approximately 250 MeV, the recoil proton
is undetectable because it does not escape the frozen-spin target
cryostat. Without that, it is problematic to separate the Compton signal from
backgrounds caused by coherent and quasi-free Compton scattering on nuclear
species in the polarized target and from the target cell wall.  One way around
this problem is to utilize a polarizable polystyrene scintillator as the
target, where the scintillation light produced in the target is used to tag a
Compton scattering event on the proton. A polarizable active target has been
developed at Mainz and shown to function well.

\begin{table}[h]
	\caption{Suggested beam parameters for proton spin polarizability.}
	\centering
	\begin{tabular}{c|c}
		\hline  Parameter & Value  \\ 
		\hline  Energy range & 150 to 300 MeV \\ 
		\hline  Flux on target & $10^9 \gamma/s$ \\
		\hline  Relative energy resolution & 2\% FWHM\\ 
		\hline  Polarization& Linear and circular, $>$95\%  \\ 
		\hline  Beam diameter & 5 mm on target \\ 
		\hline  Time structure & highest frequency available, pulse width $<$1 ns \\ 
		\hline 
	\end{tabular} 
	\label{tab:proton_spin_pol_beam}
\end{table}


\subsubsection{Neutron Scalar Polarizabilities from Elastic Compton Scattering on Deuterium}
\HIGS has recently acquired Compton-scattering data on deuterium at 65 and 85~MeV.  Extending the current \HIGS $\gamma d$ data set to energies above 100~MeV and using a significantly higher beam intensity will greatly improve the extracted polarizabilities, since the sensitivity increases at higher energies.

\begin{figure}[!htb]
	\centering
	\includegraphics[width=0.7\linewidth]
	{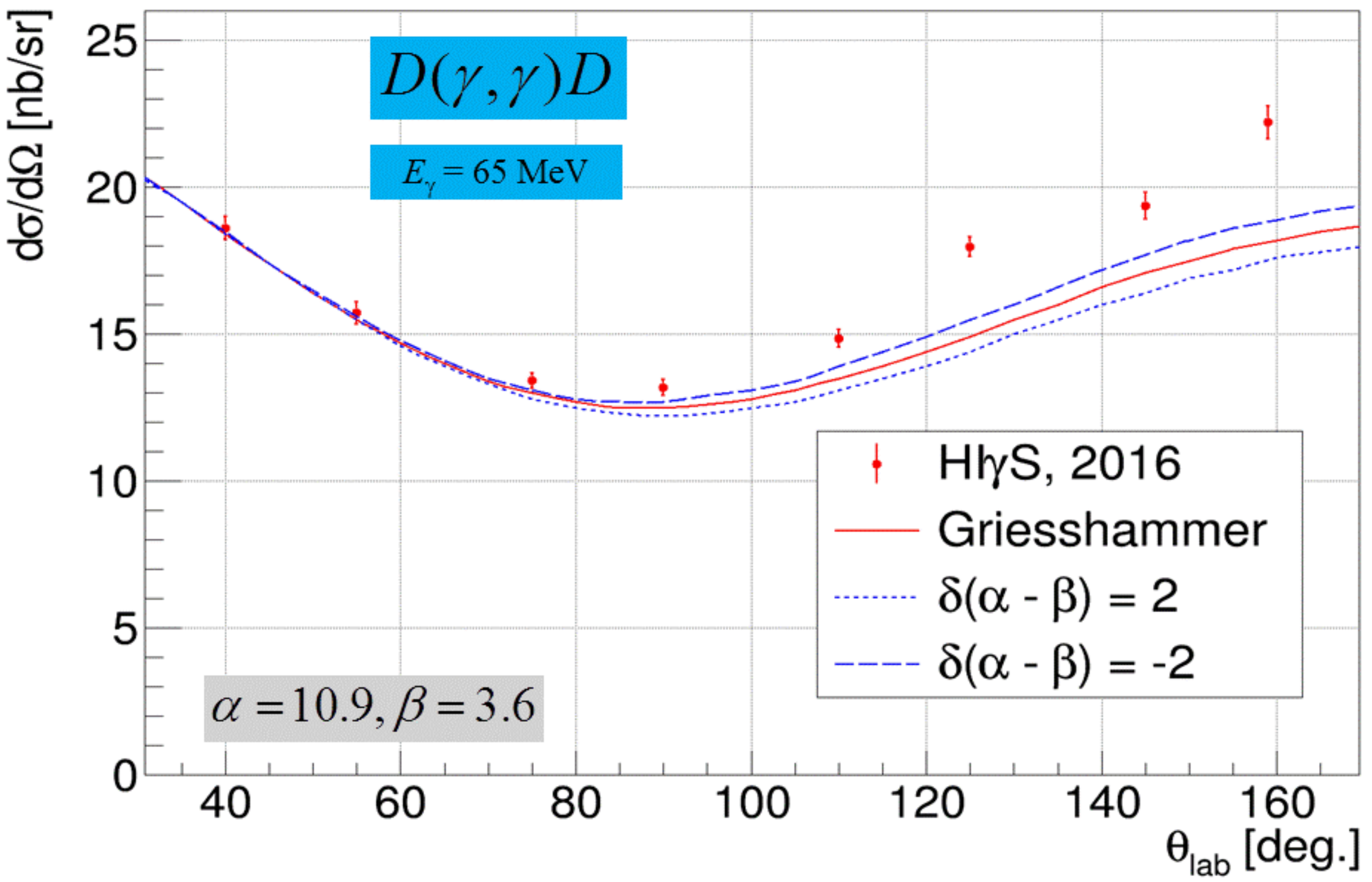}
	\caption{Differential cross section for elastic Compton scattering on
		deuterium at 65~MeV as a function of laboratory angle.  The curves
		are \ChiEFT calculations that have been varied by $\pm2$ in the
		polarizability difference
                $\alphaes-\betams$~\cite{comptonreview, Griesshammer:2013vga}. The agreement
		between data and theory is good at forward angles, but diverges at
		backward angles due to inelastic contributions.}
	\label{fig:deuteronelastic}
\end{figure}

A preliminary plot of the angular distribution of the cross section for $^2$H$(\gamma,\gamma)^2$H is shown in fig.~\ref{fig:deuteronelastic}.  The inelastic contribution that is observed at backward angles, unresolved in the \HIGS data, is the largest systematic uncertainty in the present experiment.  Part of this resolution limitation is due to the NaI detectors themselves, but part stems from poor incident beam resolution: resolution was sacrificed
in order to gain flux.
The higher intensity at the NGLCGS would permit much tighter collimation, thus enabling a vast improvement in beam energy resolution without an unacceptable reduction in beam intensity.

Nevertheless, the statistical error bars from \HIGS shown in fig.~\ref{fig:deuteronelastic} are much lower than any previous experiment.  By operating at a beam energy of 120~MeV or higher, while staying below the pion production threshold, and using several very large, high-resolution NaI detectors with FWHM values of less than 2\% (as in the MAX-Lab experiment of Refs.~\cite{Myers:2014ace,Myers:2015aba}) the new measurements would not only be more sensitive to neutron scalar polarizabilities, but would also provide much better control of the uncertainties associated with the inelastic channel.


\begin{table}[h]
	\caption{Suggested beam parameters for neutron scalar polarizabilities from deuteron.}
	\centering
	\begin{tabular}{c|c}
		\hline  Energy& 120 MeV \\ 
		\hline  Flux &$5\times 10^8$  at 1.5\% FWHM\\ 
		\hline  Polarization& Circular  \\ 
		\hline  Diameter& $<$20 mm on target \\ 
		\hline  Time Structure& TBD \\ 
		\hline 
	\end{tabular} 
\end{table}

\subsubsection{Neutron Spin Polarizabilities from Elastic Compton Scattering on Polarized \threeHe}

Polarized \threeHe approximates a polarized neutron target.
This statement can be quantified using ab initio wave functions
calculated with \ChiEFT potentials, see Ref.~\cite{Margaryan:2018opu}. Such
calculations also incorporate the meson-exchange currents that are needed for
consistency with the potential. As \threeHe is doubly charged and contains more nucleon pairs, 
interference of polarizability effects with proton Thomson terms and meson-exchange currents is larger than for the deuteron or proton,
leading to some enhanced polarizability signals for this target. 

\begin{figure}[!htb]
	\centering
	\includegraphics[width=0.6\linewidth]{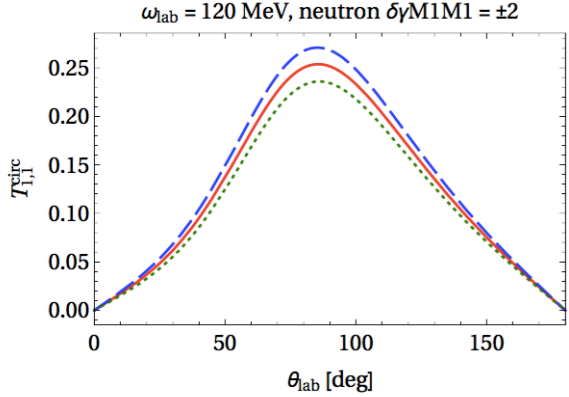}
	\caption{Predicted beam asymmetry for Compton scattering from a
		transversely polarized \threeHe target at a photon laboratory energy
		of 120 MeV. The effect of varying $\gammammn$ by 2 units is
		shown~\cite{Margaryan:2018opu}.}
	\label{fig:sigma2x3He}
\end{figure}

The calculations depicted in fig.~\ref{fig:sigma2x3He} suggest that the photon
beam asymmetry with a transversely polarized target, $\Sigma_{2x}$, is the
cleanest observable to determine a neutron spin
polarizability. It is affected by $\gammammn$ much more
strongly than it is by the scalar polarizabilities, and it is unaffected by
any proton spin polarizabilities. Other neutron spin polarizabilities affect
this observable, but with a different angular dependence. Measuring this
observable over a range of angles and energies will be crucial for checking
the \ChiEFT theory of Compton scattering from \threeHe and ensuring that
``contamination'' of the polarization observable from scalar polarizabilities
is understood and under control.

At 120 MeV, a similar sensitivity to neutron polarizabilities is predicted in the distribution of $\Sigma_{2z}$, as measured with a longitudinally polarized target. In this case, a linear combination of four spin polarizabilities could be determined from the measurements.


A high-density polarized \threeHe target has been built and tested at \HIGS. Given the demonstrated areal density of \threeHe, an experiment using the current HINDA detector array would collect on average ten elastic events per hour at each angle, assuming a beam intensity of  $5 \times 10^8\ \gamma$/s. With a target polarization of 50\%, statistical uncertainties of order 0.02 on $\Sigma_{2x}$ and $\Sigma_{2z}$ could be achieved in moderate running times. Operation at such a high beam intensity would require continued development of an effective shielding configuration against backgrounds from the target windows. Additional development work is needed to improve the capability of the HINDA array to discriminate cosmic-ray events, which contribute at the $\gamma$-ray energies of interest in this experiment.


\begin{table}[h!]
	\caption{Suggested beam parameters for neutron spin polarizabilities from polarized $^3$He.}
	\centering
	\begin{tabular}{c|c}
		\hline  Parameter & Value  \\ 
		\hline  Energy range & 100 to 140 MeV \\ 
		\hline  Flux on target & $5 \times 10^8\ \gamma/s$ \\
		\hline  Relative energy resolution & 2\% FWHM\\ 
		\hline  Polarization& Circular, polarization $>$ 95\%  \\ 
		\hline  Beam diameter & 5 mm on target \\ 
		\hline  Time structure & 20 MHz, pulse width $<$ 1 ns \\ 
		\hline 
	\end{tabular} 
	\label{tab:3He_spin_pol_beam}
\end{table}

\subsubsection{Neutron Scalar Polarizabilities from Elastic Compton Scattering on \fourHe}

\fourHe has a larger coherent Compton cross sections than any other nucleus discussed here, see fig.~\ref{fig:crosssectionplot} for the recent \HIGS data on this reaction~\cite{Sikora:2017rfk}. From a theoretical perspective, an accurate description of cross sections for ${}^4$He, which is more tightly bound than ${}^2$H or ${}^3$He, as a function of energy and angle, will challenge the $\chi$EFT description of Compton scattering on light nuclei.  Thus, although measurements just below pion threshold hold the most promise for extractions of the neutron polarizabilities, ancillary measurements at lower energies are needed to check that the theory used for polarizability extractions is sufficiently accurate.

\begin{figure}[!htb]
	\centering
	\includegraphics[width=0.8\linewidth]{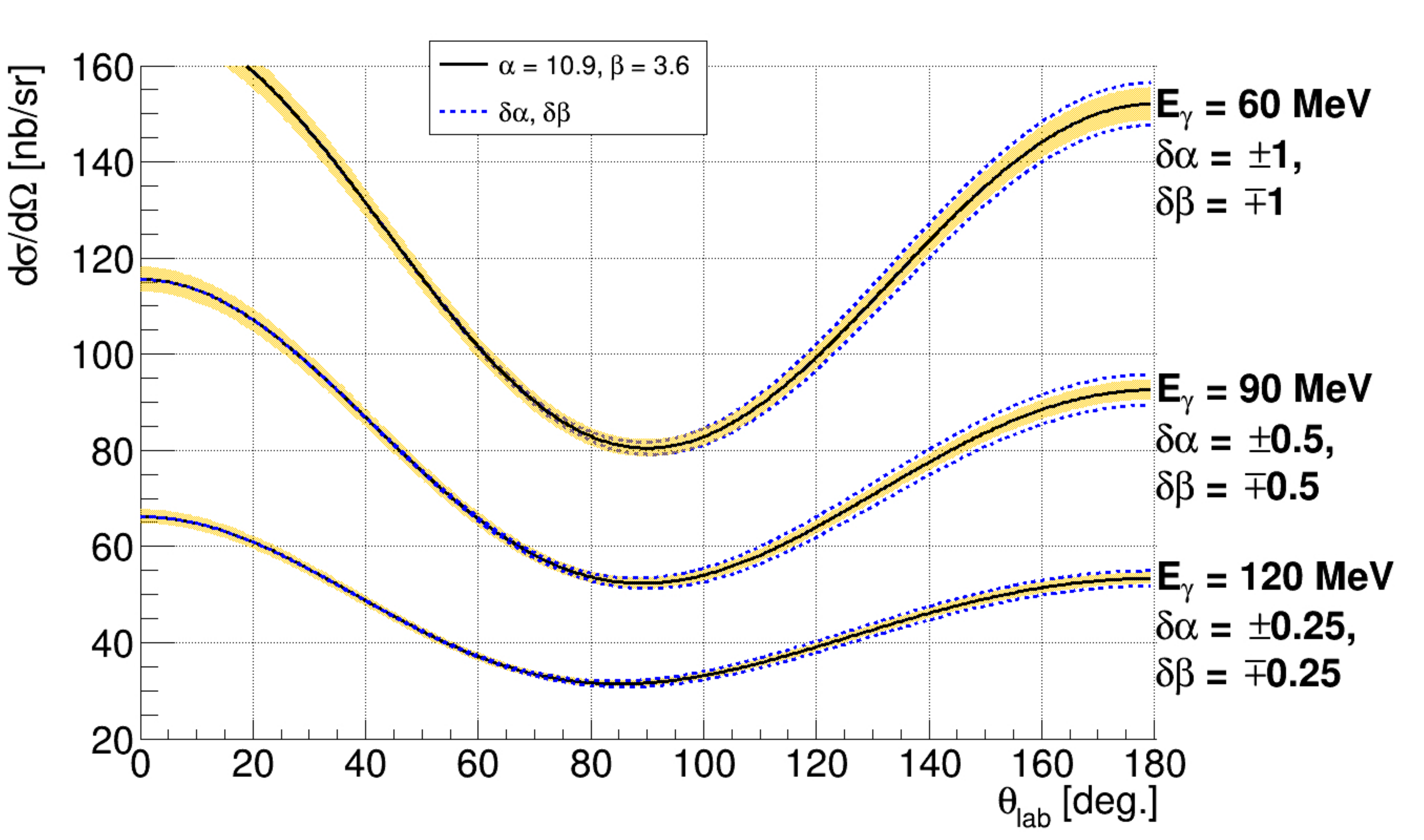}
	\caption{Predictions for the differential cross section for Compton scattering from \fourHe at photon laboratory energies of 60, 90, and 120 MeV in a phenomenological model~\cite{Wright:1985}. The effect of varying the isoscalar $\alpha_{E1}$ and $\beta_{M1}$ while keeping their sum fixed is shown by the blue dotted lines. A combined statistical and point-to-point systematic uncertainty of 2.2\% is indicated by the shaded bands.}
	\label{fig:4Heelastic}
\end{figure}

\begin{table}[h!]
	\caption{Suggested beam parameters for neutron scalar polarizabilities from $^4$He.}
	\centering
	\begin{tabular}{c|c}
		\hline  Parameter & Value  \\ 
		\hline  Energy range & 60 to 120 MeV \\ 
		\hline  Flux on target & $10^8$ to $10^9 \gamma/s$ \\
		\hline  Relative energy resolution & 3\% FWHM\\ 
		\hline  Polarization& None  \\ 
		\hline  Beam diameter & 5 mm on target \\ 
		\hline  Time structure & highest frequency available, pulse width $<$ 1 ns \\ 
		\hline 
	\end{tabular} 

	\label{tab:4He_unpol_beam}
\end{table}

The higher flux and higher energy of the new facility mean that
$\gamma$\fourHe elastic scattering could yield neutron scalar polarizabilities
of unprecedented precision. The prospects are illustrated in
fig.~\ref{fig:4Heelastic}, where the angular distributions are plotted for 60,
90 and 120 MeV.  We project that at 120 MeV the beam parameters of
Table~\ref{tab:4He_unpol_beam}, together with existing \fourHe target
technology (see Sec.~\ref{sec:unpoltargets} below), and the present HINDA
detector array, can produce a cross-section measurement with 1\% statistical
error in just 150 hours. This precision, combined with the estimated
point-to-point systematic uncertainty (2\% in solid angle) is depicted by the
shaded bands in fig.~\ref{fig:4Heelastic}. With the statistical uncertainties
at such a reduced level, we expect that systematics (including an estimated 1\% for target density and 2\% for flux) and
$\chi$EFT truncation uncertainty will ultimately limit the accuracy attained for $\alphaen$ and $\betamn$. Even so, this represents a significant advance in our knowledge of $\alphaen$ and $\betamn$. 

In addition to the cross-section advantage provided by a \fourHe target (and
also the obvious fact that the target is non-combustible), the energy
resolution of the beam and detector become significantly less critical in this
case, compared to deuterium.  Since the first excited state of \fourHe is near
20 MeV, it is possible to obtain very clean, purely elastic $\gamma$-ray
spectra with a \fourHe target.  It is no longer necessary to impose
tight beam collimation to achieve improved beam energy resolution, and greater
photon fluxes on target are possible.

A measurement of coherent Compton scattering from ${}^3$He at similar angles
and energies will provide complementary information, since it is sensitive to
a different combination of neutron and proton polarizabilities. While \threeHe
has a lower breakup threshold than \fourHe and does not offer quite the same
cross-section advantage, extractions of $\alphaen-\betamn$ from
$\gamma$\threeHe scattering will, nevertheless, be a valuable cross-check that
the theory used for polarizability extractions has nuclear effects under
control.
An experiment to measure the $\gamma$-\threeHe elastic differential cross section is approved at \HIGS and will run in 2019 at energies of 100 MeV and (later) 120 MeV~\cite{3HeHIGS}.  The higher flux and extended energy range of the NGLCGS facility could provide enhanced sensitivity of this kind of \threeHe measurement to neutron polarizabilities.

\subsubsection{Quasi-Free Neutron Compton Scattering on Light Nuclei}
Much of the experimental information we have on quasi-free (QF) Compton scattering on light systems comes from the Mainz experiment of Kossert {\it et al.} with an unpolarized photon beam and a deuterium target \cite{Kossert:2002jc}. The experiment used tagged bremsstrahlung at energies of $E_{\gamma}=200$ to 400~MeV and the large NaI(Tl) spectrometer ``CATS'' to detect the backscattered photon at 136 deg.  In addition, the ``SENECA'' array of liquid scintillators was used to detect the recoiling neutron (or proton).  The cross section peaks at around 180 nb/sr at the $\Delta$ resonance.  It rises from about 40 nb/sr to around 150 nb/sr as $E_{\gamma}$ increases from 250 to 300~MeV, so there is definite benefit if one could run at around 300 MeV. A quasi-free neutron Compton experiment at these energies with a polarized beam and a polarized \threeHe target would provide information on neutron dynamical spin polarizabilities in this energy regime. This would provide important constraints on these dynamical neutron spin-structure functions in this regime where the $\Delta(1232)$ and pion-cloud mechanisms both affect pion-production multipoles---constraints which would facilitate enhanced precision for the static values of the polarizabilities. 

\begin{figure}[!htb]
	\centering
	\includegraphics[width=0.9\linewidth]{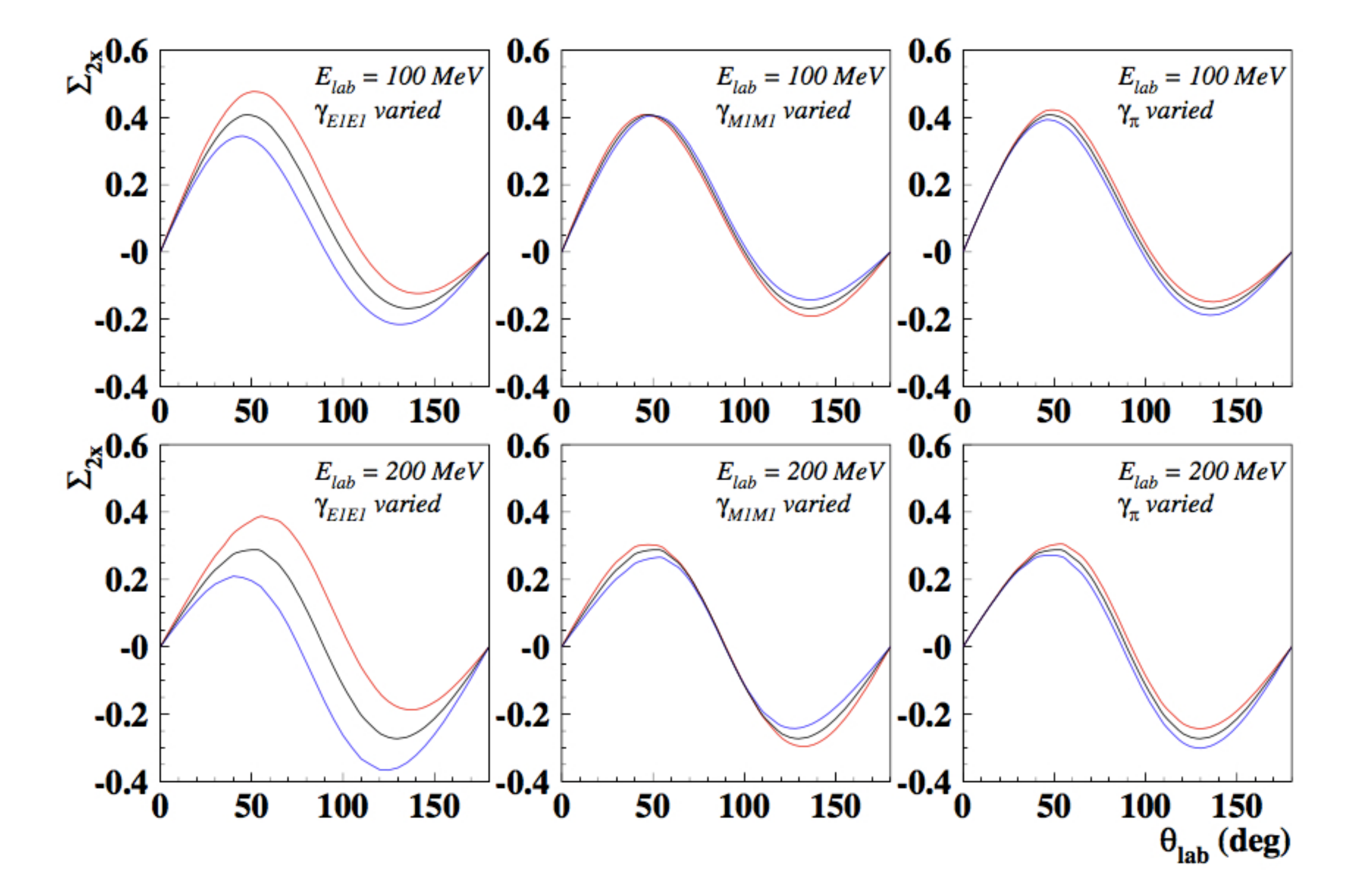}
	\caption{Predicted beam-target asymmetry for Compton scattering from a
		transversely polarized neutron at photon laboratory energies of 100
		MeV (upper row), and 200 MeV (lower row)~\cite{Pasquini:2007hf}. The results from 
		subtracted dispersion relations are obtained by using the values
		$\alphaen=12.5$, $\betamn=2.7$ and
		$\gammazeron=-0.096$, while the remaining polarizabilities are
		taken as free parameters. Left column: results for fixed
		$\gammammn$ and $\gammapin$ , and the following
		values of $\gammaeen$: -5.94 (black lines), -3.94
		(red lines), and -7.94 (blue lines); central column:
		results for fixed $\gammaeen$ and $\gammapin$, and
		the following values of $\gammammn$: 3.75 (black
		lines), 5.75 (red lines), and 1.75 (blue lines); right
		column: results for fixed $\gammaeen$ and
		$\gammammn$, and the following values of
		$\gammapin$: 13.68 (black lines), 15.68 (red
		lines), and 11.68 (blue lines).}
	\label{fig:pasquini}
\end{figure}

\begin{figure}[!htbp]
\includegraphics[width=1\columnwidth]{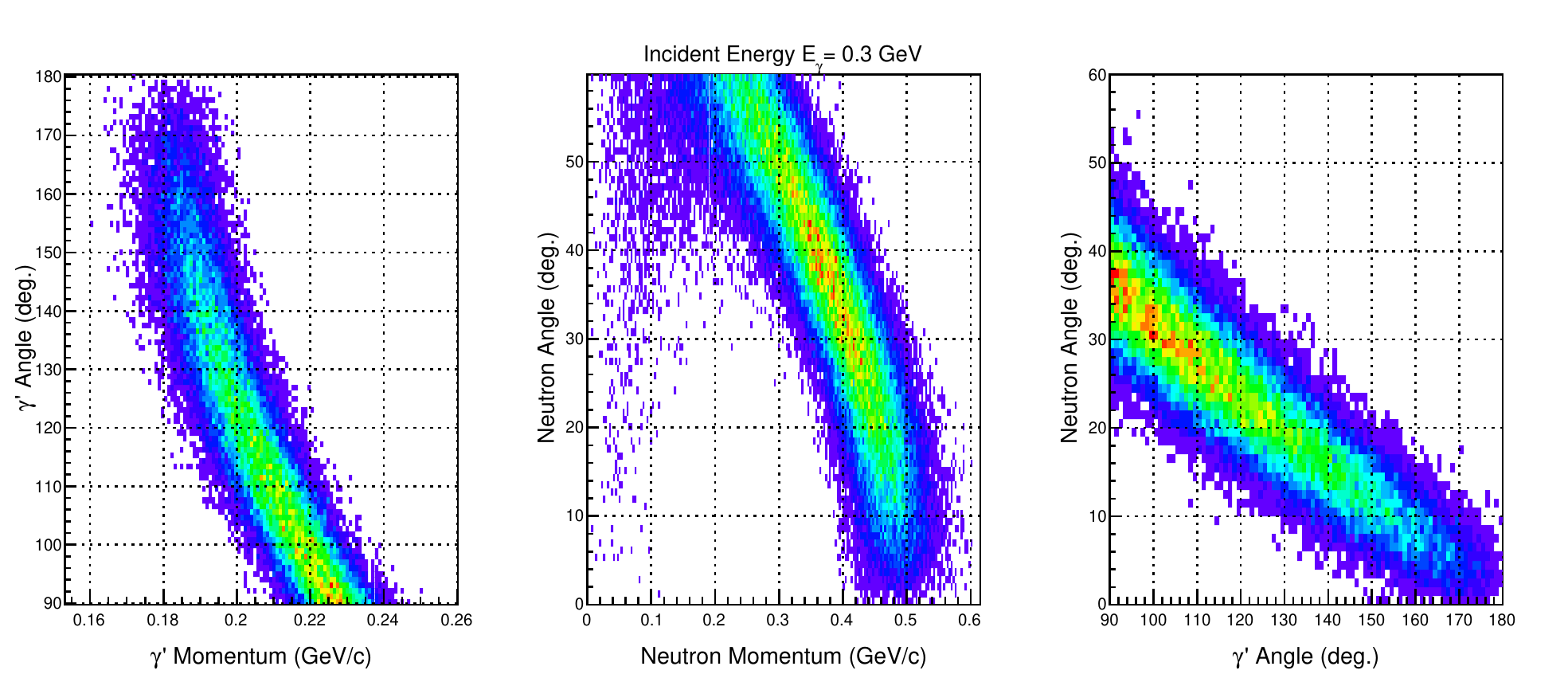}
	
	\caption{\label{fig:QF-kinem}QF kinematics at 300 MeV incident energy.}
	
\end{figure}

We note that $\Sigma_{2x}$ for Compton scattering from a neutron is mainly sensitive to $\gammaeen$, especially at forward angles, as seen in fig.~\ref{fig:pasquini}~\cite{Pasquini:2007hf,Griesshammer:2017txw}. The sensitivity to $\alphaen - \betamn$ here is quite small. The kinematics of QF Compton scattering from deuterium at an incident energy of 300~MeV is displayed in fig.~\ref{fig:QF-kinem}.

With a 10~cm liquid deuterium target, a 1~sr detector for $\gamma'$ and a photon beam intensity of $10^{9}~\gamma$/s, the Compton event rate would be about 70~Hz. Such a high rate of incident $\gamma$-rays also makes the use of a polarized $^{3}$He gas target feasible. A 30~cm, 10~bar target with a similar detector acceptance would produce a event rate of about 1.2~Hz. Given that the neutron angle is fairly tightly correlated to the $\gamma'$ angle it should be possible to design the neutron arm of the detection system to match the acceptance of the $\gamma'$-arm. 

Neutron spin polarizabilities could be accessed using two different
implementations of double-polarization: either polarized beam, polarized
target and time-of-flight neutron detector or polarized beam, unpolarized
target, and a time-of-flight polarimeter to measure the polarization transfer
to the recoiling neutron.

Typically, in a polarized experiment one measures an asymmetry in the counting rate when the beam helicity, target spin orientation, etc. are reversed. The precision in measuring a polarized observable is given by: $\delta X=\sqrt{\frac{2}{N_{inc}.F^{2}}}$, where $N_{inc}$ is the number of incident neutrons and $F^{2}$ is the experiment figure of merit. If the efficiency of the $\gamma'$ arm is close to 100\%, $F^{2}\sim\varepsilon_{n}P^{2}$, where $\varepsilon_{n}$ is the efficiency of the neutron arm and $P$ is either the target polarization or the analyzing power of the neutron polarimeter. In the polarized-beam-and-target approach, $\varepsilon_{n}\sim0.2$ and $P\sim0.9\times0.6$ are readily obtainable, so that $F^{2}\sim5.8\times10^{-2}$. To obtain a precision of $\delta X\sim0.01$, $N_{inc}$ should be about $3 \times 10^5$, which could be collected in 80~hr. 

We prefer this option to the one where the polarization of the recoiling neutron is detected. The precision of a recoil-polarization detection experiment for a given $N_{inc}$ could be similar, even though the figure of merit is lower, since a liquid target could be employed. But other complications regarding the detection of final-state neutron polarization---imprecise knowledge of the neutron analyzing power, complications in building a polarimeter that matches the solid angle of the $\gamma'$ arm---favor a polarized-target experiment. But recoil-polarization detection should continue to be considered as an option for NGLCGS experiments. 

\subsection{Target Technology}

\label{sec:targets}

History shows that polarizability measurements proceed at exactly the pace of the enabling technologies. It is telling that the first spin polarizability measurements followed the construction of a low-mass, high acceptance, frozen-spin target at Mainz, and that to drive this research further, a further extension of this device, an active polarized target, has been developed. In addition, without the construction of a modern, high-rate, highly segmented photon detector, covering a large fraction of the $4\pi$ solid angle, it will not be possible to suppress pion photoproduction backgrounds.

\subsubsection{Unpolarized Active Gas Targets}

\label{sec:unpoltargets}

In unpolarized active gas targets, a high pressure gas scintillator of \threeHe or \fourHe is generally employed with an admixture of an inert gas to shift the primary VUV scintillation into the visible region for easier detection. Xenon is a commonly used shifter, but alternatively one can use nitrogen, which gives good results at low concentrations of around 500~ppm. The scintillation at 20 bar and 500 ppm N$_{2}$ is quite fast, with rise and fall times of a few nanoseconds. This provides good timing information and allows operation at fairly high rates. Keeping the amount of high-$Z$ material in the target to a minimum reduces the sensitivity to pair production and (electron) Compton scattering by a high intensity incident photon beam.

The active target allows the recoiling charged particle from a Compton event (coherent or quasi-free) to be detected in coincidence with the scattered photon. This helps separate the weak signal from background.  Generally a gas target would be fairly long in order to achieve a reasonable thickness, so even fairly crude vertex reconstruction is advantageous in measuring angular distributions.

\subsubsection{Unpolarized Liquid Targets}

\begin{figure}[!htb]
	\centering
	\includegraphics[height=0.4\linewidth]{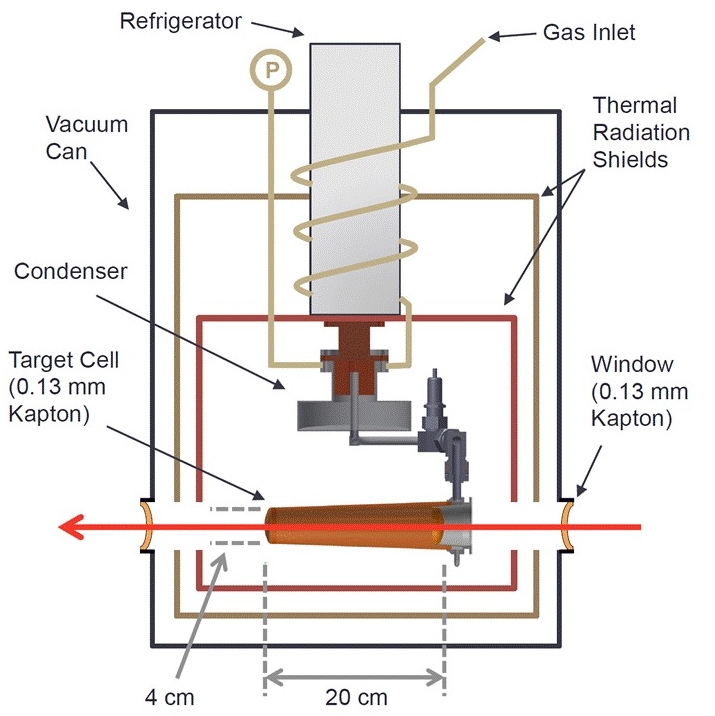}
	\hspace*{4ex}
        \includegraphics[height=0.4\linewidth]{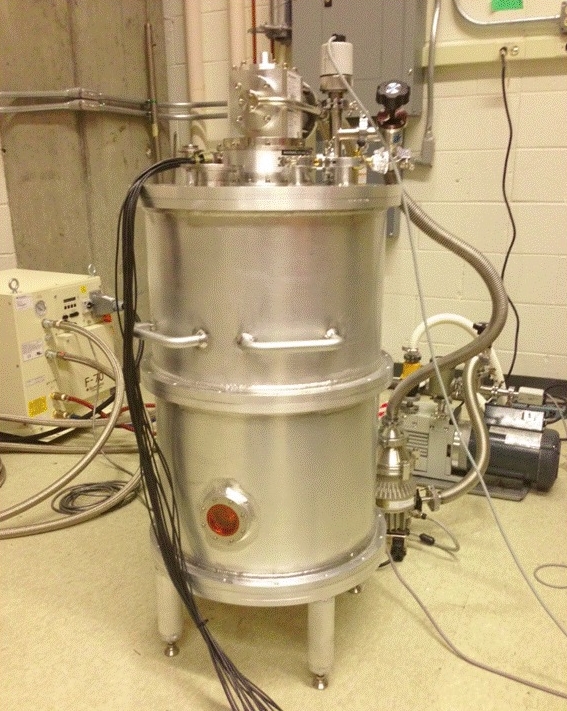}
	\caption{The cryotarget at \HIGS~\cite{Kendellen16}.}
	\label{fig:cryotarget}
\end{figure}

Liquid \threeHe and \fourHe targets are relatively standard technology. The
liquid has a density about a factor of fifty greater than a 20~bar gas
target. A cryogenic target system capable of liquefying \fourHe at 4~K, H$_2$
at 20~K, and D$_2$ at 23~K has already been developed at HI$\gamma$S
\cite{Kendellen16}; see fig.~\ref{fig:cryotarget}. The apparatus was utilized in a series of Compton scattering experiments on deuterium and \fourHe at beam energies from 60 to 85~MeV. Cooling is provided by a 4~K Gifford-McMahon cryocooler. Liquid temperatures and condenser pressures are recorded throughout each run to ensure that the target's areal density is known to about 1\%. Modifications are planned that will add the ability to liquefy \threeHe by lowering the system's base temperature to 1.5~K.

Elastic Compton scattering measurements on liquid \fourHe and deuterium targets at the NGLCGS will require a high-resolution photon detector array to identify elastic scattering unambiguously. In the case of \fourHe, the requirement is not especially stringent, since the inelastic channel is about 20~MeV below the elastic peak. The situation is considerably more challenging for deuterium targets, where the breakup channel is only 2.2~MeV below the elastic peak. A program is underway at HI$\gamma$S using the liquid deuterium target with high-resolution beams and detectors to characterize the inelastic contribution at 65 and 85~MeV.

Both liquid \threeHe and \fourHe could also be used for quasi-free Compton scattering, but the recoiling nucleon, and perhaps other fragments, would need to be detected in coincidence with the scattered photon. Time of flight to an external array could be used to measure the momentum of the recoil neutron. This would be very useful in reconstructing the QF event. If the energy is sufficiently high, similar techniques could be preformed for a recoiling proton. At lower energies a separate, active target could detect protons stopping in the target. The technique of using TOF for both neutrons and protons simultaneously could provide a tremendous advantage in the particular case of QF scattering from a deuterium target, where the proton QF measurement could provide a consistency check for the desired neutron QF result, since the two QF channels are completely symmetric. 

%

\subsubsection{Frozen-Spin Polarized Target}

The proton spin-polarizability program depends critically on developing a proton target that can be polarized in either the longitudinal or the transverse directions with high polarization and a long relaxation time. As an example, the Mainz A2 group, in collaboration with JINR Dubna, has constructed a frozen-spin butanol target ideally suited for these measurements \cite{Thomas11}.

The workhorse of the frozen-spin target is a horizontal dilution refrigerator (DR) with high cooling power designed for insertion into a nearly $4\pi$ detector. Initially, the target end of the cryostat is placed inside the warm bore of a large superconducting magnet. The target material consists of frozen beads made from butanol which has been doped with the stable free nitroxyl radical TEMPO (2,2,6,6-tetramethyl-piperidine-1-oxyl). The beads reside in the dilution refrigerator mixing chamber, which is cooled below 1~K. The electrons are readily polarized at low temperature in a 2.5~T magnetic field. Their polarization is then transferred to the protons via dynamic nuclear polarization using 70~GHz microwaves. After polarization, the microwaves are switched off and their attendant heat load is removed. The target then cools to a base temperature of around 30~mK, and a 0.6~T holding field (provided by a small superconducting coil inside the cryostat) maintains the polarization. The bulky polarizing magnet can be removed, and the target inserted into a large-acceptance detector. The Mainz frozen-spin target achieves a proton polarization of about 90\% with relaxation times greater than about 1000~h. Longitudinal or transverse directions are both available, depending upon how the internal holding coil is wound (solenoid vs. saddle). Deuterated butanol may be substituted as the target material with deuteron vector and tensor polarizations in excess of 75\% and 40\%, respectively.

\subsubsection{Active Frozen-spin Target}

As discussed in the physics justification section, there is great interest in making measurements of the proton spin polarizabilities at incident photon energies near pion threshold. However, at incident photon energies below approximately 250 MeV, recoil protons cannot escape from a frozen-spin target cryostat, and the identification of Compton scattering events from the sea of $\pi^0 \rightarrow \gamma \gamma $ events becomes problematic. At Mainz, for example, the direction of the recoil proton is used to differentiate Compton scattering events from $\pi^0$ decay events. To enable Compton scattering measurements at lower energies, where the recoil proton is presently unobserved, an active frozen-spin target has been developed. The target material is a polarizable scintillator, employing Si photomultipliers to read out the scintillation.  A similar active polarized target will be essential for this program.

\begin{figure}[!htbp]
	\centering
	\includegraphics[width=0.45\linewidth]{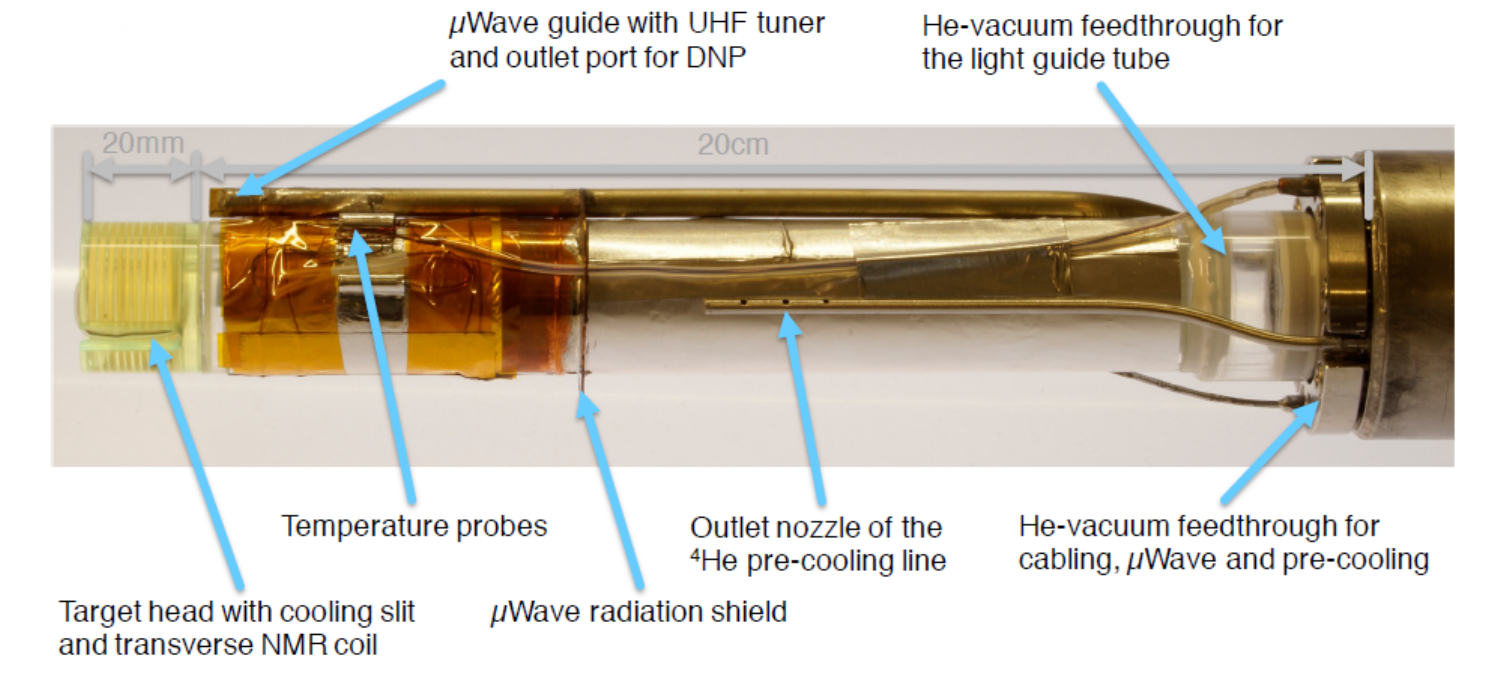}	\includegraphics[width=0.45\linewidth]{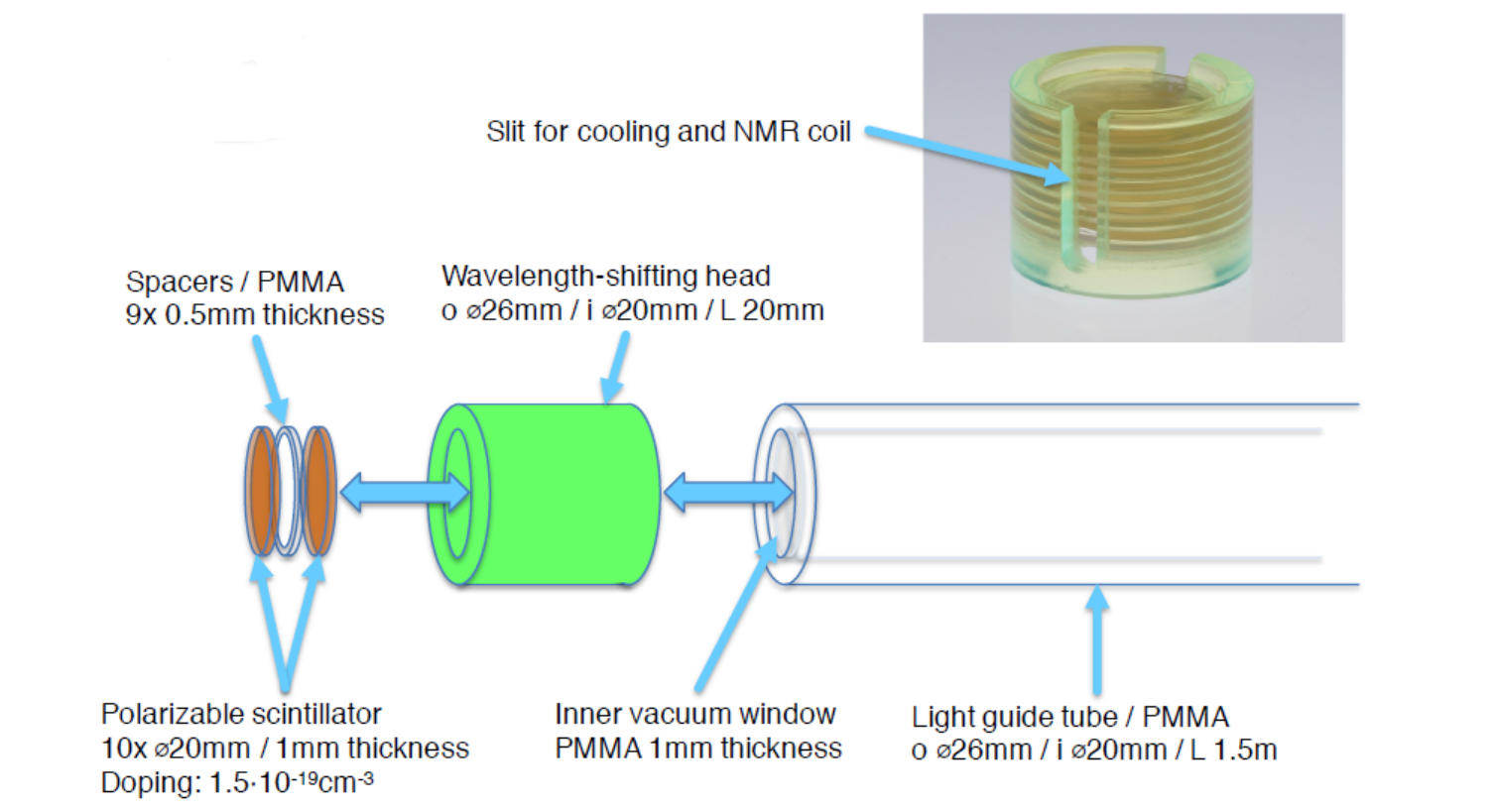}
	\caption{The Mainz Active Polarized Target \cite{Maik}.}
	\label{fig:APT}
\end{figure}

An active frozen-spin target utilizes the same \threeHe/\fourHe dilution refrigerator as a traditional frozen-spin target, with some modifications in and around the area that holds the target material. Discs of polystyrene doped with TEMPO are stacked inside the DR mixing chamber and held at 30 mK. Scintillation light from the discs is captured, redirected, and wavelenth-shifted in a containing cylinder of wavelength-shifting material \cite{Biroth:2016kye}. This cylinder is, in turn, affixed to a cylinder of Plexiglass, which transports the light to silicon photomultipliers operating at 4~K \cite{Biroth:2015lxa} (see fig \ref{fig:APT}). In a recent test at Mainz, the active target achieved 50\% polarization with a 70~h relaxation time. A target head which enables particle tracking is also under development. This could involve the use of printable scintillators in order to make a highly segmented, position-sensitive target head.

\subsubsection{Polarized \threeHe Gas Targets}

Polarized \threeHe is an effective polarized neutron target, as about 90\% of the nuclear spin can be attributed to the unpaired neutron. It is by now a quite mature, though far from trivial, technology. The technique of spin-exchange optical pumping is used to achieve high-pressure ($\approx 10$ atm) and high-polarization ($\approx 70\%$)~\threeHe targets. First, a narrow-band circularly polarized laser at a wavelength of 795 nm hitting rubidium vapor quickly polarizes those atoms. The rubidium atoms then collide with \threeHe atoms and transfer their polarization to the \threeHe nuclei. The rubidium and \threeHe mixture is maintained at about 10 atmospheres in a glass pumping cell transparent to laser light 
and at a temperature $\approx 200$ degrees C so that rubidium stays in its gaseous state. This spherical pumping cell is connected to a cylindrical target cell where the beam window made as thin as possible, so as to minimize photon interactions in the glass.  Special glass that is inert to the alkali vapor and stable in this radiation environment must be used. 

The whole target system is maintained in a very homogeneous magnetic field to ensure a long polarization lifetime. Adiabatic fast passage nuclear magnetic resonance (AFP-NMR) and electron paramagnetic resonance (EPR) are used to measure the \threeHe polarization. AFP-NMR is also used to flip the direction of polarization to reduce systematic uncertainties for asymmetry measurements.

A polarized \threeHe target of this type was developed for JLab~\cite{Zhaoetal} and achieved a world record 60\% polarization with an incident 6 GeV luminosity of $10^{36}$/(cm$^2$\,s). Meanwhile, the polarized \threeHe target system at \HIGS has reached 50\% polarization and been used successfully in several experiments~\cite{Kramer:2007zzb,Laskaris:2015wma,Ye:2009dx}. The luminosities proposed for the NGLCGS will not be an issue for this technology. R\&D on these targets is thus focused on enhancing signals with higher gas density and on reducing background from the glass cells. Note that the presence of the highly reactive alkali vapor inside the target mitigates against any sort of active target---either scintillator or drift-chamber variant.

\subsection{Detectors}

\label{sec:detectors}

An essential part of the experimental apparatus shared by many of the Compton scattering polarizability experiments is a nearly $4\pi$ crystal detector for detection of the Compton-scattered photon.  Near-$4\pi$ acceptance is desirable for two reasons: (i) to enable data taking at multiple scattering angles simultaneously, and (ii) to suppress $\pi^0 \rightarrow \gamma \gamma$ events that would show up as double neutral hits in the detector.  Surrounding the target with a charged-particle tracker---e.g., a cylindrical GEM or silicon tracker---would provide neutral/charged particle identification. For example, the tracking detector could be used to identify recoil protons, deuterons or helium nuclei from the Compton scattering reaction. 

For the case of measuring the neutron polarizabilities, detectors suitable for photons or neutrons of energy around 100 MeV will be necessary. Good detection efficiency of neutral particles requires bulk material, and scintillators are generally used.

\subsubsection{Inorganic Scintillators for Photon Detection}
\begin{figure}[!htb]
	\centering	
\includegraphics[width=0.55\linewidth]
{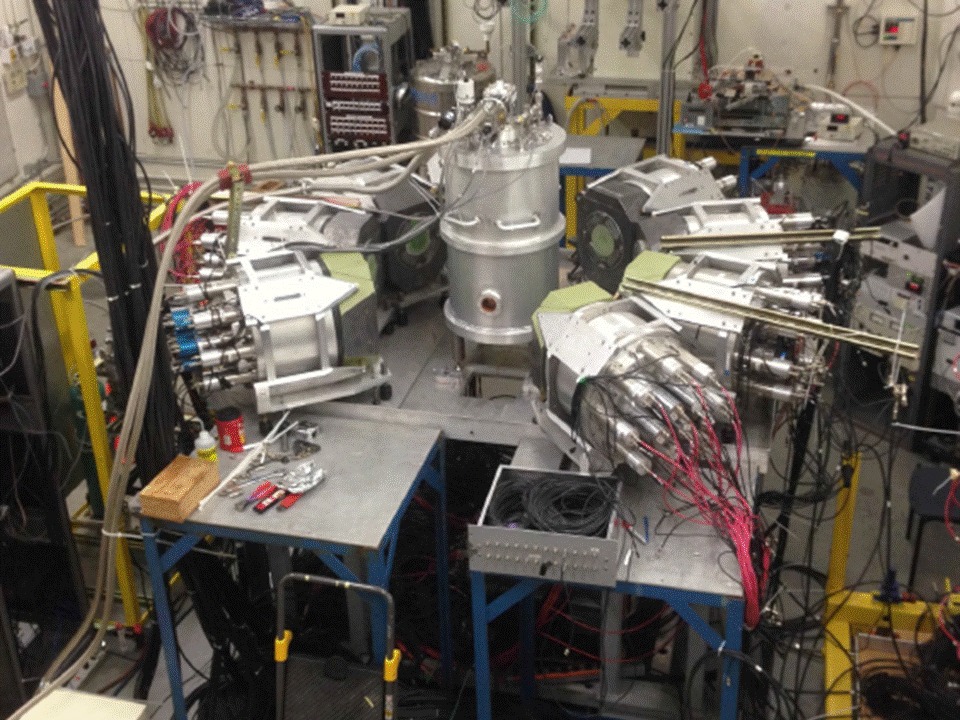}
\caption{The High Intensity Gamma Source NaI Detector Array HINDA with eight
  NaI crystals at \HIGS is an example of present-day inorganic scintillators.}
	\label{fig:HINDA}
\end{figure}

For photons one would use high-$Z$ materials which have short radiation
lengths. Since energy resolution will be important, scintillators are
preferable to Cherenkov counters. Traditionally materials such as NaI(Tl) or
CsI(Tl) have been used, like in HINDA at \HIGS; see fig.~\ref{fig:HINDA}.
However, (with thallium activation to increase light output) the scintillation
is relatively slow and timing precision for TOF measurements is
sub-optimal. Alternatively one can use materials such as:
\begin{enumerate}
	\item BaF$_2$: This has fast and slow scintillation components, with good timing properties and pulse shape discrimination (PSD) of relativistic and non-relativistic particles. The UV scintillation requires a quartz window photo-multiplier tube (PMT).
	\item PbWO$_4$: This material has quite a fast scintillation, good timing, short radiation length, and good radiation hardness. If we have a very high intensity facility, radiation hardness may become an issue in the long term.
	\item New developments in scintillator materials technology, such as the use of quantum dots. These aim to improve the light output and speed of the scintillation, and significant advances can be expected in the next few years.

\end{enumerate}
These materials are sensitive to $\gamma$-rays and neutrons above about 25 MeV as well as other generated background. For operation in intense radiation fields, the detectors should be highly segmented to keep singles rates at a manageable level. Uniformity of scintillation light collection is important if optimum energy resolution is to be obtained.

If the scintillation is in the visible region one can use silicon PMTs in place of the traditional vacuum tubes. These are relatively cheap, have very good timing properties, reasonable noise performance, and continue to improve.

\subsubsection{Organic Scintillators}
Organic scintillators are traditionally used for neutrons with energies of one to a few hundred MeV. They are also sensitive to photons, but because of their long radiation lengths, they do not have good photon energy resolution. In general they produce fast scintillations so that they are suitable for high resolution time-of-flight (TOF) measurement, which is the only way practical way to achieve some energy resolution for neutrons of around 100 MeV. Organic scintillators come in two types:
\begin{itemize}
	\item Plastic: Plastic scintillators are relatively cheap and easy to fashion into a variety of geometries. In general they do not have pulse-shape discrimination (PSD) properties, although Eljen Technology does claim PSD for one of their variants. The hydrogen-to-carbon ratio in these scintillators is generally around 1:1.
	\item Liquid: Liquid scintillators are also cheap but obviously require a container. Traditional types similar to NE213 have good PSD properties but suffer from a volatile, chemically reactive base. Mineral-oil types have varying degrees of PSD, are fast, easily contained, and some have a hydrogen-to-carbon ratio as high as 2:1.
\end{itemize}

\subsubsection{Time-of-Flight Techniques}
Typical experimental setups for time-of-flight (TOF) measurements depend on fast timing, relatively long flight paths (for good timing resolution), and segmented detector arrays (for position information). Previous TOF setups have included multi-cell liquid-scintillator arrays as well as large plastic scintillator bars that use time differences between the ends of each bar to pin down the lateral position of the neutron event.  Given the necessity of a long flight path, these arrays generally occupy a considerable area, in order to guarantee a reasonable solid-angle acceptance for the neutron arm.


\section{QCD Origins of Charge Symmetry Breaking} 
\lhead{Photomeson Production}
\rhead{Nuclear Structure and Low-Energy QCD}

The quark masses are important input parameters in the Standard Model.  They are currently determined from a variety of sources, including meson masses and lattice simulations. The currently accepted values (in the $\overline{\rm{MS}}$ scheme at a renormalization scale of 2 GeV) are~\cite{PDG17}: $$m_u \approx 2.2^{+0.6}_{-0.4} \;\;\mathrm{MeV\;\; and} \;\;  m_d \approx 4.7^{+0.5}_{-0.4} \;\; \mathrm{MeV}$$. This implies $m_d/m_u \geq 2$, which might naively imply large isospin-violation effects.  These effects are, however, efficiently masked, since chiral symmetry means that {\it all} quark-mass effects are generically suppressed in low-energy strong-interaction observables. 

Furthermore, for reactions involving only pions, effects proportional to a single power of $m_u - m_d$ are forbidden, due to G-parity. However, the presence of nucleons vitiates this theorem, and isospin violation should thus be a stronger effect in meson-nucleon interactions. This provides an important opportunity for a next-generation laser Compton $\gamma$-ray source (NGLCGS) to contribute to the determination of a fundamental parameter of the Standard Model, $m_d/m_u$. 

To be specific, pion photoproduction offers two ways of observing effects due to isospin violation.  The first consists of a precise measurement of the four pion photoproduction s-wave amplitudes: $\gamma p \rightarrow \pi^+ n$, $\gamma p \rightarrow \pi^0 p$, $\gamma n \rightarrow \pi^- p$, and $\gamma n \rightarrow \pi^0 n$. (The latter two cases would be observed via coherent production from the deuteron.)  A recent TRIUMF experiment, experiment E643, attempted to extract the s-wave amplitude for the process $\gamma n \rightarrow \pi^- p$ from the measurement of the total and differential cross sections of the inverse reaction $\pi^- p \rightarrow \gamma n$.  To be quantitative, the s-wave amplitude for the charged-particle channels should be determined to within an accuracy of 1\%, and for the neutral channels to within 5\%.  Accurate predictions for all four channels exist \cite{BKM96B} making use of the conventional isospin symmetric basis of three independent amplitudes.  The theoretical framework of consistently including operators related to the quark-mass difference and to virtual photons is currently being developed \cite{MEI95}.  A determination of the (absolute) total cross section for the charged channels with a 2\% accuracy appears to be sufficient to provide an important check on the quark-mass ratio $m_d/m_u$ extracted from mesonic processes. 

Recently, there have been two independent claims that isospin has been violated at about the 7\% level in medium energy $\pi N$ scattering \cite{GIB95}. As pointed out by Bernstein \cite{BER93}, a very precise determination of the phase of $\gamma p \rightarrow \pi^0 p$ below the secondary $\pi^+ n$ threshold would allow for a determination of the s-wave $\pi N$ scattering length $a_{\pi N}(\pi^0 p)$ via a generalized three-channel Fermi-Watson analysis.  As shown by Weinberg \cite{WEI66}, the difference $a_{\pi N}(\pi^0 p) - a_{\pi N}(\pi^0 n)$ is very sensitive to the quark-mass difference $m_u - m_d$. A measurement of Im ($E_{0+}$) is equivalent to a measurement of the corresponding $\pi N$ phase shift. It is important to map out this quantity in the region of the so-called unitary cusp (150 to 170 MeV) \cite{BER93}: the discontinuity in $E_{0+}$ which results from the fact that the $\pi^0 p$ and $\pi^+ n$ thresholds are different (a result of isospin breaking).  A measurement of Im ($E_{0+}$) on a polarized proton target using the intensity and energy resolution of the NGLCGS will yield an accuracy that easily displays isospin violation if it is present at the level claimed above. 

More generally, pion-nucleon scattering is poised for significant advances due to the advent of Roy-Steiner equations that incorporate the consequences of chiral symmetry. Threshold pion photoproduction can be used to measure phase shifts at low energies in charge states that cannot be reached in $\pi N$ scattering.  This involves measuring $\gamma p \rightarrow \pi^0 p, \pi^+ n$ and $\gamma n \rightarrow \pi^- p, \pi^0 n$ reactions. 

In the two-nucleon sector the difference between the neutron-neutron and proton-proton scattering lengths is one of the strongest signals of charge-symmetry breaking in few-nucleon systems. But a$_{nn}$ is clouded in controversy due to conflicting values from the reactions $\pi^- d \rightarrow \gamma nn$ and $n d \rightarrow p nn$. The reaction $\gamma d \rightarrow \pi^+ nn$ provides an alternative avenue for a measurement of $a_{nn}$. The NGLCGS would have the intensity and energy resolution to be well-suited for this measurement and therefore shed light on this important quantity.

\bibliographystyle{unsrt}
\renewcommand{\bibname}{References}

\chapter{Accelerator Concepts of Next Generation Laser Compton Gamma-ray Beam Facilities} 
\lhead{}
\rhead{Accelerator Concepts}
\label{Ch:Acc}

\setlength{\textwidth}{6.5in}
\setlength{\textheight}{9in}
\setlength{\topmargin}{-0.5in}
\setlength{\oddsidemargin}{-0.in}
\setlength{\evensidemargin}{-0in}
\setlength\tabcolsep{2 pt}

\section{Introduction to Compton Gamma-ray Sources}

Since the discovery of the Compton effect in the early 1920's via
the scattering of X-rays from electrons in metals~\cite{ACompton,ACompton1922secondary},
another four decades would pass before the Compton effect was recognized
as a useful mechanism to convert low energy photons to high energy
X-ray and $\gamma$-ray photons. With the development and operation
of charged particle accelerators with relativistic electron beams,
in 1963, Milburn and, independently, Arutyunian and Tumanian proposed
a method of producing very high energy $\gamma$-ray beams via Compton
back-scattering of photons from high energy electrons~\cite{RMilburn1963,FArutyunian1963}.
In the ensuing years, the first experimental demonstrations of high
energy $\gamma$-ray production using Compton scattering were carried
out by several groups around the world, including Kulikov \textit{et~al.}
with a $600$~MeV synchrotron ~\cite{OKulikov1964}, Bemporad \textit{et~al.}
with the $6.0$~GeV Cambridge Electron Accelerator~\cite{CBemporad1965},
and Ballam \textit{et~al.} with the $20$-GeV Stanford linear accelerator~\cite{JMurray1967,JBallam1969}.

While successful, the first demonstrations of $\gamma$-ray production
by Kulikov \textit{et al.}~\cite{OKulikov1964} and Bemporad \textit{et
al.}~\cite{CBemporad1965} had a very low $\gamma$-photon yield.
With much improved photon yields ($100$\textendash $500$~$\gamma$/s),
Ballam \textit{et al.}~\cite{JBallam1969} at Stanford Linac Accelerator
Center (SLAC) were able to carry out the first physics measurements
using a Compton $\gamma$-ray beam to study the photo-production
cross sections at several GeVs with a hydrogen bubble chamber. In
1978 the first Compton gamma-ray source (CGS) facility for nuclear
physics research, the Ladon project, was brought to operation in Frascati~\cite{LCasano1975,GMatone1977,LFederici1980,LFederici1980B,DBabusci1991}.
A higher $\gamma$ flux was produced by colliding the high intensity
photon beam inside a laser cavity with the electron beam in the straight
section of the $1.5$-GeV ADONE storage ring at Frascati National
Laboratory. This facility produced a polarized $\gamma$-ray beam with energies
up to $80$~MeV and  an on-target flux of up to $5\times10^{5}$~$\gamma$/s.
 Following the success of the Ladon facility
at Frascati, several more Compton gamma-ray source facilities for
nuclear physics research were brought to operation around the world starting in the 1980s, including LEGS~\cite{ASandorfi1984}
and HIGS~\cite{VLitvinenko1997} in the US, Graal in France~\cite{DBabusci1990},
ROKK-1/ROKK-2/ROKK-1M in Russia~\cite{AKazakov1984,Kezerashvili1993,Kezerashvili1995,Kezerashvili1998},
and LEPS~\cite{TNakano1998,TNakano2001} in Japan. In addition, several
other projects successfully produced Compton $\gamma$-ray beams,
including FEL-X/AIST \cite{TYamazaki1998}, UVSOR-FEL/UVSOR-II/UVSOR-III
\cite{MHosaka1997,taira2011generation,HZen2016}, NIJI-IV \cite{sei2011lasing,ogawa2012asymmetric},
SAGA-LS \cite{TKaneyasu2011}, and NewSUBARU \cite{KHorikawa2010}
in Japan, and Super-ACO \cite{DNutarelli1998,GDeNinno2001} in France.
Some specialized gamma-ray sources were developed for industrial and
national security applications \cite{barty2015advanced,albert2010isotope,albert2011design,albert2011three},
such as the MEGa-ray project \cite{gibson2010design,barty2011overview,hall2011numerical}
at Livermore National Lab, U.S., while advancing certain important
areas of accelerator and laser technologies for gamma-ray sources
\cite{anderson2011velociraptor,albert2012precision}. A few published
reviews on Compton $\gamma$-ray beams and some of the aforementioned
facilities are found in~\cite{DBabusci1996,ADAngelo2000,CSchaerf2005,weller-2009,krafft2010compton}.

Among these Compton gamma-ray source facilities, the High Intensity
Gamma-ray Source (HIGS) at Duke University is the first dedicated
Compton gamma-ray facility employing as the photon driver a high intra-cavity
power Free-Electron Laser (FEL) \cite{litvinenko1998first,litvinenko1999ok4,litvinenko2001performance,wu2006high,jia2010electron}.
The HIGS facility is a high-flux, nearly monochromatic, and highly
polarized gamma-ray source (with both linear and circular polarizations)
covering a wide range of energies from about $1$ to $100$~MeV. The HIGS has demonstrated an unprecedented
high flux operation with a maximum total flux of about $3\times10^{10}$~$\gamma$/s
around $10$~MeV, two or three orders of magnitude more flux than
what was produced by other CGS facilities ever built and operated.
The high flux capability of HIGS is made possible by the full-energy top-off booster injector and recent upgrades to the FEL optical cavity.
The wide energy range and high flux performance of the HIGS makes
it a unique and superior gamma-ray facility \cite{sun2009energy,sun2009end,sun2011theoretical},
ideal for low and medium energy nuclear physics research. It is envisioned
extensions that will first increase the energy reach of HIGS up to $130$ MeV, and then up to $150$~MeV.

During the last decade or so, while a few CGS facilities (e.g. LEGS and
Graal) ceased operation after completing their research missions,
other CGS facilities continue to flourish with accelerator and laser
system upgrades that improve beam performance and enable new capabilities.
In the meantime, a few new CGS facilities are under construction around
the world with $\gamma$-ray beam research and user operation expected to start
in the next few years. A list of
major operational laser Compton gamma-ray sources and new development
projects is provided in Table  \ref{tab:CGSFacilities}.

\begin{sidewaystable}
\caption{A list of laser Compton gamma-ray sources around the world which are
either operational or being developed for operation in the near future.
\label{tab:CGSFacilities}}
\centering{}{\small{}}%
\resizebox{\textwidth}{!}{%
\begin{tabular}{|c|c|c|c|c||c|c|c|}
\hline 
{\footnotesize{}Project Name/Parameters} & {\footnotesize{}HIGS} & {\footnotesize{}LEPS/LEPS2} & {\footnotesize{}NewSUBARU} & {\footnotesize{}UVSOR-III} & {\footnotesize{}SLEGS} & {\footnotesize{}XGLS} & {\footnotesize{}ELI-NP}\tabularnewline
 &  &  &  &  &  &  & \tabularnewline
\hline 
{\footnotesize{}Location} & {\footnotesize{}Durham, U.S.} & {\footnotesize{}Hyogo, Japan} & {\footnotesize{}Hyogo, Japan} & {\footnotesize{}Okazaki, Japan} & {\footnotesize{}Shanghai, China } & {\footnotesize{}Xi\textquoteright an, China} & {\footnotesize{}Bucharest-Magurele,}\tabularnewline
 &  &  &  &  &  &  & {\footnotesize{}Romania}\tabularnewline
\hline 
{\footnotesize{}Accelerator technology} & {\footnotesize{}Storage Ring} & {\footnotesize{}Storage ring } & {\footnotesize{}Storage ring } & {\footnotesize{}Storage ring } & {\footnotesize{}Storage ring } & {\footnotesize{}Linac} & {\footnotesize{}Storage ring }\tabularnewline
\hline 
{\footnotesize{}Laser technology} & {\footnotesize{}FEL} & {\footnotesize{}Solid state laser} & {\footnotesize{}Solid state/gas laser} & {\footnotesize{}Fiber/gas laser} & {\footnotesize{} $\text{CO}_{2}$ laser } & {\footnotesize{}Ti:Sapphire laser} & {\footnotesize{}Solid state laser}\tabularnewline
\hline 
{\footnotesize{}Collision technology} & {\footnotesize{}Intra-cavity, } & {\footnotesize{}External laser,} & {\footnotesize{}External laser,} & {\footnotesize{}External laser,} & {\footnotesize{}External laser,} & {\footnotesize{}External laser,} & {\footnotesize{}Intra-cavity,}\tabularnewline
 & {\footnotesize{}head-on} & {\footnotesize{}head-on} & {\footnotesize{}head-on} & {\footnotesize{}head-on} & {\footnotesize{}cross-angle/head-on} & {\footnotesize{}head-on} & {\footnotesize{} head-on}\tabularnewline
\hline 
{\footnotesize{}Electron energy {[}MeV{]}} & {\footnotesize{}$240$\textendash $1,200$} & {\footnotesize{}$8,000$} & {\footnotesize{}$500$\textendash $1,500$} & {\footnotesize{}$750$} & {\footnotesize{}$3,500$} & {\footnotesize{}$120$\textendash $360$} & {\footnotesize{}$234$\textendash $742$}\tabularnewline
\hline 
{\footnotesize{}Laser wavelength {[}nm{]}} & {\footnotesize{}$1,060$\textendash $190$} & {\footnotesize{}$266$ and $355$} & {\footnotesize{}$532$\textendash $10,600$} & {\footnotesize{}$1,940$ and $10,600$} & {\footnotesize{}$10,640$} & {\footnotesize{}$800$} & {\footnotesize{}$1,030$/ $515$}\tabularnewline
\hline 
{\footnotesize{}Charge and temporal structure} &  &  &  &  &  &  & \tabularnewline
{\footnotesize{}A. CW: Avg. current {[}mA{]}} & {\footnotesize{}$10$\textendash $120$} & {\footnotesize{}$100$} & {\footnotesize{}$300$} & {\footnotesize{}$300$} & {\footnotesize{}$100$\textendash $300$} &  & \tabularnewline
{\footnotesize{} $Q$ {[}nC{]} @ reprate {[}MHz{]}} & {\footnotesize{}$1.8$\textendash $22$ @$5.58$} & {\footnotesize{}$0.2$\textendash $2$ @$50$\textendash $500$} & {\footnotesize{}$0.6$ @ $500$} & {\footnotesize{}$3$ @ $90$} & {\footnotesize{}$0.28$\textendash $0.87$@$347$} &  & {\footnotesize{}$1$@ $71.4$}\tabularnewline
\cline{1-1} \cline{7-7} 
{\footnotesize{}B. Pulsed: $f_{\text{RF}}$ {[}MHz{]}} &  &  &  &  &  & {\footnotesize{}$2,856$ (s-band)} & \tabularnewline
{\footnotesize{}Pulse: $Q$ {[}nC{]} @ reprate {[}Hz{]}} &  &  &  &  &  & {\footnotesize{}$0.5$@$10$ } & \tabularnewline
{\footnotesize{}Pulse duration (full-width)} &  &  &  &  &  & {\footnotesize{}$10$ ps (micropulses)} & \tabularnewline
\hline 
\hline 
{\footnotesize{}$\gamma$-beam energy {[}MeV{]}} & {\footnotesize{}$1$\textendash $100$} & {\footnotesize{}$1,300$\textendash $2,900$} & {\footnotesize{}$1$\textendash $40$} & {\footnotesize{}$1$\textendash $5.4$} & {\footnotesize{}$0.4$\textendash $20$} & {\footnotesize{}$0.2$\textendash $3$} & {\footnotesize{}$1$\textendash $19.5$}\tabularnewline
\hline 
{\footnotesize{}Polarization} & {\footnotesize{}Lin, Cir} & {\footnotesize{}Lin, Cir} & {\footnotesize{}Lin, Cir} & {\footnotesize{}Lin, Cir} & {\footnotesize{}Lin, Cir} & {\footnotesize{}Lin, Cir} & {\footnotesize{}Lin}\tabularnewline
\hline 
{\footnotesize{}$\gamma$-beam energy resolution } & {\footnotesize{}$0.8$\% \textendash $10$\%} & {\footnotesize{}$<15$\%} & {\footnotesize{}$10$\% ($\phi=3$ mm)} & {\footnotesize{}$2.9$\% ($\phi=2$ mm)} & {\footnotesize{}$<5$\% }\textcolor{black}{\footnotesize{}($\phi=2$
mm)} & {\footnotesize{}$1.2$\%\textendash $10$\%} & {\footnotesize{}$<0.5$\%}\tabularnewline
{\footnotesize{}(FWHM)} & {\footnotesize{}collimation} & {\footnotesize{}tagging} & {\footnotesize{}collimation} & {\footnotesize{}collimation} & {\footnotesize{}collimation} & {\footnotesize{}collimation} & {\footnotesize{}collimation}\tabularnewline
\hline 
{\footnotesize{}$\gamma$-beam temporal structure} &  &  &  &  &  &  & \tabularnewline
{\footnotesize{}A. CW operation {[}MHz{]}} & {\footnotesize{}$5.58$ (typical)} & {\footnotesize{}$50$\textendash $500$} & {\footnotesize{}$500$} & {\footnotesize{}$90$} & {\footnotesize{}$347$} &  & {\footnotesize{}$71.4$}\tabularnewline
{\footnotesize{}B. Pulsed operation } & {\footnotesize{}$0.5$\textendash $1.5$ ms (FW)} &  & {\footnotesize{}$8$ ns pulse,} &  &  & {\footnotesize{}Pulsed, $10$ Hz} & \tabularnewline
 & {\footnotesize{}$2$\textendash $100$ Hz} &  & {\footnotesize{}$10$\textendash $100$ kHz,} &  &  &  & \tabularnewline
 & {\footnotesize{}Gain modulated} &  & {\footnotesize{}Q-switched lasers} &  &  &  & \tabularnewline
\hline 
{\footnotesize{}On-target flux {[}avg, $\gamma$/s{]}} & {\footnotesize{}$10^{3}$\textendash{} $3\times10^{9}$} & {\footnotesize{}$10^{6}$\textendash $10^{7}$} & {\footnotesize{}$10^{5}$\textendash $3\times10^{6}$ } & {\footnotesize{}$4\times10^{5}$} & {\footnotesize{}$10^{5}$\textendash $10^{7}$} & {\footnotesize{}$10^{6}$\textendash $10^{8}$ } & {\footnotesize{}$\sim10^{8}$}\tabularnewline
 &  &  & {\footnotesize{}($\phi=3$ mm)} & {\footnotesize{} ($\phi=2$ mm)} & \textcolor{black}{\footnotesize{}($\phi=2$ mm)} &  & \tabularnewline
\hline 
{\footnotesize{}Total flux {[}avg, $\gamma$/s{]}} & {\footnotesize{}$10^{6}$\textendash{} $3\times10^{10}$} & {\footnotesize{}$10^{6}$\textendash $10^{7}$} & {\footnotesize{}$10^{7}$\textendash $4\times10^{7}$} & {\footnotesize{}$10^{7}$} & {\footnotesize{}$10^{6}$\textendash{} $10^{8}$} & {\footnotesize{}$10^{8}$\textendash $10^{9}$ } & {\footnotesize{}$10^{11}$}\tabularnewline
\hline 
{\footnotesize{}Operation status or} & {\footnotesize{}Since 1996} & {\footnotesize{}Since 1999} & {\footnotesize{}Since 2005 } & {\footnotesize{}Since 2015 } & {\footnotesize{}Under construction,} & {\footnotesize{}Under construction,} & {\footnotesize{}Under construction,}\tabularnewline
{\footnotesize{}projected operation date} &  &  &  &  & {\footnotesize{}operation in 2022} & {\footnotesize{}operation in 2023} & {\footnotesize{}operation in 2023}\tabularnewline
\hline 
{\footnotesize{}Reference} & {\footnotesize{}\cite{weller-2009,weller2015nuclear}} & {\footnotesize{}\cite{NMuramatsu2014}} & {\footnotesize{}\cite{KHorikawa2010}} & {\footnotesize{}\cite{HZen2016}} & {\footnotesize{}\cite{WLuo2011,HXu2016}} & {\footnotesize{}\cite{chen2019commissioning}} & \tabularnewline
\hline 
\end{tabular}}{\small \par}
\end{sidewaystable}

\pagebreak{}\paragraph{A Brief Review of Compton Scattering Process}

\begin{figure}[h]

\noindent \begin{centering}
\begin{minipage}[c][1\totalheight][t]{0.6\columnwidth}%
\noindent \begin{center}
\vspace{-0.15in}
\begin{tabular}{cccc}
{\footnotesize{}\hspace{0.6in} } & {\footnotesize{}Before Collision } & {\footnotesize{}$\quad$} & {\footnotesize{}After Collision }\tabularnewline
{\footnotesize{}Electron } & {\footnotesize{}$p^{i}=({\cal E}_{e}/c,\vec{p})$ } &  & {\footnotesize{}$p'^{i}=({\cal E}'_{e}/c,\vec{p'})$ }\tabularnewline
{\footnotesize{}Photon } & {\footnotesize{}$\hbar k^{i}=(\hbar\omega/c,\hbar\vec{k})$ } &  & {\footnotesize{}$\hbar k^{\prime i}=(\hbar\omega'/c,\hbar\vec{k'})$}\tabularnewline
\end{tabular}
\par\end{center}
\noindent \begin{center}
\vspace{-0in}
\includegraphics[width=1\columnwidth]{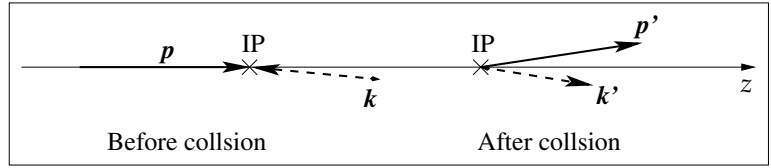} 
\par\end{center}%
\end{minipage}
\par\end{centering}
\caption{Schematic diagram of Compton scattering of an electron and a photon. IP stands for the
interaction point. \label{fig:Compton} }
\end{figure}

Compton scattering between an electron and a photon is shown in Fig.
\ref{fig:Compton}. We can list the four momenta of the electron and
photon before and after collision as the following,

Using conservation of four-momentum, we can express the energy of
the outgoing photon as: 
\begin{equation}
E_{\gamma}\equiv\hbar\omega'=\frac{\hbar\omega(1-\beta\cos\theta_{i})}{1-\beta\cos\theta_{f}+\frac{\hbar\omega}{{\cal E}_{e}}(1-\cos\theta_{ph})},\label{Eq:GammaEng1}
\end{equation}
where $\hbar$ is the reduced Planck constant, $\hbar\omega$ is the
energy of the incoming photon, $E_{\gamma}$ (or $\hbar\omega'$)
is the energy of the outgoing photon, $\theta_{i}$ and $\theta_{f}$
are the angles between the momentum of the incoming and outgoing photons
and that of the incident electron, ($\cos\theta_{i}=\hat{p}\cdot\hat{k}$
and $\cos\theta_{f}=\hat{p}\cdot\hat{k'}$), and $\theta_{ph}$ is
the angle between the two photons ($\cos\theta_{ph}=\hat{k}\cdot\hat{k'}$).

For a collision between relativistic electrons and low energy photons,
the energy of the scattered photons is peaked along the direction
of the incident electrons. The back-scattered photon has the maximum
energy in a head-on collision with $\theta_{i}=\pi$ and $\theta_{f}=0$,
\begin{equation}
\hbar\omega'=\frac{\gamma^{2}(1+\beta)^{2}}{1+R_{0}}\,\hbar\omega,\label{Eq:GammaEngHeadOn2}
\end{equation}
where $R_{0}=2\gamma^{2}(1+\beta)\hbar\omega/{\cal E}_{e}$ is the
recoil factor. When the recoil is small ($R_{0}\ll1$), the maximum
scattered photon energy is approximately, $E_{\gamma,max}\approx\gamma^{2}(1+\beta)^{2}\hbar\omega\approx4\gamma^{2}\hbar\omega,$
where the second approximation holds for ultra-relativistic electrons.
The energy boost factor, $4\gamma^{2}$, is responsible for the relativistic
Doppler effect (or relativistic blue-shift) to transform an ordinary
infrared or visible light photon into a high energy X-ray or $\gamma$-ray
photon.

At the zero-recoil limit, the total cross section is rather small
$\sigma\approx\frac{8\pi r_{e}^{2}}{3}=6.652\times10^{-29}$~m$^{2}$
(i.e., the Thomson cross section). For the collision of a relativistic
electron beam and a low-energy photon beam, scattered photons with
the highest energy are concerntrated around the direction of the incident
electron beam. This kinematic feature enables the formation of a nearly
monochromatic Compton photon beam by using a simple collimation technique.

\section{Technology for Next-Generation Compton Gamma-ray Sources}

Limited by the available flux and spectral flux of present and future
Compton gamma-ray sources, and complex event detection and beam diagnostic
techniques used for measurements, the exploration of many important
nuclear physics phenomena with small cross sections can be best realized
using a $\gamma$-ray beam with a high repetition rate. Compared with
such a CW beam, without being able to explore the multi-photon effect
due to the intensity limitation and lack of coherence, a low-repetition-rate
pulsed Compton $\gamma$-ray beam does not have practical advantages
in most cases (expect for certain experiments which can exploit extremely
fast (short) pulses). In developing a gamma-ray
source with a high average flux or spectral flux, the preferred technology
choice demands a combination of an electron accelerator and a laser
system with their respective high beam repetition rate 
matched. For the next-generation Compton gamma-ray sources,
we focus on the following optimum technology choices:
\begin{itemize}
\item Electron accelerators: a storage ring (a well-established technology)
or a super-conducting linac (a newer technology)
\item Photon beam systems: a high-finesse resonant optical cavity
\item Interaction point configuration: electron-photon collision inside
the laser cavity
\end{itemize}
Using an external laser beam in a simple optical setup for collision
has some attractive features, especially when combining a low repetition rate,
high peak power laser with a warm electron linac. However, this scheme is not
cost-effective for generating CW $\gamma$-ray beams due to its inefficiency
in utilizing the photon beam in a single pass. The use of an optical
resonator for in-cavity collision allows the electron beam to interact
with the accumulated laser power repeatedly, thereby reusing the photon beam
in a large number of passes. The in-cavity collision scheme can be
arranged in either a head-on configuration or a cross-angle configuration.
While the head-on configuration has the advantage of a higher luminosity,
it is more limiting as it requires a mechanism to guide the electron
beam into and out of the optical cavity. Meeting this requirement would limit  the geometric
design of the resonator and complicate the magnetic optics design
around the interaction point (IP). On the other hand, the cross-angle
scheme is compact and can be arranged to use a relatively short resonator.
Because of these and other benefits and shortcomings, both head-on
and crossed-angle collision configurations should be considered for
the development of the next-generation Compton gamma-ray sources.

In the following sections, a brief overview of the electron beam and
photon beam technologies is given. To mitigate many potential
risks, the focus will be on conventional, well-established, mature
or fast-maturing technologies for the next-generation Compton gamma-ray
sources which require an unprecedented level of flux, energy resolution,
stability, reliability, and precise control and manipulation. Before
going into these topics, a brief digression into other technological
developments and advancements relevant to the Compton gamma-ray sources
is in order.

\subsection{Related Technology Development}

Technological advancements relevant to the Compton gamma-ray
sources have been reviewed and summarized in the 2010 U.S. DOE Basic
Energy Sciences (BES) sponsored workshop on Compact Light Sources
\cite{osti_1291139}. This workshop examined the readiness of state of the art
technology for compact light sources (including so-called ``inverse
Compton scattering sources'') inclusing assessing cost-effectiveness, user access, availability, 
and reliability. The report also compared the performance of compact
light sources to the third-generation storage rings and Free-Electron
Lasers. While the details can be found in the report, we will briefly
describe and give updates to the two areas of technological development
that can impact the development of Compton gamma-ray sources in the future.

\subsubsection{Compton X-ray Sources}

Since mid 1990's, X-rays have been generated by means of Compton scattering
(also often referred to as ``Thomson scattering'' due to negligible
electron recoil) using low energy electron beams. The relatively low
cost and/or good availability of lower energy conventional accelerators
have allowed the exploration of laser Compton scattering with a variety
of combined accelerator and laser technologies:
\begin{itemize}
\item Combining a warm s-band linac and a pulsed laser: 

A $30$-keV, femtosecond X-ray beam was generated with a $90^{\circ}$
cross-angle at Berkeley National Laboratory \cite{leemans1995femtosecond};
a very high peak flux X-ray beam ($2\times10^{9}$ up to $10^{10}$
ph/pulse) was produced at Vanderbilt University \cite{carroll2003pulsed};
a $52$-keV X-ray beam was generated at Tsinghua University, China
with controllable polarization \cite{YDu2013,HZhang2017}; and an
$11$-keV X-ray beam was produced at CEA, France, by utilizing a ring-down
laser cavity with a folded laser path \cite{AChaleil2016};
\item With a linac FEL: tunable $7$ to $12$ keV X-rays were demonstrated
at CLIO, France, using an infrared linac FEL powered by the same electron
beam \cite{glotin1996tunable};
\item With a superconducting linac: a tunable $3.5$ to $18$ keV, high
average flux (few $10^{9}$ ph/s) X-ray beam was generated using a
kW-class infrared FEL at Jefferson Lab \cite{boyce2003jefferson};
$29$ keV, millisecond long X-ray pulses (macropulses) were demonstrated
with a four-mirror optical cavity at KEK, Japan \cite{HShimizu2015};
the generation of $7$ keV X-rays with a CW electron beam was demonstrated
with an Energy Recovery Linac and a four-mirror cavity \cite{akagi-2016};
\item Combining a storage ring and a high-finesse Fabry-Perot cavity: 

Likely the highest average flux X-ray beam was delivered by a compact
X-ray source system manufactured by Lyncean Technologies Inc. using
a dedicated storage ring and a high power, four-mirror Fabry-Perot
cavity powered by an external Nd:YAG laser \cite{eggl-2016}. Using
the same technology combinations, an even higher flux Compton X-ray
source, THOMX, is now under construction in Orsay, France \cite{variola2013thomx,variola2014thomx}.
\end{itemize}
The above brief survey of this necessarily incomplete list of Compton
X-ray source projects worldwide in the recent two decades has also
confirmed that the optimum technological combination to generate a
high average photon flux is to use a high repetition rate electron
beam to collide with a frequency matched laser beam inside an optical
resonator.


\subsubsection{Laser-plasma Accelerator Based Compton Sources}

Compton photon sources continue to be a vibrant research area. In
additional to the aforementioned sources driven by conventional accelerators,
new Compton sources based on laser-plasma accelerators \cite{leemans2009laser,esarey2009physics}
have been actively developed around the world in the recent decade
\cite{schwoerer2006thomson,phuoc2012all,chen2013mev,powers2014quasi,sarri2014ultrahigh,rykovanov2014quasi,khrennikov2015tunable,yan2017high,cole2018experimental}.
A few published reviews describing laser-plasma acceleration, ultra-fast
Compton photon generation, and applications of such light sources
can be found in \cite{hooker2013developments,corde2013femtosecond,albert2016applications}.
In these relatively compact and all-optical Compton sources, the ultra-short,
high-peak power laser plays two essential roles: (1) generating and
accelerating electrons, and (2) as an intense photon drive to collide
with the electrons. With adequate peak laser intensity the laser-plasma
Compton sources are well-suited to explore a new operational regime
of nonlinear Thomson/Compton scattering in which multiple photons
collide with a single electron to generate one Compton photon with
energy much higher than in the linear Compton scattering. 
Various experiments using different lasers and targets (solids, gases) have
proved the feasibility of this method and there is a long term path
to high beam quality. However, a specific set of beam quality at this
time, including the beam flux, energy spread, spectral flux, spectral
purity, pointing stability, etc. is less desirable compared with the
Compton sources driven by conventional accelerators.  On the other
hand, certain unique characteristics of a laser-plasma Compton source
(such as ultrafast pulses, compactness, etc.) have yet to be adequately
exploited for basic and applied science research \cite{martz2017poly,geddes2017impact}.
While providing promise for the future, at the present time, the laser-plasma
Compton systems are still at an experimental exploration stage, therefore,
not ready to implement for a user facility.

\subsubsection{New Developments}

\paragraph{ELI-NP}A major gamma-ray source under construction is
the Extreme Light Infrastructure-Nuclear Physics (ELI-NP), a European
Center of Excellence for scientific research in laser and gamma radiation. The facility is
located in the town of Magurele near Bucharest, Romania. ELI-NP's
new gamma-ray source, the Variable Energy Gamma (VEGA) System is a
dedicated system for delivering $\gamma$-ray beams to users. The
construction of the VEGA system was recently awarded to Lyncean Technologies
Inc. and its delivery, installation and acceptance are scheduled to
be completed in early 2023. 

The VEGA system will deliver $\gamma$-rays with energy continuously
variable from $1$ to $19.5$$\,$MeV, covering the energy range relevant
for low energy nuclear physics and astrophysics studies, as well as
applied research in material sciences, management of nuclear materials,
and life sciences. The beams will be quasi-monochromatic with a relative
bandwidth better than $0.5$\% (FWHM), high intensity with a spectral
density higher than $5\times10^{3}$ $\gamma$/eV/s, and a high degree
of linear polarization at more than $95$\%. With these parameters,
the VEGA system will be the most advanced gamma-ray source in the
world having about one order of magnitude higher $\gamma$-ray flux
and at least a factor of two smaller relative bandwidth than the current
state-of-the-art$\,$\cite{weller2015nuclear}. The VEGA system is
based on the use of a storage ring and a high-finesse Fabry-Perot
cavity. The parameters of the electron beam and the interaction laser
are optimized at the interaction point in a way to provide $\gamma$-rays
with the features discussed above. The electron beam system will operate
in the range of $234$ to $742$ MeV. For a given interaction laser
wavelength, this electron energy range allows at least a factor of
ten in $\gamma$-ray energy continuous tunability. Two separate optical
cavity laser systems, one at $\sim1$ $\mu$m (``IR'') and the other
at $\sim0.5$ $\mu$m (``Green'') wavelengths, will be provided
to cover the $\gamma$-ray energy range from $1$ to $10$ MeV, or
$2$ to $19.5$ MeV, respectively. The laser systems use a passive,
high-finesse optical cavity to resonantly build-up the pulsed laser
power. The optical cavity provides gains of $5,000$\textendash $10,000$
in laser power, which reduces the complexity of the interaction laser
drive system. The main parameters for the VEGA system are summarized
in Table \ref{tab:CGSFacilities}.

\paragraph{Gamma Factory at CERN}A new initiative at CERN, the Gamma
Factory (GF) \cite{krasny2015gamma}, is under rapid development in
the last two years. The Gamma Factory project is aimed at creating,
storing, and exploiting new types of relativistic atomic beams. These
atomic beams of ions with all but a few electrons stripped can be
stored in the Super Proton Synchrotron (SPS) or the Large Hadron Collider
(LHC) storage rings at very high energies ($30<\gamma<3000$), at
high bunch intensities ($10^{8}<N_{{\rm bunch}}<10^{9}$), and at
high bunch repetition rate (up to $20$$\,$MHz). With the GF approach,
a resonant atomic transition of a highly-charged relativistic ion
excited by laser light results in a spontaneously emitted photon.
Due to the relativistic Doppler effect, the photons emitted in the
direction of the ions can reach very high energies, ranging from $1$
to $400$$\,$MeV for the atomic beams stored in the LHC rings. Because
of a huge resonant photon absorption cross section compared to that
of photon scattering with a point-like electron (a factor up to $10^{9}$),
the intensity of an atomic-beam-driven light source is expected to
be several orders of magnitude higher than what is possible with Compton
gamma-ray sources driven by an electron beam, reaching a total $\gamma$-ray
flux up to $10^{17}$$\,$$\gamma$/s \cite{krasny2018cern}.

To prove experimentally the concepts underlying the Gamma Factory
proposal, feasibility tests have been and will continue to be performed
at the SPS and at the LHC. Since 2017 the experimental beam tests
have started with partially stripped $^{129}\text{Xe}{}^{39+}$ beams,
followed by dedicated SPS runs with $^{208}\mbox{Pb}^{54+}$, $^{208}\mbox{Pb}^{80+}$,
$^{208}\mbox{Pb}^{81+}$ beams.  The $^{208}\mbox{Pb}^{81+}$ beam
was injected for the first time into the LHC  and ramped to a proton
equivalent energy of $6.5$$\,$TeV in 2018 \cite{Schaumann:2019evk}.
The majority of the operation  aspects for such beams have been successfully
tested.  An important specific achievement was to demonstrate that
bunches of $10^{8}$ hydrogen-like lead atoms per bunch can be efficiently
produced and maintained at the LHC top energy with the lifetime and
intensity fulfilling  the Gamma Factory  requirements. The pivotal
concept  of the GF initiative\textemdash that relativistic atomic
beams can be produced, accelerated and stored in the existing CERN
SPS and LHC rings\textemdash has therefore already been experimentally
proven. 

The SPS and LHC beam  tests will be followed by  the Gamma Factory
 ``proof-of-principle'' SPS experiment \cite{GF_PoP_2019}.  This
experiment will study collisions of  a laser beam with the $^{208}\mbox{Pb}^{79+}$
 beam in a specially designed collision point in the SPS tunnel.   Its
results will provide a decisive proof and an experimental evaluation
  of the achievable intensities of the atomic-beam-based gamma-ray
source. 

The success of CERN's Gamma Factory initiative can have a profound
impact to the worldwide effort in developing next-generation Compton
gamma-ray sources. 

\subsection{Electron Accelerators}

\subsubsection{Electron Storage Rings}

The development and operation of a number of storage ring based third-generation
light sources since the 1980's \cite{selph1988lbl,galayda1995aps,laclare1993commissioning,zhao2009commissioning,namkung2010review}
has significantly advanced the science and technology of the electron
storage ring. In particular, advancements have been made in the following
critical areas:
\begin{itemize}
\item understanding and managing charged particle nonlinear dynamics to
realize strong focusing magnetic optics with very small transverse
beam emittance at nanometer-radians;
\item mitigating and controlling a variety of collective effects in the
storage ring to enable high current operation; and 
\item improving flexibility, stability and availability of storage ring
light sources using advanced beam diagnostics and accelerator control
systems. 
\end{itemize}
In the area of nonlinear dynamics, the more recent development of
storage ring based, diffraction limited light sources has further
advanced the control and compensation of nonlinear effects in the
storage ring lattice using multi-bend achromats (MBA), achieving an
unprecedented level of tens of picometer-radian beam emittance \cite{martensson2018saga,steier2018status,henderson2015status,raimondi2016esrf}.
The development of MBA based diffraction limited light sources has
also advanced the design and manufacture of high-field and high-quality
magnets and small vacuum chamber systems. Many effective technologies
have been developed to mitigate several important collective effects,
such as microwave instability, intra-beam scattering, and coupled-bunch
instabilities using uniform and smooth vacuum chambers, harmonic rf
cavities, and bunch-by-bunch feedback systems \cite{hettel2014challenges,borland2014lattice,byrd2000commissioning,teytelman2011overview,wu2011development}.
A variety of beam diagnostic and control techniques, such as slow/fast
orbit feedback, beam based lattice calibration, fast focusing compensation,
etc., have greatly improved the storage ring operation stability and
consistency \cite{Autin1973,Chung1993,Endo1996,Carwardine1998,Boge1999,schilcher2004commissioning,uzun2005initial,Rojsel1994,portmann1995automated,Safranek1997,Safranek2002}.
Furthermore, the development of sophisticated slow accelerator controls
has expanded the storage ring capabilities by allowing a rapid change
among a variety of operation modes with different bunch patterns,
beam currents, and beam energies \cite{Knott1993,Dalesio1994,Borland1995a,Borland1995,Corbett2003,Portmann2005,Wu2003}. 

The storage ring is a mature accelerator technology, well-suited as
the electron beam driver for the next-generation Compton gamma-ray
source. In fact, the next-generation CGS requires only a modest beam
emittance (a few to tens of nm-rad), a reasonably high current (an
average current from $100$s mA to $1$ A), and a level of beam stability
achievable using relatively long bunches (tens to hundreds of picoseconds)
(see the example gamma-ray sources below). Without a need to push
the limits of the accelerator technology, a storage ring based CGS
can be developed and constructed with a modest cost, a well-defined
schedule, and a manageable level of risk. This technology is also
being constantly improved and optimized with several existing Compton
gamma-ray sources (see Table \ref{tab:CGSFacilities}). Overall, the
electron storage ring is a highly recommended accelerator technology
for the next-generation CGS.

\subsubsection{Superconducting Linacs}

Superconducting linacs are the heart of Energy Recovery Linacs (ERLs).
ERLs can provide high brightness electron beams at high average current
(tens of millamperes). This allows them to be used in applications
where storage rings normally operate. They are well suited to CGS
applications due to the ability to tolerate relatively high beam losses
caused by the energy loss of electrons that emit high energy $\gamma$-rays. 

Superconducting linacs continue to improve in terms of  gradient, efficiency,
and current. The LCLS II program at SLAC is producing $12$-meter-long
cryomodules with over 1$28$ MV of CW accelerating voltage and the
LCLS II HE program will demand cryomodules with over 158 MV CW accelerating
voltage \cite{raubenheimer2016LCLS}. In both cases the cavity unloaded
quality factor $Q_{0}$ must be greater than $2.7\times10^{10}$ at
$2.0$ kelvin. This allows high gradients without requiring excessive
refrigeration. Source technology has also advanced both at Cornell
and Brookhaven National Lab. The LEReC program at Brookhaven has recently
used a 30 mA electron beam with excellent beam quality to cool ions
in the RHIC accelerator \cite{kayranIPAC2018_TUPMF025}. The beam
was delivered 24/7 for a user program, thus showing sufficient reliability
for a gamma source as well. The Compact ERL (cERL) at KEK in Tsukuba
Japan has recirculated $1$$\,$mA of beam current in an X-ray Compton
backscattering experiment \cite{miyajima2019}. Ultimately one would
like to use superconducting rf (SRF) cavities at $4.5$ kelvin rather
than the less efficient $2$ kelvin operation. This can be achieved
via the use of niobium-tin coated SRF cavities \cite{pudasaini2018initial}.
Recent experiments at Jefferson Lab have demonstrated over $14$ MV/m
in a 5-cell $1.497$ GHz cavity. The cavity $Q_{0}$ was $2\times10^{10}$
at $4.5$ kelvin. This effort aims at producing cavities with a $Q_{0}$
of $10^{11}$ at $20$ MV/m at $4.5$ kelvin. 

Another area of advancement is the idea of using multiple passes through
the accelerating cavities with energy recovery. For systems with a
total energy less than 1 GeV this can produce a high current, compact
accelerator. Researchers at the Daresbury Laboratory have produced
a design for a $1$ GeV recirculating linac called DIANA that would
require only two cryomodules of $165$ MV acceleration apiece. This
could be a very potent CGS \cite{williams2018}. Cornell is using
a four-pass configuration to get to $150$ MeV with one cryomodule.
The design current is $30$ mA. The machine is installed and being
commissioned \cite{gulliford2019cbeta}.

\subsection{Photon Beam Systems}

Various optical systems have been used to increase the available laser
beam intensity in Compton scattering experiments. Firstly, ``recirculator''
systems \cite{Rollason} have demonstrated effective laser beam energy
gains with an enhancement factor between $8$~\cite{AChaleil2016}
and $20$~\cite{Shverdin:10} (an enhancement factor of about $30$
is foreseen in ref. \cite{dupraz}). Secondly, a Fabry-Perot (FP)
cavity (optical resonator), a well-known optical device \cite{Kogelnik:66,arnaud,nienhuis},
can provide gain in excess of $10^{3}$. We shall concentrate here
on ``external cavities'' which are filled by external laser beams
(\textit{i.e.}, not Free-Electron Lasers). These devices can also act
as circular \cite{nous} or linear polarization filters \cite{Saraf,Liu}
by geometrical or coating designs. Optical path length inside optical
resonators ranges between micrometers up to $30$~m \cite{Ozawa}.
This latter number sets a reasonable minimum laser pulse repetition
rate at about $10$~MHz, making Fabry-Perot cavity especially useful
for CW electron beams (\textit{i.e.}, in a storage ring or superconducting
linac). Nevertheless, it has recently been demonstrated at KEK \cite{SakaueBurst}
that Fabry-Perot cavity operated in a ``burst mode'' \cite{SakaueBurst}
can also be used advantageously with a warm linac. In these devices,
laser oscillators must be tightly locked to the resonator round-trip
length. Highly efficient feedback methods are indeed well known \cite{Drever}
and have been extended to mode lock regime \cite{udem2002,JasonJones,diels,ye}.

Fabry-Perot cavities using either two mirrors or four mirrors are
of particular interest to achieve high gains. Two-mirror high-finesse
(F$\sim$$30,000$) cavities operating with CW Nd:YAG oscillators
have been used successfully for Compton polarimeters \cite{Falletto,DESY_POLAR}
and Compton laser wire \cite{Honda2005}. Moderate average power (few
kW) has been routinely obtained with these cavities. However, two
mirror cavities cannot provide stability and very strong focusing
simultaneously. In addition, the cavity length tuning required for
timing synchronization with the electron beam is not independent of
the beam focus tuning. A good alternative is four-mirror cavities.
They provide at the same time stability, strong focusing and an independent
tuning of the cavity length and beam focusing. A main drawback with
a four-mirror cavity is that the cavity modes are elliptical.

In the pulsed regime, the following state-of-the-art Fabry-Perot cavity
performance has been obtained with table-top systems:
\begin{itemize}
\item High average power: about $700$~kW ($10$ ps laser pulses) with
a Yb-doped oscillator and a laser amplifier delivering more than $400$~W
\cite{carstens-2014}. 
\item High finesse: a finesse of approximately $30,000$ (power enhancement
factor of $10,000$) using a low power Ti:Sapph ($3$ ps laser pulses)
oscillator \cite{high_finesse}. 
\end{itemize}
Combining high average power and high finesse is presently under investigation
in the context of compact Compton X-ray sources. Various optical cavities
have already been implemented with an electron storage ring, a linac
and ERL at KEK. For instance: 
\begin{itemize}
\item Non-planar four-mirror cavity ($178.5$ MHz) filled by a $10$\textendash $20$
Watts Yb-doped, picosecond oscillator/fiber-amplifier system \cite{nous2}.
Up to $50$ kW stored power was obtained routinely \cite{iryna}. 
\item Planar four-mirror cavities ($357$ MHz and $162.5$ MHz) filled by
a Nd-doped, picosecond oscillator/burst-amplifier systems \cite{fukuda2016development,Shimizu}. 
\end{itemize}
A four-mirror cavity with $70$~kW stored power (with Nd-doped $65$
MHz oscillator) is also routinely operated in a commercial Compact
X-ray source \cite{eggl-2016}.

From the above results, up to $\sim$$100$ kW average power in picosecond
regime can be obtained by careful choice of the mirror substrates/coatings
\cite{carstens-2014} and careful mechanical design. Low noise oscillators
recently available enable the use of a high-finesse cavity in order
to reduce the cost of high average power laser amplifiers. Eventually,
to avoid mirror damage and to optimize the laser beam shape at the
Compton interaction point, various cavity geometry and boundary shapes
have been proposed \cite{Skettrup,Klemz,Putnam:12,Carstens:13,DuprazABCD}.

Compared to a compact, high-power Fabry-Perot cavity driven by a multiple
MHz external laser, a larger-scale and more complex FEL oscillator
remains an attractive option as the photon beam drive for a high energy
CGS because of its demonstrated extraordinary versatility in many
areas. The oscillator FEL is well-known for its wavelength flexibility
(e.g., the Duke storage ring FEL can be operated from $2.1$ $\mu$m
to $188$ nm). More recently, the storage ring FEL has been developed
to demonstrate several new capabilities which can have an important
impact for a CGS, including simultaneous two-color lasing \cite{wu2015widely,yan2016storage}
and the use of an ``optics-free'' method to rapidly control and
manipulate the laser beam (and $\gamma$-ray beam) polarization with
an unprecedented level of precision \cite{wu2006high,yan2019precision}.

\newpage{}

\section{Next Generation Compton Gamma-ray Sources: Examples and
Capabilities}

In this section, we describe several possible ways to develop next-generation
gamma-ray sources based upon different technology choices. These example
CGSs take advantage of the most recent advancements in  laser and
accelerator technologies to achieve the key beam parameters required
by science programs. They are practical with manageable levels of
risk\textemdash without pushing multiple limits of laser and accelerator
science and technology, and these CGSs can be designed and constructed
in the next five to ten years. These sources are expected to be reliable,
especially for those adopting mature and established technological
solutions; and some are also well-suited for multi-user
operation. Depending on the $\gamma$-ray beam energy range and technology
choices, the cost for such a gamma-ray source varies significantly.

Beam performance requirements are specified by the research opportunities decribed in the previous secience sections of this report. We envision two types
of facilities needed to satisfy the user needs in the low-energy and medium-energy ranges. 
Beam requirements
are organized using a few key parameters, such as the beam energy,
intensity, energy resolution, polarization, and repetition rate in
Tables \ref{tab:HighEngBeamRequirements} and \ref{tab:LowEngBeamRequirements}. 

\begin{table}[H]
\centering{}\caption{Preliminary medium-energy $\gamma$-ray beam requirements for low energy
QCD research. Note that the energy resolution requirement can be relaxed
to $3$\% (FWHM) for higher energies ($>150$ MeV).\label{tab:HighEngBeamRequirements}}
\begin{tabular}{|c|c|c|c|c|c|}
\hline 
{\small{}Repetition rate} & {\small{}$E_{\gamma}$} & {\small{}Flux on target} & {\small{}Energy resolution } & {\small{}Total flux} & {\small{}Polarization}\tabularnewline
{\small{} (MHz)} & {\small{}(MeV)} & {\small{}($\gamma$/s)} & {\small{}(relative, FWHM)} & {\small{}Est. ($\gamma$/s)} & \tabularnewline
\hline 
{\small{}$10$} & {\small{}$60$\textendash $350$} & {\small{}$3\times10^{8}$} & {\small{}$1.5$\textendash $2$\%} & {\small{}$1$\textendash $1.5\times10^{10}$} & {\small{}Cir., flip; $10$ Hz}\tabularnewline
 &  &  &  &  & {\small{}Lin., flip; $10$ Hz}\tabularnewline
{\small{}$20$\textendash $30$} & {\small{}$100$\textendash $300$} & {\small{}$1\times10^{9}$} & {\small{}$2$\%} & {\small{}$3$\textendash $4\times10^{10}$} & {\small{}Cir;}\tabularnewline
{\small{}Up to $200$} & {\small{}$100$\textendash $300$} & {\small{}$1\times10^{9}$} & {\small{}$2$\%} & {\small{}$3$\textendash $4\times10^{10}$} & {\small{}Cir}\tabularnewline
\hline 
\end{tabular}
\end{table}

\begin{table}[H]
\centering{}\caption{Preliminary low-energy $\gamma$-ray beam requirements for nuclear
astrophysics and nuclear structure research. The nuclear structure
research requires $\gamma$-ray beams with a repetition rate not exceeding
$100$ MHz.\label{tab:LowEngBeamRequirements}}
\begin{tabular}{|c|c|c|c|c|c|}
\hline 
{\small{}Research Area} & {\small{}$E_{\gamma}$} & {\small{}Flux on target} & {\small{}Energy resolution } & {\small{}Total flux} & {\small{}Polarization}\tabularnewline
 & {\small{}(MeV)} & {\small{}($\gamma$/s)} & {\small{}(relative, FWHM)} & {\small{}Est. ($\gamma$/s)} & \tabularnewline
\hline 
{\small{}Nuclear Astrophysics} & {\small{}$2$\textendash $5$ } & {\small{}$1\times10^{11}$} & {\small{}$1$\%} & {\small{}$0.8$\textendash $1\times10^{13}$} & {\small{}Any}\tabularnewline
 & {\small{}$5$\textendash $10$ } & {\small{}$1\times10^{10}$} & {\small{}$1$\%} & {\small{}$0.8$\textendash $1\times10^{12}$} & {\small{}Any}\tabularnewline
 & {\small{}$10$\textendash $20$ } & {\small{}$5\times10^{9}$} & {\small{}$1$\textendash $2$\%} & {\small{}$2$\textendash $5\times10^{11}$} & {\small{}Any}\tabularnewline
{\small{}Nuclear Structure} & {\small{}$15$\textendash $35$} & {\small{}$1\times10^{9}$} & {\small{}$1$\textendash $2$\%} & {\small{}$4$\textendash $8\times10^{10}$} & {\small{}Lin. or Cir.}\tabularnewline
{\small{}Hadronic Parity Violation} & {\small{}} & {\small{}} & {\small{}} & {\small{}} & {\small{}}\tabularnewline
{\small{}Few-Nucleon  Nuclei} & {\small{}$2$\textendash $5$} & {\small{}$1\times10^{10}$} & {\small{}$0.5$\textendash $1$\%} & {\small{}$$ 1 $\times10^{10}$} & {\small{}Cir.}\tabularnewline
{\small{}Parity doublets in Nuclei} & {\small{}$2$\textendash $10$} & {\small{}$1\times10^{9}$} & {\small{}$0.5$\textendash $1$\%} & {\small{}$$ 1 $\times10^{10}$} & {\small{}Cir.}\tabularnewline
\hline 
\end{tabular}
\end{table}

Using two main $\gamma$-ray beam parameters, the energy range and
intensity, a comparison of the example CGSs considered for the next generation CGS are provided in Table
\ref{tab:ProjectedPerformanceSummary}. In this table, comments on
the risk level for each source are provided and several main challenges
are identified. For other beam parameters, the relevant details are
provided in the individual section for each source.

\begin{table}[H]
\caption{Comparison of several example CGSs based upon different technology
choices, in terms of their energy range, beam intensity, and risk
levels. The projected total flux is made under the assumption of a
high repetition rate beam production, and the flux on target is estimated
for a properly collimated beam with a relative energy resolution of
$2$\% (FWHM) (about $3$\% of the total flux). Note that the head-on
collision is much more efficient in producing high $\gamma$-ray beam
intensity, however, it is very challenging to simultaneously realize
this collision geometry and a very small laser beam size at the collision
point ($\sigma_{x,y}<30$ $\mu$m). \label{tab:ProjectedPerformanceSummary}}
\centering{}%
\begin{tabular}{|c|c|c|c|l|}
\hline 
 &  & {\small{}Projected } & {\small{}Flux on target} & \tabularnewline
{\small{}CGS examples} & {\small{}$E_{\gamma}$} & {\small{}total flux} & {\small{}($\frac{\Delta E_{\gamma}}{E_{\gamma}}=2$\%)} & {\small{}Risk assessment and technology}\tabularnewline
 & {\small{}(MeV)} & {\small{}($\gamma$/s)} & {\small{}($\gamma$/s)} & {\small{}requirements and challenges}\tabularnewline
\hline 
 &  &  &  & {\small{}Risk level: Low }\tabularnewline
{\small{}Storage ring based} &  & {\small{}$3\times10^{11}$} & {\small{}$\sim9\times10^{9}$} & {\small{}Modest laser power ($20$ kW);}\tabularnewline
 &  &  &  & {\small{}Capable injector; }\tabularnewline
{\small{}medium energy CGS } & {\small{}$25$\textendash $400$ } &  &  & {\small{}Small collision angle ($6^{\circ}$)}\tabularnewline
\cline{3-5} 
{\small{}($E_{e}=1.2$\textendash $3.5$ GeV)} &  &  &  & {\small{}Risk level: Medium}\tabularnewline
 &  & {\small{}$\sim2\times10^{12}$} & {\small{}$\sim6\times10^{10}$} & {\small{}High laser power ($100$ kW);}\tabularnewline
 &  &  &  & {\small{}High-charge, high-rate injector}\tabularnewline
\hline 
 &  &  &  & {\small{}Risk level: Low }\tabularnewline
{\small{}Storage ring based} & {\small{}$1$\textendash $10$ } & {\small{}$2\times10^{12}$} & {\small{}$\sim6\times10^{10}$} & {\small{}Capable injector;}\tabularnewline
 & {\small{}$10$\textendash $15$ } & {\small{}few $10^{11}$} & {\small{}$\sim3\%$ of total } & {\small{}High laser power ($100$ kW);}\tabularnewline
{\small{}low energy CGS } & {\small{}$15$\textendash $30$ } & {\small{}$10^{11}$} & {\small{}$\sim3\%$ of total } & {\small{}Small collision angle ($6^{\circ}$)}\tabularnewline
\cline{2-5} 
{\small{}($E_{e}=0.35$\textendash $0.75$ GeV)} &  &  &  & {\small{}Risk level: Medium}\tabularnewline
 & {\small{}$1$\textendash $5$ } & {\small{}$\sim3\times10^{13}$} & {\small{}$\sim9\times10^{11}$} & {\small{}Capable injector; }\tabularnewline
 & {\small{}$5$\textendash $15$ } & {\small{}few $10^{12}$ } & {\small{}$\sim3\%$ of total } & {\small{}High laser power ($100$ kW);}\tabularnewline
 & {\small{}$15$\textendash $30$ } & {\small{}few $10^{11}$\textendash $10^{12}$} & {\small{}$\sim3\%$ of total } & {\small{}Small laser beam (rms $40$ $\mu$m),}\tabularnewline
 &  &  &  & {\small{}and with head-on collision}\tabularnewline
\hline 
 &  &  &  & {\small{}Risk level: Medium}\tabularnewline
 &  &  &  & {\small{}High ERL current ($20$ mA);}\tabularnewline
{\small{}Energy Recovery Linac} & {\small{}$0.2$\textendash $30$} & {\small{}$\sim$$5\times10^{12}$} & {\small{}$\sim1.5\times10^{11}$} & {\small{}High laser power ($70$ kW);}\tabularnewline
{\small{}based low energy CGS } &  &  &  & {\small{}Small laser beam (rms $30$ $\mu$m)}\tabularnewline
{\small{}($E_{e}=0.10$\textendash $0.75$ GeV)} &  &  &  & {\small{}with head-on collision}\tabularnewline
\cline{2-5} 
 &  &  &  & {\small{}Risk level: High}\tabularnewline
 & {\small{}$0.2$\textendash $30$} & {\small{}$\sim$$3\times10^{13}$} & {\small{}$\sim9\times10^{11}$} & {\small{}Very high linac current ($40$ mA);}\tabularnewline
 &  &  &  & {\small{}High laser power ($70$ kW);}\tabularnewline
 &  &  &  & {\small{}Very small laser beam (rms $15$ $\mu$m)}\tabularnewline
 &  &  &  & {\small{}with head-on collision}\tabularnewline
\hline 
\end{tabular}
\end{table}

\newpage

\subsection{Storage Ring Based Medium Energy CGS}

\begin{figure}[h]
\noindent \begin{centering}
\begin{minipage}[t][1\totalheight][c]{0.54\columnwidth}%
\noindent \begin{center}
\includegraphics[width=1\textwidth]{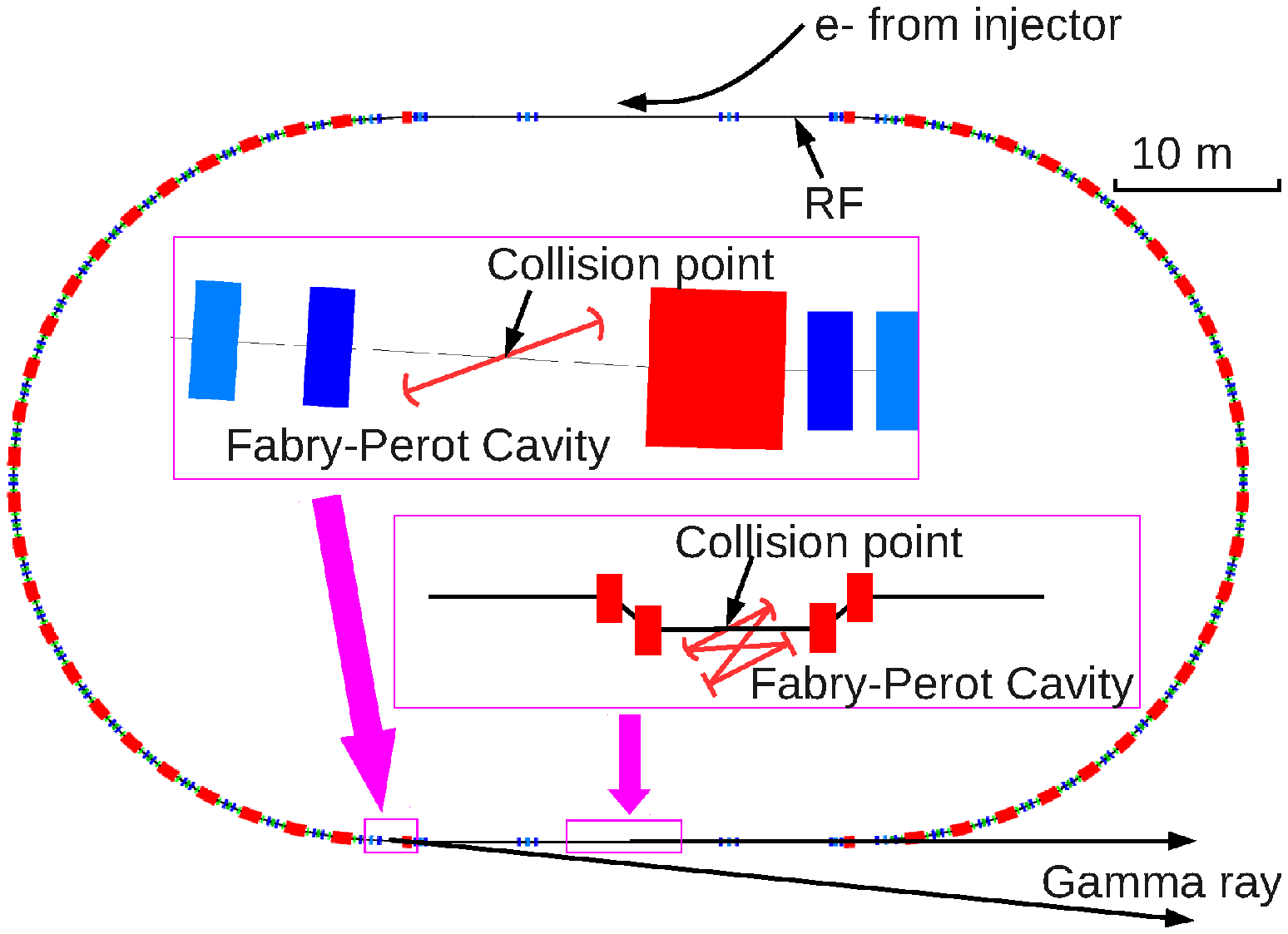}
\par\end{center}%
\end{minipage}%
\begin{minipage}[t]{0.45\columnwidth}%
\noindent \begin{center}
\begin{tabular}{ll}
\hline 
{\footnotesize{}Parameters} & {\footnotesize{}Values}\tabularnewline
\hline 
{\footnotesize{}Energy range} & {\footnotesize{}$1.2$\textendash $3.5$ GeV}\tabularnewline
{\footnotesize{}Circumference} & {\footnotesize{}$201.08$ m}\tabularnewline
{\footnotesize{}Nominal beam current} & {\footnotesize{}$500$ mA/64 bunch}\tabularnewline
{\footnotesize{}Hori. natural emittance, $\varepsilon_{x}$} & {\footnotesize{}$1.8$ nm}\tabularnewline
{\footnotesize{}Coupling, $\varepsilon_{y}/\varepsilon_{x}$} & {\footnotesize{}$0.1$}\tabularnewline
{\footnotesize{}Betatron tune, $\left(\nu_{x},\nu_{y}\right)$} & {\footnotesize{}$(23.87,9.42)$}\tabularnewline
{\footnotesize{}Main radio frequency} & {\footnotesize{}$95.42$ MHz}\tabularnewline
{\footnotesize{}Voltage of rf cavity} & {\footnotesize{}$2.5$ MV}\tabularnewline
{\footnotesize{}Natural chromaticity, $\left(\xi_{x},\xi_{y}\right)$} & {\footnotesize{}$\left(-64.7,-22.7\right)$}\tabularnewline
{\footnotesize{}Momentum compaction } & {\footnotesize{}$8.98\times10^{-4}$}\tabularnewline
{\footnotesize{}Damping time $\left(\tau_{x},\tau_{y},\tau_{s}\right)$} & {\footnotesize{}$\left(2.1,3.4,2.5\right)$ ms}\tabularnewline
{\footnotesize{}Natural bunch duration (rms)} & {\footnotesize{}$49$ ps}\tabularnewline
{\footnotesize{}Natural energy spread (rms)} & {\footnotesize{}$9.7\times10^{-4}$}\tabularnewline
{\footnotesize{}Straight section length} & {\footnotesize{}$24$ m}\tabularnewline
\hline 
\end{tabular}
\par\end{center}%
\end{minipage}
\par\end{centering}
\caption{\label{fig:HighEngStorageRingLayout}(Left) Layout and parameters
of a medium energy electron storage ring ($1.2$\textendash $3.5$ GeV).
The insets show two possible collision schemes using a two-mirror
and a four-mirror Fabry-Perot cavity, respectively (the laser cavity
length is not to scale). (Right) Main parameters for the storage ring,
where all energy-dependent parameters are shown for $3.5$ GeV operation.}
\end{figure}

The required $\gamma$-ray beam performance at higher energies (Table
\ref{tab:HighEngBeamRequirements}) can be realized with a storage
ring based Compton gamma-ray source by colliding a GeV electron beam
with a high-repetition laser beam (tens of MHz) inside a Fabry-Perot
cavity with modest intracavity power. One possible design of such
storage rings is shown in Fig. \ref{fig:HighEngStorageRingLayout}
with related machine parameters listed. This $200$-meter-long racetrack
storage ring is comprised of two multi-bend achromatic arcs with a
small emittance and a reasonably large dynamic aperture for the electron
beam injection using conventional kickers. The performance projection
of such a Compton gamma-ray source is provided below for a cross-angle
collision configuration ($\theta=6^{\circ}$). 

The storage ring should be developed to operate in a wide range of
energies using specialized equipment with a large dynamic range of
stability and reliability, including power supplies, magnets, kickers,
beam diagnostics, etc. (\cite{weller-2009,ywu-2015}). The flux performance
of this medium energy gamma-ray source will be limited by the electron
loss rate. To produce a medium energy $\gamma$-ray beam, a significant
amount of energy is transferred from an electron to the photon, causing
the electron to be lost outside the energy acceptance of the storage
ring. In this electron-loss mode, the $\gamma$-ray flux will be limited
by the electron injection rate into the storage ring. Instead of pushing
the emittance limit, this storage ring should be optimized to store
substantial beam currents and to allow flexible injection.

\subparagraph{Energy Range}

\begin{figure}[h]
\noindent \begin{centering}
\begin{minipage}[c][1\totalheight][t]{0.54\columnwidth}%
\vspace{-0.2in}
\noindent \begin{center}
\includegraphics[width=0.95\textwidth]{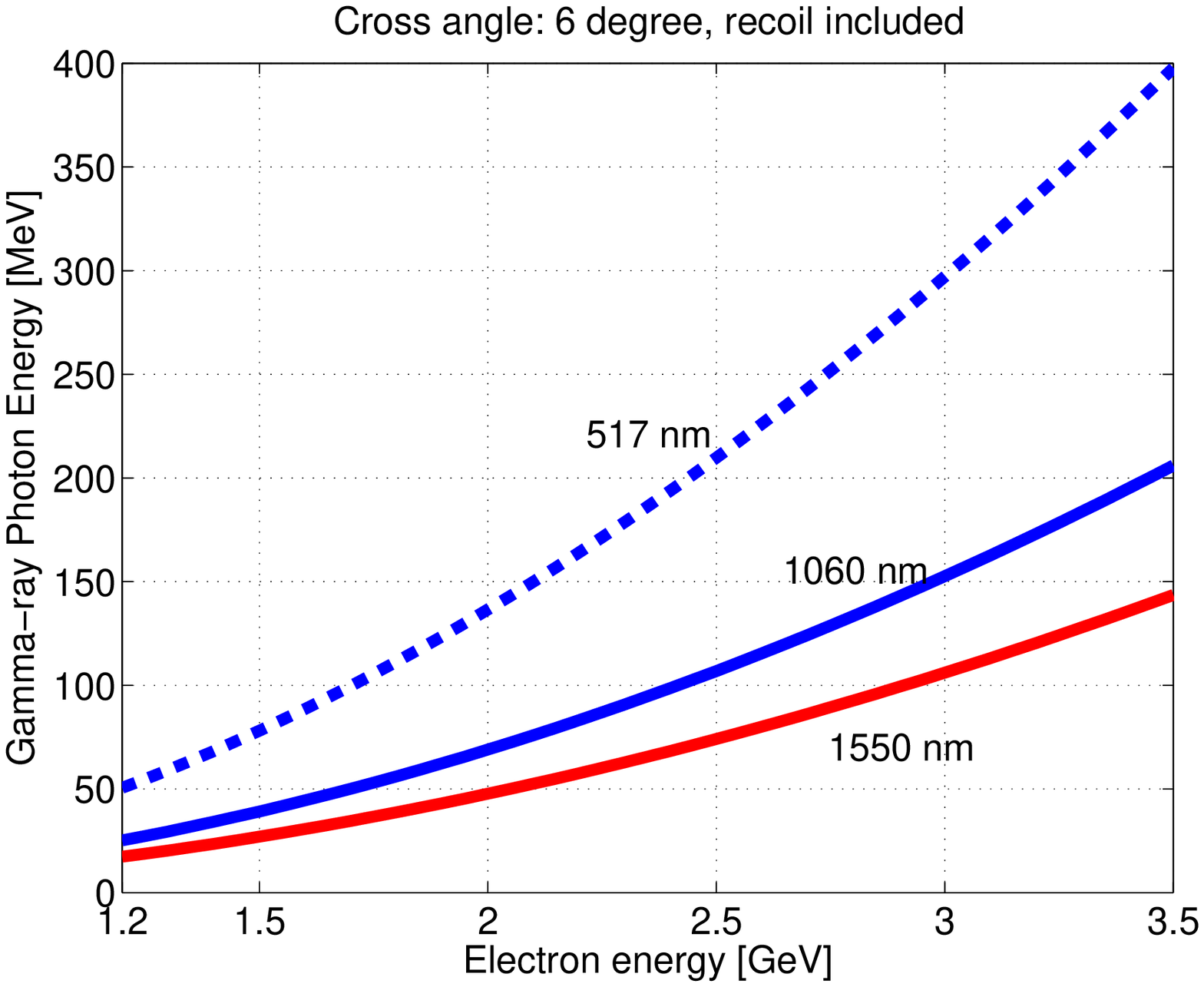}
\par\end{center}
\vspace{-0.2in}
\noindent \begin{center}
{\footnotesize{}}%
\begin{tabular}{|c||c|c|c|c|}
\hline 
{\footnotesize{}E-beam} & {\footnotesize{}$E_{\gamma}$ } & {\footnotesize{}$\lambda_{1}$ (nm) } & {\footnotesize{}$\lambda_{2}$ (nm) } & {\footnotesize{}$\lambda_{2}/3$}\tabularnewline
{\footnotesize{}(GeV)} & {\footnotesize{}(MeV)} & {\footnotesize{}$1064$ } & {\footnotesize{}$1550$ } & {\footnotesize{}$517$ }\tabularnewline
\hline 
{\footnotesize{}$1.2$ } & {\footnotesize{} $E_{\gamma,{\rm min}}$} & {\footnotesize{}$25$ } & {\footnotesize{}$17$ } & {\footnotesize{}$51$ }\tabularnewline
\hline 
{\footnotesize{}$3.5$} & {\footnotesize{}$E_{\gamma,{\rm max}}$ } & {\footnotesize{}$205$ } & {\footnotesize{}$144$ } & {\footnotesize{}$398$ }\tabularnewline
\hline 
\end{tabular}
\par\end{center}{\footnotesize \par}%
\end{minipage}%
\begin{minipage}[c][1\totalheight][t]{0.45\columnwidth}%
\begin{center}
\begin{tabular}{ll}
\hline 
\textbf{\footnotesize{}Electron beam} & \tabularnewline
{\footnotesize{}Beam energy} & {\footnotesize{}$3.27$ GeV}\tabularnewline
{\footnotesize{}Stored currents} & {\footnotesize{}$500$ mA}\tabularnewline
{\footnotesize{}Bunch filled} & {\footnotesize{}$64$}\tabularnewline
{\footnotesize{}Hori./Vert. emittance} & {\footnotesize{}$1.94/0.16$ nm-rad}\tabularnewline
{\footnotesize{}Hori./Vert. size (rms)} & {\footnotesize{}$100/26$ $\mu\mathrm{m}$}\tabularnewline
{\footnotesize{}Bunch length (rms)} & {\footnotesize{}$60$ ps}\tabularnewline
\hline 
\textbf{\footnotesize{}Laser beam} & \tabularnewline
{\footnotesize{}Wavelength} & {\footnotesize{}$517$ nm}\tabularnewline
{\footnotesize{}Intracavity power} & {\footnotesize{}$20$ kW}\tabularnewline
{\footnotesize{}Pulse length (rms)} & {\footnotesize{}$20$ ps}\tabularnewline
{\footnotesize{}Hori./Vert. size (rms)} & {\footnotesize{}$40/40$ $\mu\mathrm{m}$}\tabularnewline
\hline 
\textbf{\footnotesize{}Gamma-ray beam} & \tabularnewline
{\footnotesize{}Max. energy} & {\footnotesize{}$350$ MeV}\tabularnewline
{\footnotesize{}Collision rate} & {\footnotesize{}$95.42$ MHz}\tabularnewline
{\footnotesize{}Collision angle} & {\footnotesize{}$6^{\circ}$}\tabularnewline
{\footnotesize{}Luminosity} & {\footnotesize{}$5.7\times10^{35}\ \mathrm{cm}^{-2}\mathrm{s}^{-1}$}\tabularnewline
{\footnotesize{}Total flux (in $4\pi$ solid angle)} & {\footnotesize{}$3.4\times10^{11}$ $\gamma/\mathrm{s}$}\tabularnewline
\hline 
\end{tabular}\vspace{-0.2in}
\par\end{center}
\noindent \begin{center}
{\footnotesize{}}%
\begin{tabular}{|c|r|l|}
\hline 
{\footnotesize{}E-beam: } & \multicolumn{1}{r}{{\footnotesize{}Laser beam:}} & {\footnotesize{}$\lambda=$$517$ nm; }\tabularnewline
{\footnotesize{}$E=3.27$ GeV, $I=0.5$ A } & \multicolumn{1}{r}{{\footnotesize{}Beam size:}} & {\footnotesize{}$40/40$ $\mu\mathrm{m}$}\tabularnewline
\hline 
{\footnotesize{}FP cavity power (kW)} & {\footnotesize{}$20$ (kW)} & {\footnotesize{}$100$ (kW)}\tabularnewline
\hline 
{\footnotesize{}Tot. flux ($\gamma/\mathrm{s}$): $\mbox{\ensuremath{\theta}=}6^{\circ}$ } & {\footnotesize{}$3.4\times10^{11}$} & {\footnotesize{} $1.7\times10^{12}$}\tabularnewline
\hline 
\end{tabular}
\par\end{center}{\footnotesize \par}%
\end{minipage}
\par\end{centering}
\caption{\label{fig:SRCGSHighEngParams}(Left) Energy of the $\gamma$-ray
beam as a function of electron beam energy ($1.2$\textendash $3.5$
MeV) for lasers at $1064$ nm, $1550$ nm and its third harmonic at
$517$ nm with a $6^{\circ}$ collision angle. (Upper-Right) A table
listing the main parameters for producing a $350$ MeV $\gamma$-ray
beam using a $3.27$ GeV electron beam and a $517$ nm laser beam.
(Lower-Right) A table showing the total flux for two levels of intracavity
laser powers.}
\end{figure}

$\gamma$ rays in a wide energy range can be produced from $25$ MeV
to about $400$ MeV (see the table in Fig. \ref{fig:SRCGSHighEngParams}).
This can be done by operating the storage ring at energies between
$1.2$ and $3.5$ GeV, and by using one or more Fabry-Perot cavities
at a few selected wavelengths between $1550$ nm and $517$ nm. 

\subparagraph{Flux Performance}

The flux requirements in Table \ref{tab:HighEngBeamRequirements}
can be realized readily. A $\gamma$-ray beam with a total flux of
about $3.4\times10^{11}$ $\gamma$/s (in the $4\pi$ solid angle)
can be produced in a wide energy range. For example, a $350$ MeV
$\gamma$-ray beam of this intensity with a $95.4$ MHz repetition
rate can be produced using a $500$ mA electron beam ($3.27$ GeV)
and a modest $20$ kW laser power inside a $517$ nm FP cavity (see
the detailed beam parameters in the table in Fig. \ref{fig:SRCGSHighEngParams}).
The corresponding collimated flux for a beam with a FWHM energy resolution
of $2$\% is about $9\times10^{9}$ $\gamma$/s. For those experiments
requiring a lower repetition rate, $6$ out of $64$ rf buckets can
be filled with the same bunch charge to produce a roughly $9.5$ MHz
beam with a collimated flux of about $8\times10^{8}$ $\gamma$/s,
and higher flux is possible by increasing the bunch charge. Furthermore,
by increasing the intracavity laser power to $100$ kW to match the
capability of a high charge, high repetition rate booster injector
($32$ nC/cycle, $10$ Hz), the total $\gamma$-ray flux can be increased
to about $2\times10^{12}$ $\gamma$/s. Ultimately, the $\gamma$-ray
flux for this medium energy CGS will be limited by the electron injection
rate, not by the laser power in the FP cavity or the stored beam current
in the storage ring. 


\subparagraph*{Electron Beam}

To achieve a high average current, the electron beam will be stored
in multiple bunches. The bunch pattern should be selectable as demanded
for specific user experiments for either high flux operation at a
high repetition rate, or a modest flux operation at a lower rate.

\subparagraph*{Laser Beam}

The Fabry-Perot cavity will be powered by a commercial laser (fiber
or solid state) operated at tens to hundreds of MHz repetition rate
with a picosecond pulse duration. The commonly available wavelengths
are around $1064$ and 1550 nm; their 2nd and/or 3rd harmonics can
also be used with a lower power. 

\subparagraph*{Fabry-Perot Cavity and Collision Scheme}

For this medium energy gamma-ray source, the required intracavity laser
power is rather modest, reducing the need to push the FP cavity power
limit. Both head-on collision and small cross-angle collision can
be arranged using either a two-mirror or four-mirror cavity. The entire
$\gamma$-ray energy range can be covered using one Fabry-Perot cavity
specially designed to build up the laser power at both the fundamental
wavelength at $1550$ nm and its third harmonic at $517$ nm (see
the table in Fig. \ref{fig:SRCGSHighEngParams}). To cover a wide
intermediate energy range (from $30$ to $200$ MeV), a $1064$ nm
Fabry-Perot cavity is well suited and highly practical because the
FP cavity technology at $1064$ nm is well established and more mature
than that for $1550$ nm.

\subparagraph*{Energy Resolution}

The required $\gamma$-ray beam energy resolution ($1.5$\textendash $2$\%,
FWHM) can be met by using a narrow-band laser beam, and an electron
beam with a small energy spread and reasonably small transverse emittance
\cite{adriani-2014,Alesini-2014}. The electron beam performance is
readily achievable using the state-of-the-art storage ring magnetic
optics design, taking advantage of recent advances in developing low-emittance
storage ring based synchrotron radiation sources \cite{tarawneh-2003,farvacque-2013}. 

\subparagraph*{Beam Polarization}

A well-collimated Compton $\gamma$-ray beam driven by a polarized
laser beam can have a very high degree of polarization, either linear
or circular. The relatively slow polarization manipulation (up to
$10$ Hz), switching between two polarization states (the horizontal
and vertical, or left and right circular), can be realized using conventional
polarizing optics (including Pockels cells) outside the FP cavity
where the laser power is much lower.

\subparagraph*{Production Modes}

The medium energy gamma-ray source is likely to be operated as a single-
or few-user facility. Several Compton sources can be installed in
one or more straight sections as shown in Fig. \ref{fig:LowEngStorageRingLayout}.
When operated simultaneously, the sum of $\gamma$-ray flux from all
sources will be limited by the electron injection rate. For each gamma-ray
source, a dedicated user target room with proper shielding and a beam
dump should be constructed. Multiple target rooms are desirable as
complex mdeium energy experiments will need a dedicated staging area
to develop, test, and integrate experimental equipment while the other
target room(s) is/are being used for production data-taking.

Several challenges and important issues for this medium energy source
are summarized here:
\begin{itemize}
\item \textbf{\textit{Electron injection rate}} 

The cost-effective choice for the injection is a full-energy, top-off
booster injector. The booster injector needs to be designed to achieve
high reliability, a fast injection cycle (up to $10$ Hz), and a large
average injection rate (few $10^{11}$ to few $10^{12}$ $e^{-}$/s). 
\item \textbf{\textit{Dual wavelength Fabry-Perot cavity}}

A dual wavelength Fabry-Perot cavity system will need to be developed
to operate at both $1550$ nm and $517$ nm (the 3rd harmonic). The
cavity mirrors should have very high reflectivity at both wavelengths.
A new type of optical bench with additional third-harmonic generation
optics, two sets of laser optics, feedbacks, and diagnostics for both
wavelengths should be developed and then integrated with a shared
FP cavity.
\item \textbf{\textit{Low beam background}}

To reduce the high energy radiation background, a short interaction
region can be arranged to fully separate the $\gamma$-ray beam from
the residual gas bremsstrahlung radiation, with two possible schemes
illustrated in Fig. \ref{fig:HighEngStorageRingLayout}. This, together
with ultra-high vacuum realized in advanced storage rings \cite{eriksson-2013},
will reduce the radiation background by at least two orders of magnitude
compared to the already very low background achieved at the HIGS facility.
\end{itemize}
The following is a prioritized list of important R\&D topics for such
a medium-energy CGS:
\begin{itemize}
\item \textbf{\textit{High priority}}
\begin{itemize}
\item To develop a dual-wavelength high-finesse FP-cavity for $1550$ nm
and $517$ nm;
\item To design a reliable injector for single- and multi-bunch injection
with high charge; 
\end{itemize}
\item \textbf{\textit{Medium priority}}
\begin{itemize}
\item To design an energy-varying storage ring with small emittance and
large dynamic aperture; 
\item To design a specialized interaction region with ultra-high vacuum
either in an arc section or in a chicane to minimize radiation background;
\item To develop laser beam optics for polarization switch (circular and
linear), and for dual-wavelength operation;
\item To design radiation shielding to handle high local electron loss.
\end{itemize}
\end{itemize}

\newpage

\subsection{Storage Ring Based Low Energy CGS}

\begin{figure}[H]
\noindent \begin{centering}
\begin{minipage}[t][1\totalheight][c]{0.54\columnwidth}%
\noindent \begin{center}
\includegraphics[width=1\textwidth]{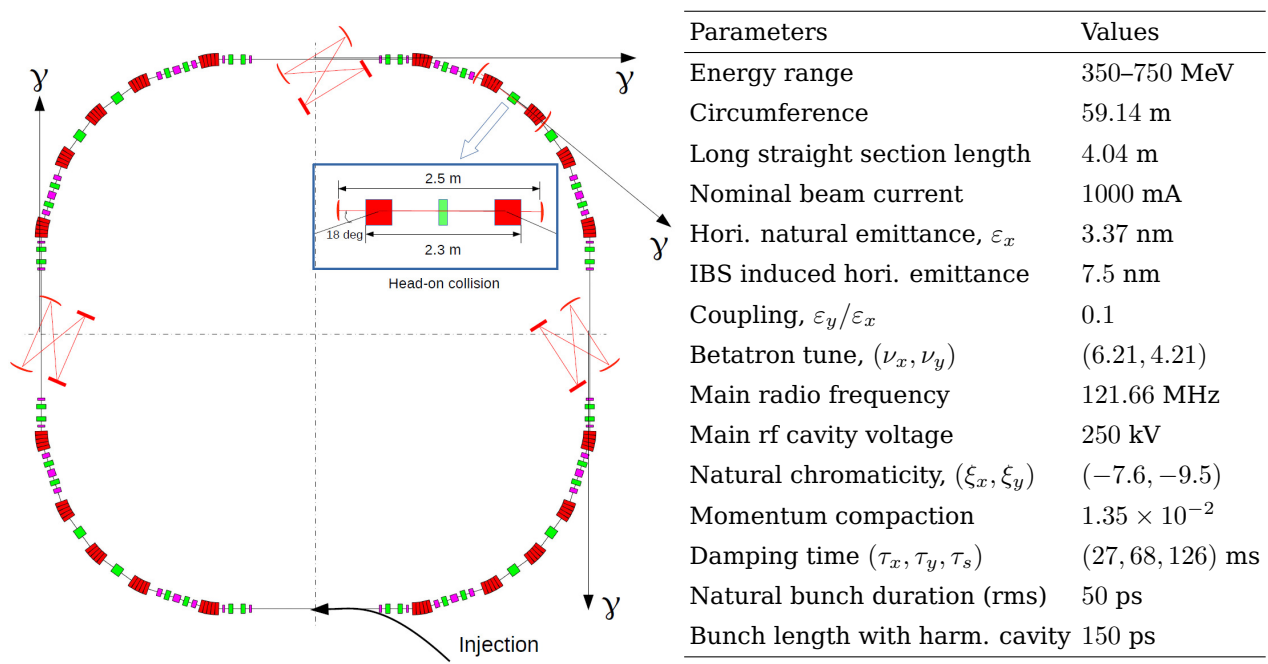} 
\par\end{center}%
\end{minipage}%
\begin{minipage}[t]{0.45\columnwidth}%
\noindent \begin{center}
\begin{tabular}{ll}
\hline 
{\footnotesize{}Parameters}  & {\footnotesize{}Values}\tabularnewline
\hline 
{\footnotesize{}Energy range}  & {\footnotesize{}$350$\textendash $750$ MeV}\tabularnewline
{\footnotesize{}Circumference}  & {\footnotesize{}$59.14$ m}\tabularnewline
{\footnotesize{}Long straight section length}  & {\footnotesize{}$4.04$ m}\tabularnewline
{\footnotesize{}Nominal beam current}  & {\footnotesize{}$1000$ mA}\tabularnewline
{\footnotesize{}Hori. natural emittance, $\varepsilon_{x}$}  & {\footnotesize{}$3.37$ nm}\tabularnewline
{\footnotesize{}IBS induced hori. emittance}  & {\footnotesize{}$7.5$ nm}\tabularnewline
{\footnotesize{}Coupling, $\varepsilon_{y}/\varepsilon_{x}$}  & {\footnotesize{}$0.1$}\tabularnewline
{\footnotesize{}Betatron tune, $\left(\nu_{x},\nu_{y}\right)$}  & {\footnotesize{}$(6.21,4.21)$}\tabularnewline
{\footnotesize{}Main radio frequency}  & {\footnotesize{}$121.66$ MHz}\tabularnewline
{\footnotesize{}Main rf cavity voltage}  & {\footnotesize{}$250$ kV}\tabularnewline
{\footnotesize{}Natural chromaticity, $\left(\xi_{x},\xi_{y}\right)$}  & {\footnotesize{}$\left(-7.6,-9.5\right)$}\tabularnewline
{\footnotesize{}Momentum compaction}  & {\footnotesize{}$1.35\times10^{-2}$}\tabularnewline
{\footnotesize{}Damping time $\left(\tau_{x},\tau_{y},\tau_{s}\right)$}  & {\footnotesize{}$\left(27,68,126\right)$ ms}\tabularnewline
{\footnotesize{}Natural bunch duration (rms)}  & {\footnotesize{}$50$ ps}\tabularnewline
{\footnotesize{}Bunch length with harm. cavity}  & {\footnotesize{}$150$ ps}\tabularnewline
\hline 
\end{tabular}
\par\end{center}%
\end{minipage}
\par\end{centering}
\vspace{-0.2in}

\caption{\label{fig:LowEngStorageRingLayout} (Left) Layout and parameters
of a low energy electron storage ring ($350$\textendash $750$ MeV).
Four gamma-ray sources are shown in two different collision configurations:
one for the head-on collision using a two-mirror Fabry-Perot cavity
in the arc and the other with a small crossing angle using a four-mirror
cavity in the straight section. (Right) Main parameters for the storage
ring, where all energy-dependent parameters are shown for $500$ MeV
operation.}
\end{figure}

The required $\gamma$-ray beam performance at lower energies (Table
\ref{tab:LowEngBeamRequirements}) can be realized with a storage
ring based Compton gamma-ray source using a low energy electron beam
($100$s of MeV) and multiple high power Fabry-Perot cavities. One
possible design of such storage rings is shown in Fig. \ref{fig:LowEngStorageRingLayout}
with related machine design parameters listed. This $60$-meter-long
storage ring is comprised of four 5-bend achromatic arcs with a small
emittance and a reasonably large dynamic aperture.

For this low energy storage ring design, the intrabeam scattering
(IBS) effect needs to be properly controlled/mitigated. Intrabeam
scattering is a process in which lossless collisions between electrons
in the same bunch lead to an increase of the electron beam distribution
in all three dimensions, which will result in the transverse beam
size growth, especially for low energy and small-emittance storage
rings. Because of a strong IBS effect, there is no need to push the
low emittance limit. One such an example storage ring design is shown
in Fig. \ref{fig:LowEngStorageRingLayout}. At $500$ MeV, this ring
has a horizontal natural emittance about $3.4$ nm-rad and the IBS
effect increases the beam emittance to about $7.5$ nm-rad with $1000$
mA beam current in $24$ bunches, and with the help of a third harmonic
rf cavity. The third-order harmonic cavity is used to lengthen the
electron bunch by a factor of three in order to increase the beam
lifetime which is mainly limited by large angle electron-electron
collisions (the so-called Touschek effect). A more complete design
of a low energy storage ring based CGS with a modest intracavity power
($20$ kW) can be found in \cite{pan2019design}. 

\vspace{-0.1in}

\begin{figure}[H]
\noindent \begin{centering}
\begin{minipage}[c][1\totalheight][t]{0.54\columnwidth}%
\noindent \begin{center}
\includegraphics[width=0.95\textwidth]{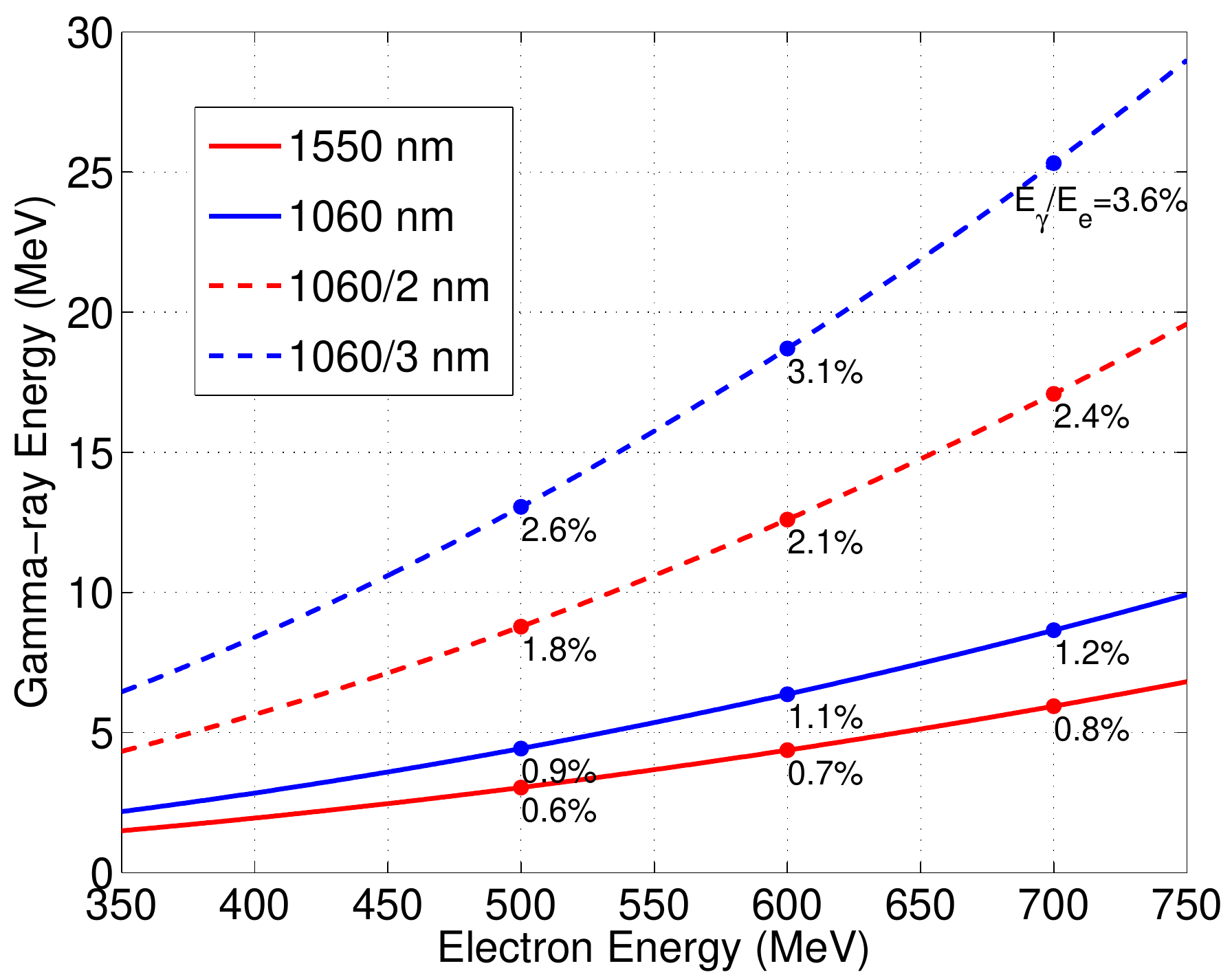}
\vspace{-0in}
\par\end{center}
\noindent \begin{center}
{\footnotesize{}}%
\begin{tabular}{|c||c|c|c|c|c|}
\hline 
{\footnotesize{}E-beam} & {\footnotesize{}$E_{\gamma}$ } & {\footnotesize{}$\lambda_{1}$ (nm)} & {\footnotesize{}$\lambda_{1}/3$} & {\footnotesize{}$\lambda_{2}$ (nm)} & {\footnotesize{}$\lambda_{2}/3$}\tabularnewline
{\footnotesize{}(MeV)} & {\footnotesize{}(MeV)} & {\footnotesize{}$1064$ } & {\footnotesize{}$355$ } & {\footnotesize{}$1550$ } & {\footnotesize{}$517$ }\tabularnewline
\hline 
{\footnotesize{}$350$} & {\footnotesize{} $E_{\gamma,{\rm min}}$} & {\footnotesize{}$2.2$ } & {\footnotesize{}$6.4$ } & {\footnotesize{}$1.5$ } & {\footnotesize{}$4.4$ }\tabularnewline
\hline 
{\footnotesize{}$750$ } & {\footnotesize{}$E_{\gamma,{\rm max}}$} & {\footnotesize{}$9.9$ } & {\footnotesize{}$29$ } & {\footnotesize{}$6.8$ } & {\footnotesize{}$20$ }\tabularnewline
\hline 
\end{tabular}
\par\end{center}{\footnotesize \par}%
\end{minipage}%
\begin{minipage}[c][1\totalheight][t]{0.45\columnwidth}%
\begin{center}
\begin{tabular}{ll}
\hline 
\textbf{\footnotesize{}Electron beam} & \tabularnewline
{\footnotesize{}Beam energy} & {\footnotesize{}$500$ MeV}\tabularnewline
{\footnotesize{}Stored currents} & {\footnotesize{}$1000$ mA}\tabularnewline
{\footnotesize{}Bunch filled} & {\footnotesize{}$24$}\tabularnewline
{\footnotesize{}Hori./Vert. emittance} & {\footnotesize{}$7.5/0.75$ nm-rad}\tabularnewline
{\footnotesize{}Hori./Vert. size (rms)} & {\footnotesize{}$212/39$ $\mu\mathrm{m}$}\tabularnewline
{\footnotesize{}Bunch length (rms)} & {\footnotesize{}$150$ ps}\tabularnewline
\hline 
\textbf{\footnotesize{}Laser beam} & \tabularnewline
{\footnotesize{}Wavelength} & {\footnotesize{}$1064$ nm}\tabularnewline
{\footnotesize{}Intracavity power} & {\footnotesize{}$100$ kW}\tabularnewline
{\footnotesize{}Pulse length (rms)} & {\footnotesize{}$20$ ps}\tabularnewline
{\footnotesize{}Hori./Vert. size (rms)} & {\footnotesize{}$40/40$ $\mu\mathrm{m}$}\tabularnewline
\hline 
\textbf{\footnotesize{}Gamma-ray beam} & \tabularnewline
{\footnotesize{}Max. energy} & {\footnotesize{}$4.43$ MeV}\tabularnewline
{\footnotesize{}Collision rate} & {\footnotesize{}$121.66$ MHz}\tabularnewline
{\footnotesize{}Collision angle} & {\footnotesize{}$6^{\circ}$}\tabularnewline
{\footnotesize{}Luminosity} & {\footnotesize{}$3.3\times10^{36}\ \mathrm{cm}^{-2}\mathrm{s}^{-1}$}\tabularnewline
{\footnotesize{}Total flux (in $4\pi$ solid angle)} & {\footnotesize{}$2.2\times10^{12}$ $\gamma/\mathrm{s}$}\tabularnewline
\hline 
\end{tabular}
\par\end{center}
\noindent \begin{center}
\vspace{-0.5in}
\par\end{center}
\noindent \begin{center}
{\footnotesize{}}%
\begin{tabular}{|c|c|c|}
\hline 
{\footnotesize{}E-beam: } & \multicolumn{1}{c}{{\footnotesize{}FP cavity:}} & {\footnotesize{}$100$ kW}\tabularnewline
{\footnotesize{}$E=500$ MeV, $I=1$ A } & \multicolumn{1}{c}{{\footnotesize{}Beam size:}} & {\footnotesize{}$40/40$ $\mu$m}\tabularnewline
\hline 
{\footnotesize{}Laser wavelength (nm)} & {\footnotesize{}$\lambda{}_{1}=1064$ } & {\footnotesize{}$\lambda_{2}=1550$ }\tabularnewline
\hline 
{\footnotesize{}Tot. flux ($\gamma/\mathrm{s}$): $\mbox{\ensuremath{\theta}=}6^{\circ}$ } & {\footnotesize{}$2.2\times10^{12}$} & {\footnotesize{} $2.8\times10^{12}$}\tabularnewline
\hline 
{\footnotesize{}Tot. flux ($\gamma/\mathrm{s}$): head-on} & {\footnotesize{} $2.4\times10^{13}$} & {\footnotesize{}$3.1\times10^{13}$}\tabularnewline
\hline 
\end{tabular}
\par\end{center}{\footnotesize \par}%
\end{minipage}
\par\end{centering}
\caption{\label{fig:SRCGSLowEngParams} (Left) Energy of the $\gamma$-ray
beam as a function of electron beam energy ($350$\textendash $750$
MeV) for lasers at $1064$ and $1550$ nm and their third harmonics.
(Upper-Right) A table listing the key operational parameters for producing
a $4.4$ MeV $\gamma$-ray beam using a $500$ MeV electron beam and
a $1064$ nm laser beam with a $6^{\circ}$ cross angle. (Lower-Right)
A table showing the total flux for $6^{\circ}$ cross-angle and head-on
collision configurations with the same electron and laser beam parameters.}
\end{figure}

\vspace{-0.1in}

\subparagraph*{Energy Range}

This storage ring based Compton source is a versatile gamma-ray factory.
It can produce $\gamma$ rays in a wide range of energies from $1.5$
to about $30$ MeV (see the table in Fig. \ref{fig:SRCGSLowEngParams}).
This is done by changing the electron beam energy from $350$ to $750$
MeV, and by using FP cavities driven by commercial lasers at multiple
wavelengths. 

\subparagraph*{Flux Performance}

At the low energy end, the $\gamma$-ray flux is determined by the
electron beam current stored in the storage ring, intracavity laser
power, and collision configuration. Below $5$ to $8$ MeV, very high
flux performance is available using a high power Fabry-Perot cavity
driven by lasers operating at their fundamental wavelengths (e.g.
$1064$ or $1550$ nm). For example, for production of $4.4$ MeV
$\gamma$ rays in a small cross-angle collision configuration ($\theta=6^{\circ}$)
using a $1000$ mA, $500$ MeV electron beam and a $100$ kW laser
beam inside a $1064$ nm FP cavity, the total $\gamma$-ray flux is
estimated to be $2.2\times10^{12}$~$\gamma$/s, and the corresponding
flux on target with $2\%$ energy resolution (FWHM) is about $6.5\times10^{10}$
~$\gamma$/s (see a list of consistent operational parameters in
the table in Fig. \ref{fig:SRCGSLowEngParams}). Using the same electron
and laser beam parameters, the total flux can be increased to $2$\textendash $3\times10^{13}$~$\gamma$/s,
assuming head-on collision can be realized. Between $15$ and $30$
MeV using an FP cavity driven by either a 2nd or 3rd harmonic laser
beam, the gamma-ray source will be operated in the electron-loss mode,
producing a lower total flux, on the order of few $10^{11}$~$\gamma$/s
as limited by the electron injection rate. Between $5$ and $15$
MeV, the total $\gamma$-ray flux will range from few $10^{11}$ to
few $10^{12}$~$\gamma$/s (or higher), determined by specific choices
of operational parameters. 

\subparagraph*{Electron Beam}

To achieve a high average current, the electron beam will be stored
in multiple bunches, typically with a vacuum clearing gap. This gap
in the $\gamma$-ray beam can be used as the start of the time trigger
for many experiments. In the example storage ring (Fig. \ref{fig:LowEngStorageRingLayout}),
when fully filled, the electron bunch rate is $122$ MHz. 

\subparagraph*{Laser Beam}

The Fabry-Perot cavities will be powered by commercial MHz lasers
(fiber or solid state) with commonly available wavelengths around
$1064$ and 1550 nm. The 2nd and/or 3rd harmonic laser beam can also
be used. The lasers with high average power and excellent beam quality
will be essential for achieving high flux.

\subparagraph*{Fabry-Perot Cavity and Collision Scheme}

Both head-on collision and small cross-angle collision can be arranged.
For the head-on collision, a simple and compact two-mirror cavity
can be used; it may be possible to install such a cavity in an arc
section between two adjacent dipole magnets as shown in Fig.~\ref{fig:LowEngStorageRingLayout}.
For the cross-angle collision, a four-mirror cavity can be implemented
in a dedicated straight section.

\subparagraph*{Energy Resolution}

A beam energy resolution of about $2$\% (FWHM) can be readily realized
by properly collimating the $\gamma$-ray beam. To achieve higher
resolution ($\le1$\%), a narrow-band drive laser will be required
and the resolution impact due to the electron beam energy spread and
angular spread should be minimized.

\subparagraph*{Beam Polarization}

Highly polarized, well-collimated Compton $\gamma$-ray beams can
be produced using a polarized drive laser beam. For certain experiments
requiring helicity switch (as fast as $50$ Hz), polarizing optics
(including Pockels cells) can be utilized outside the FP cavity.

\subparagraph*{Production Modes}

Like a storage ring based synchrotron radiation user facility, this
low energy gamma-ray factory can be operated as a multiple user facility
with several $\gamma$-ray beamlines. As shown in Fig. \ref{fig:LowEngStorageRingLayout},
several Compton sources can be installed in straight sections and
arc sections. For each source, a dedicated user target room with proper
shielding and a beam dump will be constructed. The gamma-ray facility
can be operated in the single- or multi-user mode. In the single-user
mode, the experiment on the floor can take full control of the $\gamma$-ray
beam production, including having the ability to change both electron
beam current and energy. In the multi-user mode, all gamma-ray sources
share the electron beam, therefore, individual beamlines do not have
the control of the electron beam current and energy. Beamlines can
be developed with the independent energy tuning capability\textemdash this
can be realized by constructing a special FP cavity which allows variation
of the collision angle. It is worth pointing out that the flux will
be reduced as the collision angle is increased. This gamma-ray factory
provides a full energy coverage ($1.5$ to $30$ MeV) by operating
multiple beamlines using lasers of several different wavelengths and
by running the storage ring at a few pre-determined energies.

Several challenges and important issues for this low energy source
are summarized here:
\begin{itemize}
\item \textbf{\textit{High power FP cavity design}}

The main challenge for this low energy source is to develop a very
high power FP cavity ($\ge100$ kW) operating at $50$ to $200$ MHz
with a very small beam size at the collision point (radius $\sim$$40$
$\mu$m or smaller). The smallest beam size practically realizable
will be limited by the high intracavity power.
\item \textbf{\textit{Dual wavelength FP cavity}}

Any gamma-ray source on this storage ring can greatly expand its energy
coverage by using a dual-wavelength Fabry-Perot cavity (e.g. at both
$1550$ nm and $517$ nm). 
\item \textbf{\textit{Electron injection rate}} 

There are two viable choices for the injector: (1) a low-cost, full-energy,
top-off booster injector; and (2) a full-energy room-temperature linac
injector at a higher cost. At low energies, the $\gamma$-ray flux
is not limited by the injection rate. At higher energies in the electron-loss
mode, the $\gamma$-ray flux will be limited by the electron injection
rate.
\end{itemize}
The following is a prioritized list of important R\&D topics for such
a low energy CGS:
\begin{itemize}
\item \textbf{\textit{High priority}}

\begin{itemize}
\item To develop a very high power FP cavity ($\ge100$ kW) with a small
beam size ($\le40$ $\mu$m);
\item To develop a dual-wavelength high-finesse FP-cavity (e.g. $1550$/$517$
nm);
\end{itemize}
\item \textbf{\textit{Medium priority}}

\begin{itemize}
\item To design an energy-varying storage ring with small emittance and
large dynamic aperture; 
\item To design a specialized interaction region with ultra-high vacuum
either in an arc section or in a chicane to minimize radiation background;
\item To develop laser beam optics for fast polarization switch (circular
and linear), and for dual-wavelength operation;
\item To design radiation shielding to handle large local electron loss.
\end{itemize}
\end{itemize}

\newpage

\subsection{Energy Recovery Linac Based Low-Energy CGS}

\begin{figure}[H]
\noindent \begin{centering}
\begin{minipage}[t][1\totalheight][c]{0.54\columnwidth}%
\noindent \begin{center}
\includegraphics[width=1\textwidth]{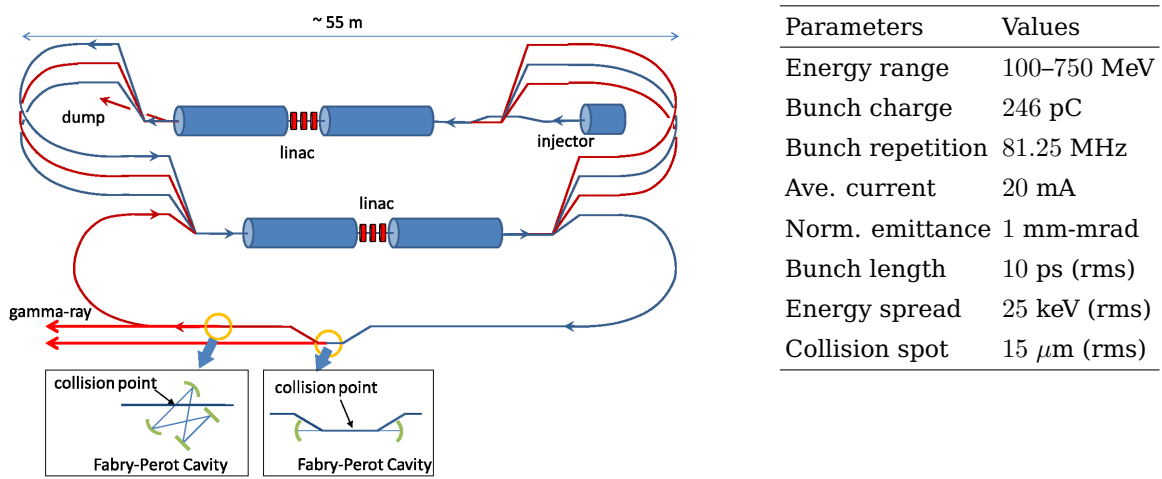} 
\par\end{center}%
\end{minipage}%
\begin{minipage}[t]{0.45\columnwidth}%
\noindent \begin{center}
\begin{tabular}{ll}
\hline 
{\footnotesize{}{}Parameters}  & {\footnotesize{}{}Values}\tabularnewline
\hline 
{\footnotesize{}{}Energy range}  & {\footnotesize{}{}$100$\textendash $750$ MeV}\tabularnewline
{\footnotesize{}{}Bunch charge}  & {\footnotesize{}{}$246$ pC}\tabularnewline
{\footnotesize{}{}Bunch repetition}  & {\footnotesize{}{}$81.25$ MHz}\tabularnewline
{\footnotesize{}{}Ave. current}  & {\footnotesize{}{}$20$ mA}\tabularnewline
{\footnotesize{}{}Norm. emittance}  & {\footnotesize{}{}$1$ mm-mrad}\tabularnewline
{\footnotesize{}{}Bunch length}  & {\footnotesize{}{}$10$ ps (rms)}\tabularnewline
{\footnotesize{}{}Energy spread}  & {\footnotesize{}{}$25$ keV (rms)}\tabularnewline
{\footnotesize{}{}Collision spot}  & {\footnotesize{}{}$15$ $\mu$m (rms)}\tabularnewline
\hline 
\end{tabular}
\par\end{center}%
\end{minipage}
\par\end{centering}
\vspace{-0in}

\caption{\label{fig:ERLBasedCGS} (Left) Schematic layout and parameters of
an ERL-based gamma source. The insets show two possible collision
schemes. (Right) Electron beam parameters for the ERL.}
\end{figure}

The required $\gamma$-ray beam performance at lower energies (Table
\ref{tab:LowEngBeamRequirements}) can also be realized with a Compton
gamma-ray source based on an Energy Recovery Linac and one or more
Fabry-Perot laser cavities. One possible design of such gamma source
is shown in Fig. \ref{fig:ERLBasedCGS} with related machine parameters
listed. This ERL is comprised of a double-sided linac with two-pass
recirculation loop and an additional loop for $\gamma$-ray beam generation.

The key components of ERL are similar to the Compact ERL \cite{akagi-2016},
in which generation of a narrow-band Compton scattered beam was recently
demonstrated. The main linac and the injector linac are both L-band
($1.3$ GHz) superconducting structures. The electron gun is a high-voltage
($500$ kV) DC gun with a multi-alkali photocathode, which can generate
a small-emittance beam, $\varepsilon_{n}\le1$~mm-mrad, at a bunch
charge of $200$\textendash $300$ pC \cite{guilliford-2015}.

\subparagraph*{Flux Performance}

\vspace{-0.1in}

\begin{figure}[H]
\noindent \begin{centering}
\begin{minipage}[c][1\totalheight][t]{0.4\columnwidth}%
\noindent \begin{center}
\includegraphics[width=1\textwidth]{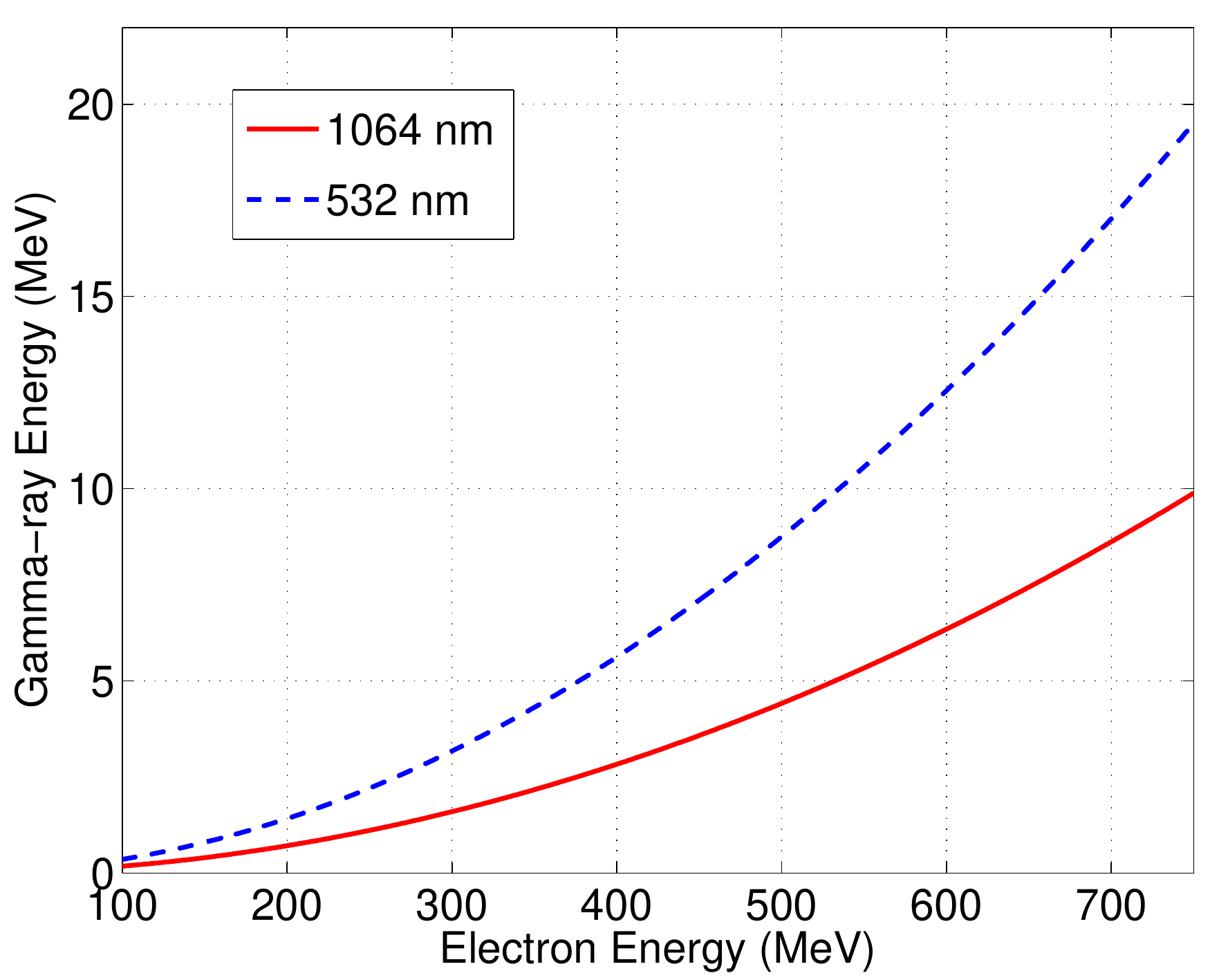} \vspace{-0in}
\par\end{center}
\noindent \begin{center}
{\footnotesize{}}%
\begin{tabular}{|c||c|c|c|}
\hline 
{\footnotesize{}E-beam} & {\footnotesize{}$E_{\gamma}$ } & {\footnotesize{}$\lambda_{1}$ (nm)} & {\footnotesize{}$\lambda_{1}/2$}\tabularnewline
{\footnotesize{}(MeV)} & {\footnotesize{}(MeV)} & {\footnotesize{}$1064$ } & {\footnotesize{}$532$ }\tabularnewline
\hline 
{\footnotesize{}$100$} & {\footnotesize{} $E_{\gamma,{\rm min}}$} & {\footnotesize{}$0.18$} & {\footnotesize{}$0.36$}\tabularnewline
\hline 
{\footnotesize{}$750$ } & {\footnotesize{}$E_{\gamma,{\rm max}}$} & {\footnotesize{}$9.9$} & {\footnotesize{}$19.5$}\tabularnewline
\hline 
\end{tabular}
\par\end{center}{\footnotesize \par}%
\end{minipage}%
\begin{minipage}[c][1\totalheight][t]{0.58\columnwidth}%
\begin{center}
\begin{tabular}{ll}
\hline 
\textbf{\footnotesize{}Electron beam} & \tabularnewline
{\footnotesize{}Beam energy} & {\footnotesize{}$500$ MeV}\tabularnewline
{\footnotesize{}Average current} & {\footnotesize{}$20$ mA }\tabularnewline
{\footnotesize{}Hori./Vert. emittance (norm.)} & {\footnotesize{}$1$mm-mrad}\tabularnewline
{\footnotesize{}Hori./Vert. size (rms)} & {\footnotesize{}$15$ $\mu$m}\tabularnewline
{\footnotesize{}Bunch length (rms)} & {\footnotesize{}$3$ ps}\tabularnewline
\hline 
\textbf{\footnotesize{}Laser beam} & \tabularnewline
{\footnotesize{}Wavelength} & {\footnotesize{}$1064$ nm}\tabularnewline
{\footnotesize{}Intracavity power} & {\footnotesize{}$70$ kW }\tabularnewline
{\footnotesize{}Pulse length (rms)} & {\footnotesize{}$10$ ps }\tabularnewline
{\footnotesize{}Hori./Vert. size (rms)} & {\footnotesize{}$15/15$ $\mu\mathrm{m}$}\tabularnewline
\hline 
\textbf{\footnotesize{}Gamma-ray beam} & \tabularnewline
{\footnotesize{}Max. energy} & {\footnotesize{}$4.41$ MeV}\tabularnewline
{\footnotesize{}Collision rate} & {\footnotesize{}$81.25$ MHz}\tabularnewline
{\footnotesize{}Collision angle} & {\footnotesize{}$6^{\circ}$}\tabularnewline
{\footnotesize{}Luminosity} & {\footnotesize{}$2.6\times10^{36}\ \mathrm{cm}^{-2}\mathrm{s}^{-1}$}\tabularnewline
{\footnotesize{}Total flux (in $4\pi$ solid angle)} & {\footnotesize{}$1.7\times10^{12}$ $\gamma/\mathrm{s}$}\tabularnewline
\hline 
\end{tabular}
\par\end{center}
\noindent \begin{center}
\vspace{-0.4in}
\par\end{center}
\noindent \begin{center}
{\footnotesize{}}%
\begin{tabular}{|c|c|c|c|}
\hline 
{\footnotesize{}E-beam current (mA) } & {\footnotesize{}$20$ } & {\footnotesize{}$20$ } & {\footnotesize{}$40$ }\tabularnewline
{\footnotesize{}FP cavity power ( kW)} & {\footnotesize{}$70$ } & {\footnotesize{}$70$ } & {\footnotesize{}$70$ }\tabularnewline
{\footnotesize{}Laser beam size ($\mu\mathrm{m}$)} & {\footnotesize{}$30/30$ } & {\footnotesize{}$15/15$} & {\footnotesize{}$15/15$}\tabularnewline
\hline 
{\footnotesize{}Tot. flux ($\gamma/\mathrm{s}$): $\mbox{\ensuremath{\theta}=}6^{\circ}$ } & {\footnotesize{}$1.1\times10^{12}$} & {\footnotesize{}$1.7\times10^{12}$} & {\footnotesize{}$3.4\times10^{12}$}\tabularnewline
\hline 
{\footnotesize{}Tot. flux ($\gamma/\mathrm{s}$): head-on} & {\footnotesize{} $5.4\times10^{12}$} & {\footnotesize{}$1.3\times10^{13}$} & {\footnotesize{}$2.7\times10^{13}$}\tabularnewline
\hline 
\end{tabular}
\par\end{center}{\footnotesize \par}%
\end{minipage}
\par\end{centering}
\caption{\label{fig:ERLCGSLowEngParams} (Left) Energy of the $\gamma$-ray
beam as a function of electron beam energy ($100$\textendash $750$
MeV) for lasers at $1064$ and $532$ nm. (Upper-Right) A table listing
the key operational parameters for producing a $4.41$ MeV $\gamma$-ray
beam using a $500$ MeV electron beam and a $1064$ nm laser beam
with a $6^{\circ}$ cross angle. (Lower-Right) A table showing the
total flux for $6^{\circ}$ cross-angle and head-on collision configurations
for two sets of values for the electron beam current and laser beam
size at IP.}
\end{figure}

The performance of an ERL with an electron beam energy of 500 MeV
and a laser wavelength of $1064$ nm is shown in a table (see Fig.
\ref{fig:ERLCGSLowEngParams}).

\subparagraph*{Energy Range}

A wide $\gamma$-ray energy range can be achieved by changing the
electron energy and laser wavelength ($1064$ nm and $532$ nm). In
this design, the electron beam energy is varied from $100$ to $750$
MeV to cover $\gamma$-ray energies from 0.18 to 20 MeV. An electron
beam below $370$ MeV can be delivered by bending the beam out after
a single turn or single linac to simplify operation.

\subparagraph*{Laser Beam}

Because the energy range of the ERL is large, the laser system does
not need to adjust to wavelengths longer than 1064 nm. To get to higher
$\gamma$-ray energies it will be necessary to frequency-double the
laser light to $532$ nm using an LBO crystal.

\subparagraph*{Fabry-Perot Cavity and Collision Scheme}

Two types of Fabry-Perot cavities can be used for the $\gamma$-ray
generation with a head-on or small cross-angle configuration as shown
in Fig. \ref{fig:ERLBasedCGS}. In order to stack laser pulses synchronizing
with the electron bunch, the cavity frequency should be equal to the
bunch repetition rate ($81.25$~MHz) or its sub-harmonic. Parameters
of laser beam and expected $\gamma$-ray flux are summarized in a
table (part of Fig. \ref{fig:ERLCGSLowEngParams}). The flux requirement,
$10^{13}$ $\gamma$/s, can be obtained with head-on collision, and
an intra-cavity laser power of $70$ kW with a tight laser focus at
IP (beam size $\sim15$$\,$$\mu$m). Separate cavities are needed
for $1064$ nm and $532$ nm operation due to the need for very high
reflectivity in the cavity.

\subparagraph*{Energy Resolution}

\vspace{-0.1in}

The ERL provides excellent beam quality so that the energy resolution
is determined by the angular acceptance of the $\gamma$-ray beam
collimation aperture. The high brightness of the electron beam allows
a very small electron beam size even for a small beta function of
$5$\textendash $10$ cm while still providing a small angular spread.
The rms energy spread is typically much less than $10^{-3}$. The
laser must be as narrow band as possible. A $\gamma$-ray beam energy
spread of $1$\textendash $2$\% (FWHM) should be easily attainable.

\subparagraph*{Beam Polarization}

The $\gamma$-ray beam polarization is determined by the laser system.
As with the storage ring design, polarization changes at a $50$ Hz
rate are attainable.

\subparagraph*{Production Modes}

An operation of low repetition rate, $10$ MHz or less, is possible
by simply reducing electron bunch repetition or reducing the collision
rate with a kicker magnet at the expense of total flux. Multiple collision
points can be placed in the recirculation loop. Since the collision
rate is small enough ($10^{-4}$ for this design), electron beam degradation
due to multiple collisions is negligible. As in the case of storage
ring sources, simultaneous experiments will be possible at some fixed
$\gamma$-ray energies unless a tunable laser or variable angle system
is developed.

The proposed gamma-ray source can be realized with existing technologies
in principle. Technical risks remaining to be resolved are summarized
as follows:

\subparagraph*{Electron Source Performance}

\vspace{-0.1in}

A photocathode used for the generation of small-emittance electron
beams has a limited life time. The Cornell injector demonstrated extraction
of a $40$ mA beam for $4$ hours without significant reduction of
quantum efficiency \cite{dunham-2013} and, using the same gun design,
the LEReC project \cite{kayranIPAC2018_TUPMF025} at Brookhaven National
Laboratory demonstrated multi-day operation at $30$ mA. The required
emittance from the injector has been demonstrated \cite{guilliford-2015}.
Operation of an $8$-mA, $115$-MeV ERL was demonstrated at JLab \cite{neil-2006}
and a proof-of-principle experiment of ERL-based Compton source was
conducted at the Compact ERL \cite{akagi-2016}.

\subparagraph*{High Current Linac Operation}

\vspace{-0.1in}

The accelerator modules for this design will see up to $80$ mA in
total with accelerating and decelerating beams. This much current
will produce copious higher order modes that must be heavily damped.
L-band cryomodules for high-current beams have been designed and fabricated
at Cornell and KEK \cite{eichhorn-2016,umemori-2014}.

\subparagraph*{Fabry-Perot Cavity Performance}

\vspace{-0.1in}

A Fabry-Perot cavity was developed at Lyncean Technologies to store
laser power of $70$~kW with a beam waist of $40$ $\mu$m \cite{eggl-2016}
and a cavity at the cERL was $10$~kW with $24$~$\mu$m (horizontal)
and $32$~$\mu$m (vertical) spot sizes \cite{akagi-2016}. In addition,
a cavity was developed for non-accelerator environment at Max-Planck
Institute to store laser power of $670$~kW with a reported rms beam
size of $13$ $\mu m$$\,\times\,$$17$ $\mu m$ at focus \cite{carstens-2014}.
A Fabry-Perot cavity to store intra-cavity laser power of $70$~kW
with a beam waist of $15$~$\mu$m must be proven. Such a cavity
requires state-of-the-art low-loss mirror coatings and thermally stable
mirror substrates.

\subparagraph*{Preservation of Injector Beam Quality}

\vspace{-0.1in}

Though the Cornell injector has demonstrated excellent performance,
it is necessary to preserve this beam quality through a large number
of long beam transport arcs. Each of these is subject to wakefields,
longitudinal space charge (LSC) forces, and coherent synchrotron radiation
(CSR) induced emittance growth. Designs have been developed to minimize
any increase in the transverse emittance in an isochronous arc. The
emittance growth can be kept to very small levels. A more serious
problem is the longitudinal phase space distortion leading to a growth
in the projected energy spread. Some of this may be compensated by
rf cavities. A harmonic rf cavity might be required to take out higher-order
curvature.

\subparagraph*{Operational Issues}

\vspace{-0.1in}

Some of the scattered electrons may be lost when operating at very
high photon energy. These losses must be controlled and understood.
The levels of loss look tolerable and many electrons will still be
energy-recovered but it is important to control where the electrons
are lost. Continuous energy tuning in ERLs has not been demonstrated,
yet. Sophisticated algorithms must be developed to tune up the beam
transport at a new energy as the energy is changed.

We recommend to conduct the following R\&D items:
\begin{itemize}
\item \textbf{\textit{High priority}}

\begin{itemize}
\item Laser stacker at $1064$ nm and $532$ nm to verify the power, beam
size, and intensity at the IP;
\item Endurance runs with CsKSb cathodes at $20$ mA;
\item Lattice designs for ERL with CGS with CSR and LSC compensation.
\end{itemize}
\item \textbf{\textit{Medium priority}}

\begin{itemize}
\item Beam breakup (BBU) and higher-order mode (HOM) loading simulations
for ERL with CGS (e.g. Cornell and KEK modules);
\item Diagnostic development for high current beams and backscattering setup. 
\end{itemize}
\end{itemize}
\vspace{-0.1in}

\newpage

\bibliographystyle{unsrt}
\addcontentsline{toc}{chapter}{References: Next Generation $\gamma$-Ray Sources}
\bibliography{cgs-all.bib}

\newpage

\pagebreak{}

\textwidth 420pt
\chapter{Appendix}
\lhead{}
\rhead{Appendix}
\section{Workshop Agenda}
\begin{table}[h!]
\hskip -1.3 cm
	\begin{tabular}{*5l}    \toprule
\multicolumn{3}{p{\dimexpr 1.00\textwidth}}{Thursday, November 17, 2016 | \textbf{Overview Sessions}, Grand Ballroom A} &  \\\midrule
		\rowcolor{blue!10}\emph{Time} & \emph{Title} & \emph{Speaker} & Affliation   \\\midrule
		8:45 &	Welcome and remarks	& Calvin Howell &	Duke U \& TUNL \\
		\toprule
		\multicolumn{3}{p{\dimexpr 1.00\textwidth}}{ \textbf{Compton Gamma-Ray Sources}} & &\\\midrule
			
9:00	& a) Global Review of Compton $\gamma$-Ray Sources 	& Ying Wu &	Duke U \& TUNL \\
9:06	& b) Facility Talks & & \\		
	       &b.1) HIGS and CGS Projects in the U.S. &	Ying Wu	& Duke U \& TUNL \\
	       &b.2) ELI-NP and CGS Projects in Europe &	Calin Ur	& ELI-NP \\
	       &b.3) Compton $\gamma$-Ray Sources in Japan	& Ryoichi Hajima	&NIRS, Japan \\
	       &b.4) Compton $\gamma$-Ray Sources in China &	Chuanxiang Tang&	Tsinghua U, China \\ 
9:54 &	c) Possible Technologies for Next-  & & \\		
        &	Generation Sources & & \\		
	    &c.1) High Power Lasers and Optics &	Fabian Zomer&	LAL, France\\
	    &c.2) Accelerator Technologies &	John Byrd	&LBNL\\
	    &c.3) Compton Source Configuration Strategies &	Ying Wu	& Duke U \& TUNL\\
	    \toprule
	    10:30 &	\textbf{Break}, Salons A\&B Foyer  & & \\		
	    	& & &  \\\bottomrule
	\end{tabular}
\end{table}
\clearpage
\begin{table}[h!]
	\hskip -1.3 cm
	\begin{tabular}{*5l}    \toprule
		\multicolumn{3}{p{\dimexpr 1.00\textwidth}}{\centering Thursday, November 17, 2016 | \textbf{Overview Sessions}, Grand Ballroom A} & &\\\midrule
		\multicolumn{3}{p{\dimexpr 1.00\textwidth}}{\textbf{Low-Energy QCD}} & &\\\midrule
		\rowcolor{blue!10}\emph{Time} & \emph{Title} & \emph{Speaker} & Affliation   \\\midrule
		
10:50	&What to Learn from  &	Harald Griesshammer	& GWU \\
	& Nucleon Polarizabilities&	 &	 \\
11:15	&Compton Experiments I	& Phil Martel &	U Mainz \& Regina\\
11:30	&Compton Experiments II	& Gerald Feldman& GWU \\
11:45	&Testing Confinement Scale QCD &	Aron Berstein &	MIT \\
			&With Low-Energy Electromagnetic  & & \\
			&Pion Production &	 & \\
			\toprule
12:35 &	\textbf{Lunch}, Salons A\&B Foyer & & \\
\midrule
			
14:00	& \textbf{Hadronic Parity Violation}	& W. Michael Snow &	Indiana U \\
			
14:25	&\textbf{Nuclear Structure} &	Deniz Savran &	GSI, Germany \\
			
14:55	& \textbf{Nuclear Astrophysics} &	Carl Brune	& Ohio U \\
\midrule
\multicolumn{3}{p{\dimexpr 1.00\textwidth}}{\textbf{Applications}} & &\\\midrule
\rowcolor{blue!10}\emph{Time} & \emph{Title} & \emph{Speaker} & Affliation   \\\midrule
15:20 &	Homeland Security &	Calvin Howell &	Duke U \& TUNL \\
15:35 &	National Nuclear Security &	Matthew Durham &	LANL\\
15:50 &	Medical Diagnosis &	Anuj Kapadia &	Duke U\\
			\toprule
16:05 &	\textbf{Charge to Working Groups} &	Calvin Howell &	Duke U \& TUNL\\
\midrule
			
16:15 &	\textbf{Break},  Grand Ballroom Salons A\&B Foyer & & \\
\toprule
16:45 &	\multicolumn{3}{p{\dimexpr 1.00\textwidth}}{\textbf{Working Group Meetings}}  &\\\midrule
\rowcolor{blue!10} & \emph{Group} & \emph{Location} &    \\\midrule
& 1) Gamma-Ray Source 	&	Grand Ballroom  B & \\
& 2.a) LE QCD:  Nucleon Polarizabilities	&	Grand Ballroom A & \\
& 2.b) LE QCD:  Photopion Physics &		Grand Ballroom A  & \\
& 2.c) LE QCD:  Hadronic Partity Violation	&	Bethesda &\\
& 3) Nuclear Structure and Astrophysics		&Chevy Chase & \\
& 4) Applications	&	Rockville & \\
\toprule
18:30 &	\textbf{Banquet} & Congressional Ballroom& \\
		& & &  \\\bottomrule
	\end{tabular}
\end{table}
\clearpage
\begin{table}[h!]
	\hskip -1.3 cm
	\begin{tabular}{*5l}    \toprule
		\multicolumn{3}{p{\dimexpr 1.00\textwidth}}{Friday, November 18, 2016 | \textbf{Working Group Sessions}} & &\\\midrule
		09:00 &	\multicolumn{3}{p{\dimexpr 1.00\textwidth}}{\textbf{Working Group Meetings}}  &\\\midrule
		\rowcolor{blue!10} & \emph{Group} & \emph{Location} &    \\\midrule
		& 1) Gamma-Ray Source 	&	Grand Ballroom  B & \\
		& 2.a) LE QCD:  Nucleon Polarizabilities	&	Grand Ballroom A & \\
		& 2.b) LE QCD:  Photopion Physics &		Grand Ballroom A  & \\
		& 2.c) LE QCD:  Hadronic Partity Violation	&	Bethesda &\\
		& 3) Nuclear Structure and Astrophysics		&Chevy Chase & \\
		& 4) Applications	&	Rockville & \\
		\toprule
		10:30 &	\textbf{Break} & Salons A\&B Foyer  & \\ \midrule
		11:00 &	Working Group Meetings & Same Locations& \\
		12:30&	Lunch	&	Salons A\&B Foyer & \\
		14:00&	Working Group meetings	 &	Same Locations & \\
        15:30&	Break		&Salons A\&B Foyer & \\
		16:00&	Information Exchange Session & Salon A & \\
        17:30&	Dinner (on your own) & & \\		\toprule
        \multicolumn{3}{p{\dimexpr 1.00\textwidth}}{Saturday, November 19, 2016 | \textbf{Working Group \& Closeout Sessions}} & &\\\midrule
        09:00 &	Working Group Meetings & Same Locations& \\
        10:00&	Break		&Salons A\&B Foyer & \\ \toprule
        \multicolumn{3}{p{\dimexpr 1.00\textwidth}}{\textbf{Closeout Sessions}} & &\\\midrule
     10:15 & 	\textbf{Working Group Summaries} | Grand Ballroom & & \\
	& 1.a) LE QCD: Nucleon Polarizabilities & & \\
	& 1.b) LE QCD:  Above the Pion-Production Threshold & & \\
	& 1.c) LE QCD:  Hadronic Partity Violation & & \\
	& 2) Nuclear Structure && \\
	& 3) Nuclear Astrophysics && \\
	& 4.a) Applications: Homeland and Nuclear Security && \\
	& 4.b) Applications: Medicine && \\
	& 5) Gamma-Ray Source  && \\
11:35 & 	\textbf{Open discussion} && \\	
12:00 & 	\textbf{Summary and Closing Remarks} &&\\
12:10 & 	\textbf{Workshop Adjourned} &&\\
				& & &  \\\bottomrule
				\hline
			\end{tabular}
		\end{table}
\clearpage
\section{Working Group Conveners}
\begin{table}[h!]
\begin{tabular}{*5l}    \toprule
	    \multicolumn{3}{p{\dimexpr 1.00\textwidth}}{\cellcolor{blue!20}\centering General Editors} & \\\midrule
	 
	 \rowcolor{blue!10}\emph{Members} & \emph{Institution} & \emph{Email}   \\\midrule
	 Mohammad Ahmed  & NCCU \& TUNL  & ahmed@tunl.duke.edu  \\
	 Harald Grie{\ss}hammer  & GWU  & hgrie@gwu.edu  \\
	 Calvin Howell  & Duke U \& TUNL  & howell@tunl.duke.edu  \\
	 Robert Janssens  & UNC-CH \& TUNL  & rvfj@email.unc.edu  \\
	 Daniel Phillips & Ohio U & phillid1@ohio.edu \\
	 Roxanne Springer & Duke U & rps@phy.duke.edu\\
	 Ying Wu  & Duke U \& TUNL  & wu@fel.duke.edu  \\
	 
	  \toprule
	  \multicolumn{3}{p{\dimexpr 1.00\textwidth}}{\cellcolor{blue!20}\centering Nucleon spin polarizabilities} & \\\midrule
	  \rowcolor{blue!10}\emph{Members} & \emph{Institution} & \emph{Email}   \\\midrule
	  Harald Grie{\ss}hammer  & GWU  & hgrie@gwu.edu  \\
	  Rory Miskimen & U Mass & miskimen@physics.umass.edu \\
	  Daniel Phillips & Ohio U & phillid1@ohio.edu \\
	 
	  \toprule
	  \multicolumn{3}{p{\dimexpr 1.00\textwidth}}{\cellcolor{blue!20}\centering Meson EM polarizabilities and QCD origin of CSB} & \\\midrule
	  \rowcolor{blue!10}\emph{Members} & \emph{Institution} & \emph{Email}   \\\midrule
	   Aron Bernstein  & MIT & bernstn@mit.edu  \\
	   Ulf-G Mei{\ss}ner & U-Bonn & meissner@hiskp.uni-bonn.de \\
	  
	  \toprule
	  \multicolumn{3}{p{\dimexpr 1.00\textwidth}}{\cellcolor{blue!20}\centering Hadronic Parity Violation} & \\\midrule
	 \rowcolor{blue!10}\emph{Members} & \emph{Institution} & \emph{Email}   \\\midrule
	  Mike Snow  & Indiana U  & wsnow@indiana.edu  \\
	  Roxanne Springer & Duke U & rps@phy.duke.edu\\
	  Matthias Schindler & U S Carolina & mschindl@mailbox.sc.edu \\
	  
	  \toprule
	  \multicolumn{3}{p{\dimexpr 1.00\textwidth}}{\cellcolor{blue!20}\centering Nuclear Structure} & \\\midrule
	  \rowcolor{blue!10}\emph{Members} & \emph{Institution} & \emph{Email}   \\\midrule
	  Ani Aprahamian   & U Norte Dame  &  aapraham@nd.edu \\
	  Deniz Savran & GSI & d.savran@gsi.edu \\

	& &  \\\bottomrule
	\hline
\end{tabular}
\end{table}
\clearpage
\begin{table}[t!]
\begin{tabular}{*5l}    \toprule
	  
	   \multicolumn{3}{p{\dimexpr 1.00\textwidth}}{\cellcolor{blue!20}\centering Nuclear Astrophysics} & \\\midrule
	   \rowcolor{blue!10}\emph{Members} & \emph{Institution} & \emph{Email}   \\\midrule
	   Carl Brune  & Ohio U   & brune@ohio.edu  \\
	   Art Champagne & UNC-CH \& TUNL & artc@physics.unc.edu \\
	   
	  \toprule
	  \multicolumn{3}{p{\dimexpr 1.00\textwidth}}{\cellcolor{blue!20}\centering Security Applications} & \\\midrule
	  \rowcolor{blue!10}\emph{Members} & \emph{Institution} & \emph{Email}   \\\midrule
	  Calvin Howell & Duke U \& TUNL  & howell@tunl.duke.edu  \\
	  Anton Tonchev & LLNL & tonchev2@llnl.gov \\
	  
	  \toprule
	  \multicolumn{3}{p{\dimexpr 1.00\textwidth}}{\cellcolor{blue!20}\centering Medical Applications} & \\\midrule
	  \rowcolor{blue!10}\emph{Members} & \emph{Institution} & \emph{Email}   \\\midrule
	  Anuj Kapadia  & Duke U  & anuj.kapadia@duke.edu  \\
	  Madan Rehani & & madan.rehani@gmail.com \\
	  
	 \toprule
	\multicolumn{3}{p{\dimexpr 1.00\textwidth}}{\cellcolor{blue!20}\centering Advanced Accelerator and Light-source Technologies} & \\\midrule
	 \rowcolor{blue!10}\emph{Members} & \emph{Institution} & \emph{Email}   \\\midrule
	 John Byrd  & LBNL  & jmbyrd@lbl.gov  \\
	  Swapan Chattopadhyay  & FNAL  & swapan@fnal.gov  \\
	    Ying Wu  & Duke U \& TUNL  & wu@fel.duke.edu  \\
	& &  \\\bottomrule
	\hline
\end{tabular}
\end{table}
\clearpage
\section{Workshop Participants}
\begin{table}[h!]
	\begin{tabular}{*5l}   \toprule
		\rowcolor{blue!10}\emph{Last Name} & \emph{First Name} & \emph{Email} & Working Group   \\\midrule
Afanasev&Andrei&afanas@gwu.edu&Nuclear Structure \\

Ahmed&Mohammad&ahmed@tunl.duke.edu&QCD origin of CSB \\

Alesini&David&david.alesini@lnf.infn.it&Accelerator \&  Light Source \\

Annand&John&john.annand@glasgow.ac.uk&Nucleon Polarizabilities \\

Aprahamian&Ani&aapraham@nd.edu&Nuclear Structure \\

Balabanski&Dimiter&dimiter.balabanski@eli-np.ro&Nuclear Structure \\

Barty&Christopher&barty1@llnl.gov&Accelerator \&  Light Source \\

Benson&Stephen&felman@jlab.org&Accelerator \&  Light Source \\

Bernstein&Aron&bernstn@mit.edu&QCD origin of CSB \\

Brune&Carl&brune@ohio.edu&Nuclear Astrophysics \\

Byrd&John&jmbyrd@lbl.gov&Accelerator \&  Light Source \\

Carlsten&Bruce&bcarlsten@lanl.gov&Accelerator \&  Light Source \\

Champagne&Art&artc@physics.unc.edu&Nuclear Astrophysics \\

Chattopadhyay&Swapan&swapan@fnal.gov&Accelerator \&  Light Source \\

Davis&David&eddaviddavis@gmail.com&Hadronic Parity Violation \\

Downie &Evie&edownie@gwu.edu&Nucleon Polarizabilities \\

Durham&J Matthew&durham@lanl.gov&  Applications \\

Feldman&Gerald&feldman@gwu.edu&Nucleon Polarizabilities \\

Friesen&Forrest&fqf@phy.duke.edu&Nuclear Structure \\

Gao&Haiyan&gao@tunl.duke.edu&Nucleon Polarizabilities \\

Geddes&Cameron&cgrgeddes@lbl.gov&Accelerator \&  Light Source \\

Griesshammer&Harald&hgrie@gwu.edu&Nucleon Polarizabilities \\

Hajima&Ryoichi&hajima.ryoichi@qst.go.jp&Accelerator \&  Light Source \\

Hao&Hao&haohao@fel.duke.edu&Accelerator \&  Light Source \\

Holstein&Barry&holstein@physics.umass.edu&Hadronic Parity Violation \\

Hornidge&David&dhornidg@mta.ca&Nucleon Polarizabilities \\

Howell&Calvin&howell@tunl.duke.edu&  Applications \\

Huffman&Paul&paul\_huffman@ncsu.edu&Hadronic Parity Violation \\

Iliadis&Christian&iliadis@physics.unc.edu&Nuclear Astrophysics \\

Isaak&Johann&jisaak@rcnp.osaka-u.ac.jp&Nuclear Structure \\
\end{tabular}
\end{table}
\begin{table}[h!]
	\begin{tabular}{*5l}    \toprule

		\rowcolor{blue!10}\emph{Last Name} & \emph{First Name} & \emph{Email} & Working Group   \\\midrule

Kapadia&Anuj&anuj.kapadia@duke.edu&  Applications \\
Kendellen&David&dpkendel@tunl.duke.edu&Nucleon Polarizabilities \\
Krasznahorkay&Attila&kraszna@atomki.hu&Nuclear Structure \\
Kovash&Micheal&kovash@pa.uky.edu&Nucleon Polarizabilities \\
Lee&Dean&djlee3@unity.ncsu.edu&Nuclear Structure \\

Lorant&Csige&csige.lorant@atomki.mta.hu&Nuclear Structure \\

Martel&Phil&martel@kph.uni-mainz.de&Nucleon Polarizabilities \\

Meissner&Ulf&meissner@hiskp.uni-bonn.de&QCD origin of CSB \\

Miskimen&Rory&miskimen@physics.umass.edu&Nucleon Polarizabilities \\

Moon&Namdoo&namdoo.moon@HQ.DHS.GOV&Government Agency \\

Mueller&Jonathan&jmmuell3@ncsu.edu&  Applications \\

Opper&Allena&aopper@nsf.gov&Government Agency \\

Pasquini&Barbara&pasquini@pv.infn.it&Nucleon Polarizabilities \\

Phillips&Daniel&phillid1@ohio.edu&Nucleon Polarizabilities \\

Pietralla&Norbert&pietralla@ikp.tu-darmstadt.de&Nuclear Structure \\

Rai&Gulshan&Gulshan.Rai@science.doe.gov&Government Agency \\

Rehani&Madan&madan.rehani@gmail.com&  Applications \\

Rhodes&William&William.Rhodes@nnsa.doe.gov&Government Agency \\

Savran&Deniz&d.savran@gsi.de&Accelerator \&  Light Source \\

Schindler&Matthias&MSCHINDL@mailbox.sc.edu&Hadronic Parity Violation \\

Sikora&Mark&msikora@tunl.duke.edu&Nucleon Polarizabilities \\

Snow&Mike&wsnow@indiana.edu&Hadronic Parity Violation \\

Springer&Roxanne&rps@phy.duke.edu&Hadronic Parity Violation \\

Stone&Terri&Terri.Stone@nnsa.doe.gov&Government Agency \\

Sun&Changchun&ccsun@lbl.gov&Accelerator \&  Light Source \\

Tang&Chuanxiang&Tang.xuh@tsinghua.edu.cn&Accelerator \&  Light Source \\

Tiburzi&Brian&btiburzi@ccny.cuny.edu&Nucleon Polarizabilities \\

Tonchev&Anton&tonchev2@llnl.gov&  Applications \\

Tornow&Werner&tornow@tunl.duke.edu&Nuclear Structure \\

Ur&Calin&calin.ur@eli-np.ro&Accelerator \&  Light Source \\

Wang&Dong&wangdong@sinap.ac.cn&Accelerator \&  Light Source \\

Weller&Henry&weller@tunl.duke.edu&QCD origin of CSB \\

Werner&Volker&vw@ikp.tu-darmstadt.de&Nuclear Structure \\

Wu&Ying&wu@fel.duke.edu&Accelerator \&  Light Source \\

Yan&Jun&junyan@fel.duke.edu&Accelerator \&  Light Source \\

Zilges&Andreas&zilges@ikp.uni-koeln.de&Nuclear Structure \\

Zomer&Fabian&zomer@lal.in2p3.fr&Accelerator \&  Light Source \\

	& & &  \\\bottomrule
	\hline
\end{tabular}
\end{table}

\end{document}